\documentclass[12pt,a4paper]{article} 
\usepackage{verbatim}
\usepackage{color}
\usepackage{rotate}
\usepackage{graphics}
\usepackage{epsf}
\usepackage{epsfig}
\usepackage{graphicx}
\usepackage{latexsym}
\usepackage{amscd}
\usepackage{amssymb}

\newcommand{\eps}{\varepsilon}

\newcommand{\beq}{\begin{equation}}
\newcommand{\eeq}{\end{equation}}
\newcommand{\beqn}{\begin{eqnarray}}
\newcommand{\eeqn}{\end{eqnarray}}
 
\newcommand\la{\langle}
\newcommand\ra{\rangle}

\newcommand{\lsi}{\raise0.3ex\hbox{$<$\kern-0.75em\raise-1.1ex\hbox{$\sim$}}}
\newcommand{\gsi}{\raise0.3ex\hbox{$>$\kern-0.75em\raise-1.1ex\hbox{$\sim$}}}

\def\inf{\int_{-\infty}^{\infty}}
\def\BA{\begin{eqnarray}}
\def\BE{\begin{equation}}
\def\EA{\end{eqnarray}}
\def\EE{\end{equation}}

\def\Ref#1{(\ref{#1})}

\begin{document}
\pagestyle{empty}

\begin{titlepage}
\begin{center} 
Dissertation\\
submitted to the \\
Combined Faculties for the Natural Sciences and for Mathematics\\
of the Rupertus Carola University of\\
Heidelberg, Germany\\
for the degree of\\
Doctor of Natural Sciences\\
\vspace*{5cm}
{\LARGE \bf QCD coherence effects in high energy reactions with nuclei}\\
\vspace*{4cm}
{\large presented by} \\[.5em]
{\large\begin{tabular}{ll}
Diplom-Physicist & J\"org Raufeisen\\
born in & Minden, Germany
\end{tabular}}
\vfill
\large Heidelberg, 12.\ Juli 2000\\[1.em]
{\large \begin{tabular}{ll}
Referees: & Prof.\ Dr.\ J\"org H\"ufner \\[.5em] 
& Prof.\ Dr.\ Andreas Sch\"afer
\end{tabular}}
\end{center}
\end{titlepage}


\begin{titlepage}
\noindent {\bf\centerline {Zusammenfassung}}
\\[.5cm]
In dieser Arbeit werden Koh\"arenzeffekte in der 
tiefinelastischen Streuung (DIS) und im Drell-Yan (DY)
Prozess an Kernen untersucht, insbesondere der
{\em Shadowing} Effekt. Es wird im Ruhesystem des Targets und in der
Farbdipol Formulierung gearbeitet. 
Die Glauber-Gribov
Theorie f\"ur Mehrfachstreuung im Kern wird so modifiziert, dass 
der Formfaktor des Kernes in allen Streutermen
ber\"ucksichtigt ist. 
Ferner wird die 
mittlere Koh\"arenzl\"ange f\"ur einen Fockzustand definiert.
Damit ist es m\"oglich abzusch\"atzen, 
dass das {\em Gluon-Shadowing} f\"ur $x_{Bj}>0.01$ 
vernachl\"assigbar ist.
Parameter freie Rechnungen werden mit Daten von NMC und
E665 f\"ur DIS und mit E772 Daten f\"ur DY verglichen.
In beiden F\"allen wird gute \"Ubereinstimmung festgestellt. 
Der von HERMES beobachtete Effekt kann jedoch nicht reproduziert werden.
F\"ur DY-Dileptonen aus Proton-Kern Kollisionen bei RHIC Energien
wird f\"ur den gesamten $x_F$ Bereich deutliches {\em
Shadowing} vorrausgesagt. Der Einfluss des Kerns auf die
Transversalimpuls-Verteilung der DY Paare wird ebenfalls untersucht.
Des weiteren  wird eine neue Parametrisierung des
Dipol-Wirkungsquerschnittes pr\"asentiert.
\\[1cm]
\noindent {\bf \centerline {Abstract}}
\\[.5cm]
In this work, coherence effects in deep inelastic scattering (DIS)
and in the Drell-Yan (DY)
process off nuclei are investigated, in particular nuclear shadowing. 
The target rest frame and the color dipole
formulation are employed. 
Multiple scatterings are treated in Glauber-Gribov theory, which is
modified to include the nuclear form factor to all orders. 
Based on the mean coherence length, which is defined in this work, 
it is estimated that
gluon shadowing is negligible at $x_{Bj}>0.01$.
Parameter free calculations are compared  
to NMC and E665 data for DIS and to E772 data for DY.  
In both cases, good agreement is found. It is  however not possible 
to reproduce the
effect observed by HERMES. 
For dileptons in proton-nucleus collisions at RHIC energies,
considerable shadowing for the whole $x_F$ range is predicted.
The influence of the nucleus on the DY transverse momentum distribution is also
studied.
Furthermore, a new
parametrization of the dipole cross section is presented.
\end{titlepage}

\clearpage

\pagestyle{headings}
\begin{flushright}
{\em ...all exact science is dominated by the idea of approximation.}\\
Bertrand Russel
\end{flushright}
\textwidth=15.1cm
\tableofcontents
\section{Introduction}\label{intro}

The use of nuclei instead of protons
in high energy scattering experiments,
like deep inelastic scattering,
provides unique possibilities to
study the space-time development of strongly interacting systems. In
experiments with proton targets the products of the scattering process can only
be observed in a detector which is separated from the reaction point by a
macroscopic distance. In contrast to this, the nuclear medium can serve as a
detector located directly at the place where the microscopic interaction
happens.
As a consequence, with nuclei one can 
study coherence effects in QCD which are
not accessible in DIS off
protons nor in proton-proton scattering. An important question that can only be
answered with the help of nuclei is for instance, how quarks and gluons evolve
from the early stages of a collision to the hadrons which are finally observed
in the detector. The large extension of the nuclear medium makes it possible to
investigate, by which time scales this hadronization process is governed.
Note that the radius of a
heavy nucleus like lead is approximately 
eight times as large ($\approx 6.8$ fm in the nuclear rest frame)
as the radius of a proton 
($\approx 0.86$ fm).

This work is mostly concerned with the theoretical analysis of coherence effects
in DIS off nuclei and in Drell-Yan (DY) dilepton production
in proton-nucleus scattering, in particular with
the phenomenon of nuclear shadowing. Before we turn to nuclear targets, we
shortly review the physics of a proton target.
 
In DIS, a lepton is  scattered off the target. This lepton radiates a virtual
photon, which can resolve the microscopic substructure of the proton.  
Therefore, the huge colliders like HERA, where such experiments
are performed, 
are big microscopes that allow to investigate, how the
proton is made up of quarks. It is well known today
that the proton contains three
quarks which are called valence quarks. These valence quarks carry the quantum
numbers of the proton and are the analog of the valence electrons which are
responsible for the chemical properties of an atom. In addition to the quarks,
there are also gluons inside the proton. Gluons mediate the color forces between
the quarks. Since gluons have no electromagnetic charge they are not directly
observable in DIS. One can however
conclude that there must be neutral partons in the proton
since quarks carry only about half of the momentum of the proton.
The missing momentum is believed to be carried by gluons.
It is natural to ask, whether quarks and gluons are elementary or whether they
have a substructure themselves. 
In order to find an answer to this question, one
has to increase the resolution of the microscope, i.e.\ build colliders with
higher energies which can measure at higher momentum transfer. 
Today, the highest energies are reached at HERA where positrons collide with
protons at a center of mass ({\em c.m.}) energy
$\sqrt{S}\approx 300$ GeV. 
As the resolution is increased, one finds that a quark, carrying the
longitudinal momentum fraction Bjorken-$x_{Bj}$ 
of the proton consists of a quark and a
gluon which carry both together the momentum of the parent quark and
therefore each 
a smaller momentum fraction of the proton. In addition, also gluons
can split into quark-antiquark ($q\bar q$) pairs. 
This is confirmed experimentally by the increase of the quark density at low
values of $x_{Bj}<0.1$.
At very low $x_{Bj}\ll 0.1$, 
the
partonic content of the proton is dominated by gluons and the photon sees
only the $q\bar q$-pairs which originate from gluon splitting. 
However, one cannot  
distinguish in DIS whether the virtual photon couples to a quark or to an
antiquark. Complementary information about the antiquark density is provided by
the Drell-Yan (DY) process. In the DY process in
proton-proton collisions, a quark from the
projectile can annihilate with an antiquark from the  target and produce a
massive photon. This photon decays into a lepton pair which can be detected. 

How does a nucleus look like at high energies, i.e.\ at low $x_{Bj}$? The answer
depends on the reference frame. 
In a frame where the nucleus is fast moving, the so-called infinite momentum
frame, the nucleus is strongly Lorentz contracted. However, the localization of
gluons which carry only a very small momentum fraction of the nucleus is
determined by the uncertainty principle. In the infinite momentum frame, 
the cloud of these
low-$x_{Bj}$ gluons extends over the whole nucleus and the nucleons are
able to communicate with each other.
The situation looks different in the rest frame of the nucleus. In the nuclear
rest frame, the nucleons are well separated from each other by a distance of
$\sim 2$ fm. How can these two pictures, infinite momentum frame and rest frame,
be reconciled with each other? Of course, all observables have to be Lorentz
invariant.

\begin{figure}[t]
  \scalebox{0.9}{\includegraphics{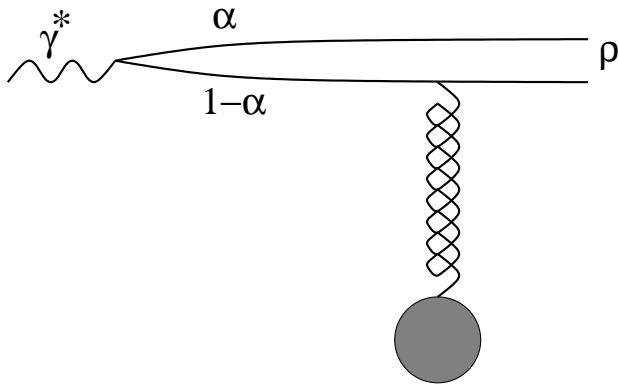}}\hfill
  \raise0.3cm\hbox{\parbox[b]{2.3in}{
    \caption{
      \label{lc1}
      At low $x_{Bj}$ and in the target rest frame,
       the virtual photon converts into a $q\bar q$-pair long
      before the target. The quark carries
      momentum fraction $\alpha$ of the $\gamma^*$, the antiquark $1-\alpha$.
      The transverse separation between the particles is denoted by $\rho$.
      The curly line represents a gluon.
    }
  }
}
\end{figure}

Note that not only the partonic structure of the nucleus is frame depended, but
also the partonic interpretation of the scattering process. 
At high energies, nuclear scattering is governed by coherence effects
which are most easily understood
in the target rest frame. In the rest frame, 
DIS looks like pair creation from a virtual
photon, see fig.\ \ref{lc1}. 
Long before the target, the virtual photon splits up into a $q\bar
q$-pair. The lifetime $l_c$ of the fluctuation 
can be estimated
with help of the uncertainty relation to be of order $\sim 1/m_N x_{Bj}$
(cf.\ section \ref{dipoledis}) where
$m_N\approx 1$ GeV is the mass of a nucleon. 
The coherence length can become
much greater than the nuclear radius at low $x_{Bj}$.
On a nuclear target, the pair will experience multiple scatterings 
off different nucleons within the
coherence length. This corresponds to the overlap of gluon clouds from different
nucleons in the infinite momentum frame. 
The long lifetime of the $q\bar q$ fluctuation, which extends over the whole
nucleus, leads to the pronounced coherence effects observed in experiment.
The measurable cross section is
independent of the reference frame, but our physical picture depends on it. 
The target rest frame is especially well suited for the study of coherence
effects.

The most
prominent example for a coherent interaction of more than one nucleon is the
phenomenon of nuclear shadowing. Naively one would expect that the cross section
for scattering a lepton off a nucleus with mass number $A$ is $A$ times as large
as the cross section for scattering the lepton off a proton. 
In experiment
it is however seen that the nuclear cross section is significantly smaller. 
Shadowing in low $x_{Bj}$ DIS and at high photon virtualities
was first observed by the EM collaboration \cite{EMC}.
The same reduction of the cross section  was measured by E772 for the
Drell-Yan (DY) process at low $x_2$ \cite{E772}.

What is the mechanism behind this suppression? 
If the coherence length is very long, as indicated in fig.\ \ref{lc1},
the $q\bar q$-dipole  undergoes multiple scatterings inside the nucleus.
The physics of shadowing in DIS is most easily understood 
in a representation, in which the pair has a
definite transverse size $\rho$. As a result of color transparency
\cite{zkl,bbgg,bm}, small size
pairs interact with a small cross section $\sigma_{q\bar q}(\rho)$, while large
pairs interact with a large cross section. 
The term "shadowing" can be taken literally in the target rest frame. The large
pairs are absorbed by the nucleons at the surface which cast a shadow on the
inner nucleons. The small pairs are not shadowed. They can propagate 
through the whole
nucleus. From these simple arguments, one can already understand the two most
important conditions for shadowing. First, the hadronic fluctuation of the
virtual photon has to interact with a large cross section and second, the
coherence length has to be long enough to allow for multiple scattering.
At $x_{Bj}\to 0$ 
the coherence length becomes infinite and
shadowing in DIS can be calculated from a formula very similar to
the Glauber-Gribov 
eikonal formula \cite{Glauber,gribov},
if one knows the cross section $\sigma_{q\bar q}(\rho)$
for scattering a $q\bar q$-pair 
with transverse size $\rho$ off a nucleon.
The application of the eikonal formula is possible 
because  at infinitely high energies 
$q\bar q$-dipoles with fixed separation in impact parameter space are 
eigenstates of the interaction
\cite{zkl,Borisold,Miettinen}.
The dipole cross section is the main nonperturbative ingredient to all
calculations in this work.

Nuclear shadowing has
received much attention within the last two decades.
A detailed review of
different approaches to nuclear shadowing
can be found in \cite{Arneodo}, see also the more recent
review \cite{PillerRev}.

Note however that shadowing has
not yet been measured
at RHIC ($\sqrt{S}\approx 200$ GeV) nor LHC ($\sqrt{S}\approx 5.5$ TeV)
energies. Furthermore, it is only known for quarks. Shadowing is also expected
for the nuclear gluon distribution. This gluon shadowing could have a strong
influence on the expected quark gluon plasma formation at RHIC and LHC and is
only poorly understood at present. Only quark shadowing is calculated in this
work, but the light-cone approach can also be extended to gluon shadowing
\cite{KST99}.
Further insight into 
the physics of nuclear parton distributions
could be provided by DIS experiments off nuclei at eRHIC or at HERA. 
Nuclei could also be included into the THERA option
at TESLA ($\sqrt{S}\approx 1.4$ TeV). The advantage of using nuclear targets is
that parton densities which could be probed only at LHC energies 
with proton targets 
are  already accessible at HERA energies. Such experiments are not only
important in view of the quark gluon plasma. They can also clarify, up to which
energies the QCD improved parton model can be applied. 

\begin{figure}[t]
  \scalebox{0.7}{\includegraphics*{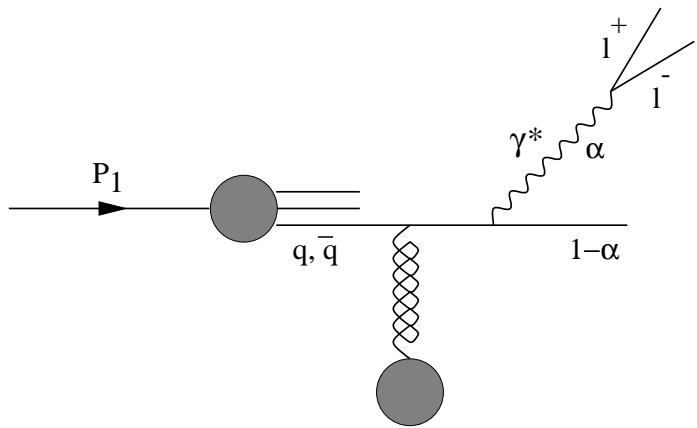}}\hfill
  \raise0cm\hbox{\parbox[b]{2.15in}{
    \caption{
      \label{bremsdy}
      In the target rest frame,
      DY dilepton production looks like bremsstrahlung. A quark 
      or an antiquark inside the
      projectile hadron scatters off the target color field and radiates a
      massive photon, which subsequently decays into the lepton pair. The photon
      can also be radiated before the quark hits the target. 
    }
  }
}
\end{figure}

While the target rest frame picture of DIS is very popular, 
the light-cone approach of Kopeliovich \cite{boris,bhq,kst}, 
which describes the DY
process in the target rest frame, is less known.
In the light-cone approach, DY
dilepton production in the rest frame of the target appears as bremsstrahlung,
see fig.\ \ref{bremsdy}. A quark from the projectile scatters off the target and
radiates a virtual photon. This photon decays into a lepton pair. Remarkably,
the DY cross section can be expressed in terms of the same dipole cross section
that appears in DIS, cf.\ section \ref{dipoledy}. 
This is a result of the QCD factorization theorem \cite{bhq}.
On a nuclear target, the quark will of course scatter several times.
The effect of multiple scattering on
bremsstrahlung is well known in QED as the
Landau-Pomeranchuk-Migdal (LPM) effect \cite{Landau,Landau2,Migdal}, see also
\cite{Ter}. 
The LPM effect leads to a reduction of the cross section due to destructive
interferences in multiple scattering within an amorphous medium. 
An electron incident
on a dense target with $n$ scattering centers, will radiate bremsstrahlung
after the first scattering. 
If the electron energy (in QED) is high enough, the
longitudinal momentum transfer in a single interaction is small and
according to the uncertainty
relation, the
electron needs a long time to recreate it's electromagnetic field.
Within this coherence time, the electron can travel
macroscopic distances. This is in complete analogy to the case of DIS.
If the electron 
is hit several times within this length, it cannot radiate again, because
it has not yet recreated it's field.
Therefore, the overall cross section is less than $n$
times the single scattering cross section. 
It is remarkable that a microscopic
process, where all momenta are
large, can exhibit a length scale of order millimeters (in QED).

Note that both, shadowing in DIS and DY, can be calculated from the Glauber
eikonal formula at infinitely high energies, where the transverse motion of the
particles in the pair, fig.\ \ref{lc1}, or the incident quark, fig.\
\ref{bremsdy}, can be neglected. In this work, a Green function approach is
developed that allows to do calculations at energies which can be reached
in experiment.

Note that shadowing in DIS can also be regarded as LPM effect for pair
production. 
The physical interpretation is however much less intuitive.
Both processes, bremsstrahlung and pair production, 
were studied in \cite{Migdal}.
After the experimental discovery of the LPM effect for bremsstrahlung
in QED
\cite{a1,a2,a3}, more than forty
years after its theoretical prediction, this effect 
again received much attention.
In particular the QCD variant of the LPM effect was extensively studied
with respect to suppression of gluon radiation and energy loss of particles
propagating in a quark gluon plasma. 
The two most important approaches are the 
path integral approach by Zakharov \cite{Zak1}-\cite{Zak7} and the
diagrammatic technique of the BDMPS
collaboration \cite{BDMPS1}-\cite{BDMPS7}.
It is demonstrated in \cite{BDMPS7} that the two formulations are equivalent. 
Further work on the LPM effect in heavy ion collisions is done in
\cite{sonst1,sonst2,sonst3}.
A review of the present theoretical understanding of the LPM  
effect can be
found in \cite{BSZ}.
The formulation of nuclear shadowing presented in this work is the one of
Zakharov since our both approaches are extensions of Glauber-Gribov theory.
The treatment of
shadowing in DIS as LPM effect was 
first suggested in \cite{zakh} and elaborated in
\cite{first,zwei}. 

This work mainly consists of two parts. 
In the first part, section \ref{proton}, 
DIS and diffraction
on a proton target and DY dilepton production in proton-proton scattering
are considered.
The deep inelastic structure functions and the definitions of kinematical
variables are introduced in section \ref{f2sec}.
In section \ref{dissec}, the parton model and the QCD improved parton model are
shortly reviewed. The first two sections can be skipped by a reader familiar
with these concepts. The space-time picture of low $x_{Bj}$ DIS 
in the color dipole formulation is discussed in
some detail in section \ref{dipoledis}. The basic concepts of the light-cone
approach and the techniques for treating 
nonperturbative effects, which are employed
throughout this work, are presented in this section, too.
Section \ref{diffractionsec} is a continuation of the preceding section and
contains a discussion of diffraction in the color dipole picture.
The sections \ref{dysec} and \ref{dipoledy} treat the DY process. While 
in section \ref{dysec} the well
known parton model of the DY process is shortly explained, section
\ref{dipoledy} introduces the color dipole formulation of the DY process. This
formulation is employed in section \ref{nucleus} 
where nuclear targets are considered.
Finally, a new parametrization of the dipole cross section is presented in
section \ref{DCS}. All results of numerical calculations 
for proton targets
and comparison with
experimental data are shown in this section.

The light-cone approach is extended to nuclear targets in 
section \ref{nucleus}. 
This section contains the main results of this work.
The connection between nuclear shadowing in DIS and diffraction is discussed in
section \ref{shadowdiffr}. In addition, the Glauber-Gribov multiple scattering
theory is shortly introduced and the physical conditions which have to be
fulfilled in order to observe shadowing are explained. 
In the sections \ref{lightcone} and \ref{derivation}, it is explained how the
nuclear form factor can be introduced in all multiple scattering terms. While
section \ref{lightcone} is intended to explain the physics of our approach, a
more formal derivation and a discussion of approximations which are employed is
given in section \ref{derivation}. The mean coherence length is defined in
section \ref{meancoh}. 
This quantity is intended to be a tool for qualitative
considerations.
In section \ref{dilepton} an estimate is given,
on how many dilepton pairs in the low mass
region are produced via the bremsstrahlungs-mechanism. 
Proton-nucleus and nucleus-nucleus collisions are considered in this section,
but no shadowing is taken into account.
Shadowing for DY dilepton production is discussed in
section \ref{dyshadow}, where also predictions for RHIC are presented. 
Section \ref{comparison} contains the comparison of our calculation with data
for shadowing in DIS and DY.

\section{Soft contributions to hard QCD reactions}\label{proton}

Deep inelastic scattering (DIS) and the Drell-Yan (DY)
process are the two classical
examples for the application of perturbative QCD (pQCD). 
The high virtuality
of the photon or the large mass of the dilepton pair, respectively, 
provides the hard scale, which is necessary 
for perturbation theory. 
Due to confinement, 
the fundamental QCD degrees of freedom, quarks and gluons
are not directly observable but occur  only
in bound
states. Therefore
all QCD reactions also
involve a soft scale, given e.\ g.\ by the radius of the hadron. 

Such two-scale processes have been extensively 
studied for more than 20 years and their theoretical treatment has reached a
high level of sophistication. Basic to the standard approach is the operator
product expansion (OPE), invented by Wilson \cite{OPE}.
Combined with asymptotic freedom, the OPE is a systematic way to separate hard
and soft scales in DIS, DY and other hard processes. The observable cross
sections can be written in factorized form, namely as a convolution 
of a hard partonic cross section and 
of the soft parton distribution of the hadrons (or of hadronic matrix elements,
more generally speaking). Contributions which do not factorize vanish in the
limit $Q^2\to\infty$.
While the partonic cross section is governed by the hard scale and can be
calculated perturbatively, the parton distributions are of completely 
nonperturbative origin and no way is known to calculate them from QCD. Even
lattice calculations are not applicable in the kinematical region of DIS and DY. 

Instead, the parton distributions have to be extracted from  experimental data. 
Their most important property is, that they are
independent of the particular hard process under consideration and depend only 
on
the hadron. This universality is crucial to make the theory predictive.
Today, several collaborations exist which provide
parametrizations of these
distributions \cite{GRV,MRST,CTEQ}.
The parton distributions are typically given
at a semihard input scale $Q_0^2$.
The evolution of the parton distributions to higher virtuality $Q^2$
is then correctly
described by the DGLAP evolution equations \cite{GL}-\cite{Dok}. This is one of
the great successes of QCD and an important argument, that QCD is the
correct field theory of the strong interaction.

One should however bear in mind, that the soft physics is completely described
in terms of fit functions and nothing is known about the nonperturbative
mechanisms, leading to the specific shape of the parton distributions. 
In particular, all coherence effects, which are so important in high energy
nuclear physics, are hidden in these parametrizations. The aim of this work is
to develop an approach which makes the physics underlying these coherence
effects transparent and intuitively understandable.

The purpose of this sections is to 
introduce the basics of the light-cone 
approach to DIS and DY before nuclear
targets are considered
in section \ref{nucleus}. 
The fundamental input to
all calculations is the cross section for scattering a quark-antiquark dipole
off a proton.
A parametrization of the dipole cross section is presented, 
which describes well all $F_2$ data from
H1 and ZEUS at $Q^2\le 35$ GeV$^2$
and also reproduces total hadronic cross sections.
We do not aim to achieve the same level of sophistication as the OPE
approach in this work. 
We rather intend to develop a good physical understanding
of high energy DIS and DY in the framework of the light-cone approach, which is
especially well suited to describe coherence effects in multiple scattering off
nuclei.  
This section also 
serves as a reminder of the
definition of structure functions and kinematical variables.

\subsection{Description of DIS in terms of structure functions}\label{f2sec}

The natural language to describe hard QCD processes 
is based on structure functions which contain all information about the
substructure of the target. These  functions are defined
by relating them to the experimentally observable lepton-target cross section
and can be expressed 
in terms of hadronic matrix elements. They
 acquire a physical
interpretation within a parton model of the structure of the target.
The definition of structure functions in DIS is explained in more detail in
standard textbooks, see e.\ g.\ \cite{Ellis,ApQCD,Muta}, we will only give a
short overlook.
Only the scattering of charged leptons is discussed in this work. Deep inelastic
neutrino scattering, as well as $Z_0$ exchange are not considered.

\begin{figure}[ht]
  \scalebox{0.94}{\includegraphics{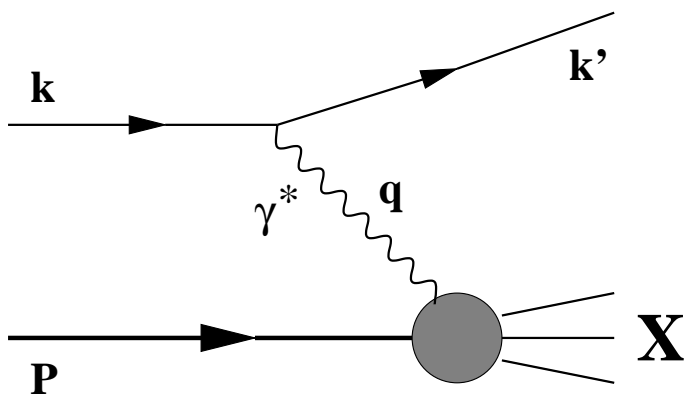}}\hfill
  \raise0mm\hbox{\parbox[b]{2.2in}{
    \caption{
      \label{dis1}
      Inclusive deep-in\-el\-as\-tic lep\-ton-pro\-ton scat\-tering. 
      A charged lepton with
      four-momentum $k^\mu=(E,\vec k)$ 
      scat\-ters off a pro\-ton with four-mo\-men\-tum $P^\mu=(E_p,\vec P)$.
      In DIS a space\-like 
      pho\-ton with mo\-men\-tum $q^2=-Q^2<0$ is ex\-changed. Only the
      scat\-tered lep\-ton with 
      mo\-men\-tum $k^{\mu\prime}$ is ob\-served in inclusive
      measure\-ments. 
    }
  }
}
\end{figure}

The kinematics for inclusive DIS is explained in fig.\ \ref{dis1}. A lepton
scatters off a proton or a nucleus. 
Due to the impact of the virtual photon, the target 
breaks up into an unobserved
final state $X$. Only the scattered lepton is observed in the final state.
We work with the standard kinematical variables. The virtuality of the photon is
denoted by
\beq
Q^2=-q^2=(k-k^\prime)^2>0
\eeq
and the mass of a proton by
\beq
m_N^2=P^2\approx 1~{\rm GeV}^2.
\eeq
Furthermore the lepton-nucleon center of mass ({\em cm}) energy
squared is defined as
\beq
S=(k+P)^2
\eeq
and the same for the $\gamma^*$-nucleon system
\beq
s=(q+P)^2.
\eeq
For convenience, we have chosen our convention slightly different from the
standard convention, where $s$ is denoted by $W^2$.
The Bjorken variable is given by
\beq
x_{Bj}=\frac{Q^2}{2P\cdot q}=\frac{Q^2}{2m_N\nu}\approx\frac{Q^2}{Q^2+s}
\eeq
and the relative energy loss of the lepton is
\beq\label{ypsilon}
y=\frac{P\cdot q}{P\cdot k}\approx\frac{Q^2+s}{S}.
\eeq
We will also frequently use the Lorentz invariant variable
\beq
\nu=\frac{P\cdot q}{m_N}.
\eeq

\begin{figure}[ht]
  \centerline{\scalebox{0.8}{\includegraphics{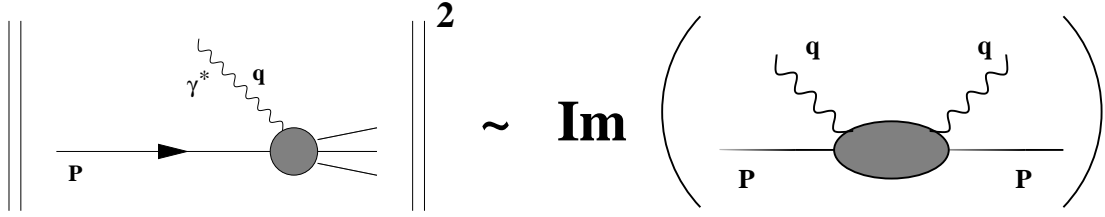}}}
    \caption{
      \label{compton}
      The hadronic tensor $W^{\mu\nu}$ is related to the imaginary part of the
      forward ($P=P^\prime\;,\;q=q^\prime$), virtual ($Q^2 >0$) 
      Compton scattering
      amplitude via the optical theorem.
    }  
\end{figure}

In one photon exchange approximation, the differential
cross section for inclusive
scattering reads
\beq
d\sigma(eP\to e^\prime X)=\frac{1}{2S}\frac{d^3k^\prime}{(2\pi)^3 2E^\prime}
\sum_X (2\pi)^4\delta^4(P+k-p_X-k^\prime)|{\cal{A}}|^2.
\eeq
The sum over final states is understood to include also the integration
over phase space. The absolute square of the matrix element reads
\beq
|{\cal{A}}|^2=\frac{4\pi\alpha_{em}}{Q^4}l_{\mu\nu}
\la P|J^{\mu\dagger}(0)|X\ra\la X|J^{\nu}(0)|P\ra
\eeq
after summation of the polarizations of the virtual photon and in one photon
exchange approximation.
Here, $J^\mu$ is the electromagnetic current.
The leptonic tensor is known as
\beq
l_{\mu\nu}=2(k^\prime_\mu k_\nu+k^\prime_\nu
k_\mu-g_{\mu\nu}k^\prime\cdot k).
\eeq
We have averaged over spins of the initial lepton and summed 
over the final spins. 
All information about the target resides now in the hadronic tensor defined by
\beqn
W^{\mu\nu} & = & \sum_X 
\la P|J^{\mu}(0)|X\ra\la X|J^{\nu}(0)|P\ra
(2\pi)^4\delta^4(P+q-p_X)\\
\label{second}
& = &
\int d^4x{\rm e}^{{\rm i}qx}\la P|J^{\mu}(x)J^{\nu}(0)|P\ra.
\eeqn
In order to obtain (\ref{second}), one has to make use of the integral
representation of the $\delta$-function and of the completeness of the final
hadronic states $X$. The hadronic tensor is then
related to the discontinuity of the forward virtual Compton scattering
amplitude $T^{\mu\nu}$ via
\beq
W^{\mu\nu}=\frac{1}{2\pi}{\rm Disc}T^{\mu\nu}
=\lim_{\eps\to 0}
\frac{1}{4\pi{\rm i}}(T^{\mu\nu}(q_0+{\rm i}\eps)-T^{\mu\nu}(q_0-{\rm i}\eps)),
\eeq
with
\beq\label{Comamp}
T^{\mu\nu}={\rm i}
\int d^4x{\rm e}^{{\rm i}qx}\la P|{\cal T}( J^{\mu}(x)J^{\nu}(0))|P\ra.
\eeq
Here, ${\cal T}$ is the time ordering operator. Note that the discontinuity of
$T$ equals the imaginary part of $T$ since the current operators $J^\mu$ are
hermitian. As illustrated in fig.\ \ref{compton}, the relation between the
hadronic tensor and the Compton amplitude is a manifestation of the optical
theorem. 

Making use of Lorentz covariance, gauge invariance and parity conservation, one
finds that the most general structure of $W^{\mu\nu}$ is
\beq
W^{\mu\nu}=\left(-g^{\mu\nu}+\frac{q^\mu q^\nu}{q^2}\right)W_1(x_{Bj},Q^2)+
\left(P^\mu+\frac{q^\mu}{2x_{Bj}}\right)\left(P^\nu+\frac{q^\nu}{2x_{Bj}}\right)
W_2(x_{Bj},Q^2).
\eeq
The dimensionless invariant structure functions for DIS are commonly
defined as
\beqn
F_1(x_{Bj},Q^2)&=&W_1(x_{Bj},Q^2),\\
F_2(x_{Bj},Q^2)&=&\nu W_2(x_{Bj},Q^2)
\eeqn
and with this notation, 
the cross section for $ep\to e^\prime X$ reads
\beq
\frac{d^2\sigma}{dx_{Bj}dQ^2}=\frac{4\pi\alpha_{em}^2}{Q^4}\left\{
\left(1-y-\frac{x_{Bj}^2y^2m_N^2}{Q^2}\right)\frac{F_2(x_{Bj},Q^2)}{x_{Bj}}
+y^2F_1(x_{Bj},Q^2)\right\}.
\eeq

Since the only purpose of the lepton is to radiate the virtual photon, it is
convenient to think about DIS as $\gamma^*p$-scattering and to define
corresponding cross sections for transverse 
and longitudinal photons. The cross section for a virtual photon
with helicity $\lambda$ can be defined as
\beq
\sigma_\lambda=\frac{4\pi^2\alpha_{em}}{K}
\epsilon_\mu(\lambda)\epsilon_\nu^*(\lambda)\,{\rm Im}T^{\mu\nu}.
\eeq
Note that the flux factor of a virtual photon is not well defined. We employ
the convention of Hand \cite{Hand}, i.\ e.\,
\beq
K=2P\cdot q-Q^2\approx 2s.
\eeq
With this convention one obtains
\beqn
\sigma_T&=&\frac{4\pi^2\alpha_{em}}{Q^2(1-x_{Bj})}2x_{Bj}F_1(x,Q^2),\\
\sigma_L&=&\frac{4\pi^2\alpha_{em}}{Q^2(1-x_{Bj})}\left[
\left(1+\frac{Q^2}{\nu^2}\right)F_2(x_{Bj},Q^2)
-2x_{Bj}F_1(x_{Bj},Q^2)\right].
\eeqn
Here, $\sigma_T$ is the cross section for transverse photons and $\sigma_L$ for
longitudinal photons. One also often introduces the helicity structure
functions,
\beqn
F_T(x_{Bj},Q^2)&=&\frac{Q^2(1-x_{Bj})}{4\pi^2\alpha_{em}}\sigma_T,\\
F_L(x_{Bj},Q^2)&=&\frac{Q^2(1-x_{Bj})}{4\pi^2\alpha_{em}}\sigma_L.
\eeqn
The invariant structure functions
$F_1$ and $F_2$ can be expressed in terms of the
helicity structure functions $F_T$ and $F_L$ and vice versa. 
In particular, one obtains in the Bjorken limit, $Q^2\to\infty$,
$x_{Bj}$ fixed,
\beqn
F_2(x_{Bj},Q^2)&=&F_T(x_{Bj},Q^2)+F_L(x_{Bj},Q^2)\\
F_1(x_{Bj},Q^2)&=&\frac{F_T(x_{Bj},Q^2)}{2x_{Bj}}.
\eeqn
There are only two independent structure functions in DIS off unpolarized
targets.

\subsection{The parton model of DIS}\label{dissec}

\begin{figure}[ht]
\centerline{
  \raise2cm\hbox{\scalebox{0.7}{\includegraphics{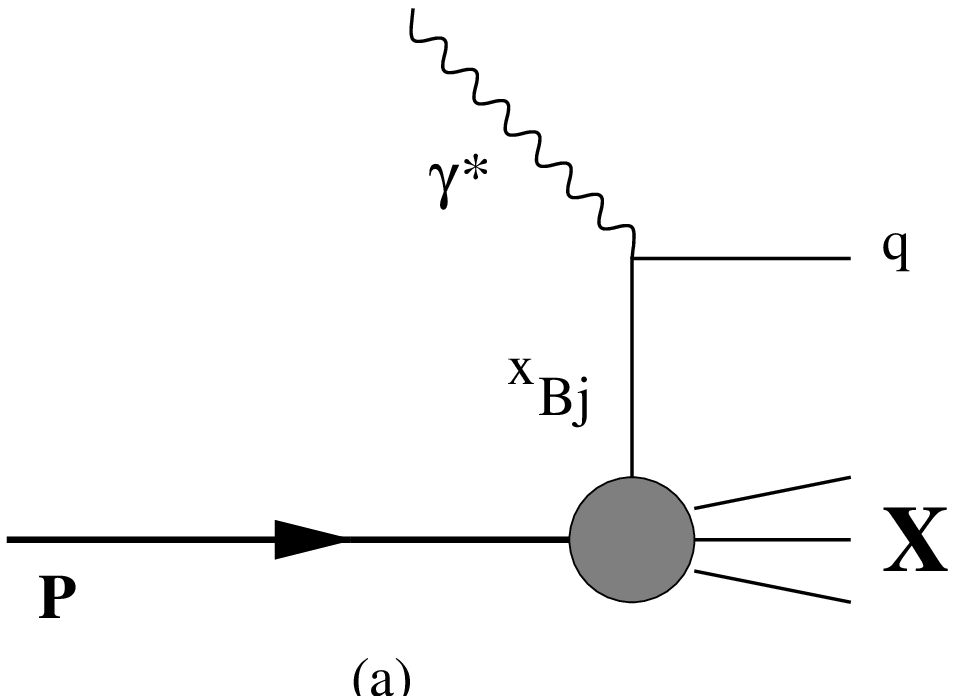}}}
  \scalebox{0.7}{\includegraphics{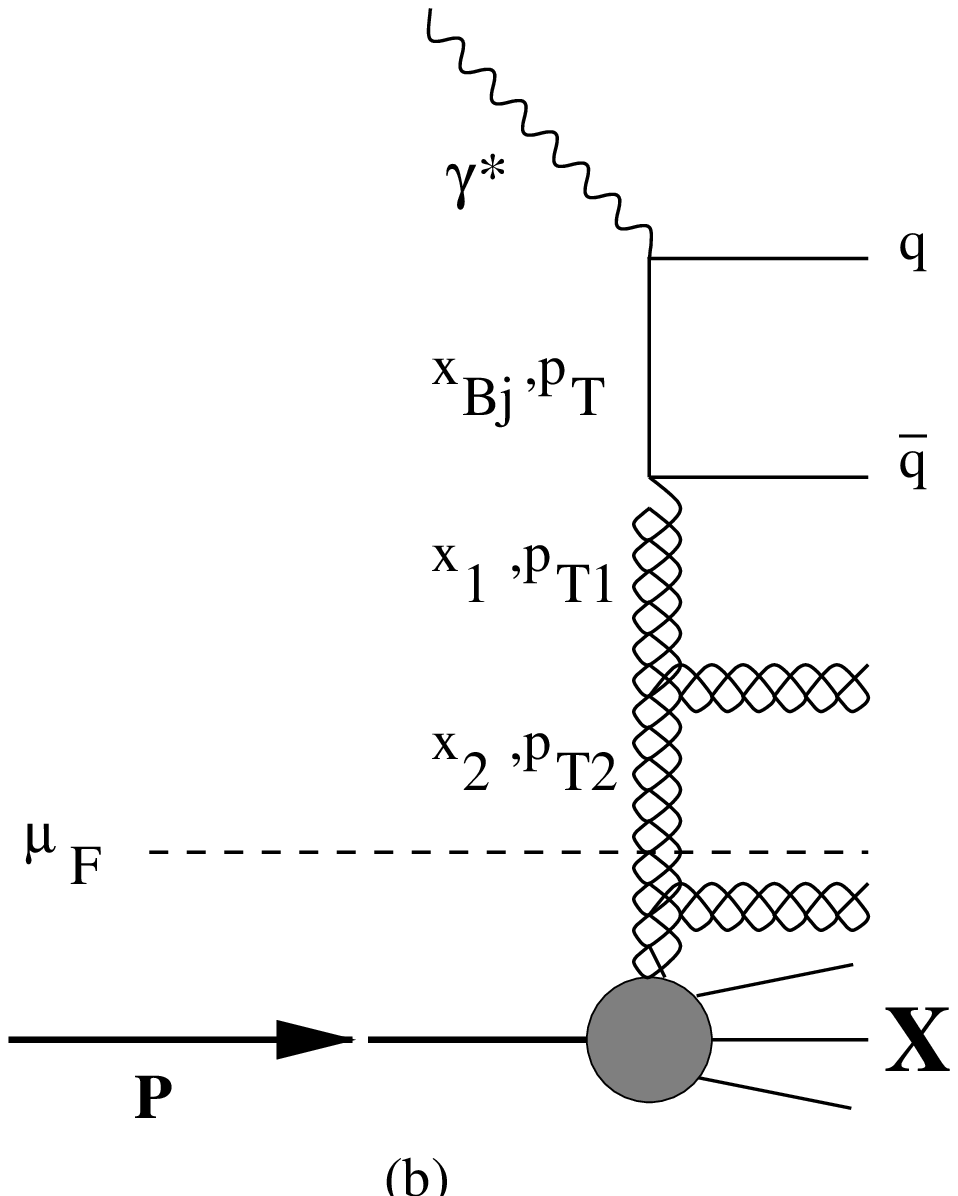}}
}
\caption{
  \label{dis2}
  The parton model for DIS (a) and its QCD improvement (b). In the naive
  parton model, the virtual photon hits a quark (q) 
  inside the target. In the QCD
  improved parton model, the quark can e.\ g.\ 
  be created from a  gluon, denoted by a
  curly line, by the splitting process $G\to q\bar q$. 
  The $x_i$ are the longitudinal momentum fractions of the particles with
  respect to the target. Therefore $x_{Bj}<x_1<x_2<\dots$.
}
\end{figure}

It came as a big surprise, when the first 
DIS measurements at SLAC
\cite{SLAC} showed that the structure function $F_2(x_{Bj},Q^2)$ 
is nearly constant as
function of $Q^2$ at fixed $x_{Bj}$.
An explanation of this phenomenon was given by Bjorken \cite{Bjorken} and by
Feynman \cite{Feynman}. Feynman's intuitive explanation is depicted in fig.\
\ref{dis2}a. He proposed that the proton is made up of pointlike charged
constituents, so called partons, 
and the total $\gamma^*p$ cross section is the incoherent sum of
photon-parton cross sections. The transverse momenta of the partons are
neglected, which is well justified in a frame where the proton is fast moving,
e.\ g.\ in the Breit frame or in the infinite momentum frame \cite {Povh}. 
In such a frame the longitudinal momenta of the partons are much larger than
the transverse ones. 

The cross sections for transverse and longitudinal
photons scattering off spin-$1/2$ partons, i.\ e.\ quarks (fig.\ \ref{dis2}a),
 are given by
\beqn\label{discross}
\sigma_T^{\gamma^*q}&=&\frac{4\pi^2\alpha_{em}Z_f^2}{Q^2(1-x_{Bj})}
\delta\left(1-\frac{x_{Bj}}{x_q}\right)\\ 
\sigma_L^{\gamma^*q}&=&0,
\eeqn
where $x_q$ is the momentum fraction of the proton carried by the struck quark
and $Z_f$ is the flavor charge in units of the elementary charge.
The $\delta$-function arises from momentum conservation and 
gives a physical meaning to the Bjorken variable. In the Breit frame,
$x_{Bj}$ is the momentum
fraction of the proton carried by the struck quark.  
For massless quarks, 
the longitudinal cross section is zero due to helicity conservation.
Introducing the density $q_f(x_{Bj})$ 
of quarks of flavor $f$, inside the
proton, one obtains a simple partonic interpretation of the structure
functions,
\beqn\label{master}
F_2(x_{Bj})&=&x_{Bj}\sum\limits_f^{}Z_f^2( q_f(x_{Bj})+\bar q_f(x_{Bj})),\\
F_L&=&0.
\eeqn
In the naive parton model, the structure functions depend only on $x_{Bj}$ and
not on $Q^2$, the longitudinal structure function vanishes and one obtains the
Callan-Gross relation \cite{CallanGross},
\beq
F_2-2x_{Bj}F_1=0.
\eeq
These equations are the basic results of the parton model and they are
approximately confirmed by experiment. 

\begin{figure}[t]
  \scalebox{0.68}{\includegraphics*{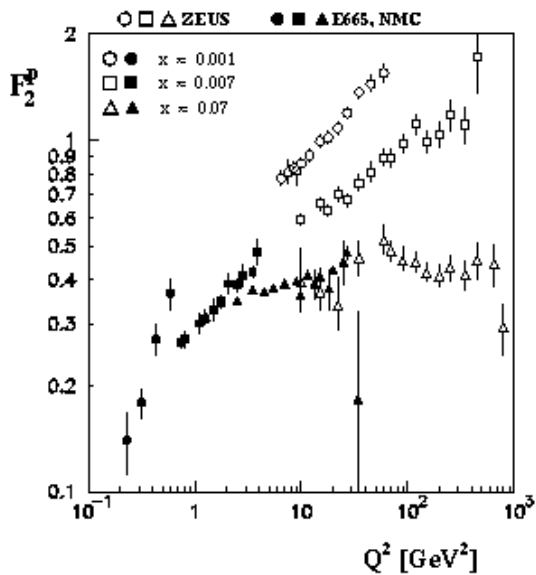}}\hfill
  \raise1.3cm\hbox{\parbox[b]{2.3in}{
    \caption{
      \label{f2pq2}
      Deviations from Bjor\-ken scal\-ing can be explained by pQCD corrections
      to the naive parton model. While Bjorken scaling is 
      approximately fulfilled
      at high $x_{Bj}$, scaling violations become stronger and stronger as
      $x_{Bj}$ decreases. At very low $x_{Bj}$, practically no scaling can be
      observed any more.   
      The 
      figure is taken from \cite{PillerRev} and the data are from
      ZEUS \cite{zeus961+2,zeus98}, E665 \cite{E665} and NMC \cite{nmc1}. 
    }
  }
}
\end{figure}

One of the great achievements 
of QCD is the successful description of the deviations
from the naive parton model seen in experiment. 
In particular at low
$x_{Bj}$
deviations from Bjorken scaling become quite pronounced, see fig.\ \ref{f2pq2}.
In  the QCD improved parton
model, perturbation theory is applied to calculate corrections to the parton
model predictions. An example of such a correction, which is important
at low $x_{Bj}$, is shown in 
fig.\ \ref{dis2}b. The quark seen by the photon is
generated perturbatively from a gluon by the splitting process $G\to q\bar q$.
Higher order corrections allow also to take into account further gluons
radiated from the gluon that splits into the $q\bar q$-pair. 

The evolution of the structure function with $Q^2$ is described by the DGLAP
equations. There are several ways to motivate these equations. 
The DGLAP evolution equations resum logarithms in $Q^2$ originating from
ladder diagrams like the one in fig.\ \ref{dis2}b. Historically, they
originate from a time even before QCD. Gribov and Lipatov \cite{GL,Lipatov}
were the first who
found that these ladder diagrams receive dominant contributions from
configurations with strong ordering in the transverse momenta, i.e.\
\beq
p_{T1}^2\gg p_{T2}^2\gg \dots.
\eeq
They obtained integrals of the form $dp_T^2/p_T^2$ which could be resummed to
all orders.
However, their calculations were done in an abelian theory. 
A very intuitive derivation in non-covariant Hamiltonian perturbation theory
was given later (in QCD) by Altarelli and Parisi \cite{AP}. 
The illustrative interpretation
of the DGLAP equations is due to them. The photon acts as a
microscope with a resolution determined by $Q^2$. As one increases $Q^2$, the
photon resolves more and more of the target substructure.  
The probability to find
a quark and an antiquark 
of flavor $f$ inside a gluon, fig.\ \ref{dis2}b, is then 
described by a
splitting function $P_{fG}$.
Similarly, there exists a splitting function which describes the probability 
to find a quark and another gluon inside the parent gluon.
Finally,
Dokshitzer \cite{Dok} derived the same equation again with the method proposed
by Gribov and Lipatov, but now in QCD. 

From the process where the quark seen by the photon is generated by the
splitting of a gluon, one obtains in perturbation theory the contribution
\beqn\nonumber\label{eins}
\frac{F_2(x_{Bj},Q^2)}{x_{Bj}}&=&\sum\limits_fZ_f^2\Bigg[
q_f\left(x_{Bj}\right)\\
&+&\frac{\alpha_s}{2\pi}\int\limits_{x_{Bj}}^1 \frac{dx_1}{x_1} 
G(x_1)\left\{P_{fG}\left(\frac{x_{Bj}}{x_1}\right)
\ln\left(\frac{Q^2}{\kappa^2}\right)
+\dots\right\}
\Bigg]
\eeqn
to the structure function. 
Here, $G(x_1)$ ist the gluon density of the target and $x_{Bj}/x_1\le 1$ is
the momentum fraction of the gluon carried away by the quark in the
ladder.
The logarithmic divergence in (\ref{eins}) 
is regularized by the cutoff $\kappa$.
The dots denote renormalization scheme dependent functions which  
can be calculated and
contain no such divergence. 
These functions are of no interest to our more qualitative discussion and we
refer to \cite{ApQCD} for more details.
The splitting function is given by
\beq\label{splitting}
P_{fG}(z)=\frac{1}{2}\left[z^2+(1-z)^2\right].
\eeq
The divergence in (\ref{eins}) is a collinear divergence,
originating from configurations with $p_T\to 0$ in fig.\
\ref{dis2}b. This limit corresponds to a long range part of the strong
interaction which cannot be calculated in perturbation theory. 
Therefore, the quark ($q_f$) and gluon ($G$)
densities in (\ref{eins}) are unmeasurable, bare parton
distributions, $q_{f0}$ and $G_0$. 
The renormalized quark density may be defined as
\beq
q_f(x_{Bj},\mu_F)=q_{f0}(x_{Bj})
+\frac{\alpha_s}{2\pi}\int_{x_{Bj}}^1 \frac{dx_1}{x_1}
G_0(x_1)\left\{P_{fG}\left(\frac{x_{Bj}}{x_1}\right)
\ln\left(\frac{\mu_F^2}{\kappa^2}\right)
+\dots\right\}.
\eeq
The collinear singularity is absorbed into the bare
quark density at the factorization scale $\mu_F$. 
With this prescription, the partonic
interpretation of $F_2$ (\ref{master}) is maintained, but the structure
function is no longer independent of $Q^2$.

The distribution $q_f(x_{Bj},\mu_F)$ cannot be calculated in pQCD, since it is
dominated by soft physics. It has to be extracted from measurements of $F_2$.
However one can calculate in pQCD, how the parton distribution runs with the
factorization scale. Since $\mu_F$ is not a physical quantity, observables
cannot depend on it. Therefore,
\beq
\frac{dF_2(x_{Bj},Q^2,\mu_F)}{d\mu_F}=0
\eeq
and 
\beq
Q^2\frac{dq_f(x_{Bj},Q^2)}{dQ^2}=
\frac{\alpha_s}{2\pi}\int_{x_{Bj}}^1 \frac{dx_1}{x_1}
P_{fG}\left(\frac{x_{Bj}}{x_1}\right)G(x_1,Q^2).
\eeq
This is one of the DGLAP equations 
which describes the evolution of the quark density. 

Note that also the gluon density depends on $Q^2$, because of pQCD 
corrections. Altogether one obtains $N_f+1$ coupled equations describing the
$Q^2$ evolution of the singlet parton densities,
\beq\label{DGLAP}
Q^2\frac{d}{dQ^2}
\left(
\begin{array}{c}
q_f(x_{Bj},Q^2)\\[1em]
G(x_{Bj},Q^2)
\end{array}
\right)
=
\frac{\alpha_s}{2\pi}\int\limits_{x_{Bj}}^1 \frac{dx_1}{x_1}
\left(
\begin{array}{cc}
P_{ff}\left(\frac{x_{Bj}}{x_1}\right)
&P_{fG}\left(\frac{x_{Bj}}{x_1}\right)\\[1em]
P_{Gf}\left(\frac{x_{Bj}}{x_1}\right)&P_{GG}\left(\frac{x_{Bj}}{x_1}\right)
\end{array}\right)
\left(
\begin{array}{c}
q_f(x_{1},Q^2)\\[1em]
G(x_{1},Q^2)
\end{array}
\right),
\eeq
How can one calculate $P_{Gf}$ and $P_{GG}$. 
The photon does not couple directly to
gluons. The most straightforward way is to consider a longitudinal photon in
fig.\ \ref{dis2}b. In this case, the $\alpha_s$ correction (\ref{eins}) does not
contain a divergence and therefore does not contribute to the running of the
quark densities. It is easy to understand intuitively, why there is no
divergence for longitudinal photons. The divergence originates from
configurations with $p_T\to 0$. This case however is exactly the naive parton
model, in which $\sigma_L=0$ due to helicity conservation.
In order to obtain $P_{Gf}$ and $P_{GG}$, 
one has to take into account at least one
rung in the ladder. The splitting function is the coefficient of the 
logarithmic divergence
occuring at $p_{T1}\to 0$. For this reason, longitudinal photons are often
called "gluonometer" \cite{Dok}.

At present the splitting functions are known to next to leading order
in pQCD. They can e.\ g.\ be found in \cite{Ellis}.
The function 
$P_{GG}(x_{Bj}/x_1)\sim 6x_1/x_{Bj}$ (at low $x_{Bj}$)
exhibits a
pole at $x_{Bj}=0$.
It is widely believed that this pole is responsible
for the steep rise of $F_2$ and the high gluon density at low $x_{Bj}$, 
which behaves
approximately like
\beq\label{asymptotic}
x_{Bj}G(x_{Bj},Q^2)\propto\exp\left(2\sqrt{\frac{N_c\alpha_s}{\pi}
\ln(1/x_{Bj})\ln(Q^2/Q^2_0)}\right)
\eeq 
for fixed strong coupling constant $x_{Bj}$.
There are however also different opinions \cite{DL,CDL}.

DGLAP is a special kind of a renormalization group equation. 
One of its most important properties is that it separates scales. All
the soft physics is contained in the parton distributions. These are of entirely
nonperturbative origin and have to be parametrized at some input scale $Q_0^2$.
Such parametrizations are provided by the three collaborations GRV, MRST and
CTEQ \cite{GRV,MRST,CTEQ} in leading and in next to leading order. With the
parton distributions as input, one can then 
calculate $F_2$ at a higher value of
$Q^2$,
because the splitting functions can be calculated in pQCD. 
The other 
essential property is, that the parton distributions are universal, i.\ e.\
they do not depend on the process under consideration, but only on the hadron
state. 

A more rigorous technique to separate scales in QCD is Wilson's operator product
expansion \cite{OPE}.
Using the OPE, the matrix of
anomalous dimensions, i.e.\ the Mellin transforms of the splitting functions,
was derived by 
Georgi and Politzer \cite{GP} and by
Gross and Wilcek \cite{GW}. 
Explaining the OPE is however not within the scope of this work and we rather
refer to standard texts \cite{Muta}. 
We only mention 
that in this method the time ordered product of currents in the virtual
Compton amplitude (\ref{Comamp}) 
is expanded near the light cone in a series of local
operators with coefficient functions which are singular on the light cone.
These coefficient functions can be calculated in pQCD and are 
related
to the splitting functions by Mellin transform.
It is interesting to see within this framework,
that (i) DIS is a long distance process receiving large
contributions from the region near the lightcone, (ii) the partonic
interpretation of $F_2$ is valid only in the infinite momentum frame and (iii)
additional terms occur, which cannot be interpreted as parton densities.
These are the so called higher twist terms which are formally expressed as
matrix elements of multiple field correlators. 
A typical example is a double scattering process.
These matrix elements are, like
the parton densities, not accessible by means of perturbation theory and have
to be adjusted to experimental data. 
The matrix elements are however universal, since
they depend only on the state in which the field correlator is evaluated.
Higher twists are typically suppressed by additional powers of $Q^2$, but can be
enhanced by a factor $A^{1/3}$ in processes involving nuclei, where $A$ is the
nuclear mass number. The $A$ enhancement was argued by Luo, Qiu and Sterman
\cite{LQS1,LQS2}.
The calculation of higher twist terms is rather involved and they are
theoretically not well under control. 

%
%
%

\subsection{The color dipole picture of low-x DIS}\label{dipoledis}

\begin{figure}[ht]
  \centerline{\scalebox{0.8}{\includegraphics*{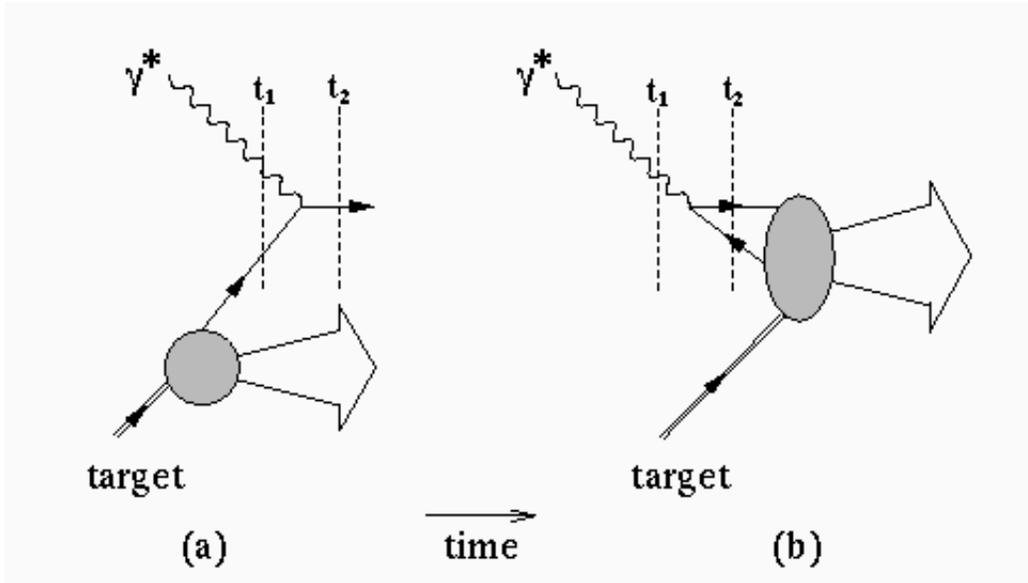}}}
    \caption{
      \label{timeord}
      The two time orderings for the interaction of a virtual photon with a
      target: (a) The photon hits a quark inside the target, (b) the photon
      splits up into a $q\bar q$-pair which subsequently scatters off the target
      color field. In the rest frame of the target, the latter contribution is
      the dominant one. The 
      figure is taken from \cite{PillerRev}.
    }  
\end{figure}

In this section we have a closer 
look at the space time picture of DIS at low $x_{Bj}$. 
The two possible time orderings for the photon-quark
interaction are depicted in fig.\ \ref{timeord}. In \ref{timeord}a, the virtual
photon penetrates the target and hits a quark. In 
\ref{timeord}b however, the $\gamma^*$ splits up into a $q\bar q$ pair that
subsequently scatters off the target color field. In covariant
Feynman-perturbation theory, both contributions are automatically taken into
account. It is however instructive to consider DIS in the target rest frame and
in a Hamiltonian, i.e.\ not manifestly covariant, picture, because this will
give us a clearer physical understanding of the dominant mechanism and in turn
we can make full use of our physical intuition.

The relative importance of the two contributions in fig.\ \ref{timeord} may be
estimated from the ratio of the energy denominators as it was done in
\cite{weise}. For large photon energies, $\nu\gg m_N$, these energy denominators
were found to be
\beqn
\Delta E_a&=&E_a(t_2)-E_a(t_1)\approx-\la p_q^2\ra^{1/2}
+\frac{\frac{2}{3}\la p_q^2\ra+Q^2}{2\nu},\\
\Delta E_b&=&E_b(t_2)-E_b(t_1)\approx
\frac{M_{q\bar q}^2+Q^2}{2\nu},
\eeqn
where $M_{q\bar q}$ 
is the invariant mass of the $q\bar q$-dipole and $\la p_q^2\ra$ is
the mean square of the quark momentum inside the target. In order
to estimate the ratio of the energy denominators, one usually sets 
$M_{q\bar q}^2\to Q^2$, since
$Q^2$ is the only large dimensionful scale available. This was also argued in
\cite{weise} with the result
\beq
\left|\frac{\Delta E_b}{\Delta E_a}\right|
\approx 2x_{Bj}\frac{m_N}{\la p_q^2\ra^{1/2}}.
\eeq
This expression makes very clear, that in the target rest frame, the process in
fig.\ \ref{timeord}a is suppressed by a factor of order $x_{Bj}\ll 1$. 
For our
physical intuition it is therefore sufficient to think of 
fig.\ \ref{timeord}b. 
Indeed, we can estimate the lifetime of the $q\bar q$
fluctuation with help of the uncertainty relation,
\beq\label{cohlength}
l_c=\frac{1}{\Delta E_b}\approx\frac{1}{2m_Nx_{Bj}},
\eeq
where we again replaced $M_{q\bar q}^2$ by $Q^2$. 
One recognizes that the lifetime or
coherence time can become very long at low $x_{Bj}$, e.\ g.\ $10^5$ fm at the
lowest values of $x_{Bj}$ accessible at HERA and $10$ fm at NMC energies
($x_{Bj}\approx 10^{-2}$). This is illustrated in fig.\ \ref{lc1}, which
corresponds to the Feynman diagram in fig.\ \ref{dis2}b, when only the gluon
($x_1,p_{T1}$) is taken into account. It is interesting to see,
that although the $\gamma^*$-proton
cross section is a Lorentz invariant quantity, the partonic interpretation of
the scattering process depends on the reference frame.

The coherence length $l_c$ is one of the key quantities in low-x DIS and we will
present a more precise calculation than the prescription 
$M^2_{q\bar q}\to Q^2$ later in
this work. Everything which happens to the photon within the length
$l_c$, is governed by coherence effects. It is therefore essential to achieve a
good physical understanding of these effects, especially in view of processes
involving nuclei \cite{PillerRev}, 
where multiple scatterings will occur. 
It is the main purpose of this work to provide insight into
the mechanisms underlying these coherence effects.

The cross sections 
for transverse and longitudinal photons are most conveniently
written in a mixed representation. The two transverse directions are treated in
coordinate space, while the longitudinal direction is described in momentum
representation. Let $\vec\rho$ be the two dimensional vector pointing from the
quark to the antiquark in the transverse plane and $\alpha$ the fraction of the
photon energy $\nu$ carried by the quark. The
momentum fraction of the antiquark is then $1-\alpha$, see fig.\ \ref{lc1}.
The cross section reads  \cite{nz91,ForshawRoss}
\beq\label{stl}
\sigma_{T,L}^{\gamma^*p}=\int_0^1d\alpha\int d^2\rho
\left|\Psi_{q\bar q}^{T,L}(\alpha,\rho)\right|^2\sigma_{q\bar q}(\rho),
\eeq
where the $\Psi_{q\bar q}^{T,L}(\alpha,\rho)$ are the light-cone (LC) wavefunctions
for the transition $\gamma^*\to q\bar q$. 
The LC wavefunctions can be calculated in perturbation
theory and read in first order in the fine structure constant $\alpha_{em}$
\beqn\label{psit}
\left|\Psi_{q\bar q}^{T}(\alpha,\rho)\right|^2&=&
\frac{2N_c\alpha_{em}}{(2\pi)^2}\sum\limits_{f=1}^{N_f}Z_f^2
\left\{\left[1-2\alpha(1-\alpha)\right]\eps^2{\rm K}^2_1(\eps\rho)
+m_f^2{\rm K}^2_0(\eps\rho)\right\},\\
\label{psil}
\left|\Psi_{q\bar q}^{L}(\alpha,\rho)\right|^2&=&
\frac{8N_c\alpha_{em}}{(2\pi)^2}\sum\limits_{f=1}^{N_f}Z_f^2
Q^2\alpha^2(1-\alpha)^2{\rm K}^2_0(\eps\rho),
\eeqn
where ${\rm K}_{0,1}$ are the modified Bessel functions of the second kind.
These functions are also called MacDonald functions \cite{Abramowitz}. 
Furthermore we have introduced the extension parameter 
\beq\label{disextension}
\eps^2=\alpha(1-\alpha)Q^2+m_f^2,
\eeq
which depends on the flavor mass $m_f$.
The cross section for scattering  a $q\bar q$-dipole off the proton is
denoted by $\sigma_{q\bar q}(\rho)$. 
In Born approximation (two gluon exchange) 
it is independent of energy and related to the two-quark form factor of the
proton via \cite{zkl}
\beq\label{borisdipole}
\sigma_{q\bar q}(\rho)=\frac{16\alpha_s^2}{3}\int d^2p_{T1}
\frac{\left[1-\la P|\exp({\rm i}\vec p_{T1}\cdot(\vec r_1-\vec r_2))|P\ra\right]
\left[1-\exp({\rm i}\vec p_{T1}\cdot\vec\rho)\right]}
{p_{T1}^4},
\eeq
where $p_{T1}$ is the transverse momentum of the exchanged gluon, fig.\
\ref{dis2}b.
In Regge phenomenology, two gluon exchange is a simple model of pomeron
exchange, cf.\ section \ref{diffractionsec}. Indeed,
the dipole cross section
takes only the pomeron contribution to the total cross section (\ref{stl}) into
account. Therefore (\ref{stl}) can be applied only at high energies.
Note the color screening factor 
$[1-\exp({\rm i}\vec p_{T1}\cdot\vec\rho)]$ in (\ref{borisdipole}), which makes 
the dipole cross section
vanish $\propto \rho^2$ at $\rho\to 0$. This salient property of the dipole
cross section is
the heart of the color transparency  phenomenon \cite{zkl,bbgg,bm}. 
In Born approximation,
the two-quark form factor is related to the differential gluon density of the
target by
\beq
\frac{\partial (x_{Bj}G(x_{Bj},p_{T1}^2))}{\partial \log(p_{T1}^2)}
=\frac{4\alpha_s(p_{T1}^2)}{\pi}\left[1-\la P|\exp({\rm i}\vec p_{T1}
\cdot(\vec r_1-\vec r_2))|P\ra\right].
\eeq
The energy dependence of the gluon density is caused by higher order QCD
corrections, i.e.\ one has to take the gluon rungs in fig.\ \ref{dis2}b into
account.
The dipole cross section reads then \cite{nzdipol}
\beq\label{nzsigma}
\sigma_{q\bar q}(x_{Bj},\rho)=\frac{4\pi}{3}
\rho^2\alpha_s(\lambda/\rho^2)
\int\frac{d^2p_{T1}}{p_{T1}^2}
\frac{[1-\exp({\rm i}\vec p_{T1}\cdot\vec\rho)]}{p_{T1}^2\rho^2}
\frac{\partial (x_{Bj}G(x_{Bj},p_{T1}^2))}{\partial \log(p_{T1}^2)},
\eeq
with $\lambda=2.25$. 
For small distances $\rho\to 0$ one can expand the color screening factor in
(\ref{nzsigma}), which leads to
\beq\label{nzsmall}
\sigma_{q\bar q}(x_{Bj},\rho)=
\frac{\pi^2}{3}\rho^2\alpha_s(\lambda/\rho^2)
\,x_{Bj}G(x_{Bj},A_\sigma/\rho^2).
\eeq
It was found in \cite{nzdipol} that $A_\sigma\approx10$. 
The dipole cross section still
vanishes quadratically at small $\rho$, up to
logarithms which originate from the gluon density.
Note that the result of Nikolaev and Zakharov (\ref{nzsmall})
coincides with the result from
\cite{dipol1,dipol2,dipol3},
\beq\label{fssigma}
\sigma_{q\bar q}(x_{Bj},\rho)=
\frac{\pi^2}{3}\rho^2\alpha_s(\lambda^\prime/\rho^2)
\,x_{Bj}G(x_{Bj},\lambda^\prime/\rho^2),
\eeq 
except for the scale, at which the strong coupling constant enters.
For a detailed derivation of (\ref{fssigma})
see \cite{dipol3}.
The gluon density in (\ref{fssigma}) is tested at the same scale as the one in
(\ref{nzsmall}), i.e.\ $\lambda\approx 10$ \cite{Koepf}.

For convenience,
we do not write out the energy dependence of the dipole cross section
explicitly, until section \ref{DCS}.
The dipole cross section is largely unknown and has to be parametrized. Several fits already
exist in the literature \cite{Wuesthoff1,Wuesthoff2,Forshaw,McDFS99}, 
we will present our own parametrization in section \ref{DCS}.
As already pointed out above, the space time picture of DIS depends on the
reference frame. In the target rest frame the virtual photon splits up into a
$q\bar q$-pair which then scatters off the target. In the infinite momentum
frame, the photon couples to the electric charge of the quarks in the target. 
Note, that there are valence and sea quarks in the target.
Since the dipole cross section is proportional to the gluon density
of the target, only quarks which are generated from gluon splitting are taken
into account by (\ref{stl}). In other words, the valence quark contribution to
the cross section is neglected and therefore (\ref{stl}) is only applicable when
sea quarks dominate, i.e.\ at low $x_{Bj}$. It is worth noting that it is
impossible to decide whether a sea quark belongs to the target and is generated
by gluon splitting or whether it is part of the hadronic structure of the
virtual photon. This depends on the reference frame.

It is important to note that while (\ref{psit}) and (\ref{psil}) are valid
only in perturbation theory, our formula (\ref{stl}) for the total cross
section is completely general and does not rely on the applicability of pQCD.
In the mixed $\rho - \alpha$-representation, 
the scattering matrix is diagonal. 
Using the standard notation for the scattering matrix, see e.g.\
\cite{ForshawRoss},
\beq
S_{fi}=\delta_{fi}+{\rm
i}(2\pi)^4\delta^{(4)}\left(\sum_fp_f-\sum_ip_i\right){\cal A}_{fi},
\eeq
where the index $f$ denotes the final state and $i$ the initial state. One also
often introduces the $T$-matrix element
\beq
T_{fi}=(2\pi)^4\delta^{(4)}\left(\sum_fp_f-\sum_ip_i\right){\cal A}_{fi}
\eeq
such that $S_{fi}=1+{\rm i}T_{fi}$. The unitarity of the $S$-matrix
leads to the optical theorem as a special case of the
Cutkosky rules \cite{Cutkosky},
\beq
2{\rm Im}{\cal A}_{ii}=K\sigma_{tot},
\eeq
where $K\approx 2s$ is the flux factor.
The eigenstates of interaction are defined by
\beqn
\sum_k|\psi_k\ra\la \psi_k|&=&1,\\
T|\psi_k\ra&=&t_k|\psi_k\ra.
\eeqn
Here, $t_k$ is the probability that the eigenstate 
$|\psi_k\ra$ scatters off
the target.
The photon can now be expanded in this basis
\cite{Borisold,Miettinen},
\beq\label{expansion}
|\gamma^*\ra=\sum_kc_k^{\gamma^*}|\psi_k\ra,
\eeq
with
\beq
\la\gamma^*|\gamma^*\ra=1.
\eeq 
For simplicity, we do not consider wavefunction renormalization and assume that
alss states are normalized to unity.
Thus one obtains
\beq
\sum_k
|c_k^{\gamma^*}|^2t_k=s\sigma^{\gamma^*p}.
\eeq
Comparing this expression with (\ref{stl}), we identify the eigenstates of the
interaction with color dipoles with fixed transverse separation and fixed
longitudinal momentum fractions.
The index $k$ labels the ($\rho,\alpha$) of the dipoles.
Furthermore, we can identify $t_k/s$  with the cross section of a dipole with
a given $\rho$ off the target, i.\ e.\ with $\sigma_{q\bar q}(\rho)$,
\beq
\sigma_{q\bar q}(\rho)=\frac{t_k}{s}.
\eeq
The
identification of interaction eigenstates with color dipoles of fixed
transverse separation in QCD was first made in \cite{zkl}.
There are of course also higher Fock-states in the photon, for example the
$q\bar qG$ state will play an important role for diffraction in DIS. This will
be discussed in section \ref{diffractionsec}.

It is a special virtue of the color dipole description of DIS 
to work with the correct degrees of freedom, which are the eigenstates of the
interaction. This will be of great use, when we consider nuclear targets in the
section \ref{nucleus}.

The $\rho$-integral in (\ref{stl})
extends from small distances, where pQCD is applicable, up to very large
distances, which are governed by infrared properties of QCD that are not
understood at present. 
The extension parameter 
(\ref{disextension}) controls which
distances $\rho$ in (\ref{stl}) contribute to the integral. Note that the
asymptotic behavior of the Bessel functions is
\beqn \label{asymptotics}
{\rm K}_0(z)=-\log(z) & {\rm for} & z\to 0, \\
{\rm K}_1(z)=\frac{1}{z} & {\rm for} & z\to 0, \\
{\rm K}_0(z)={\rm K}_1(z)=\sqrt{\frac{\pi}{2z}}\,{\rm e}^{-z} 
& {\rm for} & z\to \infty.
\eeqn
Both functions, K$_0$ and K$_1$, 
decay exponentially for large arguments and are divergent at
$z\to 0$. While ${\rm K}_0$ has only an integrable, logarithmic divergence,
${\rm K}_1$ diverges much stronger and the wavefunction is not square
integrable. Numerically, the ${\rm K}_1$-part is the dominant term.

Following \cite{boris},
we will now investigate, how these hard and soft
contributions interplay.
Special attention is paid to the question, which regions
of integration give scaling contributions to $F_2$ and which are higher
twist, i.e.\ are suppressed by additional powers of $Q^2$.

First consider transverse photons and estimate is the probability $w$, 
that the
photon splits into a small size pair $\rho^2\sim 1/Q^2$. 
This probability is of course
practically $1$, because the LC 
wavefunction (\ref{psit}) decays exponentially at large
$\eps\rho$. The largest dipoles that can contribute have a size of order
$1/\eps$.
A small size dipole interacts with a cross section roughly
proportional to its size, because of color transparency. 
As result, one finds that the contribution of small size
dipoles from transverse photons give a contribution to 
$\sigma_T^{\gamma^*p}$ that
decays as $1/Q^2$ and thus a leading twist contribution to $F_2$. 

How can the virtual photon convert into a large size dipole? The only
possibility is, that the extension parameter 
in the Bessel function (\ref{psit})
$\eps$ becomes small, i.e.\ one
needs $\alpha\to 0$ or $\alpha\to 1$.
These limits are known as Bjorken's aligned jet configurations \cite{bk}.
The phase space for these highly asymetric configurations is however very
small, of order $m^2_f/Q^2$, and thus the probability of finding a large dipole is
also small. This is however compensated by the large interaction cross section,
which is of the order of a typical hadronic cross section, $\sigma_h$. 
One therefore obtains a leading twist contribution to $F_2$ from large dipoles
that cannot be treated perturbatively. These
considerations are summarized in tab.\ {\ref{distabT}.

For longitudinal photons and for small dipoles, the situation looks like in the
transverse case and the hard part gives a leading twist contribution to $F_2$.
This is however different in the case of large dipoles. Because of the term
$\alpha^2(1-\alpha)^2$ in the longitudinal LC wavefunction
(\ref{psil}), aligned jet
configurations are suppressed by an extra power of $Q^2$ compared to the
transverse case. There are no large dipoles from longitudinal photons. Also the
large cross section $\sigma_h$ cannot compensate for this and the soft
contribution to $F_2$ is higher twist.
This is summarized in tab.\ {\ref{distabL}.

\begin{table}[t]
\begin{center}
\renewcommand{\arraystretch}{2.3}
\begin{tabular}{|c|c|c||c|}
\hline
\vphantom{XXXXXX}\hphantom{XXXX} T \hphantom{XXXX}
&\hphantom{XXXX} $w$ \hphantom{XXXX}
&\hphantom{XXXX} $\overline\sigma\hphantom{XXXX}$ 
&\hphantom{XXXX} $\sigma^{\gamma^*p}$\hphantom{XXXX}\\
\hline
hard & 1 & $\displaystyle\frac{1}{Q^2}$ &  
$\displaystyle\propto\,\frac{1}{Q^2} $ \\
\hline
soft & $\displaystyle\frac{m_f^2}{Q^2}$ & 
$\displaystyle\sigma_h$ &  $\displaystyle\propto\,\frac{1}{Q^2} $ \\
\hline
\end{tabular}
\caption{\label{distabT}
The $Q^2$ behavior of hard and soft contributions to the total photon-proton
cross section for transverse photons.
 The probability that the photon converts into a small (hard)  or
large (soft)  
dipole is denoted by $w$, $\overline\sigma$ is the order of 
magnitude of the
dipole cross section and the last column contains the $Q^2$-behavior of the
total photon-proton cross section. $\sigma_h$ is a typical hadronic cross
section, i.e.\ independent of $Q^2$.
For transverse photons both,
small and large dipoles, give leading twist
contributions to $F_2$.} 
\renewcommand{\arraystretch}{1}
\end{center}
\end{table}
\begin{table}[t]
\begin{center}
\renewcommand{\arraystretch}{2.3}
\begin{tabular}{|c|c|c||c|}
\hline
\vphantom{XXXXXX}\hphantom{XXXX} L \hphantom{XXXX}
&\hphantom{XXXX} $w$ \hphantom{XXXX}
&\hphantom{XXXX} $\overline\sigma\hphantom{XXXX}$ 
&\hphantom{XXXX} $\sigma^{\gamma^*p}$\hphantom{XXXX}\\
\hline
hard & 1 & $\displaystyle\frac{1}{Q^2}$ &  
$\displaystyle\propto\,\frac{1}{Q^2} $ \\
\hline
soft & $\displaystyle\left(\frac{m_f^2}{Q^2}\right)^2$ & 
$\displaystyle\sigma_h$ &  $\displaystyle\propto\,\frac{1}{Q^4} $ \\
\hline
\end{tabular}
\caption{\label{distabL}
The same as tab.\ \ref{distabT} but now for longitudinal photons.
In the longitudinal case, 
only small (hard) dipoles give a leading twist
contributions to $F_2$. There are no large dipoles from longitudinal photons.} 
\renewcommand{\arraystretch}{1}
\end{center}
\end{table}

Note that the integral for the transverse cross section 
in (\ref{stl})
is logarithmically
divergent (for a quadratically rising dipole cross section), 
if one would not include the quark mass $m_f$. 
This divergence originates from the end points of the integration
over $\alpha$ and is the analog of the logarithmic divergence in (\ref{eins}).
Again the reason is, that the integral extends over domains, where pQCD is not
applicable.  
In contrast to this, the longitudinal cross section is finite, even with
$m_f=0$. No collinear divergence occurs in this case, because the endpoints of
the $\alpha$-integration correspond to the naive parton model in which
$\sigma_L^{\gamma^*p}=0$. 
Therefore, no divergence can come from these configurations.
We point out, that there is a one-to-one 
correspondence between the divergences in
the standard pQCD in momentum space and the color dipole picture of DIS, which
is formulated in the mixed representation. Factorization properties in impact
parameter space are also considered in
\cite{bhq}.

For a photon, the LC wave functions, i.e.\ the coefficients $c_k^{\gamma^*}$
in (\ref{expansion}), can be calculated in perturbation theory.
For an introduction to LC wave functions and how to use them, see
\cite{BigRev}. Although they can be calculated in a manifest covariant way, the
situation is more transparent in a Hamiltonian framework.
One has in first order pQED
\beq\label{psip}
\Psi_{q\bar q}^{T,L}(\alpha,p_T)=
\frac{{\rm i}Z_f\sqrt{\alpha_{em}}}{4\pi\alpha(1-\alpha)\nu}
\frac{\bar u(p_q)\vec e_{T,L}\cdot\vec{\mathbf{\alpha}}
v(p_{\bar q})}
{\frac{p_T^2+\eps^2}{2\alpha(1-\alpha)\nu}+{\rm i}0_+},
\eeq
where $u(p_q)$ and $v(p_{\bar q})$ are Dirac spinors of the quark and the
antiquark, respectively and $\vec{\mathbf\alpha}$ is the three-vector of Dirac
matrices, not to be confused with the longitudinal momentum fraction $\alpha$.
Furthermore, $\vec e$ is the polarization vector of the photon.
The energy-denominator in (\ref{psip}) depends on
 mass $M_{q\bar q}$ of the pair.
In the mixed representation, one 
can express the LC wave function in terms of the
Green function for the propagation of the pair in vacuum, integrated over time,
or over the $z$-coordinate which is the same,
\beq
G_{q\bar q}^{vac}(\vec\rho_2,z_2|\vec\rho_1,z_1)=
\int\frac{d^2p_T}{(2\pi)}
{\rm e}^{-{\rm i}\vec p_T\cdot(\vec\rho_2-\vec\rho_1)}
\exp\left({\rm i}
\left[\frac{p_T^2+\eps^2}{2\alpha(1-\alpha)\nu}+{\rm i}0_+
\right]
(z_2-z_1)\right).
\eeq
Integrating this expression over the $z$-difference, $\Delta z=z_2-z_1$, one
obviously recovers the energy-denominator,
\beq
\int_{-\infty}^0d\Delta z G_{q\bar q}^{vac}(\vec\rho_2,z_2|\vec\rho_1,z_1)=
-{\rm i}\int\frac{d^2p_T}{(2\pi)}
\frac{{\rm e}^{-{\rm i}\vec p_T\cdot(\vec\rho_2-\vec\rho_1)}}
{\frac{p_T^2+\eps^2}{2\alpha(1-\alpha)\nu}+{\rm i}0_+}
=-{\rm i}2\alpha(1-\alpha)\nu\,{\rm K}_0(\eps(\vec\rho_2-\vec\rho_1)).
\eeq
In the last step the relation
\beq
{\rm K}_0(\eps\rho)=\frac{1}{2\pi}\int d^2p_T\frac{
{\rm e}^{-{\rm i}\vec p_T\cdot\vec\rho}}
{p_T^2+\eps^2}
\eeq
was used.
Finally, the LC wavefunctions can be written as
\beq\label{psidefinition}
\Psi_{q\bar q}^{T,L}(\alpha,\vec\rho)=\frac{{\rm i}Z_f\sqrt{\alpha_{em}}}
{4\pi\alpha(1-\alpha)\nu}
\left(\bar\chi_q\widehat{\cal  O}_{q\bar q}^{T,L}\chi_{\bar q}\right)
\int_{-\infty}^0d\Delta z G_{q\bar q}^{vac}(\vec\rho,z_2|\vec 0,z_1),
\eeq
where the $\chi$ are two component spinors and the operators 
$\widehat O_{q\bar q}^{T,L}$ 
are defined by
\beqn\label{transop}
\widehat {\cal  O}_{q\bar q}^T&=&
m_f\vec\sigma\cdot\vec e+
{\rm i}(1-2\alpha)(\vec\sigma\cdot\vec n)(\vec e\cdot\vec\nabla_\rho)
+(\vec\sigma\times\vec e)\cdot\vec\nabla_\rho\\
\label{longop}
\widehat{\cal  O}_{q\bar q}^L&=&
2Q\alpha(1-\alpha)\vec\sigma\cdot\vec n.
\eeqn
The two dimensional gradient $\vec\nabla_\rho$ acts on the transverse coordinate
$\vec\rho$, $\vec n$ is the unit vector parallel to the photon momentum and
$\vec\sigma$ is
the three vector of the Pauli spin-matrices. For a Green function
$G_{q\bar q}^{vac}$ without
interaction, one obtains of course the perturbative LC wavefunctions on page
\pageref{psit}. 
To see this, one simply notes that
\beq
{\rm K}_1(z)=-\frac{d}{dz}{\rm K}_0(z).
\eeq

Introducing a quark mass which acts like a cutoff and prevents the integrals
from diverging, is obviously a poor model of the nonperturbative effects at
large $\rho$. A better solution was suggested in \cite{KST99}, where an
interaction between the quark and the antiquark was explicitly 
introduced. For this purpose, the Green function in (\ref{psidefinition}) was
modified by introducing a harmonic oscillator potential,
\beq
V_{q\bar q}(\alpha,\rho)=\frac{a^4(\alpha)\rho^2}{2\nu\alpha(1-\alpha)}.
\eeq
The Green function including the nonperturbative interaction reads
\beqn
G^{vac}_{q\bar q}\left(\vec \rho_2,z_2;\vec \rho_1,z_1\right)
\!&=&\!\frac{a^2\left(\alpha\right)}
{2\pi\sinh\left(\omega\,\Delta z\right)}\,
\exp\left[-\frac{\varepsilon^2\,\Delta z}{2\,\nu\,\alpha(1-\alpha)}\right]
\nonumber\\ \!&\times&\!
\exp\left\{-\frac{a^2\left(\alpha\right)}{2}\left[
\left(\rho_1^2+\rho_2^2\right)\coth\left(\omega\,\Delta z\right)
-\frac{2\vec\rho_1\cdot\vec\rho_2}{\sinh\left(\omega \,\Delta z\right)}
\right]\right\},
\eeqn
where 
\beq
\omega=\frac{a(\alpha)^2}{\nu\,\alpha(1-\alpha)}\ ,
\eeq
is the oscillator frequency. It was found in \cite{KST99} that data
for diffraction dissociation and the photoabsorption 
are well reproduced with
the parametrization
\beq\label{2.1a}
a^2\left(\alpha\right)=v^{1.15}(112{\rm MeV})^2
+\left(1-v\right)^{1.15}
(165{\rm MeV})^2\alpha\left(1-\alpha\right).
\eeq
The first term in this ansatz prevents that the transverse distance becomes
arbitrarily large at the endpoint $\alpha\to 0,1$. Note 
however that the more
conventional ansatz, which results from a relativistic approach to the $q\bar
q$-bound state problem \cite{Halperin} is setting the parameter $v=0$.
Numerical results are surprisingly 
insensitive to the value of the parameter $v$,
which could not be fixed in \cite{KST99}.
We will always use the value $v=0.5$.

The nonperturbative interaction changes of course the LC wavefunctions,
which read now
\beqn
\label{psitnpt}\nonumber
\left|\Psi_{T}^{npt}(\alpha,\rho)\right|^2\!\!&=&\!\!
\frac{2N_c\alpha_{em}}{(2\pi)^2}\sum\limits_{f=1}^{N_f}Z_f^2
\left\{\left[1-2\alpha(1-\alpha)\right]\Phi^2_1(\eps,\lambda,\rho)
+m_f^2\Phi^2_0(\eps,\lambda,\rho)\right\},\\
&&\\
\label{psilnpt}
\left|\Psi_{L}^{npt}(\alpha,\rho)\right|^2\!\!&=&\!\!
\frac{8N_c\alpha_{em}}{(2\pi)^2}\sum\limits_{f=1}^{N_f}Z_f^2
Q^2\alpha^2(1-\alpha)^2\Phi^2_0(\eps,\lambda,\rho),
\eeqn
with
\beqn
\Phi_{0}\left(\varepsilon,\lambda,\rho\right)
&=&\frac{1}{2}
\int\limits_0^\infty du\frac{\lambda}
{\sinh\left(\lambda u\right)}
\exp\left[
-\frac{\lambda\varepsilon^2\rho^2}{4}\coth\left(\lambda u\right)-u\right],\\
\vec\Phi_{1}\left(\varepsilon,\lambda,\rho\right)
&=&
\frac{\vec\rho}{\rho^2}
\int\limits_0^\infty du
\exp\left[
-\frac{\lambda\varepsilon^2\rho^2}{4}\coth\left(\lambda u\right)-u\right].
\eeqn
The strength of the nonperturbative interaction is described by the
dimensionless parameter
\beq\label{2.10}
\lambda=\frac{2a^2(\alpha)}{\eps^2}.
\eeq
Compared to the expression for $\Phi_{1}$ in \cite{KST99}, we
have integrated by parts over the parameter $u$. This
considerably simplifies 
the expression.

It is worthwhile to investigate the two limiting cases of vanishing interaction,
$\lambda\to 0$, and of strong interaction, $\lambda\to \infty$.
With the useful relations \cite{gr}
\beqn
{\rm K}_0(\eps\rho)&=&\frac{1}{2}
\int\limits_0^\infty \frac{du}{u}
\exp\left[
-\frac{\varepsilon^2\rho^2}{4u}-u\right],\\
\eps\rho{\rm K}_1(\eps\rho)&=&
\int\limits_0^\infty du
\exp\left[
-\frac{\varepsilon^2\rho^2}{4u}-u\right],
\eeqn
one recognizes that in the limit $\lambda\to 0$, where the nonperturbative
interaction becomes negligible, the perturbative LC wavefunctions, (\ref{psit})
and (\ref{psil}), are recovered.
In the strong interaction limit,  $\lambda\to \infty$, the functions
$\Phi_{0,1}$ again acquire simple forms \cite{KST99},
\beqn
\Phi_{0}\left(\varepsilon,\lambda,\rho\right)\Big|_{\lambda\to \infty}
&=&\frac{1}{4\pi}{\rm K}_0\left(\frac{a^2(\alpha)\rho^2}{2}\right),\\
\vec\Phi_{1}\left(\varepsilon,\lambda,\rho\right)\Big|_{\lambda\to \infty}
&=&
\frac{\vec \rho}{2\pi\rho^2}\exp\left(\frac{a^2(\alpha)\rho^2}{2}\right).
\eeqn
This limit is appropriate in the case of real photons. The nonperturbative
interaction confines even massless quarks. However, we will include current
quark masses in all our calculations.

We point out that our approach to describe the soft contributions to DIS is
quite different from parametrizing all the unknown physics into parton
distributions. Instead we develop a model, that smoothly interpolates between
the hard and the soft part of the interaction. It will turn out later that this
way of treating the soft physics allows us to develop a better physical
understanding of coherence effects in QCD.

\subsection{Diffraction}\label{diffractionsec}

Diffraction in hadron-hadron collisions has been known for a long time. A
process is called diffractive, if only the quantum numbers of the vacuum are
exchanged in the $t$-channel \cite{Goulianos,Levy}. 
The best example for such a process is elastic
scattering, $A+B\to A+B$. In the language of Regge phenomenology, the trajectory
with the quantum numbers of the vacuum is called Pomeron trajectory, because it
was first proposed by
Pomeranchuk \cite{Pomeranchuk58}.
Elastic scattering is not the only type of diffractive processes. There is also
inelastic diffraction, where one or both colliding
particles are excited into states with
the same quantum numbers as the incoming particles. Since no quantum numbers are
exchanged, the observation of a large rapidity gap between the outgoing
particles is characteristic for a diffractive event and serves as experimental
criterion of diffraction.

An intuitive picture of diffraction was proposed by Good and Walker 
\cite{GoodWalker}, introducing an analogy between diffraction and wave optics.
The beam particle can be decomposed into a coherent sum of interaction
eigenstates.
These eigenstates scatter with different amplitudes off the target and the
coherence is destroyed. New particles are produced in the same way, as 
white light is
decomposed into different colors, by sending it through a prism. The term
"diffraction" originates from this analogy.

Typical properties of total hadronic cross sections are among others, 
(i) a slow rise 
of the total and the elastic cross section with energy, 
(ii) a large imaginary part of the forward
scattering amplitude, compared to the real part
and (iii) a small elastic cross section, compared to the total cross section.
One can argue from these observations that the
theory of strong interaction must be nonabelian. 
Indeed, if only one particle exchange is considered
in an abelian theory one would have elastic scattering in the first
order of the coupling constant and therefore a large elastic cross section and a
predominantly real forward scattering amplitude \cite{Low}.

The Pomeranchuk theorem \cite{Pomeranchuk56,Okun56}
states that any scattering process 
which involves charge exchange, must vanish at asymptotically high energies.
The converse is also true. 
Foldy and Peierls \cite{FP} have proven that,
if a cross section does not vanish asymptotically, the
process must be dominated by the exchange of vacuum quantum numbers.
A simple model of the Pomeron has been suggested by Low \cite{Low} and by
Nussinov \cite{Nussinov1,Nussinov2}. The basic idea is that elastic
hadron-hadron scattering is mediated by exchange of two gluons in a color
singlet state. This model qualitatively explains the basic 
experimental observations. One obtains a purely imaginary forward scattering
amplitude and the total cross section is independent of energy. It is argued
that the observed
increase with energy, $\sigma_{tot}\propto s^{0.08}$, can be attributed to
higher order corrections.

It came as a big surprise, when large rapidity gap events were discovered in DIS
at HERA. Naively one would expect that the virtual photon hits a quark inside
the proton and produces a jet. Due to the strong color forces, hadronic activity
is expected in the whole rapidity region between the jet and the proton
remnants. However, about 10\% of all DIS events at $x_{Bj}<0.1$ show a large
rapidity gap between an almost elastically scattered proton and a diffractively
excited state. The amount of these events shows only a weak $Q^2$ dependence,
suggesting that diffraction in DIS is a leading twist. 
The energy dependence is however much stronger than in the case of hadron-hadron
scattering, approximately $\sigma^{\gamma^*p}\propto x_{Bj}^{-0.3}$, 
which lead to the assumption that there are two Pomerons, see e.\ g.\
\cite{DL,DMP}. 
The so called soft
Pomeron shows only a weak energy dependence and is responsible for the rise of
total hadronic cross sections with energy, while the hard Pomeron is observed in
the stronger energy dependence in DIS.

A popular way to think of diffraction is to imagine that the virtual
photon scatters off a preformed color neutral cluster inside the proton, the
Pomeron. 
This picture is known as Ingelman-Schlein model \cite{Ingelman}.
The space-time picture of a diffractive event is shown in fig.\
\ref{diff}, where the Pomeron is represented by two gluons. 
The hadronic fluctuation of the photon is developed long before the target.

Diffraction in DIS requires the introduction of additional kinematical
variables. 
The quantity 
\beq
x_{\cal P}=\frac{Q^2+M_X^2}{Q^2+s^2}
\eeq
may be regarded as the momentum fraction of the proton carried by the Pomeron.
Here, $M_X$ is the mass of the diffractively excited state.
One also often uses the variable
\beq
\beta=\frac{Q^2}{Q^2+M_X^2}=\frac{x_{Bj}}{x_{\cal P}}
\eeq
instead, which can be interpreted as the momentum fraction of the struck quark
relative to the Pomeron. Furthermore, if the proton is observed in the final
state, one also has to deal with the four momentum transfer squared at the
proton vertex,
\beq
t=(P-P^\prime)^2<0,
\eeq
where $P$ and $P^\prime$ are the four momenta of the incoming and outgoing
proton, respectively.

\begin{figure}[t]
  \scalebox{0.9}{\includegraphics{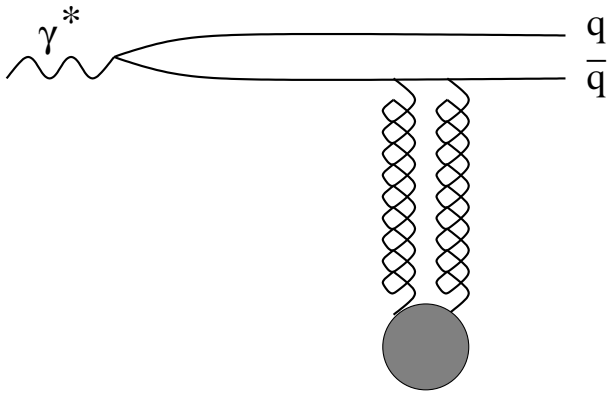}}\hfill
  \raise-0.2cm\hbox{\parbox[b]{2.4in}{
    \caption{
      \label{diff}
      Diffraction in DIS seen in the target rest frame, cf.\ fig.~\ref{lc1}. 
      Only the quantum numbers of the vacuum are exchanged in the $t$-channel.
      This can be modeled by the exchange of
      two gluons in a color singlet state. 
      The proton remains almost intact
      and a large rapidity gap is observed between the outgoing proton and the
      diffractively excited state $q\bar q$.
    }
  }
}
\end{figure}

Thinking about diffraction as DIS off the Pomeron allows to introduce parton
densities for the Pomeron and to evolve these densities with the DGLAP
evolution equations. A fit to these parton densities can be found in
\cite{Alvero1,Alvero2}. It is however only a postulate that diffractive parton
distributions evolve according to DGLAP. It is not even clear, whether these
distributions have to fulfill  a momentum sumrule. Indeed, in \cite{Alvero1} the
momentum of all partons together 
is significantly larger than 100\% and the parton
densities become even negative at small $\beta$.

The color dipole picture of diffraction does however not rely on the assumption
that the $\gamma^*$ scatters off the pomeron.  The decomposition of the virtual
photon in interaction eigenstates (\ref{expansion}) is completely independent of
this picture. Furthermore, we can not only decompose the photon, but any hadron
in such a series. However, in the photon case the LC wavefunctions, i.\ e.\ the
coefficients $c_k^{\gamma^*}$, can be calculated perturbatively at least at
small $\rho$. This is not the case for hadron wavefunctions. Proceeding like on
pp.\ \pageref{expansion}, we obtain for the mass integrated diffraction
dissociation cross section
\beqn
\int dM_X^2\frac{d\sigma^{DD}}{dM_X^2dt}\Bigg|_{t=0}
&=&\frac{\sum_k|\la\gamma^*|T|\psi_k\ra|^2}{16\pi s^2}
-\frac{d\sigma^{el}}{dt}\Big|_{t=0}\\
&=&\frac{1}{16\pi s^2}\left[\sum_k|c_{k}^{\gamma^*}|^2t_k^2-
\left(\sum_k|c_{k}^{\gamma^*}|^2t_k\right)^2\right],
\eeqn
where we have subtracted the elastic part. This part is however of order
$\alpha_{em}^2$ and will be dropped from now on.
Keeping only the $q\bar q$ Fock-component of the photon, one obtains
\beq\label{diffraction}
\int dM_X^2\frac{d\sigma^{DD}}{dM_X^2dt}\Bigg|_{t=0}
=\frac{1}{16\pi}\int_0^1d\alpha\int d^2\rho
\left|\Psi_{q\bar q}(\alpha,\rho,Q^2)\right|^2\sigma^2_{q\bar q}(\rho),
\eeq
where $\Psi$ denotes either the perturbative (\ref{psit}, \ref{psil}) or
the nonperturbative (\ref{psitnpt}, \ref{psilnpt}) LC wavefunctions of the
virtual photon. We omit the indices $T$ and $L$ from now on 
and assume summation
over all polarizations.

It is instructive to consider also off-diagonal diffraction, where the photon
goes into a final hadronic state $h$. Decomposing the hadron into interaction
eigenstates,
\beq
|h\ra=\sum_kc_k^h|\psi_k\ra,
\eeq
one obtains
\beqn
\frac{d\sigma^{DD}(\gamma^*\to h)}{dt}\Bigg|_{t=0}&=&\frac{1}{16\pi s^2}
\Big|\la h|T|\gamma^*\ra\Big|^2\\
\label{formel}
&=&\frac{1}{16\pi s^2}\Big|\sum_k\left(c_k^h\right)^*c_k^{\gamma^*}t_k\Big|^2\\
&=&\frac{1}{16\pi}\int d^2\rho\, d\alpha\Psi_h^*(\alpha,\rho)
\Psi_{q\bar q}(\alpha,\rho)\sigma^2_{q\bar q}(\rho),
\eeqn
where $\Psi_h$ is the LC wavefunction of the hadron.
It is convenient to express this result in terms of the diffractive amplitude
\beq\label{diffamp}
f(\gamma^*\to q\bar q)={\rm i}\Psi_{q\bar q}(\alpha,\rho)\sigma_{q\bar q}(\rho),
\eeq
\beq
\frac{d\sigma^{DD}(\gamma^*\to h)}{dt}\Bigg|_{t=0}=\frac{1}{16\pi}
\la f(h\to q\bar q)|f(\gamma^*\to q\bar q)\ra.
\eeq
 We see from (\ref{formel})
that the photon cannot go into an orthogonal state, if all its eigencomponents
would scatter off the target with the same amplitude $t_k=s\,\sigma_{q\bar
q}(\rho)$. Offdiagonal diffraction occurs, because the eigencomponents scatter
with different amplitudes and thus the coherence between them is disturbed. This
is exactly the picture of \cite{GoodWalker}.
We mention that the photon can 
of course also be represented as a superposition of hadronic states with the
same quantum numbers,
i.\ e.\  vector mesons, rather than interaction eigenstates. 
This is done in the vector dominance model
(VDM). A review article can be found in \cite{Shaw}.

We will now examine, like in the preceding section, from which values of $\rho$
the leading twist contribution to diffraction originates.
This was also done in \cite{boris}.
Since the LC wavefunctions are the same as in DIS, the probabilities $w$ to find
small or large dipoles does not change, when one considers diffraction.
Note however we have now $\sigma_{q\bar q}^2(\rho)$ in the
integrand (\ref{diffraction}). As a consequence, the contribution of small size
dipoles is suppressed by an extra power of $Q^2$, see tabs.\
\ref{difftabT} and \ref{difftabL}. For transverse photons, the leading twist
originates purely from soft contributions, i.\ e.\ from large dipole sizes,
making diffraction a phenomenon dominated by nonperturbative physics, in spite
of the large $Q^2$.
Furthermore, we have seen 
that in the case of longitudinal photons, the only scaling
contribution in DIS comes from small size dipoles, see tab.\ \ref{distabL}. 
This is no longer the case in diffraction and the diffractive cross section for
longitudinal photons seems to fall away as $\propto 1/Q^4$.
This is however not the case, it will be shown later, that 
the $q\bar
qG$ Fock state gives a contribution $\propto 1/Q^2$.
 
\begin{table}[t]
\begin{center}
\renewcommand{\arraystretch}{2.3}
\begin{tabular}{|c|c|c||c|}
\hline
\vphantom{XXXXXX}\hphantom{XXXX} T \hphantom{XXXX}
&\hphantom{XXXX} $w$ \hphantom{XXXX}
&\hphantom{XXXX} $\overline{\sigma^2}\hphantom{XXXX}$ 
&\hphantom{XXXX} $\la\sigma_{q\bar q}^2\ra $\hphantom{XXXX}\\
\hline
hard & 1 & $\displaystyle\frac{1}{Q^4}$ &  
$\displaystyle\propto\,\frac{1}{Q^4} $ \\
\hline
soft & $\displaystyle\frac{m_f^2}{Q^2}$ & 
$\displaystyle\sigma_h^2$ &  $\displaystyle\propto\,\frac{1}{Q^2} $ \\
\hline
\end{tabular}
\caption{\label{difftabT}
The $Q^2$ behavior of hard and soft contributions to the diffractive cross
section in DIS for transverse photons, cf.\ tab.~\ref{distabT}.
 The probability that the photon converts into a small (hard)  or
large (soft)  
dipole is denoted by $w$, $\overline{\sigma^2}$ is the order of 
magnitude of the
dipole cross section squared 
and the last column contains the $Q^2$-behavior of the
diffractive cross section. $\sigma_h$ is a typical hadronic cross
section, i.e.\ independent of $Q^2$.
The probabilities for finding small or large dipoles is the same as in
DIS. The contribution from small dipoles is higher twist in diffraction, the
leading twist originates from large, nonperturbative dipoles.}
\end{center}
\end{table}
\begin{table}[ht]
\begin{center}
\renewcommand{\arraystretch}{2.3}
\begin{tabular}{|c|c|c||c|}
\hline
\vphantom{XXXXXX}\hphantom{XXXX} L \hphantom{XXXX}
&\hphantom{XXXX} $w$ \hphantom{XXXX}
&\hphantom{XXXX} $\overline{\sigma^2}\hphantom{XXXX}$ 
&\hphantom{XXXX} $\la\sigma_{q\bar q}^2\ra$\hphantom{XXXX}\\
\hline
hard & 1 & $\displaystyle\frac{1}{Q^4}$ &  
$\displaystyle\propto\,\frac{1}{Q^4} $ \\
\hline
soft & $\displaystyle\left(\frac{m_f^2}{Q^2}\right)^2$ & 
$\displaystyle\sigma_h^2$ &  $\displaystyle\propto\,\frac{1}{Q^4} $ \\
\hline
\end{tabular}
\caption{\label{difftabL} 
The same as tab.~\ref{difftabT} but for longitudinal photons.
In the longitudinal case both, hard and
soft, contributions to the diffraction dissociation cross section are higher
twist. The leading twist contribution comes from the $q\bar qG$ Fock state of
the photon, see fig.~\ref{lc2}.} 
\renewcommand{\arraystretch}{1}
\end{center}
\end{table}

Consider the $M_X^2$ dependence of diffraction
for large $M_X^2$. First one notices that a
$q\bar q$-pair with fixed transverse separation 
does not
have a well defined mass,
because the transverse momenta are completely undetermined in coordinate space
representation. In momentum space representation one has
\beq
M_X^2=\frac{p_T^2+m_f^2}{\alpha(1-\alpha)}.
\eeq
We have seen above that the diffractive cross section for transverse photons is
dominated by 
contributions from the endpoints, $\alpha\to 0$ and $\alpha\to 1$.
These are
large size fluctuations, which have small transverse momenta.
The mass is therefore approximately $M_X^2\approx m_f^2/\alpha$ and thus
\beq
\frac{dM_X^2}{M_X^2}\approx-\frac{d\alpha}{\alpha}.
\eeq
In the large mass limit, one can write \cite{ForshawRoss} 
\beq
\frac{d\sigma^{DD}}{dM_X^2dt}\Big|_{t=0}\propto 
\frac{m_f^2}{M_X^4}\int d^2\rho |\Psi_{q\bar q}^T(\alpha,\rho)|^2
\sigma_{q\bar q}^2(\rho).
\eeq
Here we consider the limit $\alpha\to 0$. The other endpoint gives the same
contribution. Since large dipoles interact with a typical hadronic cross section
$\sigma_h$, one sees that the diffractive mass spectrum decays as $\propto
1/M_X^4$. 
This behavior is a result of the small phase space for the aligned jet
configurations
and is at variance with the experimental observation that the
mass spectrum decays only as $\propto 1/M_X^2$.

The solution to this problem is that one of the quarks can radiate a gluon
before the impact on the target, fig.\ \ref{lc2}. Then the multi parton
configuration, frozen in impact parameter space scatters off the target.
Although the gluon is radiated on the cost of an additional power in the strong
coupling constant $\alpha_s$, the spin-$1$ nature of the gluon leads to a much
weaker decay of the diffractive mass spectrum and the longitudinal cross section
for diffraction gets a leading twist contribution.
The LC-wavefunction for the transition $\gamma^*\to q\bar qG$ reads in the small
recoil approximation, $\alpha_G\to 0$,
\cite{KST99} 
\beq
\Psi_{q\bar qG}(\alpha,\alpha_G,\vec\rho_1,\vec\rho_2)=
\Psi_{q\bar q}(\alpha,\vec\rho_1-\vec\rho_2)
\left[\Psi_{qG}\left(\frac{\alpha_G}{\alpha},\vec\rho_1\right)
-\Psi_{\bar qG}\left(\frac{\alpha_G}{1-\alpha},\vec\rho_2\right)
\right],
\eeq
where $\Psi_{qG}$ is the LC wavefunction for the transition $q\to qG$. 
Here, $\alpha$ is the longitudinal momentum fraction of the photon carried by
the quark and $\alpha_G$ is the momentum fraction of the gluon with respect to
the photon. Furthermore, $\rho_1$ is the distance between the quark and the
gluon and $\rho_2$ the distance between the antiquark and the gluon.
The first
term in the square brackets corresponds to radiation of the gluon from the
quark, while in the second term the gluon is radiated from the antiquark. In
analogy to (\ref{psidefinition}), $\Psi_{qG}$ is defined as 
\beq\label{psiqGdefinition}
\Psi_{qG}(\alpha,\vec\rho)=\frac{{\rm i}\sqrt{\alpha_{s}/3}}
{2\pi\alpha(1-\alpha)\nu}
\left(\bar\chi_q \widehat\Gamma \chi_{q}\right)
\int_{-\infty}^0d(\Delta z)\, G_{qG}^{vac}(\vec\rho,z_2|\vec 0,z_1),
\eeq
where the operator
\beq\label{gluonop}
\widehat {\cal  O}_{qG}=
{\rm i}m_f\alpha^2\vec e\cdot(\vec n\times\vec\sigma)-
{\rm i}(2-\alpha)(\vec e\cdot\vec\nabla_\rho)
+\alpha\vec e\cdot(\vec\sigma\times\vec\nabla_\rho)
\eeq
comes again from the spinor structure of the vertex. 
Without a potential between the quark and the gluon, the integral over the Green
function reads
\beq
\int_{-\infty}^0d\Delta z G_{qG}^{vac}(\vec\rho,z_2|\vec 0,z_1)
={\rm -i}2\nu\alpha(1-\alpha){\rm K}_0(\tau\rho),
\eeq
where 
\beq
\tau^2=\alpha^2m_f^2
\eeq
plays now the role of the extension parameter. The LC wavefunction reads
explicitly
\beqn\nonumber
\Psi_{qG}(\alpha,\vec\rho_1)\Psi_{qG}(\alpha,\vec\rho_2)&=&
\frac{16\alpha_s}{3}\Bigg(\left[1+(1-\alpha)^2\right]
\tau^2\frac{\vec\rho_1\cdot\vec\rho_2}{\rho_1\rho_2}
{\rm K}_1(\tau\rho_1){\rm K}_1(\tau\rho_2)\\
&+&
m_f^2\alpha^4{\rm K}_0(\tau\rho_1){\rm K}_0(\tau\rho_2)\Bigg).
\eeqn

\begin{figure}[t]
  \scalebox{0.9}{\includegraphics{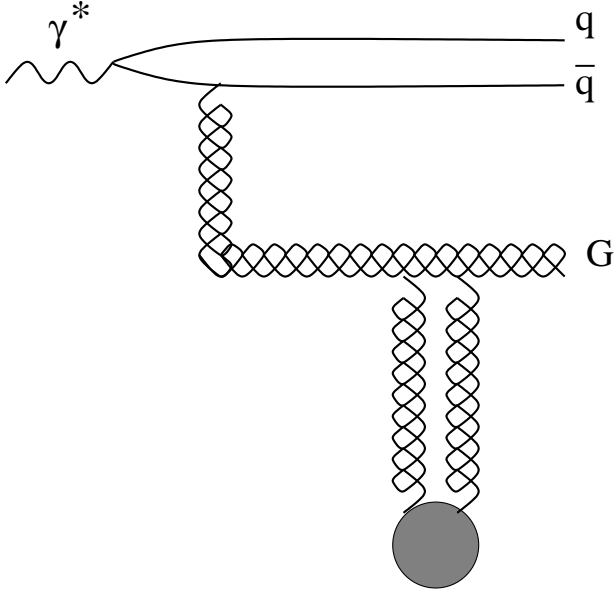}}\hfill
  \raise4.5cm\hbox{\parbox[b]{2.4in}{
    \caption{
      \label{lc2}
      The same as fig.~\ref{diff} but now for the $q\bar qG$ Fock-component of
      the virtual photon.
       For longitudinal photons, a
      leading twist con\-tri\-bution to dif\-fraction  arises
      from configurations, where the gluon is soft and far away from the $q\bar
      q$-pair.
    }
  }
}
\end{figure}

The contribution of the
$q\bar q G$ Fock state to the diffractive cross section is given by
\beq\label{qqG}
\frac{d\sigma^{DD}_{q\bar qG}}{dt}\Bigg|_{t=0}
=\frac{1}{16\pi}\int_0^1d\alpha\int \frac{d\alpha_G}{\alpha_G}
\int d^2\rho_1 d^2\rho_2
\left|\Psi_{qqG}(\alpha,\alpha_G,\vec\rho_1,\vec\rho_2)\right|^2
\sigma^2_{q\bar qG}(\vec\rho_1,\vec\rho_2).
\eeq
The cross section for scattering the $q\bar q G$ state off the proton can be
expressed in terms of the dipole cross section \cite{qqG1,qqG2},
\beq\label{sigmaqqG}
\sigma_{q\bar qG}(\vec\rho_1,\vec\rho_2)
=\frac{9}{8}\left[\sigma_{q\bar q}(\rho_1)+\sigma_{q\bar q}(\rho_2)
-\sigma_{q\bar q}(\vec\rho_1-\vec\rho_2)\right].
\eeq
The solution to the problems mentioned above lies in the integral over
$d\alpha_G/\alpha_G$. The singular behavior at $\alpha_G\to 0$ is
characteristic for the radiation of vector bosons. Note the phase
space in $\alpha_G$ is no longer suppressed as $\alpha_G\to 0$. Therefore, the
gluon can go far away from the quark and the $q\bar qG$ system scatters with a
large cross section. This makes diffraction for longitudinal photons a leading
twist effect and also leads to the asymptotic behavior
\beq
\frac{d\sigma^{DD}}{dM_X^2dt}\Bigg|_{t=0}\propto\frac{1}{M_X^2}
\eeq
for $M_X^2\gg Q^2$, as observed in experiment.
In the language of Regge theory \cite{PDBCollins}, the $q\bar qG$ component
corresponds to the triple Pomeron vertex, which dominates the large mass
behavior. In the language of pQCD, the additional gluon gives us the evolution
of the gluon density. We explained in section \ref{dissec}
why it is necessary in DIS
to take at least one gluon-rung
in the
ladder in fig.\ \ref{dis2}b into account to obtain the gluon splitting function
$P_{GG}$.

There are two interesting limiting cases of (\ref{sigmaqqG}). 
First, the separation of the
$q\bar q$ dipole is very small, of order $1/Q^2$. This 
situation is typical for
longitudinal photons. Then we obtain
\beq
\sigma_{q\bar qG}(\vec\rho_1=\vec\rho_2)=
\frac{9}{4}\sigma_{q\bar q}(\rho)=\sigma_{GG}(\rho),
\eeq
the cross section reduces to the cross section for an octet-octet, or
gluon-gluon, dipole.
Indeed, after radiation of the gluon, the $q\bar q$ dipole is with a high
probability in an octet state, as one can see from color algebra. 
If the size of this dipole is small, it appears
like a gluon and we have an effective gluon-gluon dipole.
 The factor $9/4$ is the ratio of the two Casimir operators
of the adjoint and the fundamental representation of SU(3).
In the opposite limit, where the gluon stays close to one of the quarks and the
$q\bar q$ dipole has a large separation, one obtains
\beq
\sigma_{q\bar qG}(\vec\rho_1,\vec\rho_2=\vec 0)= 
\sigma_{q\bar q}(\rho_1).
\eeq
This limit corresponds to low virtualities $Q^2$.

In the last section, it was demonstrated that the quark mass acts like an effective cutoff,
rendering all integrals finite. Also in the case of the $q\bar qG$ Fock state,
the integrals would diverge, as the gluon gets infinitely far from the quark it
has been radiated from. We do not want to introduce a gluon mass, instead we
follow \cite{KST99}, where 
an interaction between the quark and the gluon
was proposed, which is again described
by a harmonic oscillator potential,
\beq
V_{qG}(\alpha,\rho)=\frac{b^4(\alpha)\rho^2}{2\nu\alpha(1-\alpha)}.
\eeq
The ansatz for the parameter that describes the strength of the interaction is
chosen like in the $q\bar q$ case,
\beq
b^2(\alpha)=b_0^2+4b_1^2\alpha(1-\alpha).
\eeq
However, since the dominant contributions come from the region $\alpha_G\to 0$,
only 
\beq
b(0)=b_0=650{\rm MeV},
\eeq
was fixed in \cite{KST99} with help of a triple pomeron analysis of CDF data
\cite{CDF}.
For this reason, we will always use the limit
\beq\label{psigluon}
\Psi_{qG}^{npt}(\vec\rho)=\lim_{\alpha_G\to 0}\Psi_{qG}^{npt}(\alpha,\vec\rho)
=\frac{2}{\pi}\sqrt{\frac{\alpha_s}{3}}\frac{\vec e\cdot\vec\rho}{\rho^2}
\exp\left(-\frac{b_0^2\rho^2}{2}\right)
\eeq
in our calculations. We point out, that the qualitative corrections due to the
additional gluon are not a result of the ansatz for the nonperturbative
interaction, but are purely due to the spin $1$ nature of the gluon.


It is worth noting that the relatively large value of $b_0$ does not allow that
the distance between the gluon and the quark becomes as large as $1$ fm. The
maximum distance is rather about $0.3$ fm, considerably smaller than a typical
confinement radius.  
There are several phenomenological indications that this must be the case.
The small size of the gluonic fluctuation is related to the high
octet-octet string tension, $\kappa_{[8]}\approx 4$ GeV$/$fm, compared to the
triplet string tension, $\kappa_{[3]}\approx 1$ GeV$/$fm. These string tensions
are related to the slopes of the pomeron and reggeon trajectories respectively
\cite{PDBCollins}. If one thinks of diffraction in proton-proton scattering as
proton-pomeron scattering, one can extract the proton-pomeron total cross
section from CDF data. This cross section is about one order of magnitude
smaller than the total $pp$ cross section. The only explanation is that the
pomeron is a small size object compared to the proton \cite{sizes}. 
The conjecture that the pomeron has a rather small size is also supported by the
approximately fulfilled quark counting rules \cite{Gunion}. Indeed, the pomeron
seems to couple to the single valence quarks inside a hadron. A large pomeron
would couple to the whole hadron at once. Finally, we mention that a rather
short gluon correlation length, about $0.3$ fm, 
also results from lattice calculations
\cite{lattice}. These lattice calculations are used to determine the values of
the free parameters in the stochastic vacuum model of Dosch and Simonov
\cite{SVM,SVM2}.  Calculations within the SVM for soft diffraction are in good
agreement with experimental data.

\subsection{The parton model of the DY process}\label{dysec}

The concepts of the parton model, originally invented for DIS, can also be
applied to certain processes in hadron-hadron collisions. The most prominent
example for this is the Drell-Yan process \cite{DrellYan}, where vector bosons
are created in hadronic collisions. We will consider only photons. 
The model of Drell and Yan is depicted in fig.\ \ref{dyfig}. 
Two hadrons collide and a quark from one
      hadron annihilates with an antiquark from the other hadron into a timelike
      photon. This photon decays into a lepton pair that can
      be detected.

\begin{figure}[ht]
  \centerline{\scalebox{0.8}{\includegraphics{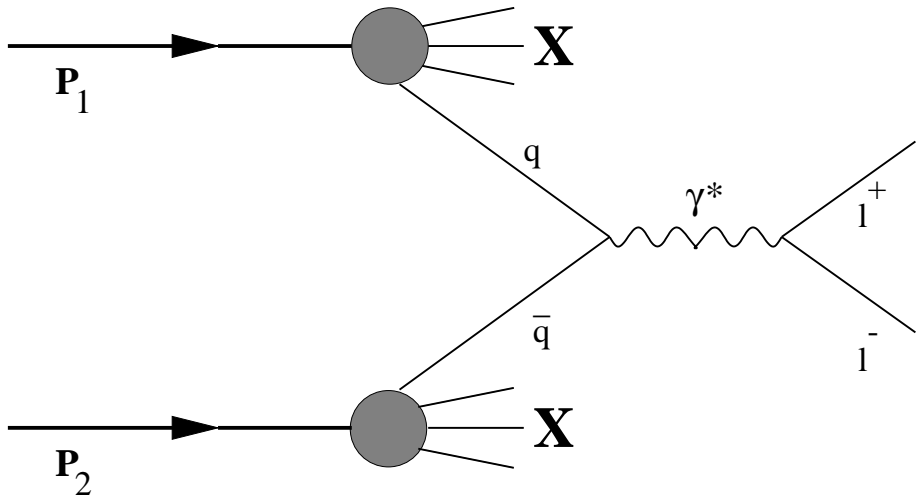}}}
    \caption{
      \label{dyfig}
      The DY process to leading order. Two hadrons collide and a quark from one
      hadron annihilates with an antiquark from the other hadron into a timelike
      photon with mass $M^2>0$. This photon decays into a lepton pair that can
      be detected.
    }  
\end{figure}

We denote the mass of the spacelike photon by
\beq
M^2=q^2>0,
\eeq
where $q^{\mu}$ is the four momentum of the virtual photon.
The square of the center of mass energy of the colliding hadrons is 
\beq
S=(P_1+P_2)^2,
\eeq
where $P_1$ and $P_2$ are the four momenta of hadron $1$ and hadron $2$,
respectively. A
convenient variable to work with is Feynman $x_F$, 
\beq
x_F=\frac{2p_L^{cm}}{\sqrt{S}}\approx x_1-x_2.
\eeq
Here, $p_L^{cm}$ is the longitudinal momentum of the dilepton in the 
hadron-hadron center of
mass frame and $x_1$ and $x_2$ are given by
\beq
x_1=\frac{2P_2\cdot q}{S}\quad,\quad x_2=\frac{2P_1\cdot q}{S}.
\eeq
These variables have the meaning of the longitudinal
momentum fractions of the quarks
participating in the hard process. The quark $q$ in fig.\ \ref{dyfig} has
longitudinal momentum $x_1P_1$ and the antiquark $x_2P_2$. 
It holds
\beq\label{scaling}
x_1x_2=\frac{M^2}{S},
\eeq
where the transverse momentum 
of the virtual photon is zero in the naive parton
model and has been neglected.
Another frequently used variable is
\beq
\tau=\frac{M^2}{S}.
\eeq

Similar structure functions as in DIS can be introduced for the DY process,
starting directly from the hadronic tensor,
\beq
W^{\mu\nu}=\int d^4x{\rm e}^{{\rm i}qx}\la P_1P_2|J^{\mu}(x)
J^{\nu}(0)|P_1P_2\ra,
\eeq
as we did in section \ref{f2sec}.
Since there are two hadrons involved in DY, there are more possible Lorentz
structures than in DIS. One can define four independent structure functions for
DY, rather than two in DIS. We do not pursue this approach in detail and refer
only to the literature \cite{LT}.

The partonic cross section for fig.\ \ref{dyfig} reads
\beq\label{dycross}
\frac{d\widehat\sigma}{dM^2}=\frac{4\pi\alpha^2_{em}Z_f^2}{3N_cM^2}
\delta(x_1x_2S-M^2).
\eeq
The factor $N_c$, number of colors, appears in the denominator,
because quark and antiquark must have the same color in order to annihilate.
Embedding the partonic cross section into the hadronic environment yields 
\beq
\frac{d\sigma}{dM^2}=\int_0^1dx_1dx_2\sum_f
\left\{q_f(x_1)\bar q_{f}(x_2)+(1\leftrightarrow 2)\right\}
\frac{d\widehat\sigma}{dM^2},
\eeq
where $q_f(x_1)$ is the probability to find a quark of flavor $f$ with 
longitudinal momentum
fraction $x_1$ in hadron $a$ and $\bar q_f$ is the analog for antiquarks.
The $\delta$-function in (\ref{dycross})
allows to write the  cross section in the scaling form
\beq\label{dylo}
M^2\frac{d\sigma}{d\tau}=\frac{4\pi\alpha^2_{em}}{3N_c}
\int_0^1dx_1\sum_fZ_f^2
\left\{q_f(x_1)\bar q_{f}(\tau/x_1)+(1\leftrightarrow 2)\right\}.
\eeq
The r.\ h.\ s.\ of (\ref{dylo}) depends only on $\tau$ and not 
separately on $M^2$ and $S$.

The observation of this
scaling property in experiment, see e.\ g.\  \cite{Ito,Moreno},
confirms that the mechanism illustrated in fig.\
\ref{dyfig} is correct. There are however features of dilepton production which
cannot be understood in the lowest order picture. 
\begin{itemize}
\item{Cross sections calculated straightforward from (\ref{dylo}) are too small
by a factor of $2-3$ compared to the measured value. This discrepancy is usually
treated by introducing a so called $K$-factor. The $K$-factor is approximately
independent of $M^2$, but it is process dependent.}
\item{Large transverse photon momenta, of order few GeV, are observed in
experiment. There are however no transverse momenta in the naive parton model.
Phenomenologically, one can introduce a primordial momentum distribution of the
quarks. One usually assumes a gaussian shape for this distribution, but the
width necessary to describe what is observed in experiment is much larger than
what one would expect from Fermi motion.}
\end{itemize}

These problems can be overcome by taking into account the first order QCD
correction, shown in fig.\ \ref{dyho}. The first row contains the virtual
corrections to the quark propagator and to the vertex. The second row shows the
so called annihilation process, where the quark or the antiquark radiates a
gluon before it annihilates with a parton from the other hadron. Due to the
radiation of the gluon, the quark acquires a transverse momentum. 
In this way, the
pQCD correction provides the missing mechanism for the production of lepton
pairs with large transverse momentum $p_T$. 
This explanation was suggested first in
\cite{fritzsch,transal,transal2}. 
The last row displays the diagrams for the QCD Compton
process, where a quark in one hadron picks up a gluon from the other hadron and
radiates a photon. This mechanism is dominant at large $p_T$ \cite{ApQCD}.

\begin{figure}[t]
  \scalebox{0.68}{\includegraphics*{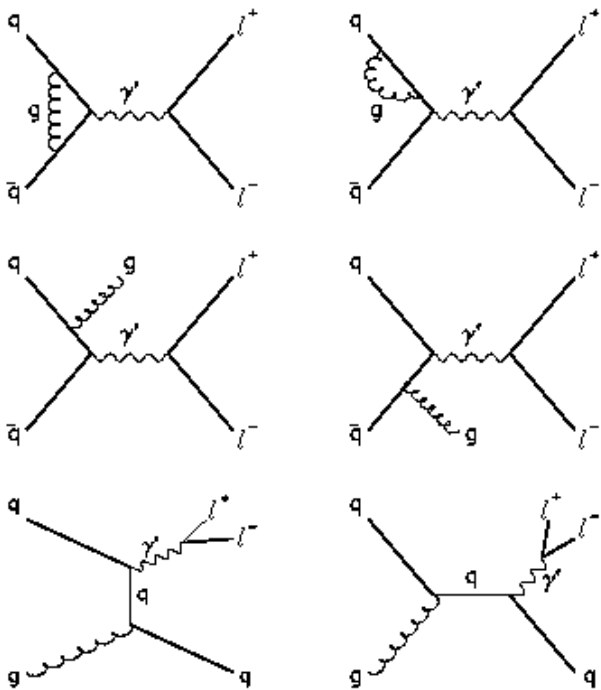}}\hfill
  \raise0.cm\hbox{\parbox[b]{2.4in}{
    \caption{
      \label{dyho}
      Higher order QCD corrections to the DY process. The upper row contains
      virtual corrections. The diagrams for the annihilation process are shown
      in the middle, where the quark or the antiquark can radiate a gluon
      before the annihilation process. The QCD Compton process is depicted in
      the last row. Here, the projectile quark scatters off a gluon from the
      target and radiates a massive photon. The Compton process is the dominant
      contribution. 
      These higher order corrections account for most of the K-factor and
      explain the occurance of large transverse momenta. The figure is taken
      from \cite{dyreport}.
    }
  }
}
\end{figure}

The graphs in fig.\ \ref{dyho} contain of course divergencies. While the
infrared divergencies cancel in the sum of virtual and real corrections, the
collinear divergencies, which occur e.\ g.\ 
when the transverse momentum $p_T$ of the intermediate quark 
in the QCD Compton process
goes to $0$, turn out to be identical to the divergencies observed in DIS, cf.\
section \ref{dissec}. Thus they can be absorbed into a redefinition of parton
densities. 
For example, the QCD Compton process gives the additional contribution
\cite{transal}
\beqn\nonumber\label{Compton}
M^2\frac{d\sigma}{d\tau}\Bigg|_{C}&=&\frac{4\pi\alpha^2_{em}}{3N_c}
\int_0^1\frac{dx_1}{x_1}\int_0^1\frac{dx_2}{x_2}\sum_fZ_f^2
\left\{(q_f(x_1)+q_{\bar f}(x_1))G(x_2)+(1\leftrightarrow 2)\right\}\\
&\times&
\Theta(x_1x_2-\tau)\frac{\alpha_s}{2\pi}P_{fG}\left(\frac{\tau}{x_1x_2}\right)
\ln\left(\frac{M^2}{\kappa^2}\right)+\dots,
\eeqn
where $\Theta$ is the step function and $P_{fG}$ is
given by (\ref{splitting}). It was first noticed in
\cite{transal}, that the whole DY cross section, including (\ref{Compton}) can
be recasted in the form (\ref{dylo}) by redefining the quark density,
\beq
q_f(x)\to q_f(x)+\frac{\alpha_s}{2\pi}\ln\left(\frac{M^2}{\kappa^2}\right)
\int_x^1\frac{dy}{y}P_{fG}\left(\frac{x}{y}\right)G(y).
\eeq
This is exactly what was done for DIS in section \ref{dissec}.
Thus the redefined parton densities in DY obey the same renormalization
group equation as the DIS parton densities, namely the DGLAP equations
(\ref{DGLAP}).

The first order correction solves most of the problems of the naive parton
model. It explains how large $p_T$-dileptons are produced and account for almost
all of the $K$-factor \cite{dyreport}. 
However, not all problems are solved by this correction.
Since it is numerically large, one has to investigate how much higher order
corrections change the result. Furthermore, the transverse momentum spectrum
is even qualitatively not well described. The theoretical result agrees only at
$p_T^2\sim M^2$ with the data and even diverges at $p_T\to 0$, 
\beq
\frac{d\sigma}{dp_T^2}\sim\frac{\alpha_s(p_T^2)}{p_T^4}
\eeq
while the experimental result is of course finite.
The reason for this behavior is that 
large logarithms
$\ln({M^2}/{p_T^2})$
occur in higher order corrections
and one has to resum all these terms.
This is possible within pQCD \cite{resumal,resumcol,Pirner}, by a resummation of
soft gluons radiated from the quark or the antiquark, respectively. The result
indicates that one needs almost no intrinsic transverse momentum and practically
all $p_T$ is generated perturbatively \cite{AEM}. 

\subsection{Light-cone approach to DY}\label{dipoledy}

Although cross sections are Lorentz invariant, the partonic interpretation of
the microscopic process depends on the reference frame. We have seen this in
section \ref{dipoledis} for the case of DIS and similar considerations can be
made for DY. It was pointed out by Kopeliovich \cite{boris} that in the target
rest frame, DY dilepton production looks like bremsstrahlung, rather than parton
annihilation. The space-time picture of the DY process in the target rest frame
is
illustrated in  
fig.\ \ref{bremsdy}. A quark (or an antiquark)
from the projectile hadron radiates a virtual
photon on impact on the target. The radiation can occur before or after
the quark scatters off the target. Only the latter case is shown in fig.\
\ref{bremsdy}. 

The cross section for radiation of a virtual photon from a quark after
scattering on a proton, can be written in factorized light-cone form
\cite{boris,bhq,kst}, 
\beq\label{dylctotal}
\frac{d\sigma(qp\to q\gamma^*p)}{d\ln\alpha}
=\int d^2\rho\, |\Psi^{T,L}_{\gamma^* q}(\alpha,\rho)|^2
    \sigma_{q\bar q}(\alpha\rho),
\eeq
similar to the DIS case (\ref{stl}).
Here $\alpha$ is the longitudinal momentum fraction of the quark, carried away
by the photon. 
The LC wavefunctions for DY can be written in the same form as in DIS
(\ref{psidefinition}),
\beq\label{psidydefinition}
\Psi^{T,L}_{\gamma^* q}(\alpha,\vec\rho)=\frac{\sqrt{\alpha_{em}}}
{2\pi}
\left(\bar\chi_q\widehat {\cal  O}_{\gamma^*q}^{T,L}\chi_{q}\right)
{\rm K}_0(\eta\rho).
\eeq
Summation over quark helicities and polarizations of the photon is understood in
(\ref{psidefinition}).
Here, the extension parameter
\beq\label{dyextension}
\eta^2 = m_f^2 \alpha^2 + M^2 \left(1-\alpha\right). 
\eeq
is the anolog of the extension parameter $\eps$ (\ref{disextension}) in DIS.
As usual, the index $T$
stands for transverse photons and $L$ for longitudinal. 
The $\chi$ are again two component spinors, but
the operators $\widehat O_{\gamma^*q}^{T,L}$ 
read now \cite{kst}
\beqn\label{transdyop}
\widehat {\cal  O}_{\gamma^*q}^T&=&
{\rm i}m_f\alpha^2\vec e\cdot(\vec n\times\vec\sigma)-
{\rm i}(2-\alpha)(\vec e\cdot\vec\nabla_\rho)
+\alpha\vec e\cdot(\vec\sigma\times\vec\nabla_\rho),\\
\label{longdyop}
\widehat{\cal  O}_{\gamma^*q}^L&=&
2M(1-\alpha).
\eeqn
The two dimensional gradient $\vec\nabla_\rho$ acts on the transverse coordinate
$\vec\rho$, $\vec n$ is the unit vector parallel to the momentum 
of the projectile quark and
$\vec\sigma$ is
the three vector of the Pauli spin-matrices.
The LC wavefunctions for the transition $q\to \gamma^* q$
read explicitly for a given flavor of unit charge
\beqn
|\Psi_{\gamma^* q}(\alpha,\rho)|^2   &=&
    |\Psi^T_{\gamma^* q}(\alpha,\rho)|^2
  + |\Psi^L_{\gamma^* q}(\alpha,\rho)|^2, \\
\label{dylct}|\Psi^T_{\gamma^* q}(\alpha,\rho)|^2 &=&
  \frac{\alpha_{em}}{\pi^2}\left\{
     m_f^2 \alpha^4 {\rm K}_0^2 \left(\eta\rho\right)
   + \left[1+\left(1-\alpha\right)^2\right]\eta^2
     {\rm K}_1^2 \left(\eta\rho\right)\right\}, \\
\label{dylcl}|\Psi^L_{\gamma^* q}(\alpha,\rho)|^2 &=&
  \frac{2\alpha_{em}}{\pi^2}
  M^2 \left(1-\alpha\right)^2 {\rm K}_0^2 \left(\eta\rho\right).
\eeqn
Comparing
(\ref{dylct}) and (\ref{dylcl}) with their DIS counterparts (\ref{psit}) and
(\ref{psil}) shows that the factor $N_c$ is no longer present 
in the $q\to \gamma^* q$ LC wavefunctions. This corresponds to the $N_c$ in the
denominator of (\ref{dycross}). Furthermore, we see that 
$|\Psi^T_{\gamma^* q}(\alpha,\rho)|^2$
 has an extra factor of $2$, because in the
DY process, one has to sum over the transverse polarizations
of the photon, rather than
average like in DIS.

For embedding the partonic cross section (\ref{dylctotal}) into the hadronic
environment, one has to note that the photon carries away the momentum fraction
$x_1=(\sqrt{x_F^2+4\tau}+x_F)/2$ from the projectile hadron. The hadronic cross
section reads then 
\beqn\label{dylctotalhadr}\nonumber
\frac{d\sigma}{dM^2dx_F}&=&\frac{\alpha_{em}}{6\pi M^2}
\frac{x_1}{x_1+x_2}\int_{x_1}^1\frac{d\alpha}{\alpha^2}
\sum_fZ_f^2\left\{q_f\left(\frac{x_1}{\alpha}\right)+
q_{\bar f}\left(\frac{x_1}{\alpha}\right)\right\}
\frac{d\sigma(qp\to q\gamma^*p)}{d\ln\alpha}\\
\\
&=&\frac{\alpha_{em}}{6\pi M^2}
\frac{1}{x_1+x_2}\int_{x_1}^1\frac{d\alpha}{\alpha}
F_2^p\left(\frac{x_1}{\alpha}\right)
\frac{d\sigma(qp\to q\gamma^*p)}{d\ln\alpha},
\eeqn
where the factor $\alpha_{em}/(6\pi M^2)$ accounts for the decay of the photon
into the lepton pair.
Remarkably, the parton density of the projectile enters just in the combination
$F_2^p$ (\ref{master}), which is the structure function of the proton. 
Therefore we did not include the fractional quark charge
$Z_f$ in the DY wavefunctions (\ref{dylct}, \ref{dylcl}). 
The structure function
$F_2^p$ is needed at large values of $x_{Bj}$. We will employ the
parametrization from \cite{Mils} in our calculations, unless mentioned
differently.

An interesting feature of the light-cone approach is the appearance 
of the dipole
cross section in (\ref{dylctotal}), although there is no physical $q\bar
q$-dipole in fig.\ \ref{bremsdy}. 
The explanation is illustrated in fig.\ \ref{dygraphs}.
The antiquark enters, when one takes the complex conjugate of the amplitude. The
dipole cross section 
appears, because the quark is displaced in the impact parameter plane after
radiation of the photon. If $\rho$ is the transverse separation between the
quark and the photon, the $\gamma^*q$ fluctuation has a center of gravity in the
transverse plane which coincides with the impact parameter of the parent quark.
The transverse separation between the photon and the center of gravity is
$(1-\alpha)\rho$ and the distance between the quark and the center of gravity is
correspondingly $\alpha\rho$. A displacement in coordinate space corresponds to
a phase factor in momentum space. The two graphs (a) and (b) in fig.\
\ref{dygraphs} have the relative phase factor 
$\exp({\rm i}\alpha\vec\rho\cdot\vec p_T)$, which produces the color screening
factor $[1-\exp({\rm i}\alpha\vec\rho\cdot\vec p_T)]$ in the dipole cross
section (\ref{borisdipole}).
Therefore, the argument of the dipole cross section is $\alpha\rho$. 

\begin{figure}[t]
  \scalebox{0.6}{\includegraphics*{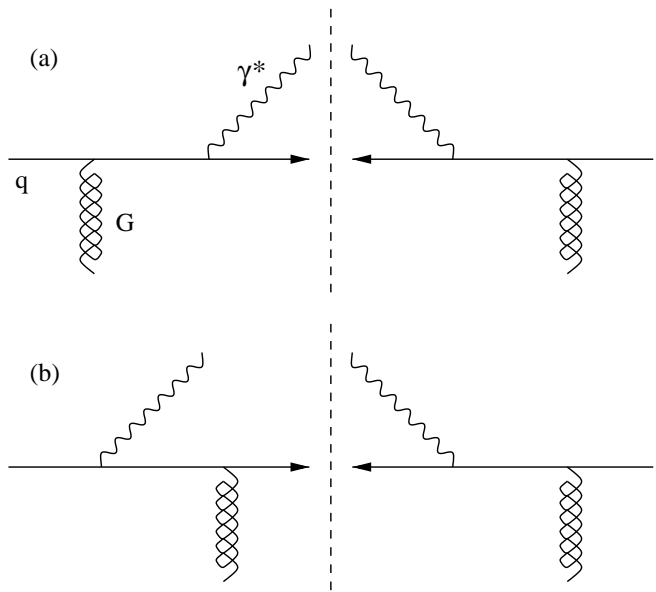}}\hfill
  \raise0.5cm\hbox{\parbox[b]{2.4in}{
    \caption{
      \label{dygraphs}
      Two out of four contributions for the DY cross section. In the first row,
      the photon is radiated on both sides of the cut
      after the quark gluon vertex. 
      The antiquark enters through the
      complex conjugate of the amplitude
      on the r.h.s.\ of the cut. The second row shows the interference term.
      The graphs in (a) and (b) have a relative phase factor $\exp({\rm
      i}\alpha\vec\rho\cdot\vec p_T)$, where $p_T$ is the transverse momentum of
      the photon. This gives 
      rise to the color screening factor
      in the dipole cross section 
      (\ref{borisdipole}) as explained in the text.
    }
  }
}
\end{figure}

The physical interpretation of (\ref{dylctotal}) is the same as for the DIS
case, cf.\ section \ref{dipoledis}. The projectile quark is expanded in the
interaction eigenstates (\ref{expansion}). We keep only the first eigenstate,
\beq
|q\ra=\sqrt{Z_2}|q_{bare}\ra+c^q_{\gamma^*}|q\gamma^*\ra+\dots,
\eeq
where, $Z_2$ is the wavefunction renormalization constant for fermions.
Different eigenstates scatter with different amplitudes off the target. This
disturbs the coherence between the eigencomponents and the photon is freed.
Since there is no real dipole in DY, we will not include a nonperturbative
interaction, as it was done 
for DIS in section \ref{dipoledis}, but always use the
perturbative LC wavefunctions (\ref{dylct}, \ref{dylcl}).

The light-cone approach to the DY process is applicable only at low $x_2$,
because the dipole cross section takes only sea quarks generated from gluon
splitting into account. Therefore, the dipole cross section is proportional to
the target gluon density. 
The statement, 
whether a sea quark belongs to the target or to the projectile, is
however frame dependent. 
If the projectile quark or antiquark becomes slow in the limit $\alpha\to 1$, in
the infinite momentum frame, it can be interpreted 
in the infinite momentum frame as anti-seaquark or seaquark
of the target which annihilates with the projectile parton. No annihilation with
valence quarks from the target is taken into account. Note however that valence
as well as sea parton distributions of the projectile are taken into account in
the light-cone approach, if one employs a parametrization of the projectile
structure function in (\ref{dylctotalhadr}). 
Therefore, the formulation of the DY process presented in this section
is not fully symmetric between
projectile and target. 
 
It is also interesting to see that with the substitution
\beq
\alpha=\frac{1}{1+\widetilde\alpha}\quad,\quad
\rho=\frac{\widetilde\rho}{\alpha},
\eeq
the transverse DY cross section (\ref{dylctotalhadr}) can be rewritten as
\beqn\label{dydis}\nonumber
\frac{d\sigma^T}{dM^2dx_F}&=&\frac{\alpha^2_{em}}{6\pi^3 M^2}
\frac{x_1}{x_1+x_2}\int^\frac{1-x_1}{x_1}_0d\widetilde\alpha\sum_fZ_f^2
\frac{q_f(x_1(1+\widetilde\alpha))+q_{\bar f}(x_1(1+\widetilde\alpha))}
{(1+\widetilde\alpha)^2}\\
&\times&
\int d^2\widetilde\rho
\left\{\left[1+2\widetilde\alpha(1+\widetilde\alpha)\right]
\widetilde\eps^2{\rm K}^2_1(\widetilde\eps\widetilde\rho)
+m_f^2{\rm K}^2_0(\widetilde\eps\widetilde\rho)\right\}
\sigma_{q\bar q}(\widetilde\rho),
\eeqn
and similarly for the longitudinal case,
where $\widetilde\eps=\widetilde\alpha(1+\widetilde\alpha)M^2+m_f^2$ is the
analog of the DIS extension parameter (\ref{disextension}). 
One immediately recognizes the similarity with the 
$q\bar q$ wavefunction (\ref{psit}) in this
equation. Indeed, the exact LC wavefunctions for DIS can be obtained by
analytic continuation from timelike to spacelike region $M^2\to -Q^2$ and
$\widetilde\alpha\to -\alpha$.

In the rewritten form (\ref{dydis}), 
our expression for the DY cross section agrees with the result of Brodsky,
Hebecker and Quack \cite{bhq}, 
where factorization in impact parameter space was considered.
Note that the discussion of hard and soft
contributions to DIS in section \ref{dipoledis} immediately carries over to DY.
Configurations with $\widetilde\alpha\to 0$ 
are the analog of the aligned jet configurations in DIS. Since
$\widetilde\alpha=p_q^{\prime\,0}/q^0$ is the energy of the quark in the final
state divided by the photon energy, large distances become important, when
$\widetilde\alpha\to 0$ and $\alpha\to 1$, respectively. 
It appears however only one limiting case in DY, whereas in DIS
(\ref{stl}), two extremely asymmetric cases allow for large distances.
 
For zero quark mass, the divergencies arising from the endpoints in the
$\alpha$-integration, correspond to the logarithmic divergencies 
occuring in the standard
approach, cf.\ sections \ref{dissec} and \ref{dysec}, which were absorbed into
the nonperturbative parton densities. A similar procedure can be performed in
impact parameter space \cite{bhq}. 
In the parton model, the differential DY cross section can be written as
\beq\label{pm}
\frac{d^2\sigma}{dx_FdM^2}=\frac{4\pi\alpha_{em}}{9M^4}\frac{x_1x_2}{x_1+x_2}
\sum_fZ_f^2\left\{q_f(x_1)\bar q_{f}(x_2)+q_f(x_2)\bar q_{f}(x_1)\right\}.
\eeq
The region $\widetilde\alpha\to 0$ is the DY analog of Bjorkens aligned jet
configurations and gives a leading twist contribution to the transverse DY cross
section. 
As explained above, a soft outgoing quark in the target rest frame can be
regarded as part of the nonperturbative sea-antiquark distribution of the
target in the infinite momentum frame. 
Neglecting the quark mass $m_f$, the divergent part
of the integral (\ref{dydis}) can be recasted into the form (\ref{pm}) and one
can read of the quark and antiquark distributions of the
target at low $x_2$. One find \cite{bhq}
\beq\label{dypart}
x_2q^{DY}_{f,\bar f}(x_2)=\frac{3M^2}{8\pi^4}\int_0^\Lambda d\widetilde\alpha
\int d^2\widetilde\rho\,
\widetilde\alpha M^2 {\rm K}^2_1(\widetilde\eps\widetilde\rho)
\sigma_{q\bar q}(\widetilde\rho),
\eeq
where $\Lambda\ll 1$ is a cutoff and the approximation 
$\widetilde\eps\to\widetilde\alpha
M^2$ for $\widetilde\alpha\to 0$ was used. 
On the other hand, in the parton model of DIS,
one has
\beq
\frac{Q^2}{4\pi^2\alpha_{em}}\sigma^{\gamma^*p}_{T}=
\sum_fZ_f^2\left\{q_f(x_{Bj})+\bar q_{f}(x_{Bj})\right\},
\eeq
and one 
can again extract the divergent part of the integral (\ref{stl}) for the
transverse cross section and read of the target parton distributions \cite{bhq},
\beq\label{dispart}
x_{Bj}q^{DIS}_{f,\bar f}(x_{Bj})=\frac{3Q^2}{8\pi^4}\int_0^\Lambda d\alpha
\int d^2\rho
\alpha Q^2 {\rm K}^2_1(\eps\rho)
\sigma_{q\bar q}(\rho).
\eeq
Note, that in DIS a slow quark ($\alpha\to 0$) seen in the target rest frame,
can be interpreted as part of the sea-antiquark distribution of target in the
infinite momentum frame.
Obviously, the equations (\ref{dypart}) and (\ref{dispart}) are identical and
therefore
the DIS parton distribution is identical to the DY parton distribution. 
This manifestation of factorization in impact parameter space was first pointed
out in \cite{bhq}.

The transverse momentum distribution of DY pairs
can also be calculated in the light cone
approach \cite{kst}. The differential cross section is given by the four-fold
Fourier integral
\beqn\nonumber\label{dylcdiff}
\frac{d\sigma(qp\to q\gamma^*p)}{d\ln\alpha d^2p_T}
&=&\frac{1}{(2\pi)^2}
\int d^2\rho_1d^2\rho_2\, \exp[{\rm i}\vec p_T\cdot(\vec\rho_1-\vec\rho_2)]
\Psi^*_{\gamma^* q}(\alpha,\vec\rho_1)\Psi_{\gamma^* q}(\alpha,\vec\rho_2)\\
&\times&
\frac{1}{2}
\left\{\sigma_{q\bar q}(\alpha\rho_1)
+\sigma_{q\bar q}(\alpha\rho_2)
-\sigma_{q\bar q}(\alpha(\vec\rho_1-\vec\rho_2))\right\}.
\eeqn
after integrating this expression over $p_T$, one obviously recovers
(\ref{dylctotal}). The negative term in the curly brackets arises from the
interference between the two graphs for bremsstrahlung. 
The expressions for the LC wavefunctions needed here are
\beqn\nonumber
\Psi^{*T}_{\gamma^* q}(\alpha,\vec\rho_1)\Psi^T_{\gamma^* q}(\alpha,\vec\rho_2)
&=& \frac{\alpha_{em}}{\pi^2}\Bigg\{
     m_f^2 \alpha^4 {\rm K}_0\left(\eta\rho_1\right)
     {\rm K}_0\left(\eta\rho_2\right)\\
   &+& \left[1+\left(1-\alpha\right)^2\right]\eta^2
   \frac{\vec\rho_1\cdot\vec\rho_2}{\rho_1\rho_2}
     {\rm K}_1\left(\eta\rho_1\right)
     {\rm K}_1\left(\eta\rho_2\right)\Bigg\},\\
 \Psi^{*L}_{\gamma^* q}(\alpha,\vec\rho_1)\Psi^L_{\gamma^* q}(\alpha,\vec\rho_2)
&=& \frac{2\alpha_{em}}{\pi^2}M^2 \left(1-\alpha\right)^2
  {\rm K}_0\left(\eta\rho_1\right)
     {\rm K}_0\left(\eta\rho_2\right). 
\eeqn
The hadronic cross section is then given by
\beq\label{dylcdiffhadr}
\frac{d\sigma}{dM^2dx_Fd^2p_T}=\frac{\alpha_{em}}{6\pi M^2}
\frac{1}{x_1+x_2}\int_{x_1}^1\frac{d\alpha}{\alpha}
F_2^p\left(\frac{x_1}{\alpha}\right)\frac{d\sigma(qp\to q\gamma^*p)}
{d\ln\alpha d^2p_T},
\eeq
in analogy to (\ref{dylctotalhadr}). Three of the four integrations in
(\ref{dylcdiff}) can be performed
analytically for arbitrary $\sigma_{q\bar q}$. This
will be done in appendix \ref{appdypp}.

\subsection{The dipole cross section}\label{DCS}

In this section we present a  parametrization of the dipole cross section
which is adjusted to DIS data and to total hadronic cross sections.
In the preceding sections we have
expressed the cross section for DIS and the
DY-process at low $x$ in terms of the dipole cross section. 
Thus the dipole cross section is the crucial missing input one needs to perform a calculation
that can be compared to experimental data. 
The dipole cross section is of interest not only for
DIS and DY. Total hadronic cross sections, like for pion-proton scattering
can also be written in terms of the dipole cross section. In this case, the pion wave-function
in coordinate representation $\Psi_{\pi}(\rho)$,
gives the probability to find a quark-antiquark
pair with a given transverse separation inside the pion. 
If only the lowest Fock state is taken into account, the total cross section
at {\em c.m.} energy $s$ can then be written as
\beq\label{pioncross}
\sigma_{\pi p}(s)=\int d^2\rho |\Psi_{\pi}(\rho)|^2
\sigma_{q\bar q}(s,\rho),
\eeq
in complete analogy to the DIS case. This is remarkable since hadronic cross
sections are governed by nonperturbative long distance physics in contrast to
DIS. Still, (\ref{pioncross}) is valid, because the decomposition into
interaction eigenstates (\ref{expansion}) is completely general and does not
rely on perturbation theory.

The dipole cross section is largely unknown, only at small distances $\rho$ it can be expressed
in terms of the gluon density (\ref{fssigma}). However, several parametrizations
exist in the literature, 
describing the whole function $\sigma_{q\bar q}(\rho)$, without 
explicitly taking
into account the QCD evolution of the gluon density. 
A very economical parametrization is provided by the saturation model of
Golec-Biernat and W\"usthoff \cite{Wuesthoff1,Wuesthoff2},
\beq\label{wuestsigma}
\sigma_{q\bar q}(x,\rho)=
\sigma_0\left[1-\exp\left(-\frac{\rho^2Q_0^2}{4(x/x_0)^\lambda}\right)\right],
\eeq
where
$Q_0=1$ GeV and the three fitted parameters are 
$\sigma_0=23.03$ mb, $x_0=0.0003$, and $\lambda=0.288$.
This dipole cross section vanishes $\propto\rho^2$ at small distances, as
implied by color transparency and levels off exponetially at large separations,
which reminds of eikonalization.
The authors are able to fit all available HERA data with 
a quite low $\chi^2$
\cite{Wuesthoff1} and can furthermore also describe diffractive HERA data
\cite{Wuesthoff2}. It is argued that saturation is necessary to obtain a nearly
$Q^2$ independent ratio of the diffractive to the total cross section, which
cannot be reproduced with a dipole cross section that rises quadratically up to infinity. 

There is however no $x_{Bj}$ in hadron-hadron scattering and
the appropriate variable to describe the energy dependence of the dipole cross section is $s$.
This is different in DIS where the virtual photon scatters off pointlike
constituents inside the proton and $x_{Bj}$ is the correct variable to use.
Parametrizing the dipole cross section as function of $s$ will automatically lead to a violation
of Bjorken scaling. This is easy to see, since $s\approx Q^2/x_{Bj}$. If the
cross section is supposed to grow like a power of $s$, it will
grow with approximately the same power in $Q^2$. 

A two pomeron model 
of the dipole cross section as function of
$s$, motivated from Regge phenomenology, 
was given in \cite{Forshaw}. It was argued in
\cite{McDFS99} that the authors of \cite{Forshaw} had to parametrize the dipole
cross section as a function of $\rho^2s$, to get approximate Bjorken scaling. 
The ansatz of \cite{Forshaw} can describe deep inelastic data up
to $Q^2=60$ GeV$^2$, with $10$ parameters.

However, as can be seen from fig.\ \ref{f2pq2}, Bjorken scaling
is quite strongly violated at low $x_{Bj}$. Since we have in mind to describe
hadron-hadron scattering and DIS with the same dipole cross section, we write the dipole cross section
as function of $s$. We are aware of the fact that our ansatz will fail at very
large values of $Q^2$. Nevertheless, practically all physics which is
interesting to us happens at quite small values of $Q^2<10$ GeV$^2$.  

An ansatz inspired by the saturation model of \cite{Wuesthoff1,Wuesthoff2}, was
presented in \cite{KST99}. It reads
\beq
  \label{sasha}
  \sigma_{q\bar q}(s,\rho) = 
      \sigma_0(s)\left[1-\exp\left(
    - \frac{\rho^2}{r_0^2\left(s\right)}\right)\right],
\eeq
with $r_0\left(s\right) = 0.88 (s/s_0)^{-0.14}~{\rm fm}$, $s_0 = %
1000~$GeV$^2$ and 
\beq\label{vorfaktor}
  \sigma_0(s) = 
     \sigma_{tot}^{\pi p}(s)\left(1
   + \frac{3r_0^2\left(s\right)}{8\left<r^2_{ch}\right>_{\pi}}\right),
\eeq
where $\sigma_{tot}^{\pi p}(s) = 23.6(s/s_0)^{0.08}~{\rm mb}$ and
$\left<r^2_{ch}\right>_{\pi} = 0.44~{\rm fm}^2$ is the pion charge radius.
This cross section is proportional to $\rho^2$ for 
$\rho\to 0$, but flattens off at large $\rho$. The energy dependence 
correlates with $\rho$. At small $\rho$ the dipole cross section rises
with a hard pomeron intercept, $0.36$, and at large separations it still depends
with a soft pomeron intercept, $0.08$, on energy. This is different in
(\ref{wuestsigma}), where the maximum value of the dipole cross section is independent of energy.
We also mention that a value of $\sigma_0=23.03$ mb is too
 small to reproduce the
pion-proton cross section. 
Note that the energy behavior of our parametrization
is also conceptually different from Regge phenomenology,
where the pomeron intercept has to be independent of the hadron. 
We think however, that the growth of the cross section with $s$ should depend on
the typical distances involved and our ansatz smoothly interpolates between
cross sections in hard QCD processes and soft hadronic scattering. The
saturation scale $r_0(s)$ decreases with increasing $s$. This means that at very
high $s$, only very small dipoles exhibit a hard pomeron behavior. At
separations $\rho\gg r_0(s)$, the dipole cross section grows only slowly with energy. The ansatz
(\ref{sasha}) ensures that the dipole cross section is a monotonous function of
$s$ and $\rho$. This is not the case for the parametrization of \cite{Forshaw},
which can decrease with increasing $\rho$. 

The energy dependent saturation value of the parametrization (\ref{sasha}) is
chosen in such a way, that the pomeron part of the
fit by Donnachie and Landshoff \cite{pipfit}
is automatically
reproduced,
if one employs a gaussian pion wavefunction,
\beq\label{pionwave}
|\Psi_\pi(\rho)|^2=\frac{3}{8\pi\la r^2_{ch}\ra_\pi}
\exp\left(-\frac{3\rho^2}{8\la r^2_{ch}\ra_\pi}\right).
\eeq
To understand the factor of $3/8$, note that it has to hold
\beq
\int d^2\rho\,\rho^2\,|\Psi_\pi(\rho)|^2=\frac{2}{3}\cdot2^2
\la r^2_{ch}\ra_\pi,
\eeq
because $\vec\rho$ is only a two dimensional vector ($2/3$) 
and it has the meaning
of a diameter, rather than a radius ($2^2$). 

The dipole cross section (\ref{sasha}) is however adjusted only by eye and no $\chi^2$ was
calculated. Furthermore, it tends to overestimate $F_2$ at $Q^2>10$ GeV$^2$.
We improve the parametrization (\ref{sasha}) in several ways. First, we use the
more recent fit by Cudell et al.\ \cite{Cudell},
\beq\label{Cudellfit}
\sigma_{\pi^-p}=26.2~{\rm mb}\,\left(\frac{s}{1~{\rm GeV}^2}\right)^{-0.357}
+7.63~{\rm mb}\,\left(\frac{s}{1~{\rm GeV}^2}\right)^{-0.56}
+12.08~{\rm mb}\,\left(\frac{s}{1~{\rm GeV}^2}\right)^{0.0933},
\eeq
for $\sqrt{s}>9$ GeV$^2$.
which has a slightly larger pomeron intercept and agrees better with 
experimental data than the old version by Donnachie and Landshoff \cite{pipfit}.
A pomeron intercept larger than $0.08$ is also suggested by recent SELEX
measurements \cite{Uwe}, fig.\ \ref{pion}. 
To improve the
agreement with data, we multiplied (\ref{sasha}) by a function that leads to a
suppression at small value of $\rho$ and becomes $1$ at large $\rho$. We have
tried several functions. 
Best agreement with HERA data is obtained with the parametrization
\beq\label{improved}
  \label{ansatz1}
  \sigma_{q\bar q}(s,\rho) = 
      \sigma_0(s)\left[1-\exp\left(
    - \frac{\rho^2}{r_0^2\left(s\right)}\right)\right]
    \left[1-0.9\exp\left(
    - \frac{\rho^2}{r_1^2}\right)\right],
\eeq
where 
\beq
  \sigma_0(s) = 
     12.08~{\rm mb}\left(\frac{s}{1~{\rm GeV^2}}\right)^{0.0933}
     \left(1
    +\frac{3r_0^2\left(s\right)}{8\left<r^2_{ch}\right>_{\pi}}\right),
\eeq
\beq
\left<r^2_{ch}\right>_{\pi} = 0.44~{\rm fm}^2,
\eeq
\beq\label{r0s2}
r_0^2(s)=(2.19~{\rm fm})^2\left(\frac{s}{1~{\rm GeV^2}}\right)^{-0.34},
\eeq
\beq\label{r12}
r_1=0.27~{\rm fm}.
\eeq

\begin{figure}[t]
\centerline{
  \scalebox{0.43}{\includegraphics*{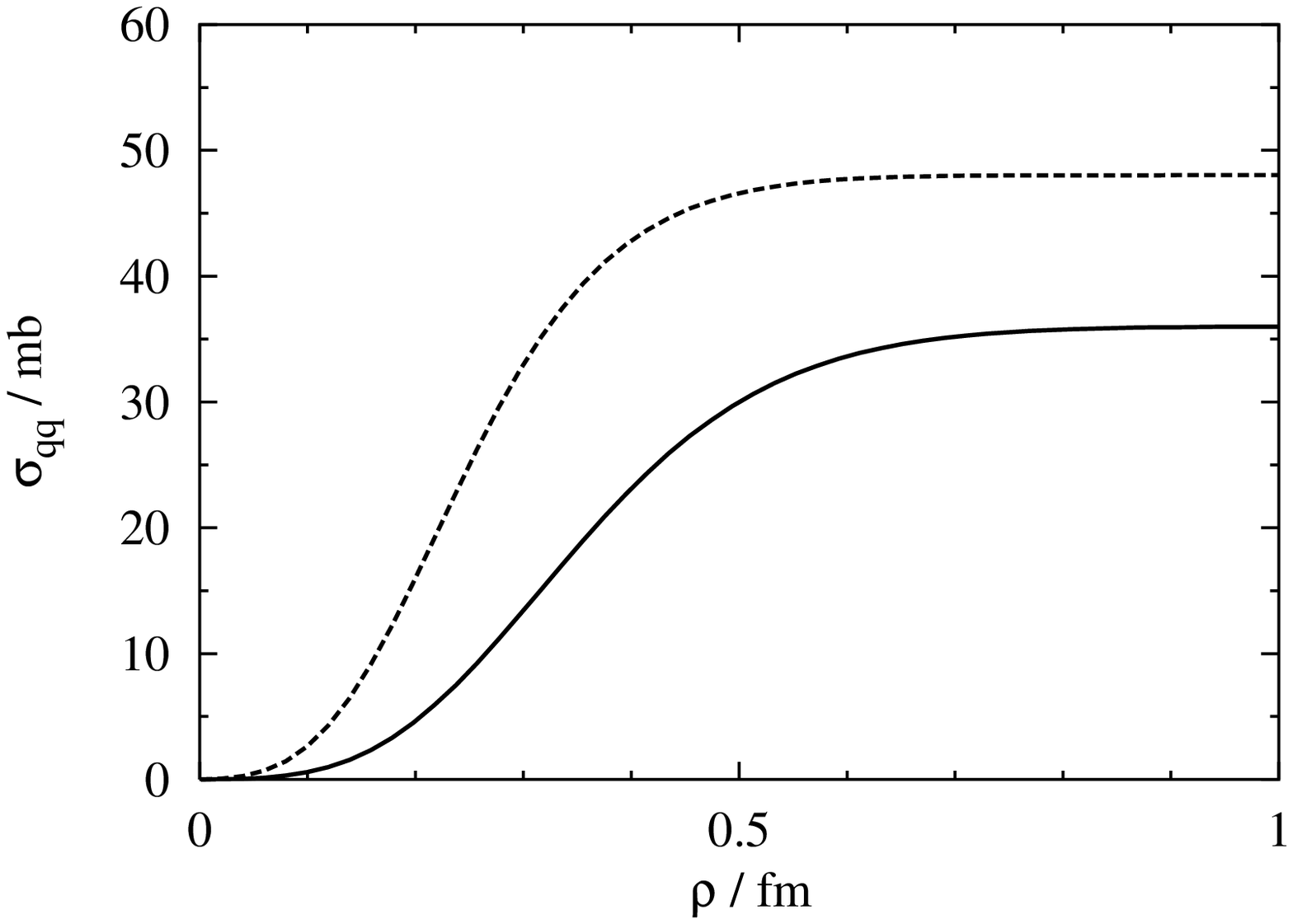}}
  \scalebox{0.43}{\includegraphics*{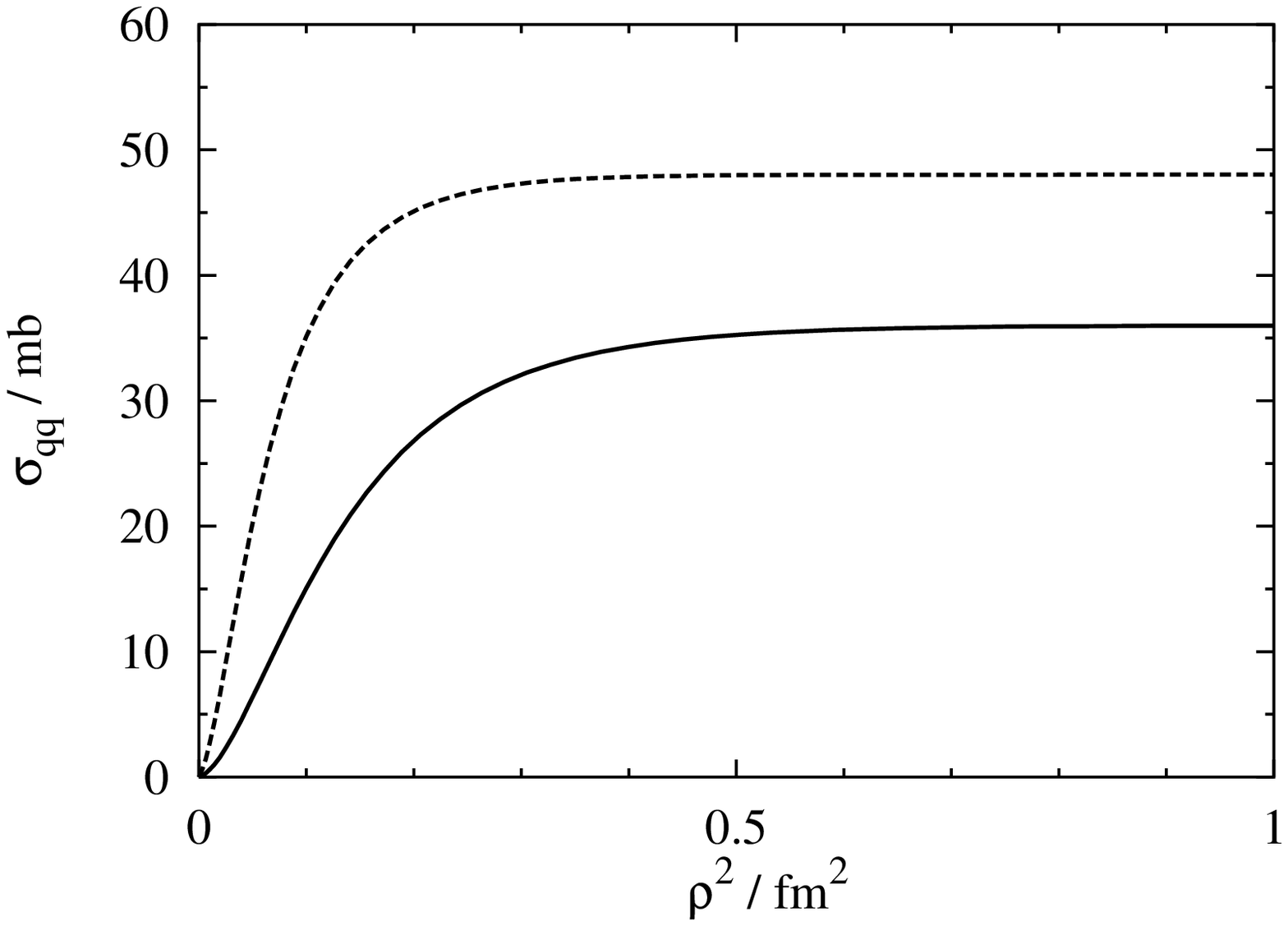}}
}
\caption{
  \label{sigma}
  The dipole cross section (\ref{improved})
  for $\sqrt{s}=200$ GeV (HERA), 
  solid line, and for $\sqrt{s}=1.4$ TeV (THERA),
      dashed line. The left figure shows the dipole cross section
       vs.\ $\rho$, while the figure
      on the right displays $\sigma_{q\bar q}(s,\rho)$ vs.\ $\rho^2$. 
      At large separations, the dipole cross section 
      becomes constant in $\rho$, but
      still  
      grows slowly with energy.  
      At small separations the dipole cross section depends quadratically on
      $\rho$.
}
\end{figure}

Our result is depicted in fig.\ \ref{sigma}.
The energy dependence of (\ref{improved}) is qualitatively
the same as discussed for (\ref{sasha}), because
$r_1$ does not depend on $s$. 
In the left figure, we plotted the dipole cross section vs.\ $\rho$. One can see how the cross
section increases from HERA to THERA energies. The figure on the right shows
$\sigma_{q\bar q}(s,\rho)$ vs.\ $\rho^2$. At small separations, the dipole cross section
increases quadratically with $\rho$.

We determined the 
$3$ parameters, $r_1$ and the two numbers in (\ref{r0s2}), 
using H1-data up
to $Q^2=65$ GeV$^2$. 
We have tried to achieve good agreement with the same
parameters for perturbative and nonperturbative wavefunctions.
A comparison with data is presented in figs.\ \ref{newh1_1}-\ref{newzeus2}.
One observes that with perturbative LC wavefunctions, (\ref{psit},\ref{psil}), the
very low $Q^2$ data is not well reproduced. The agreement can be improved by
introducing an unphysically small quark mass. This would however allow for
arbitrarily large separations of the pair, at variance with confinement. We
therefore fix the mass for $u$- and $d$-quarks at a typical inverse confinement
radius,
\beq\label{quarkmasses}
m_{u,d}=200~{\rm MeV}.
\eeq
We also include strange and charm quarks in our calculation,
\beq
m_s=350~{\rm MeV}\quad,\quad m_c=1.5~{\rm GeV}.
\eeq
With these parameters, the lowest $\chi^2$ that can be obtained with perturbative
wavefunctions is
\beq
\chi^2_{pt}({\rm H1}, Q^2\le 65~{\rm GeV}^2)=523/233~{\rm d.o.f.}
\eeq
Statistical and
systematic errors are added in quadrature. However, the dashed curves
still make a reasonable impression at not too high $Q^2$, 
when one looks at 
figs.\ \ref{newh1_1}-\ref{newzeus2}.
We will therefore perform all our calculations with perturbative wavefunctions,
too. 
Better agreement is obtained with the nonperturbative wavefunctions, 
(\ref{psitnpt},\ref{psilnpt}), which 
describe the data especially well at low $Q^2$. We obtain
\beqn
\chi^2_{npt}({\rm H1}, Q^2\le 10~{\rm GeV}^2)&=&86.4/100~{\rm d.o.f},\\
\chi^2_{npt}({\rm H1}, Q^2\le 30~{\rm GeV}^2)&=&214/182~{\rm d.o.f}.
\eeqn
For the nonperturbative wavefunctions, we fix the quark masses at $m_u=m_d=10$
MeV, $m_s=150$ MeV and $m_c=1.5$ GeV. We have also plotted the result of our
calculation for very high virtualities $Q^2=2000$~GeV$^2$,
see figs.\ \ref{newh1_2}-\ref{newzeus1}. The parametrization
(\ref{improved}) overshoots the data at large $Q^2$.

\begin{figure}[ht]
  \centerline{\scalebox{0.77}{\includegraphics{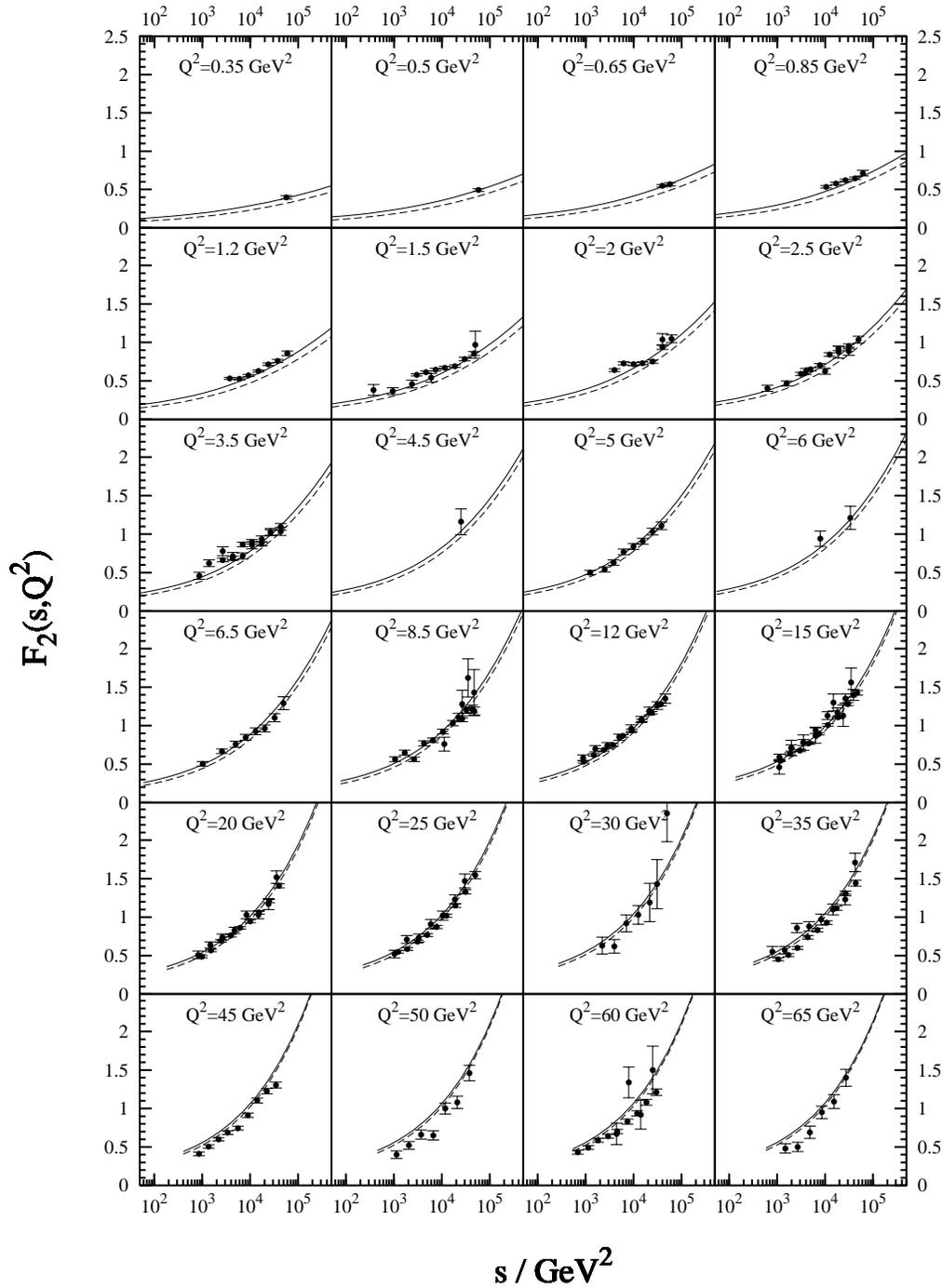}}}
    \caption{
      \label{newh1_1}
      The points show H1 data  \cite{h1931}-\cite{h1962} for the proton
      structure function $F_2$ vs.\ $s$. Only statistical errors are shown.
      The solid curves are calculated with the parametrization 
      (\ref{improved}) of
      the dipole cross section and with the nonperturbative LC wavefunctions
      (\ref{psitnpt},\ref{psilnpt}). The dashed curves are the same
       but calculated with the perturbative
      LC wavefunctions (\ref{psit},\ref{psil}). The parameters in
      (\ref{improved}) are adjusted to this set of data. See text for more
      details. 
    }  
\end{figure}

\begin{figure}[ht]
  \centerline{\scalebox{0.77}{\includegraphics*{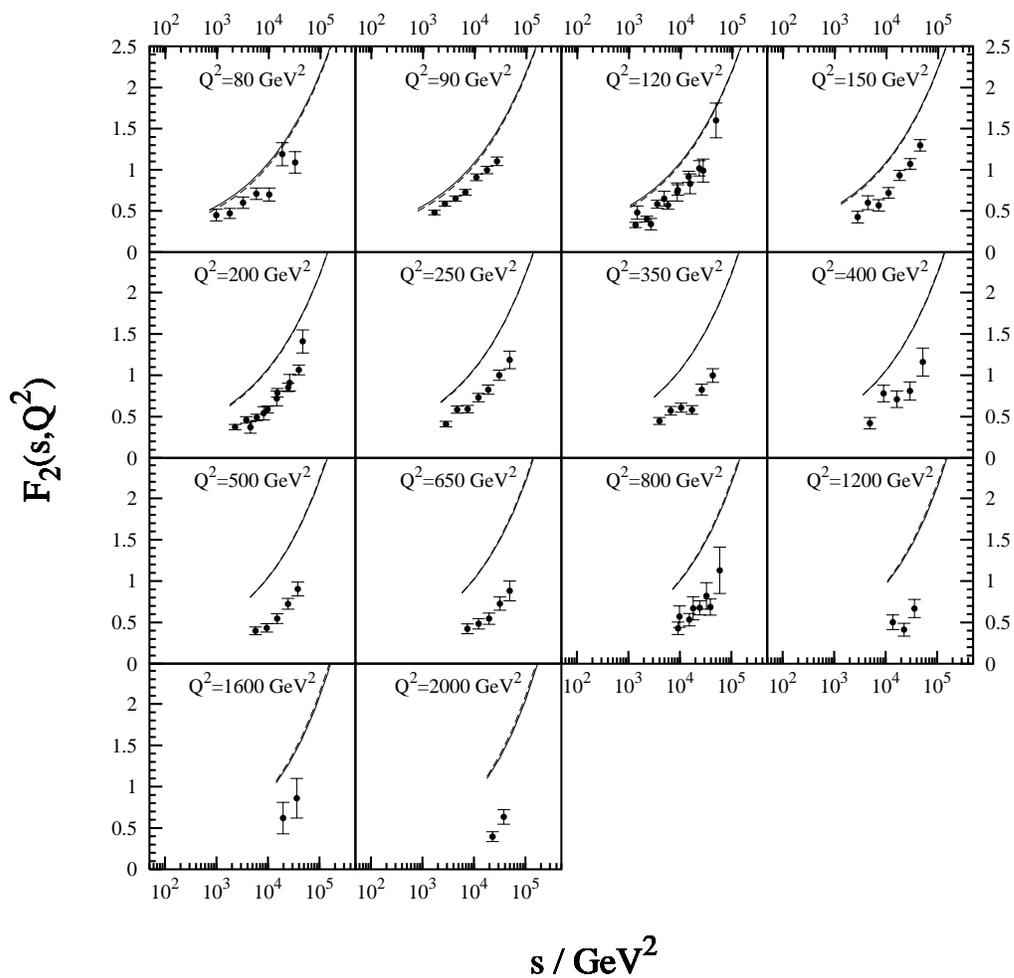}}}
    \caption{
      \label{newh1_2}
      The same as fig.~\ref{newh1_1}. The points shown here are not used to
      determine the parameters in (\ref{improved}). 
    }  
\end{figure}

\begin{figure}[ht]
  \centerline{\scalebox{0.77}{\includegraphics{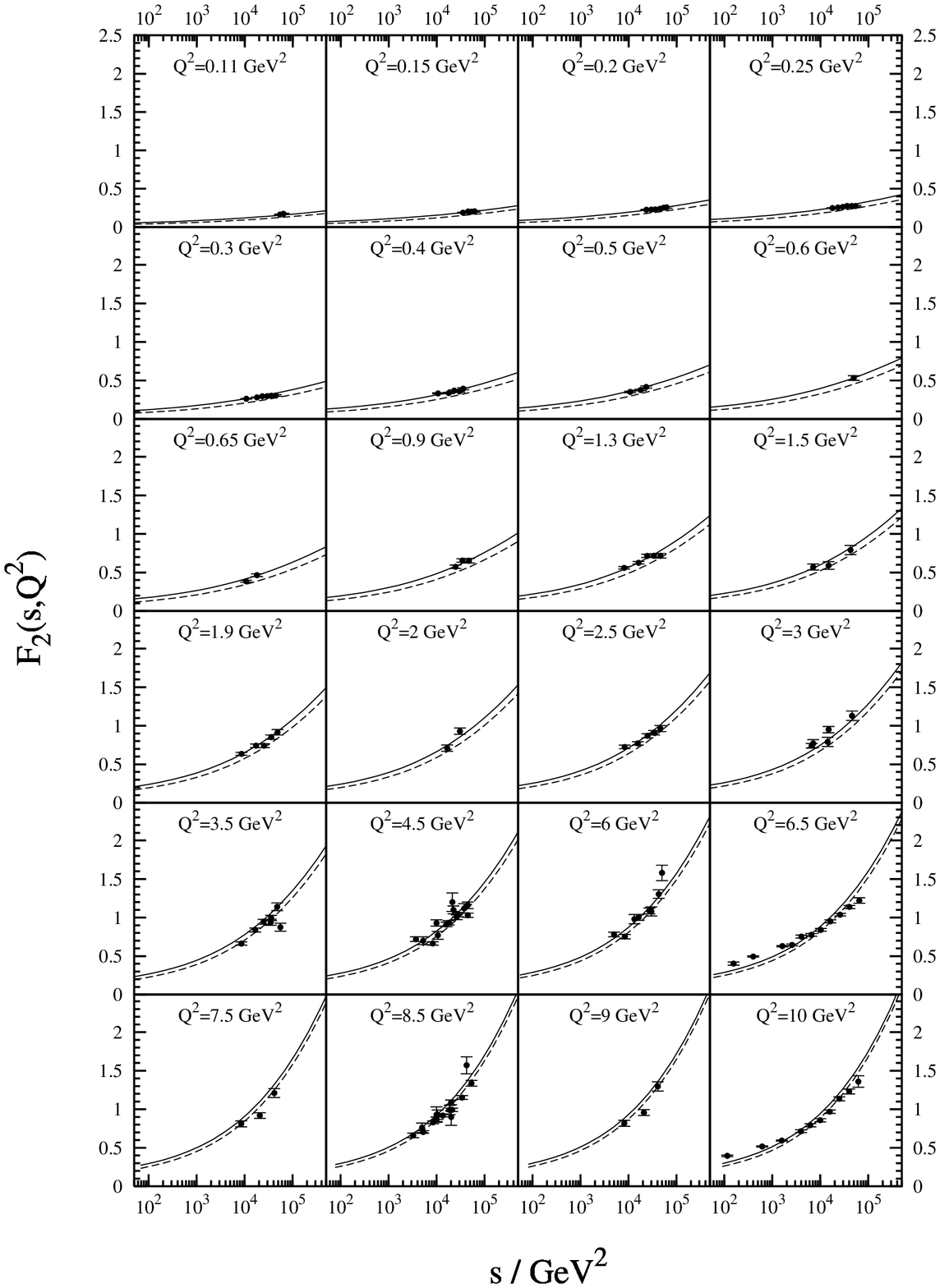}}}
    \caption{
      \label{newzeus1}
      The same as fig.~\ref{newh1_1}
      but the points are from ZEUS
      \cite{zeus961+2,zeus98,zeus93}-\cite{zeus97}. 
      These points are not used to
      determine the parameters in (\ref{improved}).  
    }  
\end{figure}

\begin{figure}[ht]
  \centerline{\scalebox{0.77}{\includegraphics{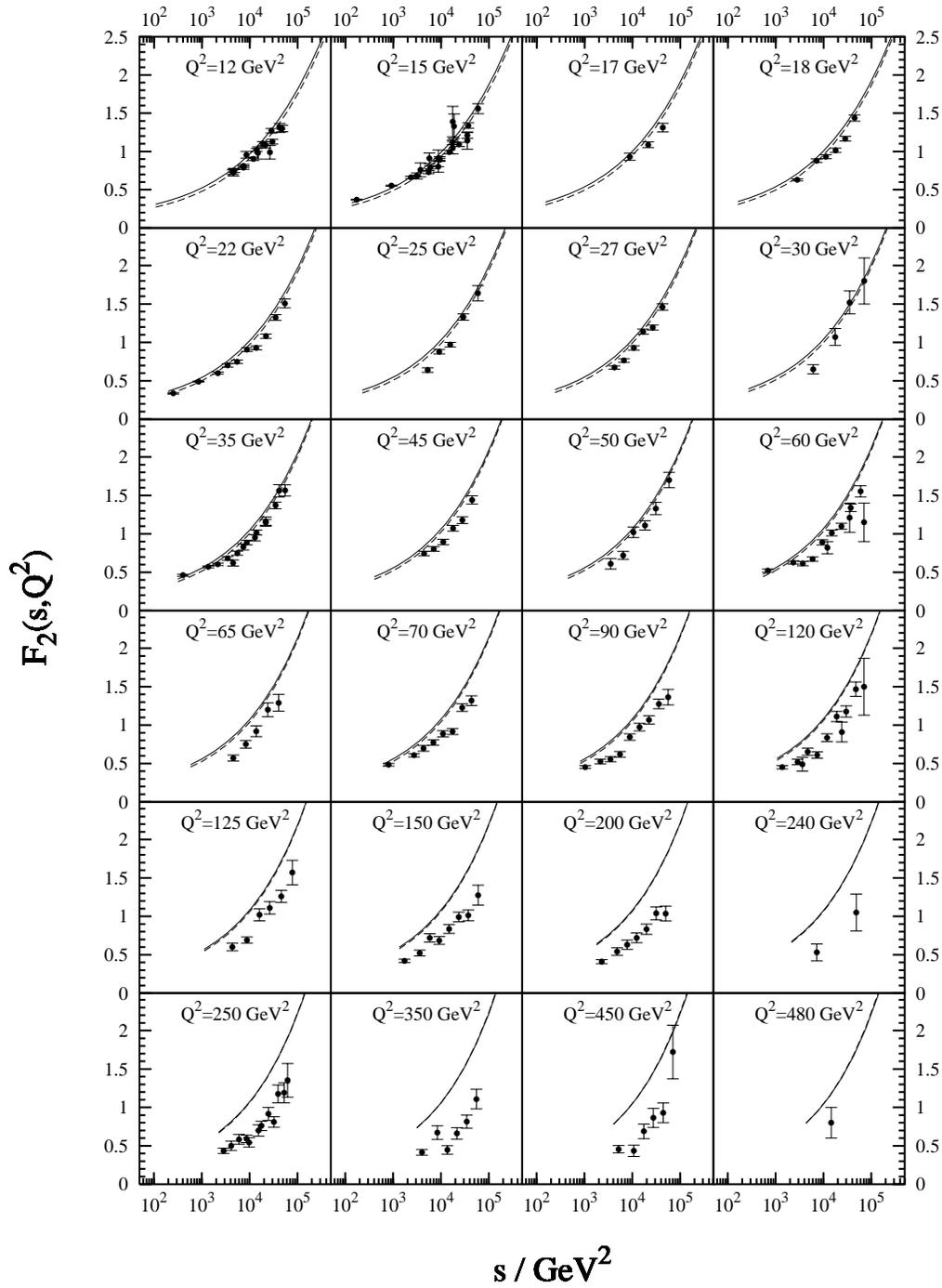}}}
    \caption{
      \label{newzeus2}
      The same as fig.~\ref{newh1_1}
      but the points are from ZEUS
      \cite{zeus961+2,zeus98,zeus93}-\cite{zeus97}. 
      These points are not used to
      determine the parameters in (\ref{improved}).    
    }  
\end{figure}

\clearpage

To see in more detail, 
how the nonperturbative interaction modifies the LC wavefunctions and
which distances contribute to $F_2$, we calculate the
following three quantities. First, we integrate the LC wavefunctions
over the longitudinal momentum fraction $\alpha$ and over the angle,
\beq\label{psialpha}
A_0(\rho)=\frac{2\pi\rho}{\alpha_{em}}
\int_0^1d\alpha|\Psi^{T,L}_{q\bar q}(\alpha,\rho)|^2.
\eeq
The result is shown in fig.\ \ref{psiz}. The results for our calculation of the
wavefunctions integrated over $\alpha$ and angle weighted with the dipole cross section,
\beq\label{psialphasig}
A_1(\rho)=\frac{2\pi\rho}{\alpha_{em}}\sigma_{q\bar q}(s,\rho)
\int_0^1d\alpha|\Psi^{T,L}_{q\bar q}(\alpha,\rho)|^2,
\eeq
can be seen in fig.\ \ref{psizsig}. This quantity is important for the structure
function $F_2$. Finally, we weight the integral with $\sigma^2_{q\bar
q}(s,\rho)$
\beq\label{psialphasig2}
A_2(\rho)=\frac{2\pi\rho}{\alpha_{em}}\sigma_{q\bar q}^2(s,\rho)
\int_0^1d\alpha|\Psi^{T,L}_{q\bar q}(\alpha,\rho)|^2,
\eeq
which is important for diffraction. The result is depicted in fig.\
\ref{psizsig2}. 
In all three cases, the factor $2\pi\rho$ arises from the area element. It has
to be included, if one wants to get a correct impression of the distances
contributing to $F_2$.
For convenience, we divide  $A_0$--$A_2$ by the fine structure constant
$\alpha_{em}$. All calculations are done 
at $\sqrt{s}=200$ GeV$^2$ and
for two values of $Q^2$, namely for
$Q^2=0.5$ GeV$^2$ and for $Q^2=10$ GeV$^2$.

\begin{figure}[t]
\centerline{
  \scalebox{0.43}{\includegraphics*{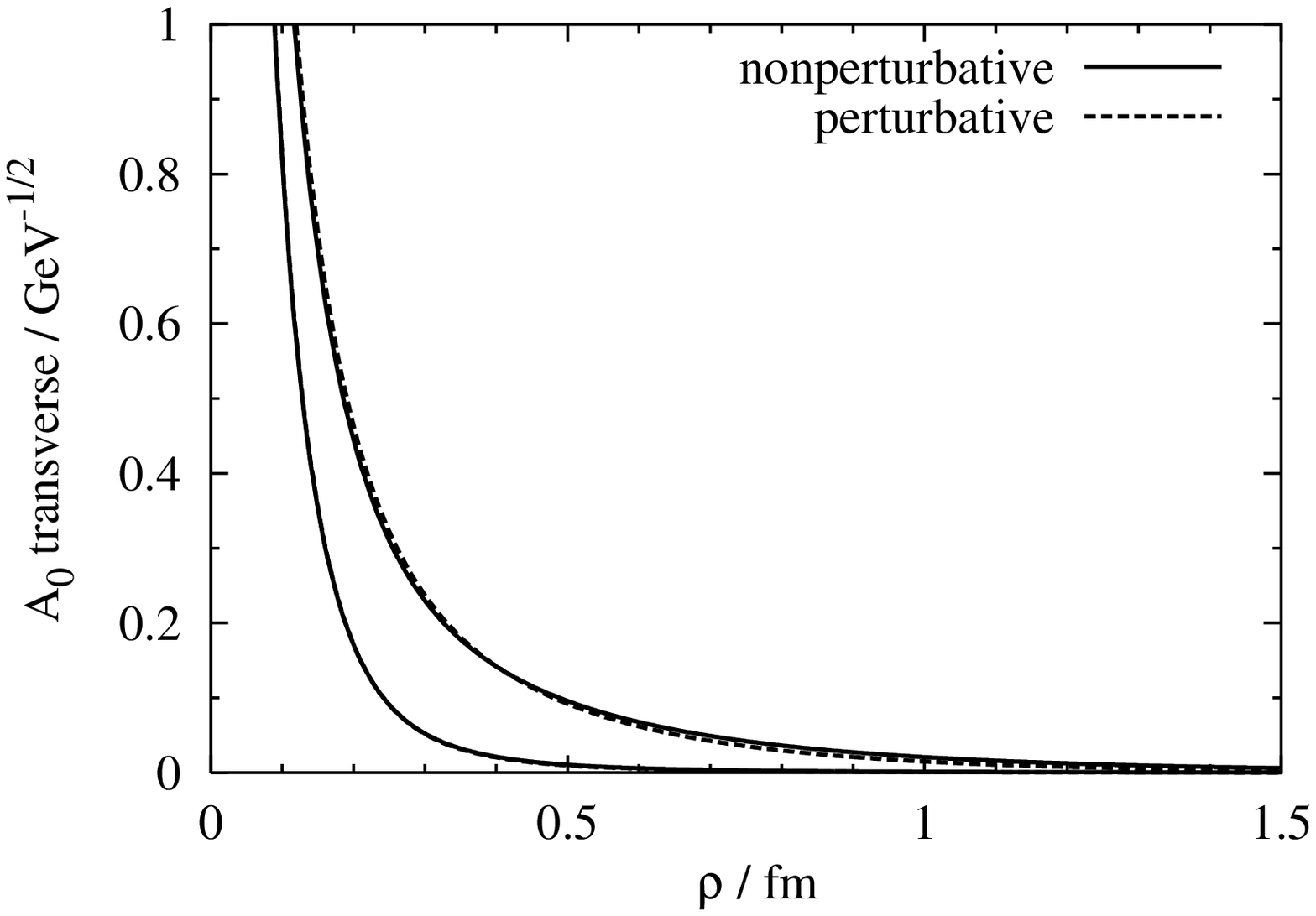}}
  \scalebox{0.43}{\includegraphics*{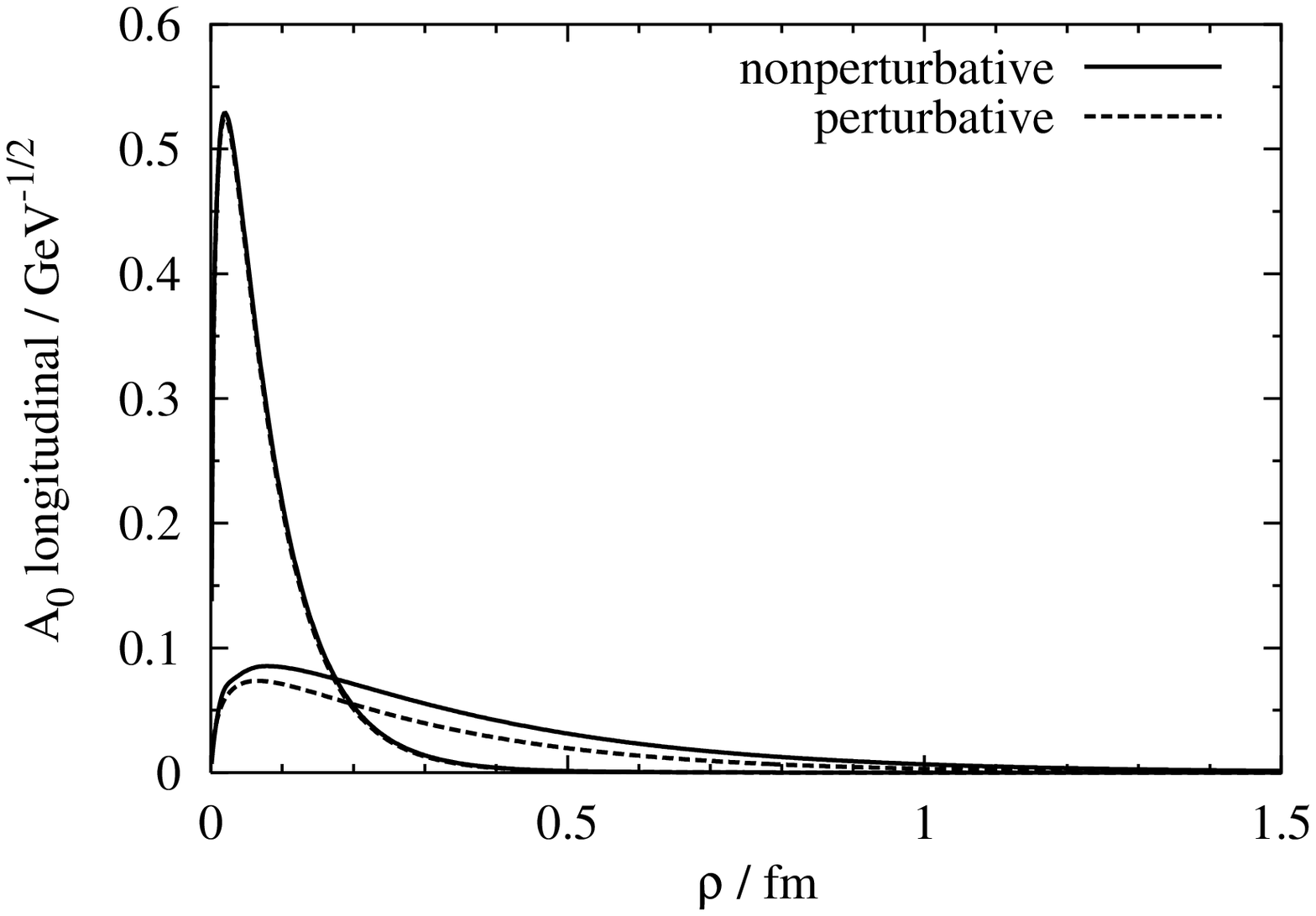}}
}
\caption{
  \label{psiz}
  The LC wavefunctions integrated over $\alpha$, (\ref{psialpha}), at
  $\sqrt{s}=200$ GeV. The solid curves are calculated with the nonperturbative
  wavefunctions, (\ref{psitnpt},\ref{psilnpt}), while the dashed curves
  are calculated with the perturbative ones, 
  (\ref{psit},\ref{psil}). The lower two curves (at large $\rho$)
  are for $Q^2=10$ GeV$^2$
  and the upper ones for $Q^2=0.5$ GeV$^2$. 
}
\end{figure}

It is hard to see in fig.\ \ref{psiz}, how the nonperturbative interaction
modifies the transverse LC wavefunctions. Even at $Q^2=0.5$ GeV$^2$, the curves
are almost identical. The transverse part diverges $\propto 1/\rho$ at $\rho\to
0$. The wavefunction is not normalizable. The curves rapidly decay as $\rho$
becomes larger, cf.\ (\ref{asymptotics}). The situation is different in the
longitudinal case. Here, $A_0$ goes to $0$ at $\rho\to 0$. The
longitudinal wavefunction is square integrable. At high $Q^2$ the integral $A_0$
has a peak at small $\rho\approx 1/Q$ 
and then decreases exponentially. At low virtuality,
this peak is much less pronounced and $A_0$ has quite a long tail into the large
distance region. Somewhat surprisingly, the longitudinal LC wavefunctions are
stronger affected by the nonperturbative interaction at low $Q^2$
than the transverse.
At $Q^2=10$ GeV$^2$, no effect of the nonperturbative interaction is visible any
more.

\begin{figure}[t]
\centerline{
  \scalebox{0.43}{\includegraphics*{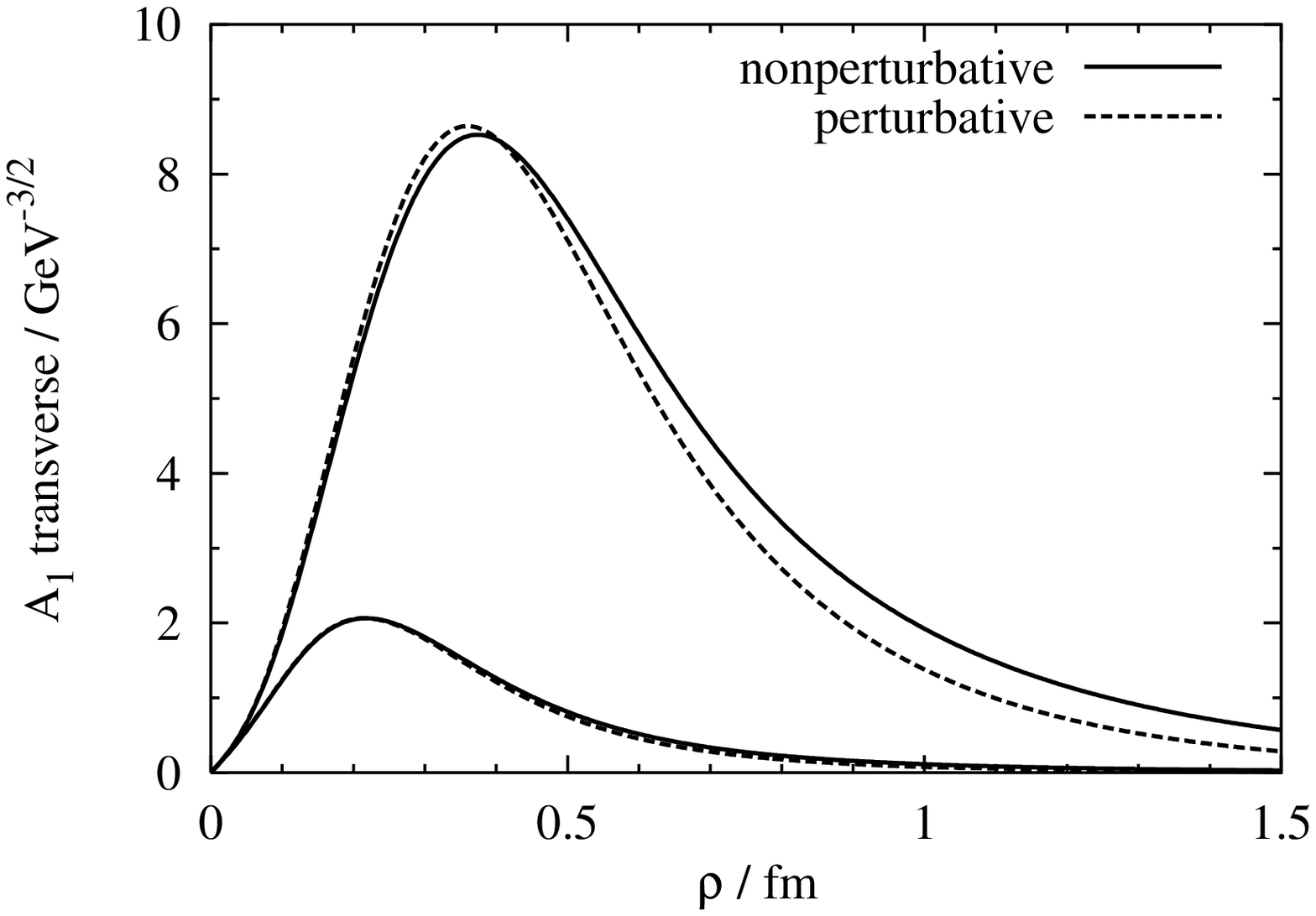}}
  \scalebox{0.43}{\includegraphics*{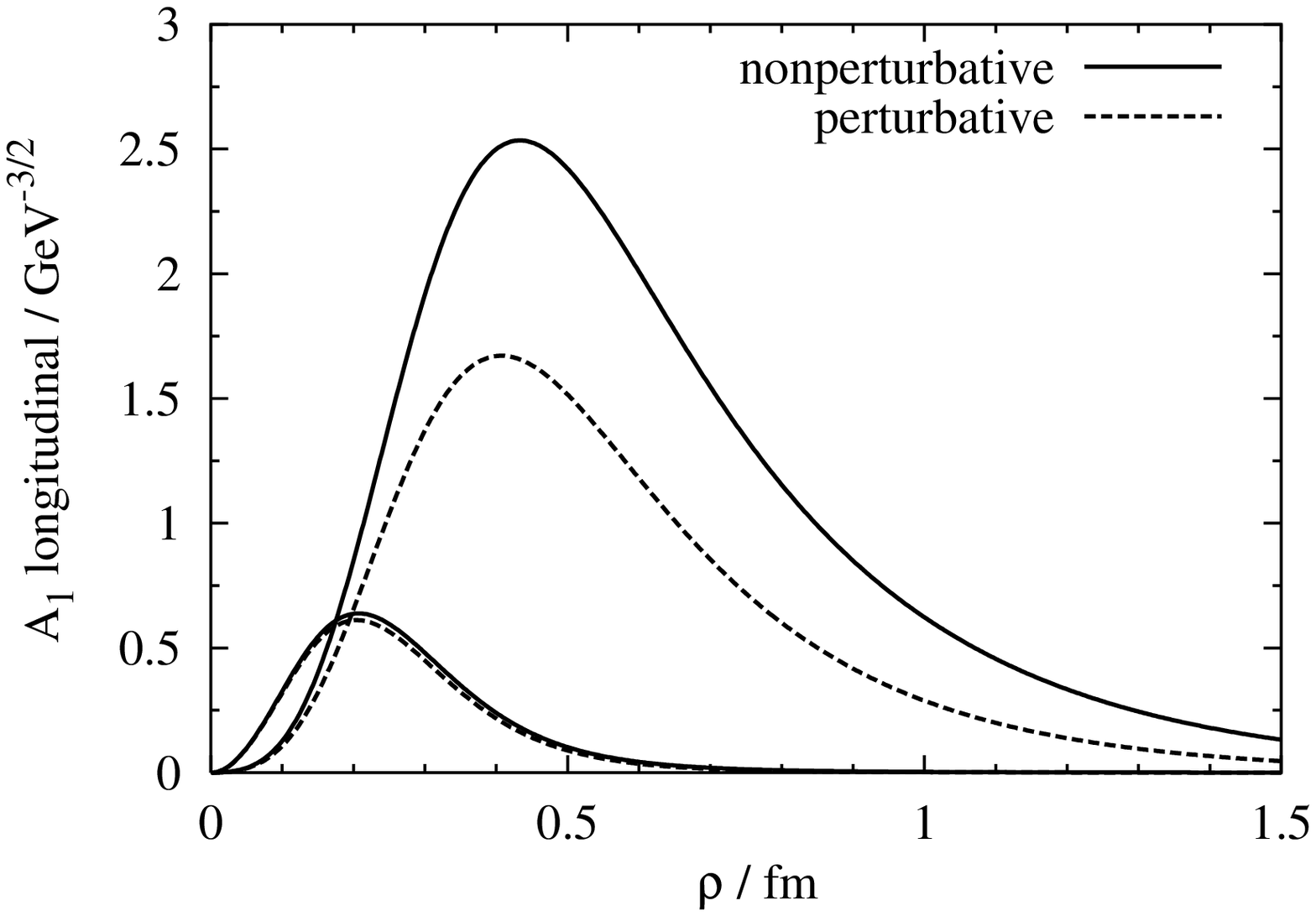}}
}
\caption{
  \label{psizsig}
  The LC wavefunctions integrated over $\alpha$ and weighted with
  $\sigma_{q\bar q}(s,\rho)$, (\ref{psialphasig}), at
  $\sqrt{s}=200$ GeV. See fig.~\ref{psiz} for more explanations.
}
\end{figure}

The quantity $A_1$
displayed in fig.\ \ref{psizsig} integrated over $\rho$ is nothing
but the structure function $F_2$, up to a factor $Q^2/(4\pi^2)$. 
The region of small $\rho$ is suppressed compared to fig.\ \ref{psiz}, because of
an extra factor $\rho^2$ from the dipole cross section. Therefore, larger distances as one would
expect contribute. Both, for low and high $Q^2$, the curves have a
long tail into the large distance domain. Again, there is practically no
influence of the nonperturbative interaction at high virtuality, but at low
$Q^2$ and large $\rho$, the perturbative part is slightly suppressed compared to
the nonperturbative. This means that our choice of the quark mass, squeezes the
wavepacket somewhat more than the interaction. This can also be seen in figs.\ 
\ref{newh1_1}-\ref{newzeus2}, where the perturbatively calculated curve is
always below the nonperturbative. At very high $Q^2$, the curves become
indistinguishable. We see again, that the longitudinal part is more strongly
influenced  by the interaction, than the transverse. 
Note also, that the longitudinal part is much smaller than the transverse,
namely of order 20\%.
 
\begin{figure}[t]
\centerline{
  \scalebox{0.43}{\includegraphics*{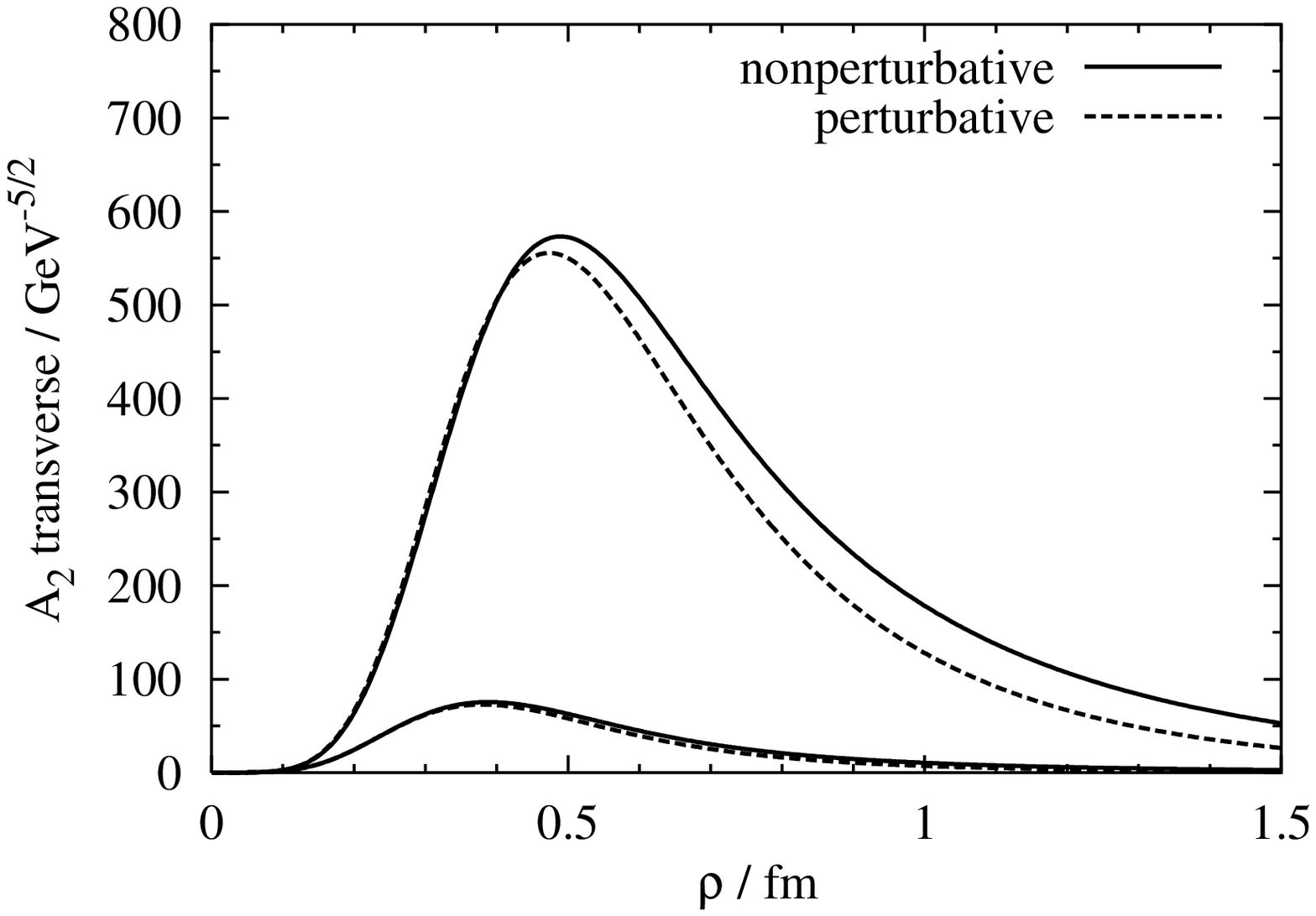}}
  \scalebox{0.43}{\includegraphics*{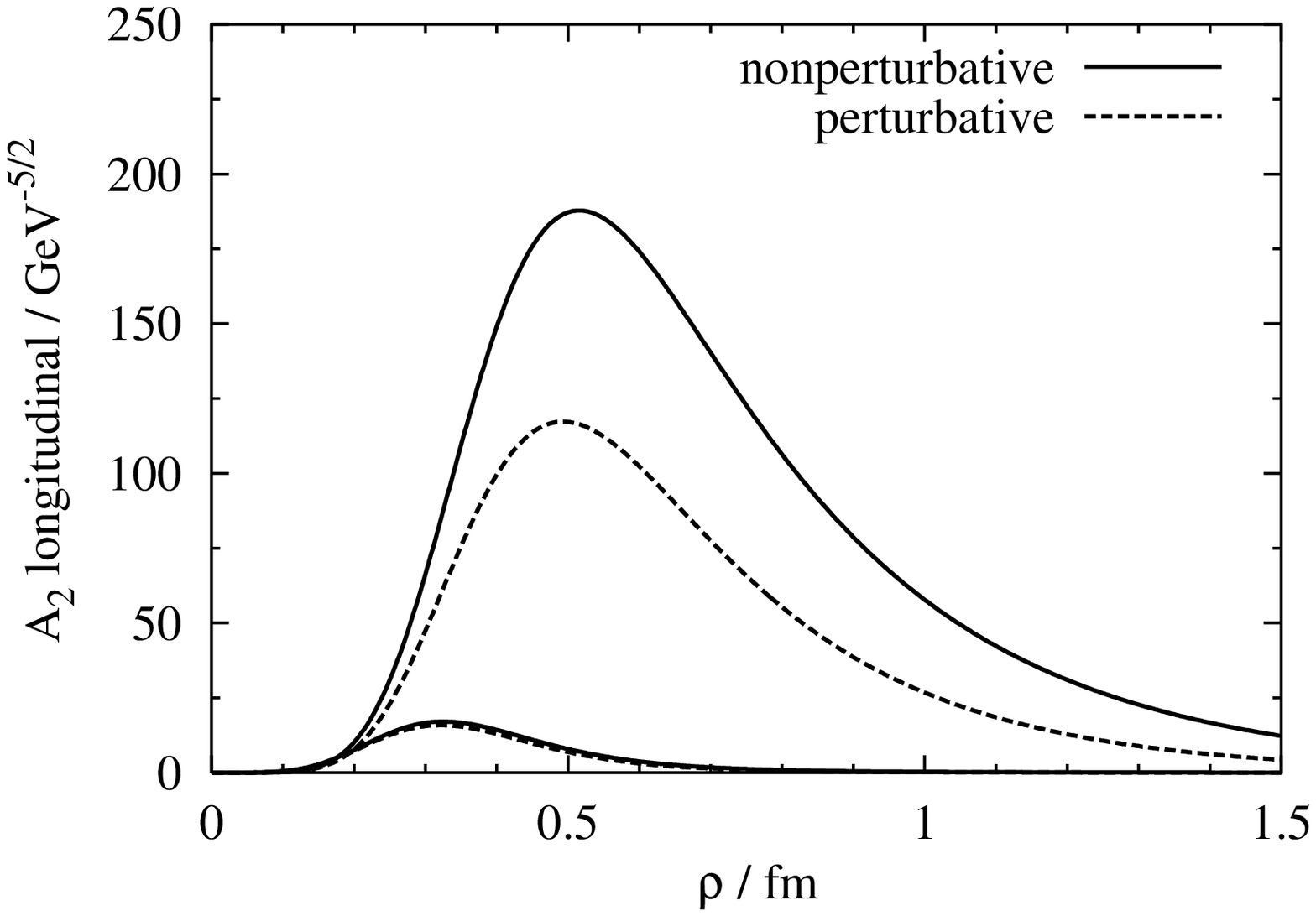}}
}
\caption{
  \label{psizsig2}
  The LC wavefunctions integrated over $\alpha$ and weighted with
  $\sigma^2_{q\bar q}(s,\rho)$, (\ref{psialphasig2}), at
  $\sqrt{s}=200$ GeV. See fig.~\ref{psiz} for more explanations.
}
\end{figure}

We also plot the $\alpha$-integrated LC wavefunction, weighted with
$\sigma^2_{q\bar q}(s,\rho)$. In fig.\ \ref{psizsig2}, we see that the small
distance region is suppressed by an additional factor $\rho^4$, compared to
fig.\ \ref{psiz}. Small distances play numerically no role, if one would
integrate $A_2$ over $\rho$. This illustrates our discussion in section
\ref{diffractionsec}, where we argued that diffraction is dominated by soft 
physics, despite the large virtuality. 

\begin{figure}[ht]
  \centerline{\scalebox{0.8}{\includegraphics{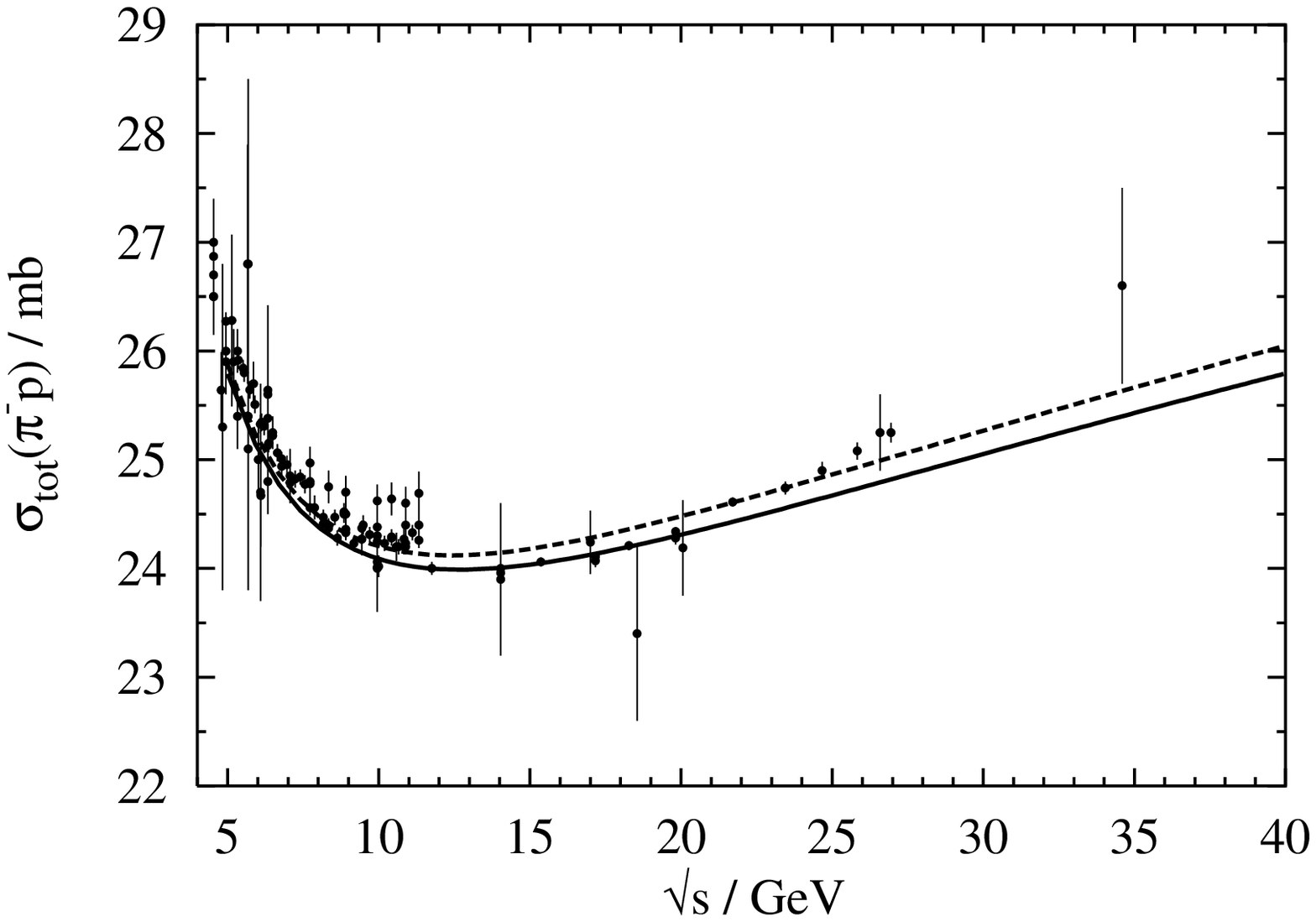}}}
    \caption{
      \label{pion}
      The points show the world data \cite{PDG}
      for the $\pi^-p$ total cross section. 
      The point at $\sqrt{s}\approx 35$ GeV
       is from the
      SELEX collaboration \cite{Uwe}. Only statistical errors are shown.
      The solid
      curve is calculated with the dipole cross section 
      (\ref{improved}), while
      the dashed curve is the fit (\ref{Cudellfit}) from \cite{Cudell}.
    }  
\end{figure}

While the original parametrization of the dipole cross section from \cite{KST99} reproduced
identically the pomeron part of the Donnachie-Landshoff fit, 
the improved parametrization (\ref{improved}) does not have this property.
We therefore check, how well we can reproduce the total $\pi^-p$ cross section
with the wavefunction (\ref{pionwave}). 
Of course we have to add the reggeon exchange contribution from 
(\ref{Cudellfit}),
since the dipole cross section describes only the pomeron part.
The result is shown in fig.\ \ref{pion}.
We see that the cross section is not identically reproduced, as could have been
expected. Our result is slightly smaller than the fit (\ref{Cudellfit}), because
we multiplied the old parametrization (\ref{sasha}) with a function that is
always smaller than $1$. Nevertheless, the pion-proton cross section is
reasonably well reproduced. The cross section calculated from our
parametrization (\ref{improved}) grows with the same power of $s$ at high
energy.
One gets the impression from fig.\ \ref{pion} that the
data grow even faster. At present, there is however no data
available at higher energies than $\sqrt{s}=35$.

\begin{figure}[ht]
\centerline{
  \scalebox{0.43}{\includegraphics{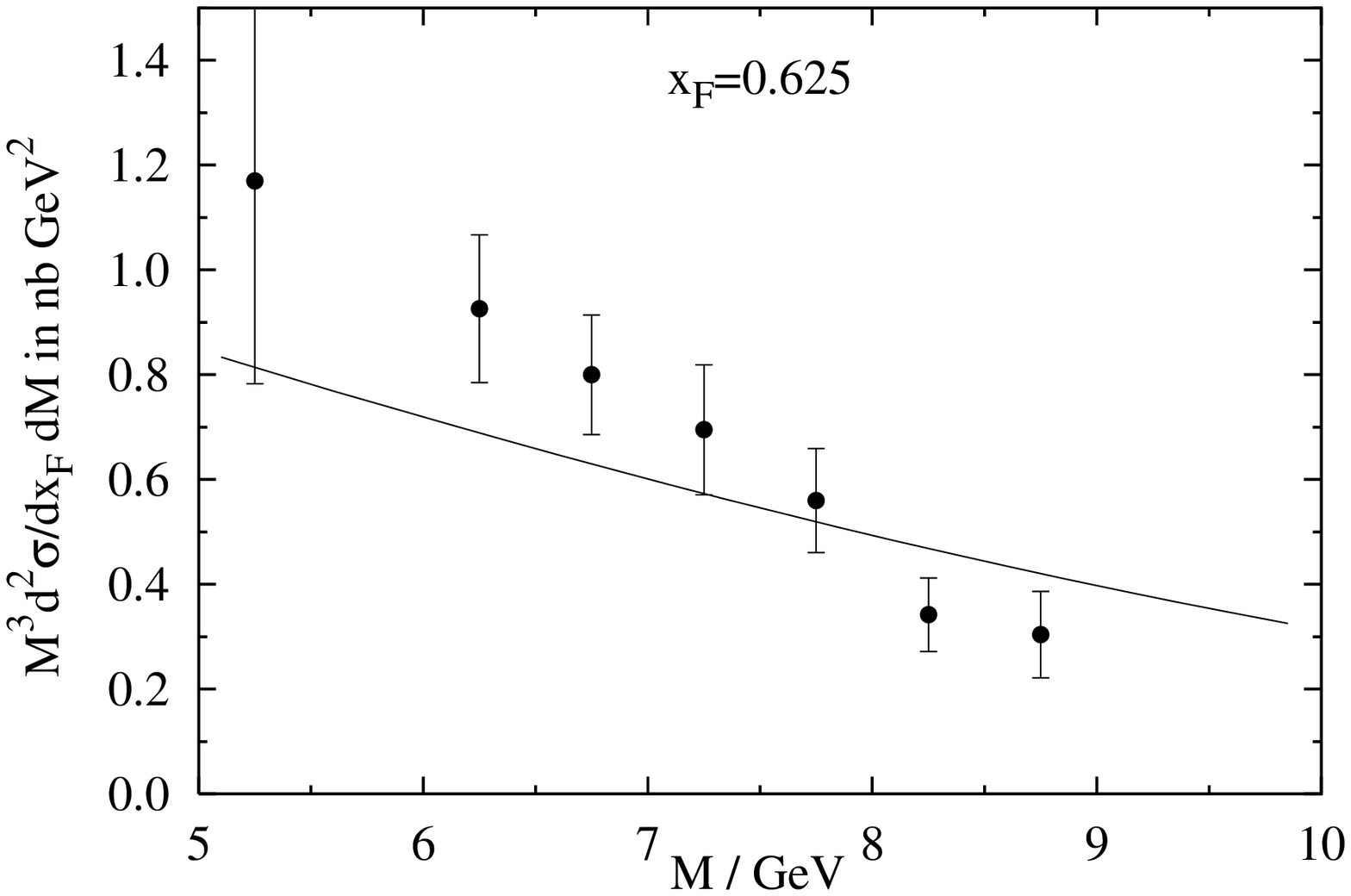}}
  \scalebox{0.43}{\includegraphics{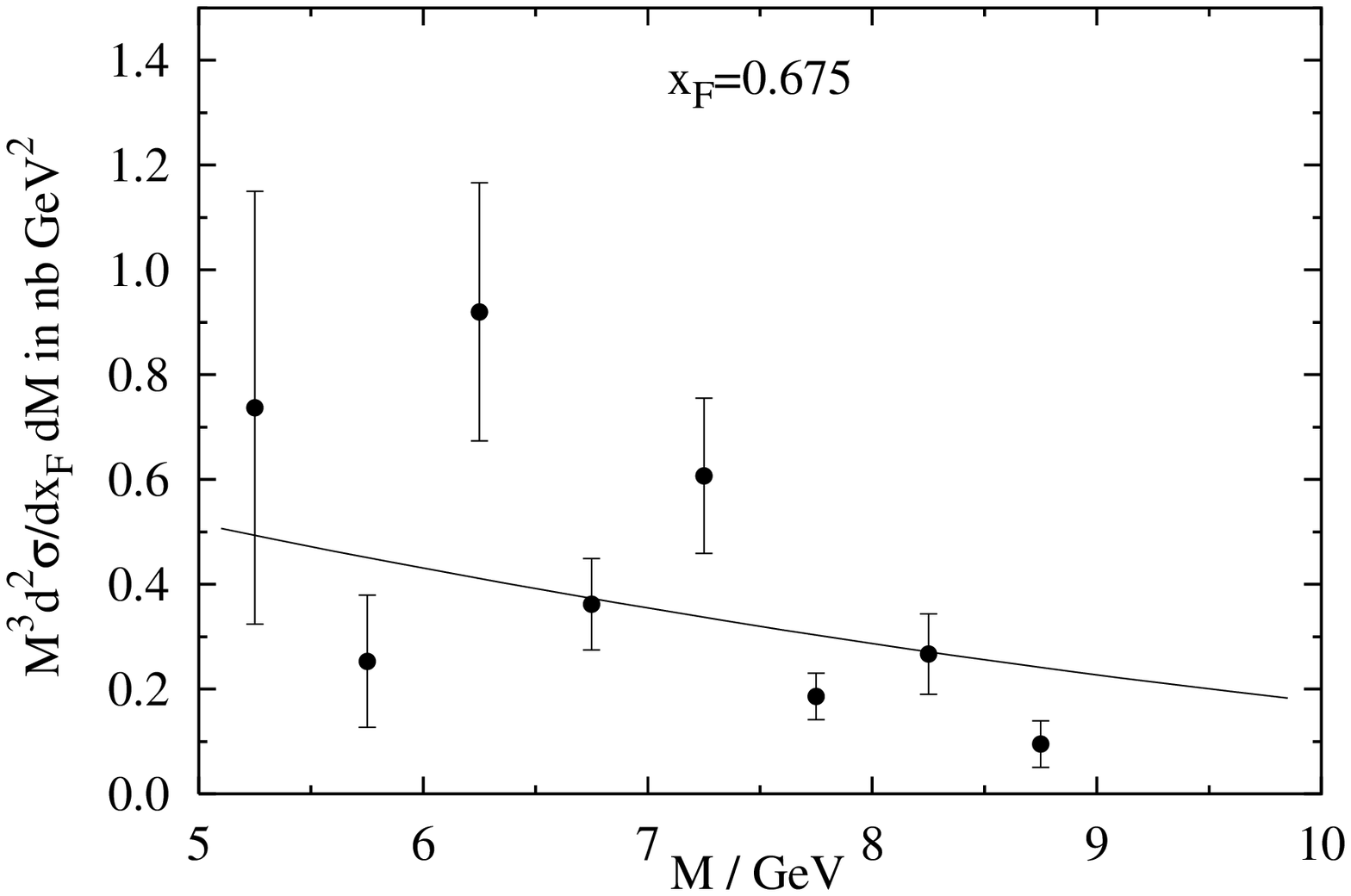}}
}
\caption{
  \label{dytotal}
  The points represent the measured DY cross section from \cite{dydata}. Only
  statistical errors are shown. The curves are calculated with the dipole cross
  section (\ref{improved}) without any further
  fitting procedure.
  In order to make $x_2$
  small, $x_F$ has to be large. In these two figures $0.01\le x_2\le 0.1$.
}
\end{figure}
\begin{figure}[ht]
  \centerline{\scalebox{0.8}{\includegraphics{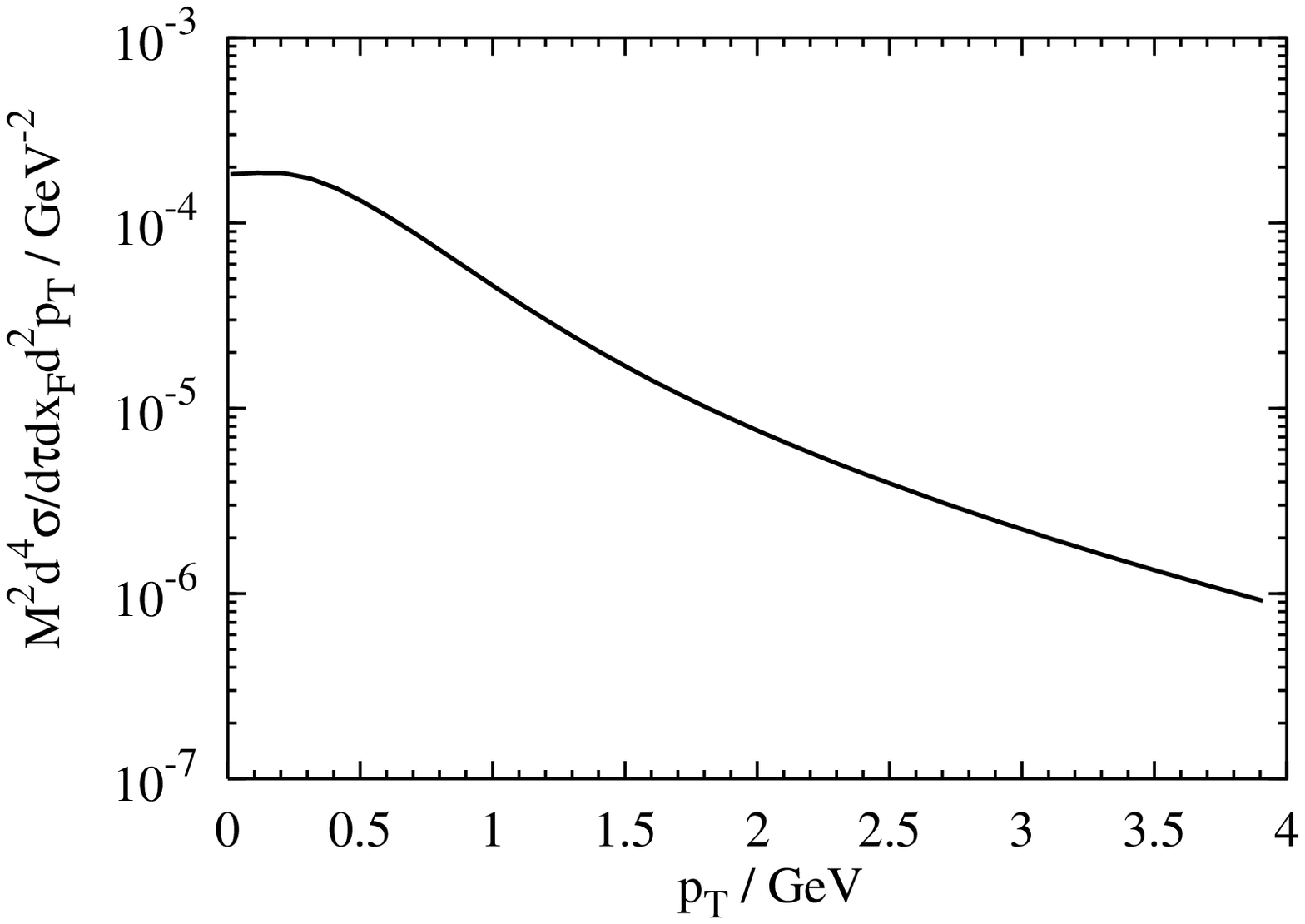}}}
    \caption{
      \label{dyperp}
      The transverse momentum distribution for DY pairs calculated from 
      (\ref{dylcdiffhadr}) with the dipole cross section (\ref{improved}). 
    }  
\end{figure}

We can now proceed and investigate how well our parametrization describes DY
data. The light-cone approach is applicable only at low $x_2$, because it
neglects valence quarks in the target. At present, there are however not many
data for DY cross sections at low $x_2=(\sqrt{x_F^2+4\tau}-x_F)/2$ or at large
$x_F$, respectively. Only the data at the two largest values of $x_F$ from the
E772 collaboration, correspond to values of $0.01\le x_2\le 0.1$. The result of
our calculation, using (\ref{dylctotalhadr}), 
is shown in fig.\ \ref{dytotal}. Although the data scatter a lot, one observes
that the $M$ dependence is only approximately reproduced. We believe that this
can be improved by modifying the dipole cross section. Remarkable is also that the absolute
magnitude of the cross section is quite well reproduced, without introducing a
$K$-factor. Indeed, we believe that it is not legitimate to use a $K$ factor in
our approach, since the dipole cross section is supposed to parametrize all higher order
corrections.

Furthermore, we also calculate the transverse momentum distribution of DY
dilepton pairs from (\ref{dylcdiffhadr}). The result is depicted in fig.\
\ref{dyperp}. The DY cross section is finite at $p_T=0$, in contrast to
the result of first order pQCD. 
We do not compare to data in fig.\ \ref{dyperp}, because all available data are
integrated over $x_F$ and are therefore contaminated by valence quark
contributions. 
An extraction of the low-$x$ part of the transverse momentum distribution
is in progress \cite{private}. Note however, that the differential cross section
is usually parametrized as \cite{Kaplan,Moreno}
\beq\label{familiar}
\frac{d\sigma}{dp_T^2}\propto\frac{1}{[1+(p_T/p_0)^2]^6},
\eeq
while our result decays much weaker and behaves approximately like
\beq
\propto\frac{1}{[1+(p_T/1~{\rm GeV})^2]^2}.
\eeq
The behavior $\propto p_T^4$ is expected from pQCD at large $p_T^2>M^2$ and the
finiteness at $p_T\to 0$ looks encouraging. It is at present not clear to us,
whether a modified dipole cross section could reproduce the familiar shape (\ref{familiar}), or
whether the resummation of soft gluon radiation from the projectile quark 
requires
additional modifications of the approach.

To summarize, our parametrization reproduces well the proton
structure function $F_2$ from the lowest values of $Q^2\approx0.1$ GeV$^2$ 
up to
$Q^2\approx 30$ GeV$^2$, although it does not respect Bjorken scaling.
The low $Q^2$ data is much better described if one employs the nonperturbative
wavefunctions. Also at $Q^2>30$ GeV$^2$, the data is quite well reproduced if
one looks at the plot only by eye.
Since we have parametrized the dipole cross section as function of $s$, rather than $x_{Bj}$, we
can also calculate total meson-proton cross sections, given an appropriate meson
wavefunction. The total $\pi^-p$ cross section is reasonably well reproduced and
also the few available DY data can be described without invoking a $K$-factor.
Only the DY transverse momentum distribution looks different from what one would
expect. 
At present, all data on the transverse momentum distribution is
however integrated over $x_F$, making a comparison with calculations impossible.
The differential
cross section could give much more detailed information about
the dipole cross section than the presently
availabe total cross sections.  
Certainly (\ref{improved}) is not the final answer and has to be improved
further and compared to more data. Especially diffractive DIS data which are
more sensitive to the large distance region  have not been
taken into account so far. We will however compare our predictions for shadowing
with data and nuclear shadowing is closely related to diffraction, as will
be explained in the following section.
  
\section{Nuclear shadowing in DIS and for DY}\label{nucleus}

The typical energy and momentum scales in nuclear physics are
much lower than the virtualities reaches in DIS. One could therefore expect that
using nuclei instead of protons as target would only increase the cross section.
There are however very pronounced nuclear effects in DIS. At intermediate values
of $x_{Bj}$ one observes a suppression of the structure function ratio, fig.\
\ref{scheme}, known as the EMC effect \cite{Arneodo}. 
This effect was unexpectedly discovered
at CERN. It is described theoretically in terms of new degrees of freedom in
nucleus, i.e.\ others than nucleons, see review \cite{fs}. 
We do not discuss the
EMC effect any further and turn our attention to lower values of $x_{Bj}$.
After a small enhancement in the region of $x_{Bj}\approx 0.1$, which is called
antishadowing, one reaches the shadowing region, where
$\sigma^{\gamma^*A}<A\sigma^{\gamma^*p}$ ($A$ is the nuclear mass number), 
i.e.\ 
a considerable depletion of the nuclear structure function is observed.
During the last decade, 
nuclear shadowing has been measured
at CERN (EMC, NMC), 
at Fermilab (E665) and recently also at DESY (HERMES). Values of
$x_{Bj}<0.001$ can be reached, but most of the data are at higher values.  

\begin{figure}[ht]
  \centerline{\scalebox{0.6}{\includegraphics*{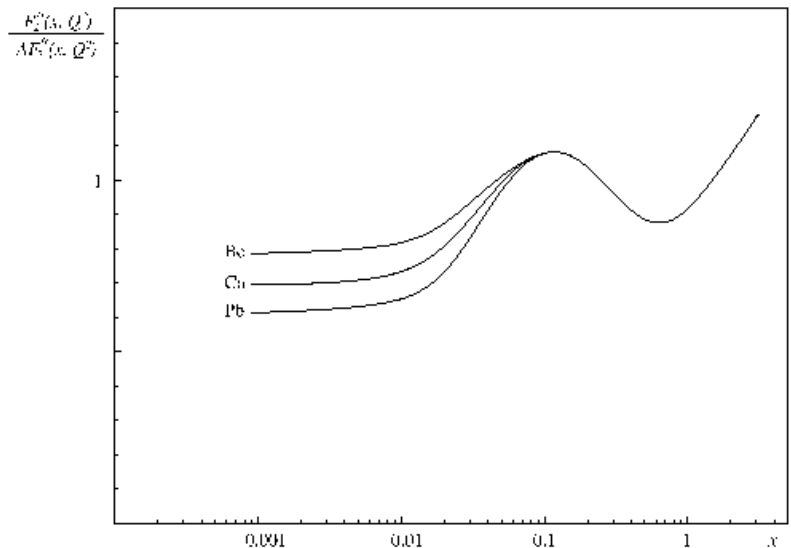}}}
    \caption{
      \label{scheme}
      Schematic plot of nuclear effects in the ratio of the structure functions
      $F_2^A$ of a nucleus and the proton structure function $F_2^N$ vs.\
      $x=x_{Bj}$. 
      The suppression, $F_2^A/AF_2^N<1$, 
      at $x_{Bj}<0.1$ is called nuclear shadowing. See text for
      more details.
      The figure is taken from
      \cite{Povhtalk}.
    }  
\end{figure}

After the pioneering work of Gribov \cite{gribov}, where shadowing
arises from the interaction of virtual photons at the surface of the nucleus,
this effect was not
unexpected.  
Nuclear shadowing 
is of high theoretical interest, because it 
can be understood in terms of the fundamental degrees of
freedom of QCD, quarks and gluons. In the preceding sections the
ground for an investigation of nuclear shadowing in terms of the color dipole
picture was prepared. 
In this part of the work, an improved version of the
Glauber-Gribov theory is developed, 
which includes the nuclear form factor in the usual
Glauber series. The interplay of multiple scattering and coherence are discussed
in detail. 

\subsection{Shadowing and Diffraction}\label{shadowdiffr}

The partonic interpretation of shadowing depends on the reference frame. In the
infinite momentum frame of the target, shadowing is due to parton fusion
\cite{kancheli} - \cite{q}, which
leads to a reduction of the parton density at low $x_{Bj}$.
This can be understood intuitively in the following way \cite{Levintalk}:
In DIS a picture of the microscopic substructure of the target is taken. 
However,
this cannot be done in an infinitely short time. The uncertainty relation 
tells us that the time needed increases as $x_{Bj}$ becomes small. This can be
seen from the energy denominator in (\ref{cohlength}) on page
\pageref{cohlength}. More precisely, 
the role of time is played by the logarithm
of $x_{Bj}$, as can be seen from the double leading log DGLAP equation,
\beq
\frac{\partial^2x_{Bj}G(x_{Bj},Q^2)}{\partial\ln(1/x_{Bj})\partial\ln Q^2}
=\frac{N_c\alpha_s}{\pi}x_{Bj}G(x_{Bj},Q^2).
\eeq
For fixed strong coupling constant $\alpha_s$, the solution behaves
asymptotically like (\ref{asymptotic}). While a fast moving nucleus is
contracted to a pancake shape, the cloud of partons with very small momentum
fraction is much less contracted and parton clouds from different nucleons will
overlap. As the density becomes higher and higher with decreasing $x_{Bj}$, the
partons start to feel the neighborhood of other partons and fusion processes
like $GG\to G$ and $GG\to q\bar q$ will occur. These fusion processes reduce the
parton density in a nucleus compared to a free nucleon.

A very intuitive picture arises in the rest frame of the nucleus,
where the same phenomenon looks like
nuclear
shadowing
of hadronic fluctuations of the virtual
photon
\cite{PillerRev,weise,nz91},\cite{bauer} - \cite{kp2}.
Shadowing occurs, because of multiple scattering of the hadronic fluctuation
inside the target. 
The nucleons at the surface of the nucleus cast a shadow on the inner nucleons,
which then have less chances to interact with the photon. Multiple scattering 
at high energies is commonly described in the optical model of Glauber-Gribov
theory. The basics of this approach have been formulated a long time ago by
Glauber \cite{Glauber} within nonrelativistic quantum mechanics. The optical
model is usually derived in eikonal approximation with the assumption that the
phases obtained in different scatterings are additive. Of course, one would like
to have a connection to quantum field theory and the Feynman diagram technique,
which allows to include relativistic kinematics. This has been elaborated 
by Gribov \cite{gribov}. However, the relation between Glauber theory and
Feynman diagrams has been fully explored only for the double scattering term.
The theory has been developed by many authors since then. In \cite{Bertocchi},
Gribovs approach is generalized to non forward scattering and an expression is
obtained which also includes inelastic transitions. For a discussion of
Glauber-Gribov theory in the context of Regge phenomenology see for instance
\cite{Weis}.

We give a short summary of the Glauber-Gribov theory, following
\cite{Bertocchi}. For simplicity, only forward scattering is considered, 
although
the approach can be generalized to include also small momentum transfers. 
All correlations between the
nucleons will be neglected in the following. 
The total amplitude for scattering off a nucleus is written as a sum
over $n$-fold scattering amplitudes.
\beq
F(s)=\sum_{n=1}^AF^{(n)}(s),
\eeq
where $s$ is the photon-nucleon {\em c.m.} 
energy squared and $F^{(n)}(s)$ is the amplitude
for an $n$-fold scattering. It can be written as an integral over impact
parameter ($b$) space and over the longitudinal coordinates $z_i$ of the
scatterings,
\beqn\nonumber\label{mscatamp}
F^{(n)}(s)&=&\left(\frac{{\rm i}}{K}\right)^{n-1}
\int d^2b
\int\limits_{-\infty}^{\infty}dz_1\rho_A(b,z_1)
\int\limits_{z_1}^{\infty}dz_{2}\rho_A(b,z_2)
\dots
\int\limits_{z_{n-1}}^{\infty}dz_{n}\rho_A(b,z_{n})\\
&\times&\sum_{\{h\}}
f_{\gamma^*\to h_1}e^{-{\rm i}q_{L\gamma^*\to h_1}(z_2-z_1)}
f_{h_1\to h_2}e^{-{\rm i}q_{Lh_1\to h_2}(z_2-z_3)}
\dots f_{h_{n-1}\to\gamma^*}.
\eeqn
Here, $K\approx 2s$ is the flux factor.
The expression (\ref{mscatamp}) 
is illustrated in fig.\ \ref{mscat}. The nuclear density 
$\rho_A$, which is normalized to $A$, enters
after integration over the wavefunction of the nucleus, neglecting correlations.
It is assumed that the impact parameter $b$ does not change in a collision. The
limits of the integrations over the longitudinal coordinates $z_i$ originate
from the condition that the $(i+1)$th collision
has to take place after the $i$th one. After the first collision, the virtual
photon converts into a multiparticle hadronic state $h_1$. 
The amplitude for this process is denoted by $f_{\gamma^*\to h_1}$. Furthermore,
the longitudinal momentum transfer $q_{L\gamma^*\to h_1}$ appears as an
oscillating phase factor for each intermediate state. One has
\beq
q_{L\gamma^*\to h_1}=\frac{Q^2+M_{h_1}^2}{2\nu},
\eeq
where $M_{h_1}$ is the invariant mass of the hadronic state $h_1$.
Finally one has to sum over all states $h_i$ that can be reached by multiple
scattering. According to the optical theorem, 
the total cross section is then given by
\beq
\sigma^{\gamma^*A}(s)=\frac{2F(s)}{K}.
\eeq

\begin{figure}[ht]
  \centerline{\scalebox{0.6}{\includegraphics{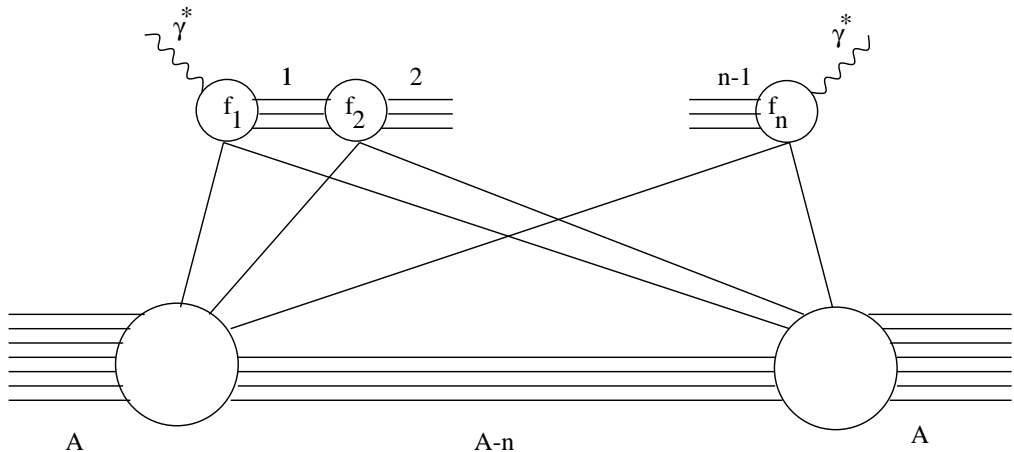}}}
    \caption{
      \label{mscat}
      Illustration of the $n$-fold scattering amplitude $F^{(n)}(s)$
      (\ref{mscatamp})
      with intermediate hadronic states
      $1\dots n-1$. 
      The upper part of the figure corresponds to the photon, which converts in
      different hadronic states due to multiple rescattering. The lower part
      corresponds the nucleus, where each line represents a nucleon.
      The amplitude for the transition $i\to j$ is denoted by
      $f_j$ in this figure.  
    }  
\end{figure}

Straightforward calculations with (\ref{mscatamp}) are however not possible,
because the intermediate states which can occur in an $n$-fold scattering are
not known.
Therefore, 
one is forced to introduce additional approximations. The idea to represent
the virtual photon as superposition of hadrons with the same quantum numbers,
namely vector mesons, leads to (generalized) vector dominance model (G)VDM
\cite{gribov,Shaw}. The advantage of this approach is that the masses of vector
mesons are well defined, in contrast to the mass of a $q\bar q$ dipole in the
mixed representation. The masses of the intermediate states have to be known,
since they enter the longitudinal momentum transfers $q_L$. On the other hand it
is not known, how to calculate scattering amplitudes, higher than the
double scattering term, because vector mesons are no eigenstates of the
interaction. A possible prescription would be to omit all off diagonal
transitions which is called the diagonal approximation. Note that in the 
VDM, which is the diagonal approximation of the GVDM,
one has to choose a vector meson-nucleon cross section that behaves as
$\sigma_{Vp}\propto 1/M^2_V$, where $M_V$ is the mass of the vector meson,
in order 
to reproduce the Bjorken scaling of the proton structure function. With a 
nearly
constant $\sigma_{Vp}$, $F_2^p$ would increase linearly with $Q^2$.
However, with such a fast decreasing meson-nucleon cross section, it 
turns out that nuclear shadowing is a higher twist effect.
It was indeed believed for quite a long time that shadowing vanishes at large
$Q^2$, until it was shown experimentally that this is not the case \cite{EMC}.
This phenomenon is sometimes called the Gribov paradox \cite{paradox}.
The solution is that one also has to take into account off-diagonal transitions,
which correspond to diffraction dissociation. These off-diagonal transitions
interfere destructively with the diagonal ones and one can reproduce the Bjorken
scaling with an $M_V$ independent $\sigma_{Vp}$
\cite{Schild}. The diagonal VDM is applicable only at low $Q^2<1$ GeV$^2$, where
the proton structure function indeed rises approximately linearly, see fig.\
\ref{f2pq2}.

The close connection between shadowing and diffraction becomes most transparent
in the formula derived by Karmanov and Kondratyuk \cite{kk}. In the double
scattering approximation, the shadowing correction can be related to the
diffraction dissociation spectrum, integrated over the mass,
\beqn\nonumber
{\sigma^{\gamma^*A}}
&\approx& A\sigma^{\gamma^*p}-
{8\pi}\,
\int d^2b
\int\limits_{-\infty}^{\infty}dz_1\rho_A(b,z_1)
\int\limits_{z_1}^{\infty}dz_{2}\rho_A(b,z_2)\\
&&\qquad\times
\int dM^2_X\,\left.\frac{d\sigma(\gamma^*N\to XN)}
{dM^2_X\,dt}\right|_{t\to 0}\cos\left(-{\rm
i}\frac{Q^2+M_X^2}{2\nu}(z_2-z_1)\right)
\label{KKK}\\
&=& A\sigma^{\gamma^*p}-{4\pi}\,
\int d^2b
\int dM^2_X\,\left.\frac{d\sigma(\gamma^*N\to XN)}
{dM^2_X\,dt}\right|_{t\to 0}F_A^2(l_c,b).
\eeqn
Note that the kinematical limit for $t$ is slightly larger than zero.
This result is easily obtained from (\ref{mscatamp}), when one neglects the
small real part of the scattering amplitude $F(s)$.
Here
\beq\label{ff}
F_A^2(l_c,b)=\left|\int\limits_{-\infty}^{\infty}dz\rho_A(b,z)
{\rm e}^{{\rm i}z/l_c}\right|^2
\eeq
is the formfactor of the nucleus. 
Again the coherence length 
\beq\label{length}
l_c=\frac{2\nu}{Q^2+M_X^2}
\eeq
emerges.
This length is of
course already present in (\ref{mscatamp}).
The Karmanov-Kondratyuk equation is illustrated in fig.~\ref{vmd}. The first
term in  (\ref{KKK}) is the single scattering term, which is proportional to
$A$.
The second term represents the double scattering contribution and 
is responsible for shadowing.  
To model the influence of higher order rescattering terms, the shadowing
correction is often multiplied by the term
\beq
\exp\left(-\frac{\sigma_{eff}}{2}\int dz\rho_A(b,z)\right),
\eeq
with the effective absorption cross section $\sigma_{eff}\sim 20$ mb for quark
shadowing.
Note that at very high energies, the
coherence length becomes infinite and the oscillating exponential 
in the nuclear form factor is practically
unity which corresponds to maximum shadowing. 
Of course, the shadowing term 
still grows with energy, because of the diffractive cross
section. 
As the energy becomes lower, the
exponential oscillates more and more rapidly and the shadowing correction
vanishes. This observation makes the importance of $l_c$ very clear. 

\begin{figure}[t]
  \centerline{\scalebox{0.68}{\includegraphics*{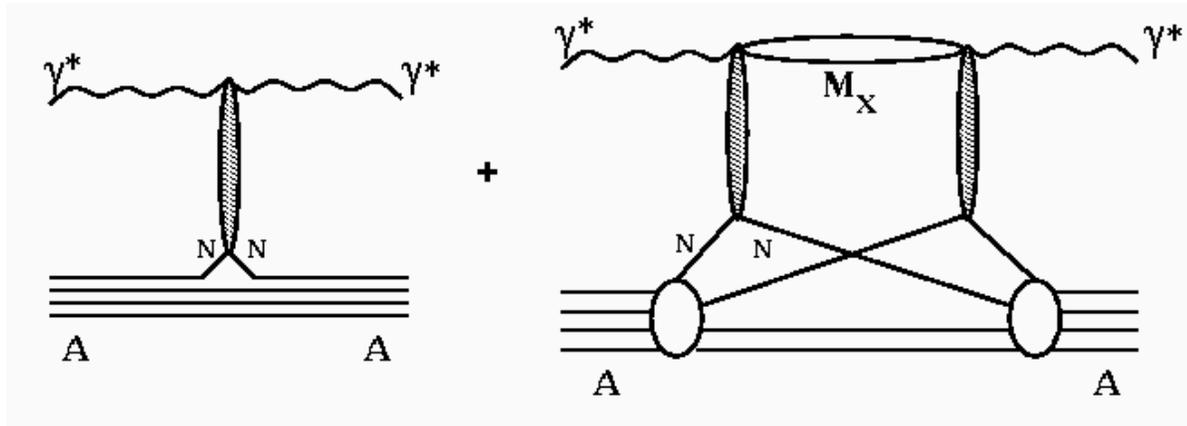}}}
    \caption{
      \label{vmd}
      Illustration of the Karmanov-Kondratyuk formula (\ref{KKK}).
      The left figure shows the single scattering contribution to DIS off
      nuclei. The double scattering term, shown on the right, 
      gives the driving contribution to
      nuclear shadowing. It scales like $A^\frac{4}{3}$.  
    }  
\end{figure}

A 
simple estimate that demonstrates
the relation between shadowing and
diffraction is performed in \cite{Ferreira}, see also \cite{PillerRev}.
In \cite{Ferreira}, an infinite coherence
length is assumed and the diffractive cross section is parametrized in the
way,
\beq
\int dM^2_X\,\left.\frac{d\sigma(\gamma^*N\to XN)}
{dM^2_X\,dt}\right|_{t\to 0}
=B\sigma^{DD}_{\gamma^*p}.
\eeq
The diffractive cross section  
$\sigma^{DD}_{\gamma^*p}\approx 0.1\sigma^{\gamma^*p}$ makes about 10\% of the
total cross section and the slope parameter which describes the $t$-dependence
of diffraction is adjusted to
$B\approx 8$ GeV$^{-2}$.
Although no higher order scattering terms are taken into account, the
authors of \cite{Ferreira}  obtain good agreement  
with experimental data at very low $x_{Bj}$. 
This analysis confirms the physical
picture that shadowing in DIS is governed by coherent interaction of
diffractively produced states in the nucleus, fig.\ \ref{vmd}. 

From this discussion, two fundamental requirements emerge, which have to be
fulfilled for shadowing.
\begin{itemize}
\item{The mean free path of the hadronic fluctuation is long enough to have
multiple scatterings. More quantitatively it must hold
\beq
\frac{1}{\rho_A\sigma_{eff}}<R_A,
\eeq
where $R_A$ is the nuclear radius.}
\item{The coherence length $l_c$ has to be bigger than the mean inter-nucleon
distance, $l_c>2$ fm. This condition ensures that the virtual photon coherently
scatters  off the nucleons.}
\end{itemize}

One can also estimate, which masses $M_X$ can contribute to shadowing. Note that
since
\beq
\frac{d\sigma(\gamma^*N\to XN)}
{dM^2_X\,dt}\Big|_{t\to 0}\propto\frac{1}{M_X^2}
\eeq
for large masses, one formally has to integrate the spectrum up to the
kinematical limit.
However, due to coherence length effects, large masses cannot contribute to
shadowing. This can be seen in the following way. The coherence length
can be written as
\beq
l_c=\frac{\beta}{m_Nx_{Bj}}=\frac{1}{m_Nx_{\cal P}}
\eeq
with
\beq
\beta=\frac{Q^2}{Q^2+M_X^2}.
\eeq
One finds from the second condition that only masses
\beq\label{estimate}
M_X^2<0.1\,s
\eeq
are relevant for shadowing. Note that this quantity can become much larger than
$Q^2$ at low $x_{Bj}$. It is however small compared to $s$.

In the VDM and for the Karmanov-Kondratyuk formula, the multiple scattering
amplitude (\ref{mscatamp})
is evaluated in a hadronic basis. The disadvantage of using
this basis is that one cannot calculate higher order scattering terms in an
unambiguous way. 
We can however make use of our knowledge of the interaction eigenstates, namely
partonic configurations frozen in impact parameter space. No off-diagonal
transitions occur in this basis. One encounters however another severe problem, 
the mass of the eigenstates, which one has to know for the coherence length,
 is undefined in the mixed representation and one cannot give an 
explicit expression for the scattering amplitude. Only in the limit
$l_c\to\infty$, it is possible to resum the whole multiple scattering series
\cite{Bertocchi} in an eikonal-formula,
\beqn\nonumber
\sigma^{\gamma^*A}&=&2\int d^2b\int_0^1 d\alpha\int d^2\rho
\left|\Psi(\alpha,\rho)_{q\bar q}\right|^2\left(
1-\exp\left(-\frac{\sigma_{q\bar q}(s,\rho)}{2}T(b)\right)\right)\\
\nonumber
&+&2\int d^2b\int_0^1d\alpha\int \frac{d\alpha_G}{\alpha_G}
\int d^2\rho_1 d^2\rho_2
\left|\Psi_{q\bar qG}(\alpha,\alpha_G,\vec\rho_1,\vec\rho_2)\right|^2\\
&&\qquad\qquad\times
\left(
1-\exp\left(-\frac{\sigma_{q\bar qG}(s,\vec\rho_1,\vec\rho_2)}{2}
T(b)\right)\right).
\eeqn
Summation over photon polarizations is understood.
The nuclear thickness function 
$T(b)=\int_{-\infty}^{\infty}dz\,\rho_A(b,z)$ is the
integral of
nuclear density over longitudinal
coordinate $z$ and depends on the
impact parameter $b$.
The light-cone wavefunctions are given by (\ref{psit}) and (\ref{psil})
on page \pageref{psit} for the free case and by (\ref{psitnpt}) 
and (\ref{psilnpt})
on page \pageref{psitnpt} including the nonperturbative interaction.
The LC wavefunction 
for the gluonic component is given 
by (\ref{psigluon}) on page \pageref{psigluon}.
The cross section for the three body system $q\bar qG$ 
(\ref{sigmaqqG}) can be expressed in terms
of the dipole cross section.
In the following more quantitative discussion, the 
$q\bar qG$-term is dropped and 
for the $q\bar q$-part the shorthand notation
\beq\label{eikonalappr}
\sigma^{\gamma^*A}=\left< 2\int d^2b\left(
1-\exp\left(-\frac{\sigma_{q\bar q}(s,\rho)}{2}T(b)\right)\right)\right>
\eeq
is introduced.
The condition $l_c\gg R_A$
insures that the $r$ does not vary during propagation
through the nucleus (Lorentz time dilation) and one can apply the eikonal
approximation.
Note that the averaging of the whole exponential
in
(\ref{eikonalappr})
makes this expression different from the Glauber eikonal
approximation where $\sigma_{q\bar q}(s,\rho)$ is averaged in
the exponent,
\beq
2\int d^2b\left(
1-\exp\left(-\frac{\sigma^{\gamma^*p}}{2}T(b)\right)\right).
\eeq
The difference is known as Gribov's inelastic
corrections
\cite{gribov}.
In the case of DIS the Glauber approximation does not
make sense and the whole cross section is due to
the inelastic shadowing. Indeed, $\sigma^{\gamma^*p}$ is at most of order $100$
$\mu$b, for real photons, which
 corresponds to a mean free path in nuclear matter of 
$l_f\approx 600$
fm. This means that the bare photon does practically not interact and therefore
no multiple scattering can occur. The photon interacts only via its hadronic
components.

The condition $l_c\gg R_A$ is however not fulfilled in most of the experiments.
Only for the lowest values of $x_{Bj}$ in the E665 experiment, 
this condition is met.
For the case $l_c \sim R_A$, one has to take into account the
nuclear form factor, i.e.\ the variation of
$\rho$ during the propagation of the $q\bar q$ fluctuation through
the nucleus. This can be done for the $q\bar q$-component of the photon and in
double scattering approximation \cite{kp,kp2},
\beq\label{doublescat}
\sigma^{\gamma^*A}
\approx A\sigma^{\gamma^*p}
-\frac{1}{4}\,
\la\sigma^2(s,\rho)\ra
\int
d^2b\,
F_A^2(l_c,b).
\eeq
It is not well defined what one has to take as argument $l_c$ of the formfactor
(\ref{ff}). The usual prescription is to 
set $M_{q\bar q}^2=Q^2$ which leads to
$l_c=1/(2m_Nx_{Bj})$, cf.\
discussion on page \pageref{cohlength}. 
In the limit $l_c\to\infty$, (\ref{doublescat}) can be obtained from expanding
the exponential in (\ref{eikonalappr}). 

Like (\ref{KKK}), (\ref{doublescat}) also resembles the close connection between
shadowing and diffraction. 
The connection between these two equations is given by (\ref{diffraction}) on
page \pageref{diffraction}.
In section \ref{diffractionsec} it 
is discussed in detail that only large size pairs give a leading
twist contribution to the diffractive cross section. Similarly, 
also shadowing is
a soft reaction, although $Q^2$ is large. The leading twist contribution to
shadowing for longitudinal photons comes from the $q\bar qG$-Fockstate. The
$q\bar q$ contribution vanishes with a higher power of $Q^2$.
The fact that only large size configurations contribute to shadowing is the
solution of the Gribov paradox in the dipole language. 
It follows from the first condition for shadowing
that the pair has to scatter at least with a cross
section of $10$ mb to be shadowed in a large nucleus. This
quantity can be several times larger for a light nucleus, like carbon.
In the VDM all hadronic
fluctuations of the photon interact with a large cross section, which leads to a
structure function $F_2$ that is proportional to $Q^2$. However, VDM cannot be
applied at large $Q^2$, since then most 
of the $q\bar q$
fluctuations interact with small cross
sections and this produces the scaling of $F_2$. Only the very unlikely Bjorken
aligned jet configurations make shadowing a leading twist effect.

To summarize, two problems remain under discussion
\begin{itemize}
\item{How can the nuclear formfactor can be included 
in the higher order scattering terms which are of great importance for heavy
nuclei?  For instance, the shadowing term in (\ref{eikonalappr}) for
lead is half as big as $A\sigma^{\gamma^*p}$, so the need of the
higher order terms is obvious.}
\item{
Even for the double scattering term
it is still unclear which argument should 
enter the formfactor.  Indeed, the effective mass of the $q\bar q$
fluctuation needed for the coherence length in (\ref{length})
cannot be defined in a representation with a definite
$q\bar q$ separation. On the other hand, (\ref{KKK})
exhibits an explicit dependence on $M_X$ and the
longitudinal momentum transfer is known. However, unknown
in this case is the effective cross section $\sigma_{eff}$.}
\end{itemize}
A solution to these problems is proposed
in the next subsection.

\subsection{Light-cone approach to nuclear shadowing}\label{lightcone}

The goal of this subsection is to give a physical explanation of the 
light-cone
approach to nuclear shadowing which includes the nuclear form factor in all
multiple scattering terms. A more formal derivation is given in 
section \ref{derivation}. 
In order to study the difference between the correct quantum
mechanical treatment of nuclear shadowing and known approximations, we restrict
ourselves to the $q\bar q$ Fock component of the photon, neglecting gluons.
The nuclear antishadowing effect is omitted as well, since we believe it is
beyond the shadowing dynamics. It might e.g.\ 
be caused by bound nucleon swelling. 
Like in (\ref{eikonalappr}) the total cross section 
is represented in the form
\beq\label{form}
\sigma^{\gamma^*A}=A\sigma^{\gamma^*p}-\Delta\sigma,
\eeq
where $\Delta\sigma$ is the shadowing correction,
\beqn\nonumber\label{correction}
\Delta\sigma&=&\frac{1}{2}{\rm Re}\int d^2b
\int\limits_{-\infty}^{\infty} dz_1\rho_A(b,z_1)
\int\limits_{z_1}^{\infty} dz_2\rho_A(b,z_2)\\
&\times&
\int_0^1d\alpha\int d^2\rho_2
\Psi_{q\bar q}^*(\vec\rho_2,\alpha)\sigma_{q\bar q}(s,\rho_2)
A(\vec\rho_2,z_1,z_2,\alpha),
\eeqn
with
\beq\label{propagation}
A(\vec\rho_2,z_1,z_2,\alpha)=\int\,d^2\rho_1\,
W(\vec \rho_2,z_2|\vec \rho_1,z_1)\,
{\rm e}^{-{\rm i}q_L^{min}(z_2-z_1)}\,\sigma_{q\bar q}(s,\rho_1)\,
\Psi_{q\bar q}(\vec \rho_1,\alpha).
\eeq
Here,
\beq
q_L^{min}=\frac{Q^2\alpha(1-\alpha)
+m_f^2}{2\nu\alpha(1-\alpha)}
\eeq
is the minimal longitudinal momentum transfer when the photon splits into the
$q\bar q$ dipole. These equations 
were first suggested by Zakharov in the context
of the LPM effect \cite{zakh}. Indeed, 
as already mentioned in the introduction,
shadowing in DIS can be regarded in the target rest frame
as the LPM effect for pair production.

\begin{figure}[ht]
\scalebox{0.9}{\includegraphics{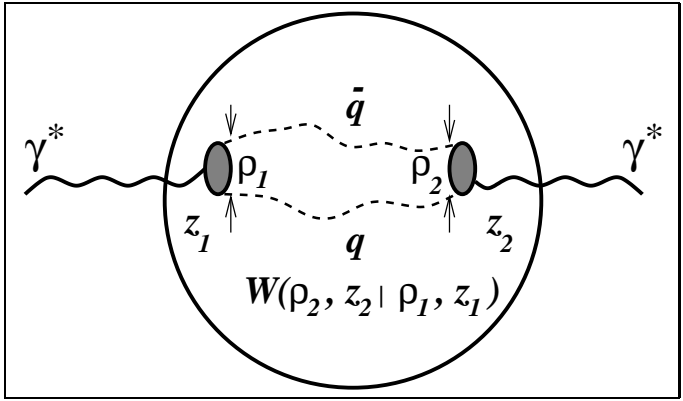}}\hfill
  \raise25mm\hbox{\parbox[b]{2.3in}{
\caption{\label{propag}
Propagation of a $q\bar q$-pair through a nucleus. Shown is the case of
a finite coherence length, where the transverse motion is described by the
Green function $W\left(\vec\rho_2,z_2|\vec\rho_1,z_1\right)$.}}}
\end{figure}

The shadowing term in (\ref{form}) is
illustrated
in fig.\ \ref{propag}.  
At the point $z_1$ the photon diffractively produces the $q\bar q$
pair ($\gamma^*N\to q\bar qN$) with a transverse separation $\vec\rho_1$.
The pair propagates through the nucleus along arbitrarily curved
trajectories, which are summed over, and arrives at the point
$z_2$ with a separation $\vec\rho_2$.  The initial and the
final separations are controlled by the light-cone wavefunction 
$\Psi_{q\bar q}^{T,L}(\vec \rho,\alpha)$.  While passing the nucleus
the $q\bar q$ pair interacts with bound nucleons
via the cross section $\sigma_{q\bar q}(s,\rho)$ which depends on the local
separation $\vec\rho$.  The Green function 
$W(\vec\rho_2,z_2|\vec\rho_1,z_1)$ describes the propagation of the pair from
$z_1$ to $z_2$, see (\ref{propagation}).
One recognizes the diffractive amplitude (\ref{diffamp})
\beq
f(\gamma^*\to q\bar q)={\rm i}\Psi_{q\bar q}(\vec \rho_1,\alpha)
\sigma_{q\bar q}(s,\rho_1).
\eeq
At the position $z_2$, the result of the propagation is again projected onto the
diffractive amplitude. 
The Green function
includes that part of the phase shift between the
initial and the final photons which is due to transverse
motion of the quarks, while the longitudinal motion is included in
(\ref{propagation}) via the exponential.

Thus, (\ref{correction}) does not suffer from either of the two
problems of the approximations (\ref{KKK}), (\ref{eikonalappr})
and (\ref{doublescat}). The longitudinal momentum transfer is known and all the
multiple interactions are included.

The Green function $W(\vec \rho_2,z_2;\vec\rho_1,z_1)$ in
(\ref{propagation}) satisfies the two dimensional Schr\"o\-din\-ger equation,
\beqn\label{schroedinger}
{\rm i}\,\frac{{\partial}W(\vec\rho_2,z_2|\vec\rho_1,z_1)}{{\partial}z_2}
&=&
-\frac{\Delta(\rho_2)}{2\nu\alpha(1-\alpha)}\,
W(\vec\rho_2,z_2|\vec\rho_1,z_1)\nonumber\\
&-&{{\rm i}\over2}\,\sigma(s,\rho_2)\,\rho_A(b,z_2)\,
 W(\vec\rho_2,z_2|\vec\rho_1,z_1)
\eeqn
with the boundary condition $W(\vec\rho_2,z_1|\vec\rho_1,z_1)
=\delta^{(2)}(\vec \rho_2-\vec \rho_1)$.  The Laplacian $\Delta(\rho_2)$ acts
on the coordinate $\vec\rho_2$.   
The kinetic term $\Delta/[2\nu\alpha(1-\alpha)]$ in this Schr\"odinger equation
takes care of the varying effective mass of the $q\bar q$ pair
and provides the proper phase shift. 
Indeed, the term $\nu\alpha(1-\alpha)$ can be regarded as 
the reduced mass of the pair.
The role of time is played by the
longitudinal coordinate $z_2$.
The imaginary part of the optical potential describes the absorptive process.
The nonrelativistic appearance of (\ref{schroedinger}) comes from the fact that
the energy in light cone coordinates can be written as
\beq
P^-=\frac{P_\perp^2+m^2}{P^+}.
\eeq
One recognizes that the two dimensional Laplacian originates from the
$P_\perp^2$ and that $P^+/2$ plays the role of the mass in this two dimensional
Schr\"odinger equation.

The Green function method contains the eikonal
approximation (\ref{eikonalappr}) and the Karmanov-Kondratyuk formula 
(\ref{KKK}) as limiting cases.

In order to obtain the eikonal approximation, one has to take the limit
$\nu \to\infty$. In this case, the kinetic term in
(\ref{schroedinger}) can be neglected and one obtains
\beq
\left.W(\vec\rho_2,z_2|\vec\rho_1,z_1)\right|_{\nu\to
\infty}=\delta^{(2)}(\vec\rho_2-\vec\rho_1)\,
\exp\left[-\frac{\sigma(s,\rho_2)}{2}
\int\limits_{z_1}^{z_2}dz\,\rho_A(b,z)\right].
\eeq
When this expression is
substituted into (\ref{propagation}) and with $q_L^{min}\to 0$ one arrives 
after a short calculation at the result (\ref{eikonalappr}).
 
One can also recover the Karmanov-Kondratyuk formula, 
when one neglects 
the absorption of the $q\bar q$  pair in the medium, i.e.\  
the imaginary potential in (\ref{schroedinger}) is omitted. 
Then $W$ becomes the  Green function of a free motion,
\beq\label{free}
\left.W(\vec\rho_2,z_2|\vec\rho_1,z_1)\right|_{\sigma\to 0}=
\int \frac{d^2p_T}{2\pi}\,
\exp\left[-{\rm i}\vec p_T\cdot(\vec\rho_2-\vec\rho_1)+
\frac{{\rm i}p_T^2(z_2-z_1)}{2\nu\alpha(1-\alpha)}\right],
\eeq
where $\vec p_T$ is the transverse momentum of the quark.
In this limit, the shadowing term in (\ref{form}) reproduces the second term in
(\ref{KKK}). To see this, note that the Fourier transform of the diffractive
amplitude reads
\beq
f_{DD}(p_T)=
\int \frac{d^2\rho}{2\pi}
 \Psi_{q\bar q}^{T,L}(\vec\rho,\alpha)\,
 \sigma_{q\bar q}(\rho)\,e^{{\rm i}\vec p_T\cdot\vec\rho}.
\eeq
In momentum space, the mass of the pair is well defined,
\beq
M_X^2=\frac{p_T^2+m_f^2}{\alpha(1-\alpha)}.
\eeq
Then one obtains (\ref{KKK}) with this $M_X^2$.

We calculate nuclear shadowing for calcium and lead with the Green function
method and compare to the approximations
(\ref{KKK}), 
 and (\ref{doublescat}), where only the double scattering term is
taken into account.  As was mentioned before,
only the valence $q\bar q$-part of the virtual photon
is taken into account, but the higher Fock
components,
containing gluons and sea quarks, are neglected as well as
the effect of anti-shadowing.  
The dipole cross section
is approximated by the
 form $\sigma_{q\bar q}(\rho)=C\rho^2$, $C\approx 3$ \cite{kp,kp2}, which
works remarkably well, even at large separations and 
is sufficient for our purpose.
A uniform density $\rho_A=0.16$ fm$^{-3}$
is used for all nuclei, 
and the quark masses are fixed at $m_u=m_d=300 $ MeV, $m_s=450$ MeV and 
$m_c=1.5$ GeV.
Within these approximations it is possible to solve
(\ref{schroedinger}) analytically.  The solution is the harmonic
oscillator Green function with a complex frequency \cite{kz},  
\begin{equation}\label{erg} 
W\left(\vec\rho_2,z_2;\vec\rho_1,z_1\right)
=\frac{a}{2\pi\sinh\left(\omega\Delta
z\right)}
\exp\left\{-\frac{a}{2}\left[\left(\rho_2^2+\rho_1^2\right)
\coth\left(\omega\Delta z\right)-\frac{2\vec\rho_2\cdot\vec\rho_1} 
{\sinh\left(\omega\Delta z\right)} \right]\right\},
\end{equation}  
where  
\begin{eqnarray}
\Delta z& = & z_2-z_1, \\ 
\omega^2 & =&
{\rm i}\;\frac{C\rho_A}{\nu\alpha\left(1-\alpha\right)},\\ 
a^2 & = & -{\rm i}\;C\rho_A\nu\alpha\left(1-\alpha\right).
\end{eqnarray}
This formal solution properly accounts for all
multiple scatterings and the finite lifetime of the hadronic
fluctuations of the photon, as well as for fluctuations of
the transverse separation of the $q\bar q$ pair.

Because of all these approximations, the calculation is not compared  
to data, but only to the standard
approximations
(\ref{KKK}), 
(\ref{eikonalappr}) and (\ref{doublescat}).
The comparison with data will be presented below in section \ref{comparison}.

The results  are shown in
fig.~\ref{bild}.
The dashed curves show the predictions of (\ref{doublescat}) which is
called standard approach.  The mean values of $\sigma^2$ and
$\sigma$ are calculated using the perturbative LC wavefunctions (\ref{psit}) and
(\ref{psil}) given on page \pageref{psit}.
The intermediate state mass is fixed at $M^2=Q^2$. At low $x_{Bj}<
0.01$ shadowing saturates because $q=2m_Nx_{Bj} \ll
1/R_A$. The
thin solid curve also corresponds to a double scattering
approximation, but now only the imaginary potential in (\ref{schroedinger}) was
set to zero.  The formfactor is treated
properly, i.e.\ the kinetic term that describes
the relative transverse motion of the $q\bar q$
pair, correctly reproduces the phase shift.  The difference
between the curves is substantial.  The thin solid curve
does not show full saturation even at $x=0.001$.
We conclude that the prescription $M_{q\bar q}^2=Q^2$ 
does not work well for shadowing in
the transition region, where $x_{Bj}\to 0.1$ and
formfactor effects are of crucial importance.
This could have been expected from the estimate (\ref{estimate}) which shows
that fluctuations with quite large masses participate in shadowing.

\begin{figure}[ht]
  \centerline{\scalebox{0.8}{\includegraphics*{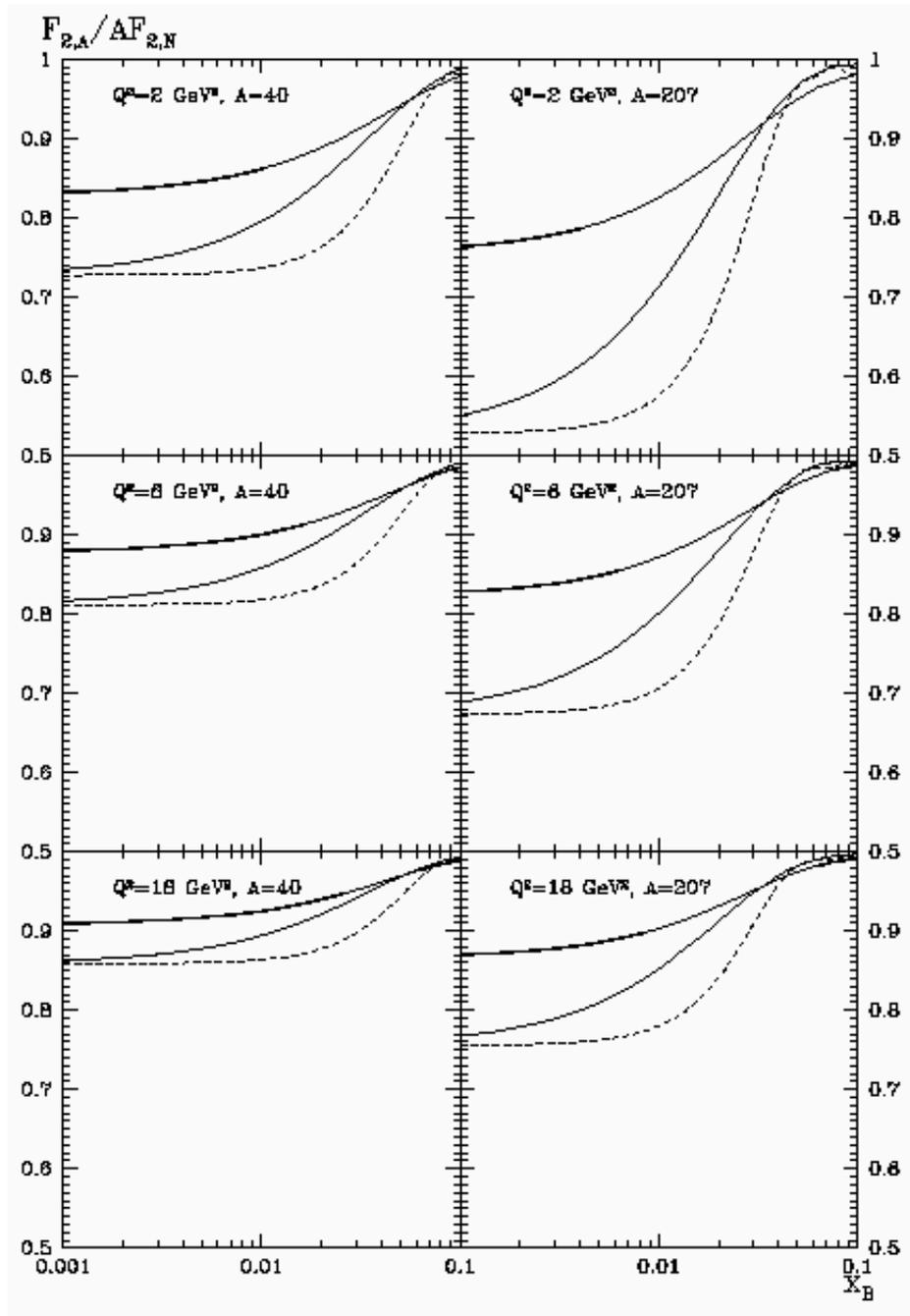}}}
    \caption{
      \label{bild}
      Nuclear shadowing for calcium and lead at different values of $Q^2$. 
      The dashed
        curve is calculated in the standard approach (\ref{1.3}), i.e.\ with
        coherence length $l_c=1/2m_Nx_{Bj}$ and in double scattering
        approximation.
        The thin solid curve corresponds
        to the double scattering approximation with the free Green
        function, (\ref{free}), and the thick solid
        curve shows the full calculation, (\ref{erg}).} 
\end{figure}

\clearpage

The next step is to do the full calculation and to study the
importance of the higher order rescattering terms in
nuclear shadowing.  The results are shown by the thick solid
curves.  Higher order scattering brings another substantial
deviation (especially for lead) from the standard approach.
At very low $x_{Bj}$ the curves saturate at the level given
by (\ref{eikonalappr}).
Note that the NM collaboration \cite{NMCfit} was indeed unable to fit the $A$
dependence of the shadowing ratio at low $x_{Bj}$ with an ansatz linear in the
nuclear density, indicating the importance of higher order rescattering terms.  

However, calculations with (\ref{schroedinger}) turn out to be numerically quite
involved, if one wants to use a realistic parametrization of the DCS and a
realistic nuclear density. The approximations introduced in this section
significantly simplify the calculations, because one can find an analytic
expression for the Green function $W$. 
We are forced to apply simplifying approximations also in our further
calculations, but  will use an effective $C$ in the dipole cross section in
order to account for the correlation between the energy and $\rho$ dependence,
as will be described in section \ref{comparison}.
Then one can still make use of the solution (\ref{erg}) of the Schr\"odinger
equation. This approach can be generalized to include also higher Fock
components of the photon \cite{KST99}. 

\subsection{Derivation of the formula 
for nuclear shadowing in DIS}\label{derivation}

Before we start with the derivation, we briefly sketch the idea of 
the calculation. The applied approximations are summarized at the end of this
subsection.
For a nuclear target, the total cross section 
may be written in eikonal form (\ref{eikonalappr}),
if the transverse separation $\rho$ is frozen, i.e.\ at very small $x_{Bj}$.
The eikonal formula can be obtained in exactly the same way as by Glauber
\cite{Glauber} in nonrelativistic quantum mechanics, because of the
nonrelativistic structure of the light-cone Hamiltonian.
In the eikonal approximation, 
one splits a fast oscillating phase factor from the 
wavefunction $\varphi(\vec r)=e^{ikz}\tilde\varphi(\vec r)$
and obtains for the slowly varying part $\tilde\varphi$
an equation of the form
\beq\label{SGL}
{\rm i}\frac{\partial}{\partial z}\tilde\varphi=
\left(-\frac{\Delta_\perp}{const.}+V\right)\tilde\varphi.
\eeq 
In the target rest frame,
the interaction is given by a color-static potential $V$.
However, anticipating that all dependence of the interaction will be absorbed into the
dipole cross section $\sigma_{q\bar q}(\rho)$, we use
an abelian potential. Strictly speaking, this is justified only in the case of
an electron-positron pair propagating in a condensed medium, 
but 
since the influence of the potential will be expressed in terms of scattering
amplitudes and because
our final result reproduces well known approximations in limiting cases,
we believe, that the results holds also for the case
of a nonabelian potential. 

The Laplacian acts only on the transverse coordinates and is omitted in the
eikonal approximation.
Then, (\ref{SGL}) is easily integrated.
The multiple scattering series is sum\-med like in Glauber theory \cite{Glauber}.
This is possible, because the typical distance between two nucleons inside a
nucleus is roughly $2$ fm, while the gluon correlation length is much smaller,
presumably $\sim 0.3$ fm. 
After averaging over the medium, one 
obtains the eikonal expression (\ref{eikonalappr}).

The idea is now to keep this Laplacian, because it
describes the transverse motion of the particles in the pair. 
Taking into account this motion, one can correctly describe the effective
mass of the fluctuation in a coordinate space representation.
The phase shift
function has then to be replaced by the Green function for (\ref{SGL}).
We write this Green function as a path integral and averaging over all
scattering centers yields an effective Green function $W$ which fulfills the two
dimensional Schr\"odinger equation (\ref{schroedinger}) 
with the optical potential  
$V_{opt}=-{\rm i}\sigma_{q\bar q}\left(\rho\right)\rho_A\left(b,z\right)/2$.

In order to derive explicit expressions for the terms in  (\ref{form}), we
start from the general expression for the cross section for production of a
$q\bar q$-pair,
\beqn\label{ds}
d\sigma^{\gamma^*A} & = & Z_f^2\alpha_{em}\,\left|M_{fi}\right|^2\,
\delta\left(\nu-p_{q}^0-p_{\bar{q}}^0\right)
\frac{d^3p_{q}\,d^3p_{\bar{q}}}{\left(2\pi\right)^4\nu\,2p_{q}^0\,
2p_{\bar{q}}^0}\\
\label{ds2}
& = & Z_f^2\alpha_{em}\,\left|M_{fi}\right|^2\,
\frac{d^2p_{\perp,q}\,d^2p_{\perp,\bar{q}}\,d\lambda}{\left(2\pi\right)^4
4\nu^2\,
\lambda\left(1-\lambda\right)}
\eeqn
where $Z_f^2$ is the flavor charge, 
and $\alpha_{em}=1/137$. Note that we prefer to call the longitudinal momentum
fraction of the quark by $\lambda$, because $\alpha$ is already used 
to denote many
other quantities.
The matrix element is given by
\beq\label{matrix}
M_{fi}=\int d^3r\, \Psi_{q}^\dagger\left(\vec r\right)\,
\vec{\bf \alpha}\cdot\vec\epsilon\;\Psi_{\bar{q}}\left(\vec r\right)
e^{{\rm i}kz}.
\eeq
The $z$-axis is chosen in direction of the propagation of the photon. The
photon's momentum is denoted by $k$ and it's energy by $\nu$, $p_q$ is
the momentum of the quark and $p_{\bar q}$ the momentum of the antiquark. 
In (\ref{ds2}) we have introduced the energy fraction $\lambda=p_q^0/\nu$.
In the
ultrarelativistic case which is considered, pair production takes place
predominantly in forward direction and therefore we distinguish between the
longitudinal direction ($z$-direction) and the transverse directions.
The longitudinal momenta are large compared to the flavor masses, $p_z\gg m_f$, 
and
to the perpendicular momenta, $\left|\vec p_{\perp}\right|^2\sim m_f^2$. In
our approximation, terms of order $m_f/p_z$ are kept only in exponentials
and neglected otherwise. All higher
order terms are omitted.

Note, that the pair is created electromagnetically in a color singlet state,
therefore  the factor $Z_f\alpha_{em}$ appears in (\ref{ds}), but 
the quarks interact
with the nucleons as described by a phenomenological, scalar potential
$\phi\left(\vec r\right)$. 
The particles in the pair move in a potential 
$U\left(\vec r\right)$ that is a superposition of the
potentials of all nucleons,
\beq\label{super}
U\left(\vec r\right)=\sum\limits_{j=1}^A\phi\left(\vec r-\vec r_j\right).
\eeq
The vector $\vec r_j$ runs over all positions of the nucleons.

The 
wavefunction of the quark fulfills the Dirac equations and is an
eigenstate with positive energy, while the antiquark is represented by an
eigenstate with negative energy, 
\beqn
\left(p_{q}^0-U\left(\vec r\right)-m_f{\bf \beta}
+{\rm i}\vec{\bf\alpha}\cdot\nabla\right)\Psi_{q}\left(\vec r\right) & = & 0,\\
\left(-p_{\bar{q}}^0-U\left(\vec r\right)-m_f{\bf \beta}
+{\rm i}\vec{\bf\alpha}\cdot\nabla\right)\Psi_{\bar{q}}\left(\vec r\right) & = & 0.
\eeqn
No interaction between the quark and the antiquark is taken into account and
therefore, the two equations decouple.
The wavefunction of the quark $\Psi_{q}\left(\vec r\right)$ 
contains an outgoing plane wave and an
outgoing spherical wave in it's asymptotic
form, while the wavefunction of the antiquark contains an incoming  
spherical wave and an incoming plane wave. 
These equations are transformed into second order
equations by applying the operator 
$\left(p_{q}^0-U\left(\vec r\right)+m_f{\bf \beta}
-{\rm i}\vec{\bf\alpha}\cdot\nabla\right)$ on the first equation, and the
corresponding operator on the second equation, as described in \cite{LL}.
 When  the term quadratic
in the potential is omitted, one obtains
\beqn
\left(\Delta+\left|\vec p_{q}\right|^2-2p_{q}^0U\left(\vec r\right)
+{\rm i}\vec{\bf\alpha}\cdot\left(\nabla U\left(\vec r\right)\right)
\right)\Psi_{q}\left(\vec r\right) & = & 0, \\
\left(\Delta+\left|\vec p_{\bar{q}}\right|^2+2p_{\bar{q}}^0U\left(\vec r\right)
+{\rm i}\vec{\bf\alpha}\cdot\left(\nabla U\left(\vec r\right)\right)
\right)\Psi_{\bar{q}}\left(\vec r\right) & = & 0. 
\eeqn
The solutions may approximately 
be written as Furry 
\cite{Furry,SM}
type
wavefunctions,
\beqn\label{sm1}
\Psi_{q}\left(\vec r\right) & = &
e^{{\rm i}\vec p_{q}\cdot\vec r}
\left(1-\frac{{\rm i}\vec{\bf\alpha}}{2p_{q}^0}\cdot\nabla
\right)\,
F_{q}\left(\vec r\right)\,
u\left(p_{q},\lambda_{q}\right), \\
\label{sm2}
\Psi_{\bar q}\left(\vec r\right) & = &
e^{-{\rm i}\vec p_{q}\cdot\vec r}
\left(1+\frac{{\rm i}\vec{\bf\alpha}}{2p_{\bar q}^0}\cdot\nabla
\right)\,
F_{\bar q}\left(\vec r\right)\,
v\left(p_{\bar q},\lambda_{\bar q}\right).
\eeqn
Here, $u\left(p_{q},\lambda_{q}\right)$ is the free spinor with positive energy
and pol\-ari\-zation $\lambda_q$ which fulfills the equation
$\left(p_q\!\!\!\!\!/-m_f\right)u\left(p_{q},\lambda_{q}\right)=0$ and similarly
$\left(p_{\bar q}\!\!\!\!\!/+m_f\right)v\left(p_{q},\lambda_{q}\right)=0$. 
In Dirac representation the spinors read
\beq\label{Dirac}
u\left(p_{q},\lambda_{q}\right)=
\sqrt{p^0_q+m_f}\left(
\begin{array}{c}
\displaystyle{\chi_q } \\
\displaystyle{\frac{\vec\sigma\cdot\vec p_q}{p^0_q+m_f}\;\chi_q} 
\end{array}\right)
\;\;,\;\;
v\left(p_{q},\lambda_{q}\right)=
\sqrt{p^0_{\bar q}+m_f}\left(
\begin{array}{c}
\displaystyle{\frac{\vec\sigma\cdot\vec p_{\bar{q}}}{p^0_{\bar q}+m_f}\;
\chi_{\bar q}} \\ 
\displaystyle{\chi_{\bar q} }
\end{array}\right).
\eeq
The three Pauli spin matrices are denoted by $\vec\sigma$ and the Pauli spin
state referred to the rest frame of the particle is $\chi_q$, or
$\chi_{\bar q}$ respectivly. This means explicitly 
$\vec s\cdot\vec\sigma\,\chi_q=\lambda_{q}\chi_q$ and 
$\vec s\cdot\vec\sigma\,\chi_{\bar q}=-\lambda_{\bar q}\chi_{\bar q}$ 
with a spin vector $\vec
s$ normalized to unity and $\lambda_q,\lambda_{\bar q}=\pm 1$.
The
functions $F_{q}$ and $F_{\bar q}$ have no spinor structure any more.
They contain all dependence of the potential and have to be calculated for a
given $U\left(\vec r\right)$.
They fulfill the equations \cite{LL}
\beqn
\label{erste}
\left(\Delta
+2{\rm i}\vec p_{q}\cdot\nabla
-2p_{q}^0U\left(\vec r\right)
\right)F_{q}\left(\vec r\right)
& = & 0,\\
\label{zweite}\left(\Delta
-2{\rm i}\vec p_{\bar q}\cdot\nabla
+2p_{\bar q}^0U\left(\vec r\right)
\right)q_{\bar f}\left(\vec r\right) & = & 0,
\eeqn
with boundary conditions $F\rightarrow 1$ for the quark and for the antiquark as
$z \rightarrow \infty$. Note, that it is essential to take the correction
proportional to $\vec\alpha$ in (\ref{sm1}) and (\ref{sm2}) into account,
although these terms seem to be suppressed by a factor of $1/p^0$. It turns out,
that when one calculates the matrix element of the current operator,
$\left(\vec\alpha\cdot\vec\epsilon\,\right)_{fi}$, between free spinors,
 the large part cancels and one is
 left with a contribution of the same order as produced by the correction
term. Thus, the terms proportional to $\vec\alpha$ may not be neglected in the
matrix element, although they give only small corrections to the wave functions.

In order to remove the dependence on the transverse momenta from the phase
factors in eq.~(\ref{sm1}) and (\ref{sm2}), we rewrite
the solutions in the  form
\beqn\label{ansatz}
\Psi_{q}\left(\vec r\right) & = &
e^{{\rm i} \left|\vec p_q\,\right|z}
\left(1-\frac{{\rm i}\vec{\bf\alpha}}{2p_{q}^0}\cdot\nabla
-\frac{\vec{\bf\alpha}_\perp}{2p_{q}^0}\cdot\vec p_{\perp,q}
+\frac{{\bf\alpha}_z}{2p_{q}^0}
\left(\left|\vec p_{q}\,\right|-p_{z,q}\right)
\right)\,
\psi_{q}\left(\vec r\right)\,
u\left(p_{q},\lambda_{q}\right),\nonumber\\
 \\
\label{ansatz2}
\Psi_{\bar q}\left(\vec r\right) & = &
e^{-{\rm i} \left|\vec p_{\bar q}\,\right|z}
\left(1+\frac{{\rm i}\vec{\bf\alpha}}{2p_{\bar q}^0}\cdot\nabla
-\frac{\vec{\bf\alpha}_\perp}{2p_{\bar{q}}^0}
\cdot\vec p_{\perp,\bar{q}}
+\frac{{\bf\alpha}_z}{2p_{q}^0}
\left(\left|\vec p_{\bar q}\,\right|-p_{z,\bar q}\right)
\right)\,
\psi_{\bar q}\left(\vec r\right)\,
v\left(p_{\bar q},\lambda_{\bar q}\right),\nonumber\\
\eeqn
with
\beqn\label{psidef1}
\psi_{q}\left(\vec r\right) & = &
e^{{\rm i}\vec p_{\perp,q}\cdot\vec r_\perp}
e^{-{\rm i}\left(\left|\vec p_{ q}\,\right|-p_{z,q}\right)z}
F_{q}\left(\vec r\right), \\
\label{psidef2}\psi_{\bar q}\left(\vec r\right) & = &
e^{-{\rm i}\vec p_{\perp,\bar{q}}\cdot\vec r_\perp}
e^{{\rm i}\left(\left|\vec p_{\bar q}\,\right|-p_{z,\bar q}\right)z}
F_{\bar q}\left(\vec r\right),
\eeqn
and $
\left|\vec p\,\right|=\sqrt{p_0^2-m_f^2}$
for the quark and the antiquark respectively. In 
the following, we neglect the terms with $\alpha_z$ in
eq.~(\ref{ansatz}) 
and (\ref{ansatz2}), because they are of order $O\left(1/p^0_qp^0_{\bar
q}\right)$. The functions $\psi_{q}\left(\vec r\right)$ and 
$\psi_{\bar q}\left(\vec r\right)$ will play the role of effective wave
functions for the quarks.

The phase factors 
combine in the matrix element (\ref{matrix}) 
to the minimal longitudinal momentum transfer, 
\beq
q_L^{min}=
k-\left|\vec p_{q}\,\right|-\left|\vec p_{\bar q}\,\right|
\approx\frac{Q^2}{2\nu}
+\frac{m_f^2}{2p_{0,q}}
+\frac{m_f^2}{2p_{0,\bar{q}}},
\eeq
and one obtains
\beqn\nonumber
\lefteqn{M_{fi} \; =\;\int d^3 r e^{{\rm i}q_L^{min}z} }\\
\nonumber &\times&{ 
\left[u^\dagger\left(p_{q},\lambda_{q}\right)
\left(\,\vec{\bf \alpha}\cdot\vec\epsilon\;
+\vec{\bf \alpha}\cdot
\frac{{\rm i}\nabla\left(\vec r_{q}\right)-\vec{p}_{\perp,q}}{2p_q^0}\,
\vec{\bf \alpha}\cdot\vec\epsilon\;
+\vec{\bf \alpha}\cdot\vec\epsilon\,\vec{\bf \alpha}\cdot
\frac{{\rm i}\nabla\left(\vec r_{\bar{q}}\right)
-\vec{p}_{\perp,\bar{q}}}{2p_{\bar{q}}^0}
\right)
v\left(p_{\bar{q}},\lambda_{\bar{q}}\right)\right]} \\
& \times &
\psi_{q}^*\left(\vec r_{{q}}\right)
\psi_{\bar{q}}\left(\vec r_{\bar{q}}\right)
\Big|_{\vec r=\vec r_{{q}}=\vec r_{\bar{q}}}.
\eeqn
Here, the coherence length, $l_c^{max}=1/q_L^{min}$, enters as an
oscillating phase factor. However, $l_c^{max}$ does not depend on the
transverse momenta. Their influence on the cross section is encoded in the
rest of the wavefunctions, (\ref{ansatz}) and (\ref{ansatz2}).
The operator $\nabla\left(\vec r_{q}\right)$ acts only on the variable
$\vec r_{q}$ and the operator 
$\nabla\left(\vec r_{\bar{q}}\right)$ only on
$\vec r_{\bar q}$.
After the derivatives have been performed, the whole integrand 
has to be evaluated at 
$\vec r=\vec r_{{q}}=\vec r_{\bar{q}}$.

With the representation (\ref{Dirac}) one obtains after some algebra
within the demanded accuracy
\beqn\nonumber\label{me}
\lefteqn{M_{fi} \; =\;\inf dz e^{{\rm i}q_L^{min}z} \int d^2 r_\perp
\frac{1}{\sqrt{\lambda\left(1-\lambda\right)}}}
\\
\nonumber &\times& \chi_q^\dagger
\displaystyle{\Bigg\{ }
m_f\vec\sigma\cdot\vec\epsilon_T
+{\rm i}\left(1-\lambda\right)\vec\sigma\cdot\vec e_z
\vec\epsilon_T\cdot\nabla\left(\vec r_{\perp,\bar{q}}\right)
+{\rm i}\lambda\vec\sigma\cdot\vec e_z
\vec\epsilon_T\cdot\nabla\left(\vec r_{\perp,{q}}\right)\\
\nonumber & & 
+\left(1-\lambda\right)\left(\vec e_z\times\vec\epsilon_T\right)
\cdot\nabla\left(\vec r_{\perp,{\bar q}}\right)
-\lambda\left(\vec e_z\times\vec\epsilon_T\right)
\cdot\nabla\left(\vec r_{\perp,{q}}\right)\\
\nonumber & & 
+\;2\,Q\,\lambda\left(1-\lambda\right)
\Bigg\}\chi_{\bar q}
 \\
& \times &
\psi_{q}^*\left(\vec r_{\perp,{q}},z\right)
\psi_{\bar{q}}\left(\vec r_{\perp,\bar{q}},z\right)
\Big|_{\vec r_{\perp,{q}}=\vec r_{\perp,\bar{q}}}.
\eeqn
Some details of the calculation can be found in \cite{OM}. The unit vector in
$z$-direction is denoted by $\vec e_z$.
The polarization vector $\vec\epsilon_T$ corresponds to transverse
states of the $\gamma^*$, while the last term in the curly brackets
is due to longitudinal polarization.
There are two
remarkable aspect concerning this last equation. First, it does not contain any
derivative with respect to $z$ any more, because of the transverse nature of the
polarization vector $\vec\epsilon_T$ and of $\vec e_z\times\vec\epsilon_T$. 
Second, all dependence on the transverse momenta in the spinor part 
has canceled.

From (\ref{erste}) and (\ref{zweite}) and from the definition of $\psi$,
(\ref{psidef1}) and (\ref{psidef2}), an equation for $\psi$ is obtained.
Assuming that these functions are only slowly 
varying with $z$, 
we omit the longitudinal part of the Laplacian and get
\beqn\label{Schroe1}
{\rm i}\frac{\partial}{\partial z}
\psi_{q}\left(\vec r_\perp,z\right)
 & = & 
\left(-\frac{\Delta_\perp}{2p_{{q}}^0}
+U\left(\vec r_\perp,z\right)\right)
\psi_{q}\left(\vec r_\perp,z\right), \\
\label{Schroe2}
{\rm i}\frac{\partial}{\partial z}
\psi_{\bar{q}}\left(\vec r_\perp,z\right)
& = & 
\left(\frac{\Delta_\perp}{2p_{\bar{q}}^0}
+U\left(\vec r_\perp,z\right)\right)
\psi_{\bar{q}}\left(\vec r_\perp,z\right).
\eeqn
Our ansatz yields two dimensional Schr\"odinger equations, where the
$z$-coordinate plays the role of time and the mass is given by the
energy. 
The Laplacian $\Delta_\perp$ acts on the transverse
coordinates only. The functions $\psi_{q}\left(\vec r_\perp,z\right)$ 
and $\psi_{\bar q}\left(\vec r_\perp,z\right)$ 
become two dimensional plane waves for $z\rightarrow \infty$, up to a phase
factor that cancels in the square of the matrix element.
It should be mentioned that the kinetic energy for the
antiquark has a negative sign.
This results from the fact that solutions of the Dirac equation with negative
energy propagate backwards in time.
The Laplacian in (\ref{Schroe1}) and (\ref{Schroe2}), 
account for the transverse motion of the pair in which we are
especially interested.
The functions $\psi_{\bar q}\left(\vec r_\perp,z\right)$ and
$\psi_{q}\left(\vec r_\perp,z\right)$ in the matrix element (\ref{me})
may now be expressed in terms of the Green functions for 
(\ref{Schroe1}) and (\ref{Schroe2}) and their asymptotic behavior,
\beqn
\psi_{q}\left(\vec r_{\perp,2},z_2\right) & = &
\int d^2 r_{\perp,1}\,G_{q}\left(\vec r_{\perp,2},z_2\,
|\,\vec r_{\perp,1},z_\infty\right)
e^{{\rm i}\vec p_{\perp,q}\cdot\vec r_{\perp,1}}
e^{-{\rm i}\left(\left|\vec p_{ q}\,\right|-p_{z,q}\right)z_\infty},\\
\psi_{\bar q}\left(\vec r_{\perp,2},z_2\right) & = &
\int d^2 r_{\perp,1}\,G_{\bar q}\left(\vec r_{\perp,2},z_2\,
|\,\vec r_{\perp,1},z_\infty\right)
e^{-{\rm i}\vec p_{\perp,\bar q}\cdot\vec r_{\perp,1}}
e^{{\rm i}\left(\left|\vec p_{\bar q}\,\right|-p_{z,\bar q}\right)z_\infty}.
\eeqn

Now one uses the
expression for $d\sigma^{\gamma^*A}$, (\ref{ds}), 
and with the matrix element (\ref{me})
and the last two relations, we obtain
\beqn
\nonumber d\sigma^{\gamma^*A} & = & Z_f^2\alpha_{em}
\;\inf dz \inf dz^\prime e^{{\rm i}q_L^{min}\left(z-z^\prime\right)} 
\int d^2 r_\perp\,d^2\tau_{q}\,d^2\tau_{\bar{q}}\,
\int d^2 r_\perp^{\,\prime}\,d^2\tau_{q}^\prime\,d^2\tau_{\bar{q}}^\prime\, \\
\nonumber & &\;\;\;\times\;\;
{\cal O}\left(\vec r_{\perp,{q}},\vec r_{\perp,\bar{q}}\right)
G_{q}^*\left(\vec r_{\perp,{q}},z\,|\,\vec\tau_{q},z_{\infty}\right)\,
G_{\bar{q}}\left(\vec r_{\perp,\bar{q}},z\,
|\,\vec\tau_{\bar{q}},z_{\infty}\right)
\Big|_{\vec r_{\perp}=\vec r_{\perp,{q}}=\vec r_{\perp,\bar{q}}}
\\
\nonumber & &\;\;\;\times\;\;
{\cal O}^*\left(\vec r^{\,\prime}_{\perp,{q}}
,\vec r^{\,\prime}_{\perp,\bar{q}}\right)
G_{q}\left(\vec r_{\perp,{q}}^{\,\prime},z^{\,\prime}\,
|\,\vec\tau_{q}^{\,\prime},z_{\infty}\right)\,
G_{\bar{q}}^*\left(\vec r_{\perp,\bar{q}}^{\,\prime},z^{\,\prime}\,
|\,\vec\tau_{\bar{q}}^{\,\prime},z_{\infty}\right)
\Big|_{\vec r_{\perp}^{\,\prime}=\vec r_{\perp,{q}}^{\,\prime}
=\vec r^{\,\prime}_{\perp,\bar{q}}}
\\
\nonumber & &\;\;\;\times\;\;
e^{{\rm i}\left(\vec\tau_{q}^{\,\prime}-\vec\tau_{q}\right)\cdot\vec p_{\perp,q}}
\,e^{{\rm i}\left(\vec\tau_{\bar{q}}^{\,\prime}-\vec\tau_{\bar{q}}\right)
\cdot\vec p_{\perp,\bar{q}}}\\
& &\;\;\;\times\;\;
\frac{d^2p_{\perp,q}\,d^2p_{\perp,\bar{q}}\,d\lambda}
{\left(2\pi\right)^4\,\lambda^2\left(1-\lambda\right)^2 4\nu^2}.
\eeqn
For convenience  the operator
\beqn\nonumber
{\cal O}\left(\vec r_{\perp,{q}}
,\vec r_{\perp,\bar{q}}\right) & = &
\chi_q^\dagger
\displaystyle{\Bigg\{ }
m_f\vec\sigma\cdot\vec\epsilon_T
+{\rm i}\left(1-\lambda\right)\vec\sigma\cdot\vec e_z
\vec\epsilon_T\cdot\nabla\left(\vec r_{\perp,\bar{q}}\right)
+{\rm i}\lambda\vec\sigma\cdot\vec e_z
\vec\epsilon_T\cdot\nabla\left(\vec r_{\perp,{q}}\right)\\
 & + & 
\left(1-\lambda\right)\left(\vec e_z\times\vec\epsilon_T\right)
\cdot\nabla\left(\vec r_{\perp,{\bar q}}\right)
-\lambda\left(\vec e_z\times\vec\epsilon_T\right)
\cdot\nabla\left(\vec r_{\perp,{q}}\right)\\
 & + & 
2\,Q\,\lambda\left(1-\lambda\right)\vec\sigma\cdot\vec e_z
\Bigg\}\chi_{\bar q}
\eeqn
is introduced.
In order to obtain the total cross section, we integrate over $p_{\perp,q}$ and
$p_{\perp,\bar{q}}$. The exponential factors give a
$\delta$-function which makes it possible 
to perform the integrations over all the
$\tau$s. Note, that ${\cal O}$ does not depend on $\tau$. The result is
\beqn
\nonumber \sigma_{tot}^{\gamma^*A} & = & Z_f^2\,
\frac{\alpha_{em}}{\nu^2}\,2{\rm Re}\int\limits_{0}^1 
\frac{d\lambda}{\lambda^2\left(1-\lambda\right)^2}
\;\int\limits_{-\infty}^{\infty} dz \int\limits_{z}^{\infty} dz^\prime 
e^{{\rm i}q_{min}\left(z-z^\prime\right)}
\int d^2 r_\perp\,\int d^2 r_\perp^{\,\prime} 
 \\ 
\nonumber & &\;\;\;\times\;\;
{\cal O}\left(\vec r_{\perp,{q}},\vec r_{\perp,\bar{q}}\right)
{\cal O}^*\left(\vec r^{\,\prime}_{\perp,{q}}
,\vec r^{\,\prime}_{\perp,\bar{q}}\right)
\\
\label{stot} & &\;\;\;\times\;\;
G_{q}\left(\vec r_{\perp,q}^{\,\prime},z^{\,\prime}\,
|\vec r_{\perp,q},z\right)\,
G_{\bar{q}}^*\left(\vec r_{\perp,\bar q}^{\,\prime},z^{\,\prime}\,
|\vec r_{\perp,\bar q},z\right)
\Big|_{\vec r_{\perp}=\vec r_{\perp,{q}}=\vec r_{\perp,\bar{q}}\,;\,
\vec r_{\perp}^{\,\prime}=\vec r_{\perp,{q}}^{\,\prime}
=\vec r^{\,\prime}_{\perp,\bar{q}}}.
\eeqn
Instead of four propagators one is 
left with only two, because of the
convolution relation
\beq
G\left(\vec r_{\perp,2},z_2\,|\,\vec r_{\perp,1},z_1\right)=
\int d^2 r_\perp\;
G\left(\vec r_{\perp,2},z_2\,|\,\vec r_\perp,z\right)
G\left(\vec r_\perp,z\,|\,\vec r_{\perp,1},z_1\right).
\eeq

In order to derive an expression that is convenient for numerical calculations,
one makes use of the path-integral 
 representations of the propagators in (\ref{stot}). They read 
\beqn\label{path}\lefteqn{
G\left(\vec r_{\perp}^{\,\prime},z^{\,\prime}\,
|\vec r_{\perp},z\right)
= \!\!\int{\cal D}\vec \tau\,\exp\left\{
{\rm i}\int\limits^{z^{\,\prime}}_{z}d\xi\left(\pm\frac{p^0}{2}
\dot{\vec \tau}^{\,2}-
U\left(\vec \tau,\xi\right)\right)\right\}}\\
\label{path2}& \approx &\!\! \int{\cal D}\vec \tau\,\exp\left\{
{\rm i}\!\int\limits^{z_2}_{z_1}\pm \frac{p^0}{2}
\dot{\vec \tau}^{\,2}d\xi-{\rm i}
\sum\limits_{j=1}^A\mbox{X}
\left(\vec \tau\left(z_j\right)-\vec r_{j,\perp}\right)
\Theta\left(z^{\,\prime}-z_j\right)\Theta\left(z_j-z\right)
\right\},\nonumber \\
\eeqn
where the upper sign corresponds to the quark and the lower to the antiquark.
In this expression, $\vec\tau$ is a function of $\xi$. The derivative with
respect to $\xi$ is denoted by $\dot{\vec\tau}$.
Furthermore, the condition
\beq
G\left(\vec r_{\perp}^{\,\prime},z^{\,\prime}\,
|\vec r_{\perp},z\right)\Big|_{z^{\,\prime}=z}
=\delta^{(2)}\left(\vec r_{\perp}^{\,\prime}-\vec r_{\perp}\right)
\eeq
has to be fulfilled together with
$\vec \tau\left(z\right)=\vec r_{\perp}$ and
$\vec \tau\left(z^{\,\prime}\right)=\vec r_{\perp}^{\,\prime}$.
In (\ref{path2}), the phase shift function 
$\mbox{X}\left(\vec \tau-\vec r_{j,\perp}\right)$ is introduced. 
The step function 
$\Theta\left(z\right)$ is $0$ for $z<0$ and $1$ for $z>0$.
As
mentioned before, $U$ is the superposition of all potentials 
of the nucleons, see eq.~(\ref{super}). Let
$\phi$ be the potential of a single nucleon in the target. 
The position of the nucleon with number $j$ is denoted by the transverse vector
$\vec r_{j,\perp}$ and the longitudinal coordinate $z_j$.
If the range
of interaction is much smaller than the distance 
$z^{\,\prime}-{z}$, the potential is practically zero outside the domain of
integration over $\xi$ and the phase shift function 
is  given by
\beq
\mbox{X}\left(\vec r_\perp-\vec r_{j,\perp}\right)
=\inf
d\xi
\,\phi\left(\vec r_\perp-\vec r_{j,\perp},\xi-z_j\right).
\eeq
Note that $\vec\tau\left(\xi\right)$ was replaced by 
$\vec\tau\left(z_j\right)$. This means,
only an average
value of the transverse coordinate is used
for calculating the phase shift for
scattering on a single nucleon.

The two path integrals (\ref{path2}) for the quark ($+$) and the antiquark ($-$)
sum over all possible trajectories of the two particles.
In order to calculate the cross section for pair production in the nuclear
medium, one has to average over all nucleons.
One obtains with the path-integral (\ref{path2})
\beqn
\nonumber
\lefteqn{
\left<G_{q}\left(\vec r_{\perp,q}^{\,\prime},z^{\,\prime}\,
|\vec r_{\perp,q},z\right)\,
G_{\bar{q}}^*\left(\vec r_{\perp,\bar q}^{\,\prime},z^{\,\prime}\,
|\vec r_{\perp,\bar q},z\right)\right>}
\\
& = &\left< \int{\cal D}\vec\tau_q\,\int{\cal D}\vec\tau_{\bar q}\,
\exp\left\{
{\rm i}\int\limits^{z^{\,\prime}}_{z}\left(
\frac{p_{q}^0}{2}\dot{\vec\tau}_q^{\,2}
+\frac{p^0_{\bar{q}}}{2}\dot{\vec\tau}_{\bar q}^{\, 2}\right)d\xi
\right.\right.\\
& & \left. \left. 
+{\rm i}
\sum\limits_{j=1}^A
\Theta\left(z^{\,\prime}-z_j\right)\Theta\left(z_j-z\right)\Bigl(
\mbox{X}\left(\vec \tau_{\bar q}\left(z_j\right)-\vec r_{j,\perp}\right)
-\mbox{X}\left(\vec \tau_{q}\left(z_j\right)-\vec r_{j,\perp}\right)
\Bigr)\right\}\right>,
\eeqn
with boundary conditions $\vec\tau_q\left(z\right)=\vec r_{\perp,q}$,
$\vec\tau_q\left(z^{\,\prime}\right)=\vec r_{\perp,q}^{\,\prime}$,
$\vec\tau_{\bar q}\left(z\right)=\vec r_{\perp, \bar q}$ and
$\vec\tau_{\bar q}\left(z^{\,\prime}\right)=\vec r_{\perp,\bar q}^{\,\prime}$.
The averaging procedure is similar to the one described in \cite{Glauber}.
All correlations between the nucleons are neglected and  the 
average
nuclear density ${\rho_A}$ is introduced, which is normalized to $A$,
(\ref{density}). Then, the whole expression may be
written as an exponential, if $A$ is large enough,
\beqn
\nonumber\lefteqn{
\left<\exp\left({\rm i}
\sum\limits_{j=1}^A
\Theta\left(z^{\,\prime}-z_j\right)\Theta\left(z_j-z\right)\Bigl(
\mbox{X}\left(\vec \tau_{\bar q}\left(z_j\right)-\vec r_{j,\perp}\right)
-\mbox{X}\left(\vec \tau_{q}\left(z_j\right)-\vec r_{j,\perp}\right)
\Bigr)
\right)\right>}&&\\
\nonumber\label{density}& = & \Biggl\{1-\frac{1}{A}\int d^2s
\int\limits^{z^{\,\prime}}_{z}d\xi^{\,\prime}
{\rho_A}\left(\vec s,\xi^{\,\prime}\right)\\
&\times&\Biggl.\left(1-\exp\Biggl({\rm i}
\Bigl(
\,\mbox{X}\left(\vec\tau_{\bar q}\left(\xi^{\,\prime}\right)-\vec s\right)
-\,\mbox{X}\left(\vec\tau_q\left(\xi^{\,\prime}\right)-\vec s\right)
\Bigr)\Biggr)\right)
\Biggr\}^A
\\
\nonumber\label{app}
& \approx &
\exp\left\{-\int\limits^{z^{\,\prime}}_{z} d\xi^{\,\prime}
{\rho_A}\left(b,\xi^{\,\prime}\right)\right.\\
& &\;\;\times\left.\;\;
\int d^2s
\left(1-
\exp\Biggl({\rm i}
\Bigl(
\,\mbox{X}\left(\vec\tau_{\bar q}\left(\xi^{\,\prime}\right)
-\vec s\right)
-\,\mbox{X}\left(\vec\tau_q\left(\xi^{\,\prime}\right)
-\vec s\right)\Bigr)\Biggr)
\right)\right\}.
\eeqn
When 
${\rho_A}$ is varying
fairly smoothly inside the nucleus, 
one can replace its dependence of $\vec s$, the transverse distance between the
pair and the scattering nucleon, by the impact parameter $b$,
eq.~(\ref{app}).
Since we have a short ranged interaction, it is reasonable to choose as impact
parameter $\vec b=\left(\vec r_\perp-\vec r_\perp^{\,\prime}\right)/2$.
This way one finds the forward scattering amplitude 
for a dipole scattering on a single
nucleon, see eq.~(\ref{app}). 
The corresponding cross section,
\beq
\sigma_{q\bar{q}}
\left(\rho\right)
=2{\rm Re}\int d^2s
\left(1-\exp\Biggl({\rm i}
\Bigl(
\mbox{X}\left(\vec\rho-\vec s\right)
-\mbox{X}\left(\vec s\right)\Bigr)\Biggr)\right),
\eeq
 appears as imaginary
potential in the propagator. At this point one notices 
that the forward scattering
amplitude in the abelian model is predominantly real, while in a nonabelian
theory it has to be imaginary. Nevertheless, we believe that the final result
holds also for a nonabelian theory, because 
all dependence on the potential of the nucleons,
$\phi\left(\vec r\right)$,
is absorbed into $\sigma_{q\bar{q}}$, for which we will use a
parametrization. 
One finally arrives at
\beqn
\nonumber
\lefteqn{
\left<G_{q}\left(\vec r_{\perp,q}^{\,\prime},z^{\,\prime}\,
|\vec r_{\perp,q},z\right)\,
G_{\bar{q}}^*\left(\vec r_{\perp,\bar q}^{\,\prime},z^{\,\prime}\,
|\vec r_{\perp,\bar q},z\right)\right>}
\\
& = &\!\!\!
\int{\cal D}\vec\tau_q\int{\cal D}\vec\tau_{\bar q}
\exp\left\{
{\rm i}\int\limits^{z^{\,\prime}}_{z}d\xi\left(
\frac{p_{q}^0}{2}\dot{\vec\tau}_q^{\,2}
+\frac{p^0_{\bar{q}}}{2}\dot{\vec\tau}_{\bar q}^{\, 2}
+\frac{{\rm i}}{2}{\rho_A}\left(b,\xi\right)\sigma_{q\bar{q}}
\left(\left|\vec\tau_q-\vec\tau_{\bar q}\right|\right)
\right)\right\}.
\eeqn

It is
convenient to introduce center of mass coordinates, 
$\vec\tau_{rel}=\vec\tau_{\bar q}-\vec\tau_q$ and
$\vec\tau_{cm}=\left(1-\lambda\right)\vec\tau_{\bar q}+\lambda\vec\tau_q$, 
and to
express the sum of the two kinetic energies
as the sum of the kinetic energy of the relative motion
and the center of mass kinetic energy,
\beqn
\nonumber
\lefteqn{
\left<G_{q}\left(\vec r_{\perp,q}^{\,\prime},z^{\,\prime}\,
|\vec r_{\perp,q},z\right)\,
G_{\bar{q}}^*\left(\vec r_{\perp,\bar q}^{\,\prime},z^{\,\prime}\,
|\vec r_{\perp,\bar q},z\right)\right>}
\\
\label{third}& = & \int{\cal D}\vec\tau_{cm}\,\int{\cal D}\vec\tau_{rel}\,
\exp\left\{
{\rm i}\int\limits^{z^{\,\prime}}_{z}d\xi\left(
\frac{\mu}{2}\dot{\vec\tau}_{rel}^{\,2}
+\frac{\nu}{2}\dot{\vec\tau}_{cm}^{\,2}
+\frac{{\rm i}}{2}{\rho_A}\left(b,\xi\right)\sigma_{q\bar{q}}
\left(\tau_{rel}\right)
\right)\right\}\\
\label{last}\nonumber & = & 
\frac{\nu}{2\pi {\rm i}\left(z^{\,\prime}-z\right)}
\,\exp\left({\rm i}\frac{\nu}{2}
\frac{\left(
\lambda\left(\vec r_{\perp,q}^{\,\prime}-\vec r_{\perp,q}\right)
+\left(1-\lambda\right)
\left(
\vec r_{\perp,\bar q}^{\,\prime}-\vec r_{\perp,\bar q}
\right)
\right)^2}{z^{\,\prime}-z}\right)\\
& &\;\;\;\times\;\;
\int{\cal D}\vec\tau_{rel}\,
\exp\left\{
{\rm i}\int\limits^{z^{\,\prime}}_{z}d\xi\left(
\frac{\mu}{2}\dot{\vec\tau}_{rel}^{\,2}
+\frac{{\rm i}}{2}{\rho_A}\left(b,\xi\right)\sigma_{q\bar{q}}
\left(\tau_{rel}\right)
\right)\right\}.
\eeqn
We have introduced the
reduced mass
\beq
\frac{1}{\mu}=\frac{1}{p_{q}^0}+\frac{1}{p_{\bar{q}}^0}
=\frac{1}{\nu\lambda\left(1-\lambda\right)}.
\eeq
Since the imaginary potential depends
only on the relative coordinate, the center of mass propagates freely and what
remains is the effective propagator
\beq\label{prop}
W\left(\vec\rho^{\,\prime},z^{\,\prime}\,|\,\vec\rho,z\right)
=\int{\cal D}\vec\tau_{rel}\,
\exp\left\{
{\rm i}\int\limits^{z^{\,\prime}}_{z}d\xi\left(
\frac{\mu}{2}\dot{\vec\tau}_{rel}^{\,2}
-V_{opt}\left(b,\tau_{rel},\xi\right)\right)\right\},
\eeq
with $\vec \rho^{\,\prime}=\vec r_{\perp,\bar q}^{\,\prime}
-\vec r_{\perp,q}^{\,\prime}$ and
$\vec \rho=\vec r_{\perp,\bar q}-\vec r_{\perp,q}$ and the optical potential
\beq
V_{opt}\left(b,\rho,z\right)=
-\frac{{\rm i}}{2}{\rho_A}\left(b,z\right)\sigma_{q\bar{q}}
\left(\rho\right).
\eeq
It fulfills the equation
\beq
\left[{\rm i}\frac{\partial}{\partial z^{\,\prime}}
+\frac{\Delta_\perp\left(\rho^{\,\prime}\right)}
{2\nu\lambda\left(1-\lambda\right)}
-V_{opt}\left(b,\rho^{\,\prime},z^{\,\prime}\right)\right]
W\left(\vec\rho^{\,\prime},z^{\,\prime}\,|\,\vec\rho,z\right)=
{\rm i}\delta\left(z^{\,\prime}-z\right)
\delta^{\left(2\right)}\left(\vec\rho^{\,\prime}-\vec\rho\right).
\eeq
The propagator for the center of mass coordinate produces a $\delta$-function
and thus, one obtains from (\ref{stot}) after averaging over the medium
\beqn\label{stotal}
\nonumber \sigma_{tot}^{\gamma^*A} & = & \frac{Z_f^2\alpha_{em}}{4\nu^2}\,
\int d^2 b\;2{\rm Re}\int\limits_{0}^1 
\frac{d\lambda}{\lambda^2\left(1-\lambda\right)^2}
\;\int\limits_{-\infty}^{\infty} dz \int\limits_{z}^{\infty} dz^\prime 
e^{{\rm i}q_L^{min}\left(z-z^\prime\right)} \\
& & \;\;\;\times\;\;
{\cal O}_{q\bar q}\left(\vec \rho\right)
{\cal O}_{q\bar q}^*\left(\vec \rho^{\,\prime}\right)\;
W\left(\vec\rho^{\,\prime},z^{\,\prime}\,|\,\vec\rho,z\right)
\Big|_{\vec\rho^{\,\prime}=\vec\rho=\vec 0},
\eeqn
with
\beq\label{opo}
{\cal O}_{q\bar q}\left(\vec \rho\right)
  = 
{\cal O}_{q\bar q}^T\left(\vec \rho\right)+
{\cal O}_{q\bar q}^L\left(\vec \rho\right),
\eeq
where the transverse part of the operator is
\beq\label{opot}
{\cal O}_{q\bar q}^T\left(\vec \rho\right)
  = 
\chi_q^\dagger
\displaystyle{\Bigg\{ }
m_f\vec\sigma\cdot\vec\epsilon_T
+{\rm i}\left(1-2\lambda\right)\vec\sigma\cdot\vec e_z
\vec\epsilon_T\cdot\nabla\left(\vec\rho\right)
+
\left(\vec e_z\times\vec\epsilon_T\right)
\cdot\nabla\left(\vec\rho\right)
\Bigg\}\chi_{\bar q}
\eeq
and the longitudinal part
\beq\label{opol}
{\cal O}_{q\bar q}^L\left(\vec \rho\right)
  = 
\chi_q^\dagger
2\,Q\,\lambda\left(1-\lambda\right)\vec\sigma\cdot\vec e_z
\chi_{\bar q}=2\,Q\,\lambda\left(1-\lambda\right)
\delta_{\lambda_q,\lambda_{\bar q}}.
\eeq
These operators are already known from the definition of the LC wavefunctions in
section \ref{dipoledis}, see (\ref{transop},\ref{longop}).
Equation (\ref{stotal}) is the central result of this section. It is the total
cross section for production of a $q\bar q$-pair from a virtual photon
scattering on a nucleus. We have not summed over the spins of the quark and the
antiquark and not averaged over the polarizations of the photon. We have not
summed over the different flavors, either. 
The expression for the operator ${\cal O}$, (\ref{opo}),
depends on the spin vector of the
quark and the antiquark. The directions of these vectors may be fixed
arbitrarily.

In order to represent the result in the form (\ref{form}), $W$ and
it's complex conjugate is rewritten in an
expansion. The results can be combined in the following way:
\beqn\nonumber\label{expand}
W\left(\vec\rho^{\,\prime},z^{\,\prime}\,|\,\vec\rho,z\right) & = &
W_0\left(\vec\rho^{\,\prime},z^{\,\prime}\,|\,\vec\rho,z\right)\\
\nonumber
& + & {\rm i}\int\limits_{z}^{z^{\,\prime}}dz_1\int d^2\rho_1
W_0\left(\vec\rho^{\,\prime},z^{\,\prime}\,|\,\vec\rho_1,z_1\right)\,
V_{opt}\left(b,\rho_1,z_1\right)\,
W_0\left(\vec\rho_1,z_1\,|\,\vec\rho,z\right)\\
\nonumber
& - &\int\limits_{z}^{z^{\,\prime}}dz_1\int d^2\rho_1
\int\limits_{z}^{z_1}dz_2\int d^2\rho_2
W_0\left(\vec\rho^{\,\prime},z^{\,\prime}\,|\,\vec\rho_1,z_1\right)\,
V_{opt}^*\left(b,\rho_1,z_1\right)\,\\
& & \;\;\times\;\;
W\left(\vec\rho_1,z_1\,|\,\vec\rho_2,z_2\right)
V_{opt}\left(b,\rho_2,z_2\right)\,
W_0\left(\vec\rho_2,z_2\,|\,\vec\rho,z\right).
\eeqn
Here, $W_0$ is the propagator corresponding to (\ref{prop}), when the potential
is absent. The first term gives a divergent contribution, which is the
wave-function renormalization for the photon. The second term leads to the first
contribution in (\ref{form}). The
operators ${\cal O}_{q\bar q}^{T,L}$ applied to the free propagator $W_0$
give the light-cone wave functions
$\Psi^{T,L}_{q\bar q}\left(\rho,\lambda\right)$, 
up to a constant overall factor. 
The third term in the above expansion is the
interference term. Since this term contains the full propagator, there are no
higher terms in this expansion. 

As an example for further calculation, consider the integral
\beqn
\nonumber
\cal{I} & = & -
\int\limits_{-\infty}^{\infty} dz \int\limits_{z}^{\infty} 
dz^\prime 
{\rm e}^{{\rm i}q_L^{min}\left(z-z^\prime\right)} 
\int\limits_{z}^{z^{\,\prime}}dz_1\int d^2\rho_1
\int\limits_{z}^{z_1}dz_2\int d^2\rho_2\\
\nonumber &&\;\;\times\;\;
V_{opt}^*\left(b,\rho_1,z_1\right)\,
V_{opt}\left(b,\rho_2,z_2\right)
\\
\label{integral}& & \;\;\times\;\;
W_0\left(\vec\rho^{\,\prime},z^{\,\prime}\,|\,\vec\rho_1,z_1\right)\,
W\left(\vec\rho_1,z_1\,|\,\vec\rho_2,z_2\right)
W_0\left(\vec\rho_2,z_2\,|\,\vec\rho,z\right),
\eeqn
which is needed to calculate the interference part
\beq\label{inter}
\sigma_{tot}^{int}  =  \frac{Z_f^2\alpha_{em}}{4\nu^2}\,
\int d^2 b\;2{\rm Re}\int\limits_{0}^1 
\frac{d\lambda}{\lambda^2\left(1-\lambda\right)^2}
{\cal O}\left(\vec \rho\right)
{\cal O}^*\left(\vec \rho^{\,\prime}\right)
\;{\cal I}\,
\Big|_{\vec\rho^{\,\prime}=\vec\rho= \vec 0}.
\eeq

With the new variable $\varepsilon^2=\lambda\left(1-\lambda\right)\,Q^2+m_f^2$,
which is also defined in \ref{disextension},
one finds
\beq
q_L^{min}=\frac{\varepsilon^2}{2\nu\lambda\left(1-\lambda\right)},
\eeq
and because of the relation
\beq
\left[{\rm i}\frac{\partial}{\partial z^{\,\prime}}
+\frac{\Delta_\perp\left(\rho^{\,\prime}\right)-\varepsilon^2}
{2\nu\lambda\left(1-\lambda\right)}
\right]
\left(W_0\left(\vec\rho^{\,\prime},z^{\,\prime}\,|\,\vec\rho,z\right)
e^{-{\rm i}q_L^{min}\left(z^{\,\prime}-z\right)}\right)=
{\rm i}\delta\left(z^{\,\prime}-z\right)
\delta^{\left(2\right)}\left(\vec\rho^{\,\prime}-\vec\rho\right),
\eeq
one can write the propagator in the form
\beq\label{propagator}
W_0\left(\vec\rho^{\,\prime},z^{\,\prime}\,|\,\vec\rho,z\right)
e^{-{\rm i}q_L^{min}\left(z^{\,\prime}-z\right)}=
\int\frac{d^2 l_\perp}{\left(2\pi\right)^2}
\int\limits_{-\infty}^{\infty}\frac{d \omega}{2\pi}\;
\frac{\exp\left\{-{\rm i}\omega\left(z^{\,\prime}-z\right)
+{\rm i}\vec l_\perp\cdot\left(\vec\rho^{\,\prime}-\vec\rho\right)
\right\}}{
{\omega-\displaystyle{
\frac{\vec{l}_\perp^2+\varepsilon^2}{2\nu\lambda\left(1-\lambda\right)}}
+{\rm i}0_+}}.
\eeq
The $0_+$-prescription for the pole in the complex $\omega$-plane ensures
that $W_0\left(\vec\rho^{\,\prime},z^{\,\prime}\,|\,\vec\rho,z\right)=0$ for
$z>z^{\,\prime}$. Putting (\ref{propagator}) 
into (\ref{integral}) yields after a short
calculation
\beqn
\nonumber
\cal{I} & = & 
-\frac{4\nu^2\lambda^2\left(1-\lambda\right)^2}{\left(2\pi\right)^2}
\int\limits_{-\infty}^{\infty} dz_1 \int\limits_{-\infty}^{z_1} 
dz_2
e^{{\rm i}q_L^{min}\left(z_2-z_1\right)} 
\int d^2\rho_1
\int d^2\rho_2\\
\nonumber& & \;\;\times\;\;
V_{opt}^*\left(b,\rho_1,z_1\right)\,
V_{opt}\left(b,\rho_2,z_2\right)
\\
\label{integral2}& & \;\;\times\;\;
{\rm K}_0\left(\varepsilon\left|\vec\rho^{\,\prime}-\vec\rho_1\right|\right)
W\left(\vec\rho_1,z_1\,|\,\vec\rho_2,z_2\right)
{\rm K}_0\left(\varepsilon\left|\vec\rho_2-\vec\rho\right|\right).
\eeqn
K$_0$ is the MacDonald function of zeroth order.
We have used the relation
\beq
{\rm K}_0\left(\varepsilon\rho\right)=
\frac{1}{2\pi}\int d^2l_\perp\frac{{\rm e}^{{\rm i}\vec
l_\perp\cdot\vec\rho}}{l_\perp^2+\varepsilon^2}.
\eeq
The next step is to insert (\ref{integral2}) into (\ref{inter})
and to calculate the contribution from the second term in the expansion
(\ref{expand})
in a similar way. One obtains for the total cross section
\beqn
\nonumber\label{cross}
\sigma_{tot}^{\gamma^*\,A} & = & 
A\,Z_f^2\,\frac{\alpha_{em}}{\left(2\pi\right)^2}
\int\limits_0^1d\lambda\int d^2\rho_1
\,\sigma_{q\bar{q}}\left(\rho_1\right)\,
\left|{\cal O}_{q\bar q}\left(\vec\rho\right)
{\rm K}_0\left(\varepsilon\left|\vec\rho-\vec\rho_1\right|\right)
\Big|_{\vec\rho= \vec 0}
\right|^2
\\
\nonumber
& - & Z_f^2\,\frac{\alpha_{em}}{\left(2\pi\right)^2}\,
2{\rm Re}\int d^2b
\int\limits_{-\infty}^{\infty} dz_1 \int\limits_{z_1}^{\infty} dz_2
\int\limits_0^1d\lambda\int d^2\rho_1\int d^2\rho_2\,
e^{-{\rm i}q_L^{min}\left(z_2-z_1\right)} \\
& & \;\;\times \;\;
V_{opt}^*\left(b,\rho_1,z_1\right)\,
V_{opt}\left(b,\rho_2,z_2\right)
\left({\cal O}_{q\bar q}^*\left(\vec\rho^{\,\prime}\right)
{\rm K}_0\left(\varepsilon\left|\vec\rho^{\,\prime}-\vec\rho_1\right|\right)
\Big|_{\vec\rho^{\,\prime}= \vec 0}\right)
\nonumber \\
& & \;\;\times \;\;
W\left(\vec\rho_2,z_2\,|\,\vec\rho_1,z_1\right)\,
\left({\cal O}_{q\bar q}\left(\vec\rho\right)
{\rm K}_0\left(\varepsilon\left|\vec\rho-\vec\rho_2\right|\right)
\Big|_{\vec\rho= \vec 0}\right).
\eeqn 

With help of
the relation
\beq
\nabla\left(\vec\rho\right){\rm K}_0\left(\varepsilon\rho\right)=
-\varepsilon\frac{\vec\rho}{\rho}{\rm K}_1\left(\varepsilon\rho\right)
\eeq
one finds the light-cone wave functions
\beqn
\Psi^T_{q\bar q}\left(\alpha,\vec\rho\right) & = & 
Z_f\,\frac{\sqrt{\alpha_{em}}}{2\pi}
{\cal O}_{q\bar q}^T\left(\vec\rho\right)
{\rm K}_0\left(\varepsilon\rho\right)\\
\nonumber& = &
Z_f\,\frac{\sqrt{\alpha_{em}}}{2\pi}
\Bigg\{m\,{\rm K}_0\left(\varepsilon\rho\right)\,
\delta_{\lambda_q,\lambda_{\bar q}}
\,\delta_{\lambda_q,\lambda_{\gamma}}\\
& &
+\Bigl({\rm i}\lambda_q\left(2\lambda-1\right)\vec\epsilon_T\cdot\vec e_\rho
+\left(\vec\epsilon_T\times\vec e_z\right)\cdot\vec e_\rho
\Bigr)\varepsilon {\rm K}_1\left(\varepsilon\rho\right)
\delta_{\lambda_q,-\lambda_{\bar q}}\Bigg\}
\eeqn
and
\beqn
\Psi^L_{q\bar q}\left(\alpha,\vec\rho\right) & = & 
Z_f\,\frac{\sqrt{\alpha_{em}}}{2\pi}
{\cal O}_{q\bar q}^L\left(\vec\rho\right)
{\rm K}_0\left(\varepsilon\rho\right)\\
& = &
Z_f\,\frac{\sqrt{\alpha_{em}}}{2\pi}
2\,Q\,\lambda\left(1-\lambda\right)
\,{\rm K}_0\left(\varepsilon\rho\right)\,
\delta_{\lambda_q,\lambda_{\bar q}},
\eeqn
with the Kronecker-$\delta$. 
The unit vector in $\vec\rho$-direction is denoted by $\vec e_\rho$.
As spin vector we have chosen the unit
vector in $z$-direction and the parameter $\lambda_q$ takes the value $+1$ for
spin in positive $z$-direction and the value $-1$ otherwise. For the antiquark
it is vice versa. For the photon, $\lambda_\gamma=+1$ for positive helicity and
$\lambda_\gamma=-1$ for negative helicity. The transverse light cone wave
function has one part that depends on K$_0$ and another part dependent on K$_1$,
the MacDonald function of first order. Note the different spin structures of
these parts. In the K$_0$-part, the spins of the quarks add up to the spin of
the photon, but in the K$_1$-part of the transverse LC wave function, 
the spins of the quarks add to $0$ and the
pair gets an orbital angular momentum.
We finally sum over all flavors, colors, helicities and spin states and get 
the following expression:
\beqn
\nonumber\label{final}
\bar\sigma_{tot}^{\gamma^*\,A} & = & A\int\limits_0^1d\lambda\int d^2\rho
\,\sigma_{q\bar{q}}\left(\rho\right)\,
\left(\left|\Psi_{q\bar q}^{T}\left(\alpha,\rho\right)\right|^2
+\left|\Psi_{q\bar q}^{L}\left(\alpha,\rho\right)\right|^2\right)\\
\nonumber
& - & \frac{3\alpha_{em}}{\left(2\pi\right)^2}
\sum\limits_{f=1}^{N_f}Z_f^2\,{\rm Re}\int d^2b
\int\limits_{-\infty}^{\infty} dz_1 \int\limits_{z_1}^{\infty} dz_2
\int\limits_0^1d\lambda\int d^2\rho_1\int d^2\rho_2\,
e^{-{\rm i}q_L^{min}\left(z_2-z_1\right)} \\
\nonumber& & \;\;\times \;\;
{\rho_A}\left(b,z_1\right)\,{\rho_A}\left(b,z_2\right)\,
\sigma_{q\bar{q}}\left(\rho_2\right)\,
\sigma_{q\bar{q}}\left(\rho_1\right)
\\
& & \;\;\times \;\;
\left\{\left(1-2\lambda\left(1-\lambda\right)\right)\varepsilon^2
\frac{\vec\rho_1\cdot\vec\rho_2}{\rho_1\rho_2}
{\rm K}_1\left(\varepsilon\rho_1\right)
{\rm K}_1\left(\varepsilon\rho_2\right)\right.\\
&&\nonumber\qquad\qquad\Biggl.
+\left(m_f^2+4Q^2\;\lambda^2\left(1-\lambda\right)^2\right)
{\rm K}_0\left(\varepsilon\rho_1\right){\rm K}_0\left(\varepsilon\rho_2\right)
\Biggr\}
W\left(\vec\rho_2,z_2\,|\,\vec\rho_1,z_1\right).
\eeqn 
Here, $|\Psi_{q\bar q}^{T,L}\left(\varepsilon\rho\right)|^2$ 
are the absolute
squares of the transverse and the longitudinal light-cone wavefunctions
summed over all flavors, see
(\ref{psit}) and (\ref{psil}) on page \pageref{psit}.
This form is more convenient for numerical calculations than
(\ref{stotal}) and was used in
\cite{first} for a calculation of nuclear shadowing.
(\ref{final}) was for the first time
suggested in a paper by Zakharov
\cite{zakh}.

In the rest of this subsection, 
the assumptions and approximations entering this derivation
are summarized.
The starting point is the Dirac equation with an abelian potential.
This
simplification is used in anticipation
that all dependence of this potential will be absorbed
into the dipole cross section.
Because the final result contains known approximations as limiting cases,
see section \ref{lightcone}, we believe that it holds also for the case
of a nonabelian interaction. 

Furthermore, no interaction between 
the quark and the antiquark is taken into account and
therefore, the two Dirac equations decouple.
The Furry-approximation \cite{Furry,SM} of the wavefunctions is employed
which is known to be a good
approximation to the continuous spectrum of the Dirac equation 
for high energies. In the next step a two
dimensional Schr\"odinger equation for a scalar function which 
may be regarded as
an effective wavefunction is derived. The $z$-co\-or\-din\-ate
plays the role of time, since the particles move almost with the velocity of
light. This Schr\"odinger equation is solved in terms of it's Green function.
Averaging over all scattering centers in the nucleus yields an optical potential
that is proportional to the total cross section for scattering a $q\bar q$-pair
off a nucleus. The real part of the forward 
scattering amplitude is omitted.
All dependence on the potential is absorbed into the dipole  cross
section. 

The averaging procedure and the summation of the multiple scattering
series is similar to the one in Glau\-ber theory
\cite{Glauber} and most of the approximations come in at this point.
First,  all correlations between the nucleons are neglected. 
Then, the influence of
the potential is described by a phase shift
function.
It is also assumed, that the interaction is short ranged
and the pair interacts
only with one nucleon at a given time. 
The phase shift for scattering 
a particle in the pair
off a
single nucleon is calculated for 
an average value of the transverse coordinate of the particle. 
This means, the
transverse coordinates 
should not vary too rapidly within a longitudinal distance of the order
of the interaction range.
In order to obtain an exponential from the averaging procedure, 
the nuclear
mass number $A$ has to be large enough.
Further approximations are, that both particles 
in the pair see the same nuclear
density. The value of the 
density in the middle between the quark and the antiquark
is used in the calculation. Furthermore, the motion 
of the center of mass of the pair is approximated
by a free motion, since the pair is scattered
predominantly in forward direction.

One finally arrives at the result (\ref{stotal}). This formula allows to
calculate the cross section $\sigma^{\gamma^*A}_{tot}$ 
for arbitrary polarization of
the photon and the pair. However, this equation is not convenient for numerical
calculations and is modified by introducing the light-cone 
wavefunctions (\ref{cross}). Finally, one sums  over all helicity
and spin states, arriving at (\ref{final}).

\subsection{The mean coherence length}\label{meancoh}

We emphasize again that
nuclear shadowing is controlled by the interplay between
two fundamental quantities.
\begin{itemize}
\item{The lifetime of photon fluctuations, or coherence time.
Namely, shadowing is possible only if the coherence 
time exceeds the mean internucleon spacing in nuclei,
and shadowing saturates (for a given Fock component)
if the coherence time substantially exceeds the nuclear 
radius.}
\item{Equally important for shadowing is the transverse
size of the hadronic fluctuation of the virtual photon. 
In order to be shadowed 
the fluctuation  has to interact with
a large cross section. 
As a result of color transparency \cite{zkl,bm,bbgg}, small size
configurations interact only weakly and 
are therefore less shadowed. For the $q\bar q$ Fock component,
the dominant
contribution to shadowing comes from  the large
aligned jet configurations
\cite{bk,fs} of the pair.}
\end{itemize}

In this section a definition of the mean coherence length 
or the mean fluctuation lifetime of a Fock state, relevant for shadowing,
is proposed.  
The mean coherence time for the $q\bar q$ Fock 
state is evaluated using the perturbative (\ref{psit}, \ref{psil})
and nonperturbative (\ref{psitnpt}, \ref{psilnpt})
wavefunctions. It is observed that 
the coherence length is substantially  longer 
for longitudinal than for transverse photons. At the same time, both
are different from the usual prescription 
$l_c=(2m_Nx_{Bj})^{-1}$. 
At high $Q^2$ one approximately has
\beqn
l_c^T&\approx&\frac{2}{5m_Nx_{Bj}},\\
l_c^L&\approx&\frac{4}{5m_Nx_{Bj}},
\eeqn
for transverse and longitudinal photons respectively.
The coherence length is found to 
vary steeply with $Q^2$ at fixed $x_{Bj}$ and small $Q^2$.
The coherence length for the $|q\bar qG\ra$ Fock component, which controlles
nuclear shadowing for gluons, is also calculated.
The latter turns out to be much shorter than for $|q\bar q\ra$ components,
see fig.\ \ref{l-Q2}.
Therefore, the onset of gluon shadowing is expected at smaller $x_{Bj}$
than for quarks.

A photon of virtuality $Q^2$ and energy $\nu$
can develop a hadronic fluctuation for a lifetime,
\beq
l_c=\frac{2\,\nu}{Q^2+M_{q\bar q}^2}=
\frac{P}{x_{Bj}\,m_N}\ ,
\label{1.1}
\eeq
where 
$M_{q\bar q}$ is the effective mass of the fluctuation,
and the factor $P^{-1}=(1+M_{q\bar q}^2/Q^2)$. 
The usual approximation is to assume that
$M_{q\bar q}^2 \approx Q^2$ since $Q^2$ is the only large 
dimensional scale available. In this case $P=1/2$.

The effective mass of a noninteracting $q\bar q$-pair 
is well defined, $M_{q\bar q}^2=(m_f^2+p_T^2)/\alpha(1-\alpha)$,
where $p_T$ and $\alpha$ are the transverse
momentum and fraction of the
light-cone momentum of the photon carried by the quark, respectively.
Therefore, $P$ has a simple form,
\beq
P(k_T,\alpha)=\frac{Q^2\,\alpha\,
(1-\alpha)}{p_T^2+\eps^2}\ ,
\label{1.2}
\eeq
where
\beq
\eps^2 = \alpha(1-\alpha)Q^2 + m_f^2\ .
\label{1.3}
\eeq
To find the mean value of the fluctuation lifetime in vacuum 
one should average (\ref{1.2}) over $p_T$ and $\alpha$ 
weighted with the wavefunction squared of
the fluctuation,
\beq
\la P\ra_{vac}=
\frac{\Bigl\la \Psi_{q\bar q}
\Bigl|P(k_T,\alpha)\Bigr|
\Psi_{q\bar q}\Bigr\ra}
{\Bigl\la \Psi_{q\bar q}\Bigl|
\Psi_{q\bar q}\Bigr\ra}\ .
\label{1.3a}
\eeq

The normalization integral in the denominator in the
r.h.s. of (\ref{1.3a})
diverges at large $p_T$ for transversely polarized photons, 
therefore one arrives at the unexpected result
$\la P^T\ra_{vac}=0$.
The divergence is related to the wavefunction renormalization constant
\beq
\Bigl\la \Psi_{q\bar q}\Bigl|
\Psi_{q\bar q}\Bigr\ra=1-Z_3
\eeq
and can be interpreted as a result
of overwhelming the fluctuations of a transverse photon by
heavy $q\bar q$ pairs with very large $p_T$. Such heavy fluctuations
indeed have a very short lifetime. However, they also have a vanishing
transverse size $\rho\sim 1/p_T$ and interaction cross section.
Therefore, such fluctuation cannot be resolved by the interaction and do
not contribute to the DIS cross section.
To get a sensible result one should properly define the averaging
procedure. One is interested in the fluctuations which contribute
to nuclear shadowing, i.e.\ they have to interact at least twice.
Correspondingly, the averaging procedure 
has to be redefined as,
\beq
\la P\ra_{shad}=
\frac{\Bigl\la f(\gamma^*\to q\bar q)
\Bigl|P(p_T,\alpha)\Bigr|
f(\gamma^*\to q\bar q)\Bigr\ra}
{\Bigl\la f(\gamma^*\to q\bar q)\Bigl|
f(\gamma^*\to q\bar q)\Bigr\ra}\ ,
\label{1.6a}
\eeq
where $f(\gamma^*\to q\bar q)$ is the amplitude of diffractive
dissociation of the virtual photon on a nucleon 
$\gamma^*\,p\to q\bar q\,p$, (\ref{diffamp}).

Therefore, $P$ has to be weighted with the 
interaction cross section squared $\sigma^2_{q\bar q}(s,\rho)$
in the averaging procedure.
Then, the mean value of factor $P(\alpha,p_T)$ reads,
\beq
\left\la P^{T,L}\right\ra=\frac{\int_0^1d\alpha\int\! d^2\rho_1d^2\rho_2
\left[\Psi_{q\bar q}^{T,L}\left(\vec \rho_2,\alpha\right)\right]^*\!\!
\sigma_{q\bar q}\left(s,\rho_2\right)
\widetilde P\left(\vec \rho_2-\vec \rho_1,\alpha\right)
\Psi_{q\bar q}^{T,L}\left(\vec \rho_1,\alpha\right)
\sigma_{q\bar q}\left(s,\rho_1\right)}
{\int_0^1d\alpha\int d^2\rho
\left|\Psi_{q\bar q}^{T,L}\left(\vec \rho_,\alpha\right)
\sigma_{q\bar q}\left(s,\rho\right)\right|^2}
\label{1.7}
\eeq
with
\beq
\widetilde P\left(\vec \rho_2-\vec \rho_1,\alpha\right)=
\int\frac{d^2p_T}{\left(2\pi\right)^2}\,
{\exp\left(-{\rm i}\,\vec p_T\cdot\left( \vec \rho_2-\vec \rho_1\right)\right)}
{P\left(\alpha,\rho\right)}.
\label{1.8}
\eeq

Using expression (\ref{1.2}) one obtains for
a non interacting  $q\bar q$-pair,
\beq
\widetilde P\left(\vec \rho_2-\vec \rho_1,\alpha\right)=
\frac{Q^2\alpha\left(1-\alpha\right)}
{2\pi}{\rm K}_0\left(\varepsilon\left|\vec \rho_2
-\vec \rho_1\right|\right).
\label{1.9a}
\eeq

As a simple estimate for the mean value 
(\ref{1.7}) one can use the small-$\rho$
approximation for the dipole cross section 
$\sigma_{q\bar q}(s,\rho)=C(s)\,\rho^2$. 
The
Factor $C(s)$ does not enter the
result since it cancels in (\ref{1.7}).
One obtains for transverse and longitudinal photons
with perturbative wavefunctions (\ref{psit}, \ref{psil})
\beq
\left\la P^T\right\ra =
\frac{2\,Q^2}{3}\,\,
\frac{\int\limits_{0}^{1}
d\alpha\,
(1-\alpha)\,\alpha\,
\Bigl(
\Bigl[\alpha^2+(1-\alpha)^2\Bigr]\,
\Bigl/\,\varepsilon^6
+\frac{7}{8}\,m_f^2(1-\alpha)\,\alpha\,\Bigl/\,\varepsilon^8
\Bigr)
}
{\int\limits_{0}^{1}
d\alpha\,
\Bigl(
\Bigl[\alpha^2+(1-\alpha)^2\Bigr]\,
\Bigl/\,\varepsilon^4\
+\frac{2}{3}\,m_f^2\Bigl/\,\varepsilon^6
\Bigr)}
 ;
\label{1.10}
\eeq
\beq
\left\la P^L\right\ra =
\frac{7\,Q^2}{8}\,\,
\frac{\int\limits_{0}^{1}
d\alpha\,
(1-\alpha)^3\,\alpha^3\,
\Bigl/\,\varepsilon^8}{\int\limits_{0}^{1}
d\alpha\,
(1-\alpha)^2\,\alpha^2\,
\Bigl/\,\varepsilon^6}\ ,
\label{1.10a}
\eeq
respectively.

The factor $\la P^{T,L}\ra$ is calculated as function of $Q^2$ 
at $x_{Bj}=0.01$ from (\ref{1.10}) and (\ref{1.10a}).
The results depicted in fig.\ \ref{l-Q2} by dotted lines are 
quite different from the naive estimate $P^{T,L}=1/2$. 
Besides, $P^L$ turns out to be substantially longer than $P^T$.
This indicates that a longitudinally polarized photon develops
lighter fluctuations than a transverse one. Indeed, the effective mass
$M_{q\bar q}$ is maximal for asymmetric pairs, i.\ e.\ when
$\alpha$ or $1-\alpha$ are small. However, such fluctuations
are suppressed in longitudinal photons by the
wavefunction (\ref{psil}).

\begin{figure}[t]
  \centerline{\scalebox{0.8}{\includegraphics{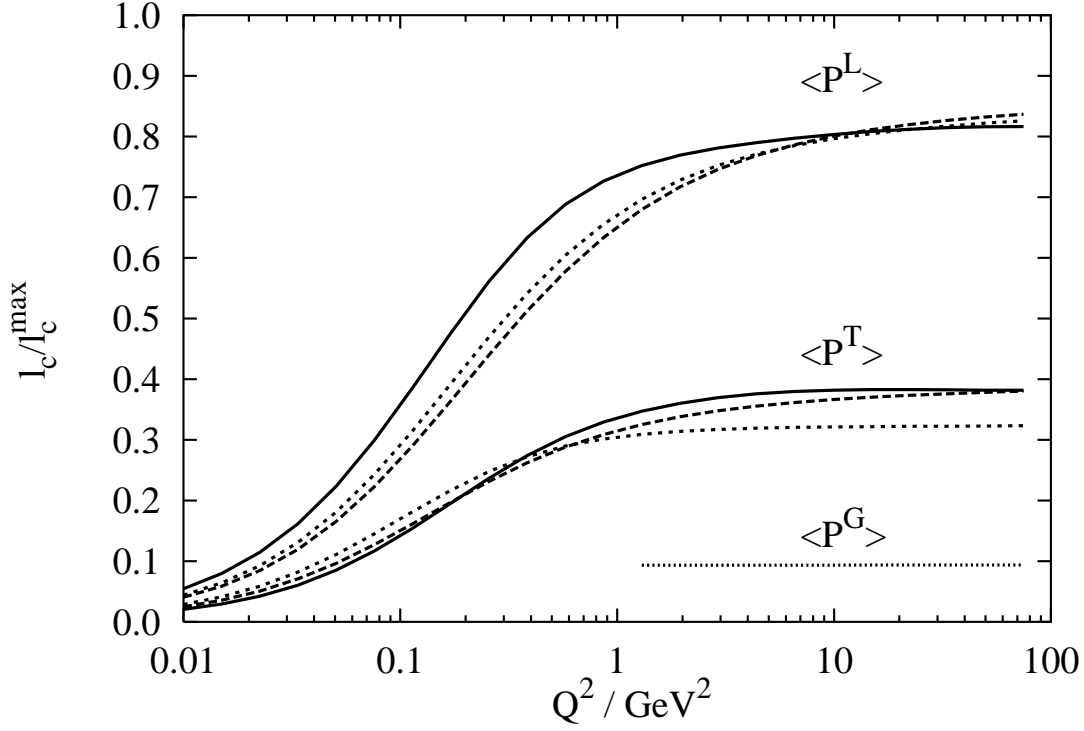}}}
    \caption{
      \label{l-Q2}
      The dependence of $\left\la P\right\ra=
        l_c/l_c^{max}$, defined in (\ref{1.1}),
        on $Q^2$
         at $x_{Bj}=0.01$. 
         Here, $l_c^{max}=1/m_Nx_{Bj}$ is the largest possible coherence 
         length.
         The curves are
        for $q\bar q$ fluctuations of transverse 
        and longitudinal
        photons, and for $q\bar qG$ fluctuation, from the top to 
        bottom, respectively. Dot\-ted curves correspond to
        calculations with perturbative wavefunctions 
        (\ref{psit},\ref{psil})
        and an approximate dipole
        cross section $\propto \rho^2$. Dashed curves are the same, except
        the realistic parameterization (\ref{improved}) is employed.
        The solid curves are calculated with 
        the nonperturbative wavefunctions (\ref{psitnpt},\ref{psilnpt}). 
        Note that as $Q^2$ becomes to low at fixed $x_{Bj}$, no high energy
        approximation can be applied any more.
    }  
\end{figure}

The dependence of $\left\la P^{T,L}\right\ra$ on $x_{Bj}$ 
depicted in fig.\ \ref{l-x} for $Q^2=4$ GeV$^2$ and 
$Q^2=40$ GeV$^2$
is rather smooth. Therefore, the coherence length 
varies approximately as $l_c\propto 1/x_{Bj}$.
\begin{figure}[t]
  \centerline{\scalebox{0.8}{\includegraphics{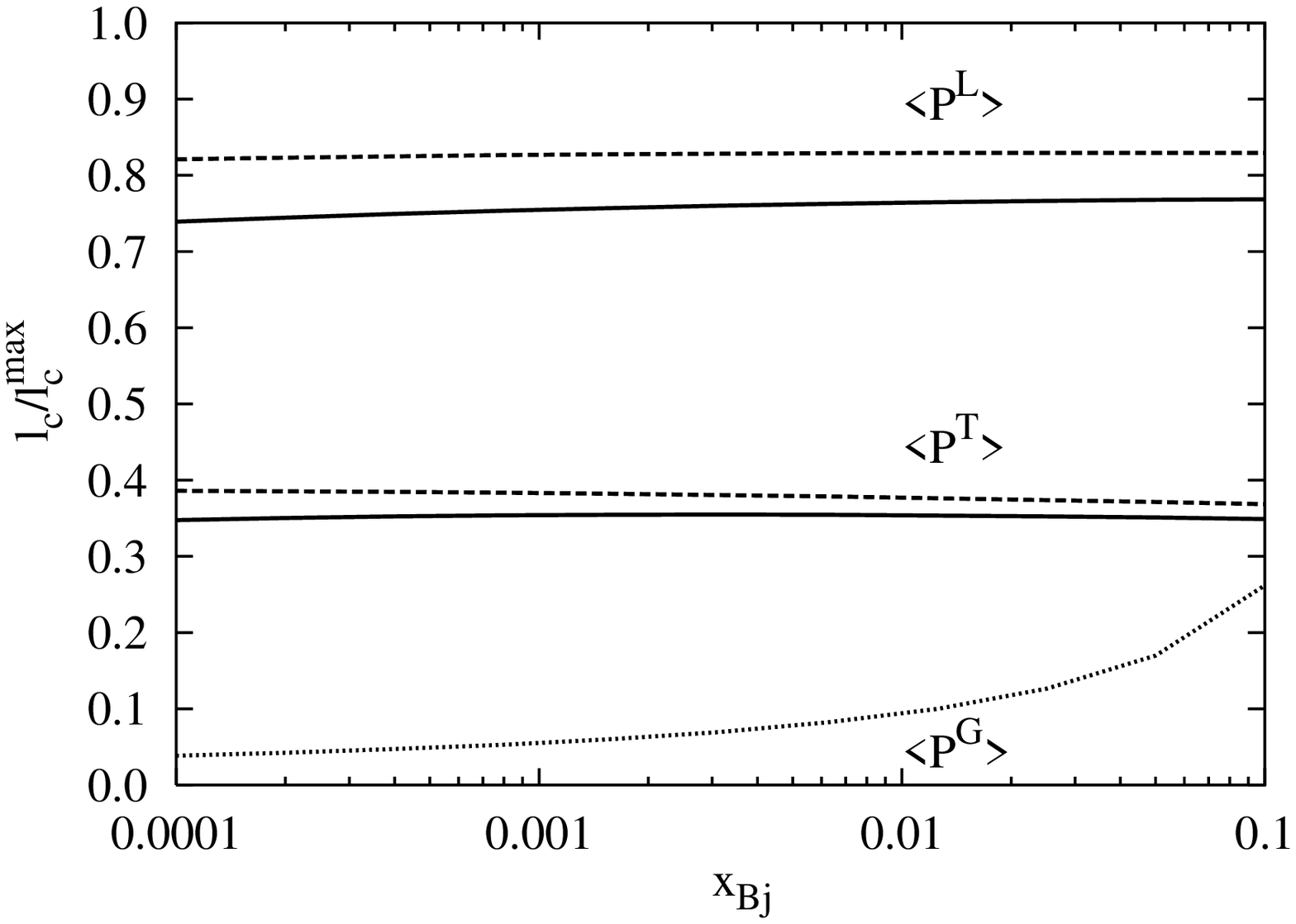}}}
    \caption{
      \label{l-x}
      $x_{Bj}$ dependence of  $\left\la P\right\ra=l_c/l_c^{max}$
        defined in (\ref{1.1}), corresponding to the coherence length 
        for shadowing of transverse and longitudinal
        photons and gluon shadowing, respectively. 
        Here, $l_c^{max}=1/m_Nx_{Bj}$ is the largest possible coherence length. 
        Solid and dashed curves
        correspond to $Q^2=4$ and $40$ GeV$^2$
        and are calculated with the
        nonperturbative wavefunctions (\ref{psitnpt},\ref{psilnpt}). The
        realistic parametrization of the dipole cross section (\ref{improved})
        is used. The dotted curve shows the factor $\left\la P\right\ra$ for the
        $q\bar qG$ Fock-state. In this case the perturbative LC wavefunction 
        (\ref{psil}) and an approximate
         dipole cross section $\propto \rho^2$ are used. The dotted curve is
         independent of $Q^2$.
    }  
\end{figure}

The simple approximation $\sigma_{q\bar q} \propto \rho^2$ is not
realistic since nonperturbative effects affect the large-$\rho$ behavior.
A more realistic parametrization of the dipole cross section (\ref{improved})
is introduced in section \ref{DCS}.

 Note that (\ref{1.7}) can be represented in the form,
\beq
\la P^{T,L}\ra =
\frac{N^{T,L}}{D^{T,L}}\ ,
\label{1.12}
\eeq
The angular integrations in (\ref{1.7}) for the denominators 
$D^{T,L}$ are trivial and for the
numerators $N^{T,L}$ one
uses the relation \cite{gr},
\beq
{\rm K}_0\left(\varepsilon\left|\vec \rho_1
-\vec \rho_2\right|\right)={\rm K}_0\left(\varepsilon \rho_>\right)
{\rm I}_0\left(\varepsilon
\rho_<\right)+2\sum_{m=1}^\infty {\rm e}^{{\rm i}m\phi}
{\rm K}_m\left(\varepsilon \rho_>\right){\rm I}_m\left(\varepsilon
\rho_<\right),
\label{1.12a}
\eeq
where $\rho_>=\max\left(\rho_1,\rho_2\right)$, 
$\rho_<=\min\left(\rho_1,\rho_2\right)$,
$\cos\phi=\vec \rho_1\cdot\vec \rho_2/(\rho_1\rho_2)$ and 
${\rm I}_m(z)$ are the
modified Bessel functions of first kind 
(Bessel function of imaginary variable). 
It is clear from this relation
 that after angular integration only one term in the sum gives a
non-vanishing contribution. One finally obtains for transverse photons

\beqn\nonumber\label{nt}
N_p^{T}&=&2\,Q^2\int\limits_0^1
d\alpha\,\alpha\,\left(1-\alpha\right)
\int\limits_0^\infty d\rho_2\ \rho_2\int\limits_0^{\rho_2}
d\rho_1\ \rho_1\left\{m_f^2\,
{\rm K}_0^2\left(\varepsilon \rho_2\right)\,
{\rm K}_0\left(\varepsilon \rho_1\right)
\,{\rm I}_0\left(\varepsilon \rho_1\right)\right.\\
&+&\left.\left[\alpha^2+
\left(1-\alpha\right)^2\right]\,\varepsilon^2\,
{\rm K}_1^2\left(\varepsilon \rho_2\right)\,
{\rm K}_1\left(\varepsilon \rho_1\right)
\,{\rm I}_1\left(\varepsilon \rho_1\right)\right\}
\,\sigma_{q\bar q}\left(s,\rho_1\right)\,
\sigma_{q\bar q}\left(s,\rho_2\right),
\label{1.13}
\eeqn
\beq\label{dt}
D_p^{T}=\int\limits_0^1 d\alpha
\int\limits_0^\infty d\rho\ \rho\left\{
m_f^2\,
{\rm K}_0^2\left(\varepsilon \rho\right)
+\left[\alpha^2+\left(1-\alpha\right)^2\right]\varepsilon^2
\,{\rm K}_1^2\left(\varepsilon \rho\right)\right\}
\sigma_{q\bar q}^2\left(s,\rho\right),
\label{1.14}
\eeq
and for longitudinal photons,
\beqn\nonumber\label{nl}
N_p^{L}&=&2\,Q^2\int d\alpha\,\alpha^3\,\left(1-\alpha\right)^3
\int\limits_0^\infty d\rho_2\rho_2\int\limits_0^{\rho_2} d\rho_1\rho_1\\
&\times&
{\rm K}_0^2\left(\varepsilon \rho_2\right)
{\rm K}_0\left(\varepsilon \rho_1\right)
{\rm I}_0\left(\varepsilon \rho_1\right)
\sigma_{q\bar q}\left(s,\rho_1\right)\sigma_{q\bar q}\left(s,\rho_2\right),
\label{1.15}
\eeqn
\beq\label{dl}
D_p^{L}=\int d\alpha
\int\limits_0^\infty d\rho \rho
\alpha^2\left(1-\alpha\right)^2
{\rm K}_0^2\left(\varepsilon \rho\right)
\sigma_{q\bar q}^2\left(s,\rho\right).
\label{1.16}
\eeq

The factor $\la P^{T,L}(x,Q^2)\ra$ calculated in this way 
is depicted by dashed lines in fig.~\ref{l-Q2} as
function of $Q^2$ at $x_{Bj}=0.01$. 
It is not much different from the previous simplified
estimate demonstrating low sensitivity to the form
of the dipole cross section.

Although the quarks should be treated perturbatively
as nearly massless, at the endpoints
$\alpha$ or $1-\alpha \to 0$ the mean ${q\bar q}$ transverse separation
${\rho}\sim 1/\eps$ 
becomes huge $\sim 1/m_f$ and nonperturbative effects are important. 
A more sophisticated treatment of these effects is developed in \cite{KST99},
where an interaction between the particles in the pair is introduced, as
explained in section
\ref{dipoledis}.
In the following, we will perform the same calculations again, but now with the
nonperturbative LC wavefunctions.

Note that in $p_T$ representation the free Green function
$G^0_{{q\bar q}}(z_1,\vec \rho_1;z_2,\vec \rho_2)$ integrated over longitudinal 
coordinate is simply related to the coherence length (\ref{1.1}), if one
performs an analytic continuation to imaginary time, $z\to -{\rm i}z$
\beq
\int\limits_{z_1}^{\infty}d\,z_2\,
G^0_{{q\bar q}}(z_1,\vec \rho_1;z_2,\vec \rho_2)=
\,\int \frac{d^2p_T}{(2\,\pi)^2}\ 
{\rm exp}\Bigl[-{\rm i}\,\vec p_T\cdot(\vec \rho_2-\vec \rho_1)\Bigr]
\,l_c(p_T,\alpha)\ .
\label{2.3}
\eeq
This is easily generalized to include the nonperturbative
interaction. Then,
making use of this relation one can switch in (\ref{1.7}) to
${\rho}$ representation and express the mean coherence
length via the Green function. One arrives at new
expressions for the functions $N^{T,L}$ and $D^{T,L}$ in
(\ref{1.12}),
\beqn
N^{T,L}&=& \int\limits_0^1 d\alpha
\int d^2\rho_1\,d^2\rho_2\,
\left[\Psi_{{q\bar q}}^{T,L}
\left(\vec \rho_2,\alpha\right)\right]^*\,
\sigma_{{q\bar q}}(s,\rho_2)\,
\left(\int\limits_{z_1}^\infty dz_2\, 
G^{vac}_{q\bar q}\left(\vec \rho_1,z_1;\vec \rho_2,z_2\right)\right)
\nonumber\\
&\times&\,\Psi_{{q\bar q}}^{T,L}
\left(\vec \rho_1,\alpha\right)\,
\sigma_{{q\bar q}}(s,\rho_1),
\label{2.4}
\\
D^{T,L}&=& \int_0^1d\alpha\int d^2r\,
\left|\Psi_{q\bar q}^{T,L}\left(\vec r_,\alpha\right)
\sigma_{q\bar q}\left(r,s\right)\right|^2\ ,
\label{2.5}\ 
\eeqn
where the nonperturbative ${q\bar q}$ wave functions are given 
by (\ref{psitnpt}) and (\ref{psilnpt}).

For a harmonic oscillator potential the Green function is known
analytically,
\beqn
G^{vac}_{q\bar q}\left(\vec \rho_2,z_1;\vec \rho_1,z_2\right)
&=&\frac{a^2\left(\alpha\right)}
{2\pi\sinh\left(\omega\,\Delta z\right)}\,
\exp\left[-\frac{\varepsilon^2\,\Delta z}{2\,\nu\,\alpha(1-\alpha)}\right]
\\ \nonumber&\times&\,
\exp\left\{-\frac{a^2\left(\alpha\right)}{2}\left[
\left(\rho_1^2+\rho_2^2\right)\coth\left(\omega\,\Delta z\right)
-\frac{2\vec \rho_1\cdot\vec \rho_2}{\sinh\left(\omega \,\Delta z\right)}
\right]\right\},
\label{2.6}
\eeqn
where $\Delta z=z_2-z_1$ and
\beq
\omega=\frac{a(\alpha)^2}{\nu\,\alpha(1-\alpha)}\ ,
\label{2.7}
\eeq
is the oscillator frequency, cf.\ section \ref{dipoledis}.

Now we have all ingredients which are necessary to calculate (\ref{1.12}).
Two from the eight remaining integrations, over the angles, can
be performed analytically. The final result is 
\beqn\label{ntnpt}
N^T&=&m_N\,x_{Bj}
\int_0^1d\alpha\int_0^\infty d\rho_1\,\rho_1\,d\rho_2\,\rho_2
\int_0^\infty d\Delta z\,
\left[\Psi_{q\bar q}^{T}\left(\varepsilon,\lambda,\vec \rho_2\right)\right]^*
\Psi_{q\bar q}^{T}\left(\varepsilon,\lambda,\vec \rho_1\right)
\nonumber\\ &\times&\,
\sigma_{q\bar q}\left(\rho_2,s\right)\,
\sigma_{q\bar q}\left(\rho_1,s\right)\,
\frac{a^2\left(\alpha\right)}
{\sinh\left(\omega\,\Delta z\right)}\ 
\exp\left[-\frac{\varepsilon^2\,\Delta z}
{2\,\nu\,\alpha(1-\alpha)}\right]
\nonumber\\ &\times&\,
{\rm I}_1\left[\frac{a^2\left(\alpha\right)\rho_1\rho_2}
{\sinh\left(\omega\,\Delta z\right)}\right]\ 
\exp\left[-\frac{a^2\left(\alpha\right)}{2}
\left(\rho_1^2+\rho_2^2\right)
\coth\left(\omega\,\Delta z\right)\right],
\label{2.13}
\eeqn
\beqn\label{nlnpt}
N^L&=&m_N\,x_{Bj}
\int_0^1d\alpha\int_0^\infty d\rho_1\,\rho_1\,d\rho_2\,\rho_2
\int_0^\infty d\Delta z\,
\left[\Psi_{q\bar q}^{T}\left(\varepsilon,\lambda,\vec \rho_2\right)\right]^*
\Psi_{q\bar q}^{T}\left(\varepsilon,\lambda,\vec \rho_1\right)
\nonumber\\ &\times&\,
\sigma_{q\bar q}\left(\rho_2,s\right)\,
\sigma_{q\bar q}\left(\rho_1,s\right)\,
\frac{a^2\left(\alpha\right)}
{\sinh\left(\omega\,\Delta z\right)}\
\exp\left[-\frac{\varepsilon^2\,\Delta z}
{2\,\nu\,\alpha(1-\alpha)}\right]
\nonumber\\ &\times&\,
{\rm I}_0\left[\frac{a^2\left(\alpha\right)\,\rho_1\,\rho_2}
{\sinh\left(\omega\,\Delta z\right)}\right]\,
\exp\left[-\frac{a^2\left(\alpha\right)}{2}
\left(\rho_1^2+\rho_2^2\right)\coth\left(\omega\,\Delta z\right)
\right],
\label{2.14}
\eeqn
\beq\label{dlnpt}\label{dtnpt}
D^{L,T}=
\int_0^1d\alpha\int_0^\infty dr\,r\,
\left|
\Psi_{q\bar q}^{T,L}\left(\varepsilon,\lambda,\vec r\right)\,
\sigma_{q\bar q}\left(r,s\right)\right|^2.
\label{2.15}
\eeq
For a dipole
cross section that levels off  at large separations
like the older parametrization
(\ref{sasha}) the
integrations over $\rho_1$ and $\rho_2$ can also be done analytically.
However, we prefer to work with the more general expressions that hold
for arbitrary $\sigma_{q\bar q}\left(s,\rho\right)$,
as long as it depends only  on the modulus of $\rho$.
The remaining integrations are performed 
numerically. The results for $l_c^{T,L}(x,Q^2)$ are shown by solid
curves in figs.\ \ref{l-Q2} and \ref{l-x}.

It is instructive to compare our calculations with the VDM 
which is usually supposed to dominate
at small $Q^2 \leq m_{\rho}^2$. The corresponding coherence length
$l_c^{VDM}$ is given by (\ref{1.1}) with $M_{q\bar q}=m_{\rho}$.
The ratio of $l_c^T$ calculated with the nonperturbative wave function
to $l_c^{VDM}$ as function of $Q^2$ is shown by solid curve in fig.\
\ref{vdm}.
It demonstrates an unexpectedly strong deviation from the VDM expectation
at quite low $Q^2$. We also calculated $l_c^T$ with
the perturbative wavefunctions and a  quark mass of $m_f=200$ MeV.
This choice
mimics the nonperturbative effects quite well, as one can see from 
fig.\ \ref{vdm}. 
\begin{figure}[t]
  \centerline{\scalebox{0.8}{\includegraphics{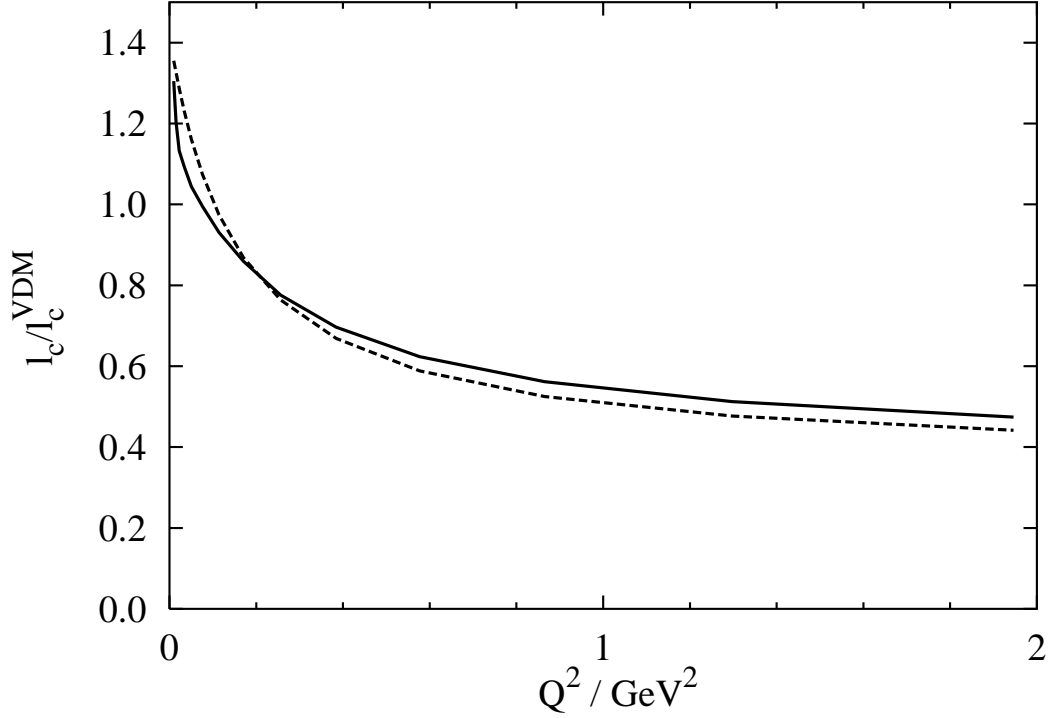}}}
    \caption{
      \label{vdm}
        $Q^2$ dependence of ratio of $\la l_c^T\ra$  calculated
        with (\ref{1.7}) and $m_f=200$ MeV to
        $l_c^{VDM}$ calculated with (\ref{1.1}) and $M=m_{\rho}$.
    }  
\end{figure}

Shadowing in the nuclear gluon distributing function at small 
$x_{Bj}$ which looks like gluon fusion $GG\to G$ in the infinite 
momentum frame of the nucleus, 
should be treated in the rest frame of the nucleus as
shadowing for the Fock components of the photon 
containing gluons. 

The lowest gluonic Fock component is the $|{q\bar q}G\ra$ state. 
The coherence length relevant to
shadowing depends according to (\ref{1.1}) on the effective mass
of the $|{q\bar q}G\ra$ which should be expected to be 
higher than that for the $|{q\bar q}\ra$ state.  Correspondingly,
the coherence length $\la l^G_c\ra$ should be
shorter and a onset of gluon shadowing  
is expected to start at smaller $x_{Bj}$.

For this coherence length one can use 
the same formula as in the $q\bar q$ case (\ref{1.1}), 
but with the effective mass,
\beq\label{mqqg}
M^2_{{q\bar q}G}=\frac{p_T^2}{\alpha_G(1-\alpha_G)}+
\frac{M_{{q\bar q}}^2}{1-\alpha_G}\ ,
\label{3.1}
\eeq
where $\alpha_G$ is the fraction of the photon momentum 
carried by the gluon, and
$M_{{q\bar q}}$ is the effective mass of the ${q\bar q}$ pair.
However, (\ref{mqqg}) is valid only in the perturbative limit.
It is modified by the nonperturbative interaction which is much stronger for
gluons than for quarks \cite{KST99}. The nonperturbative interaction in the
$q\bar qG$ state is discussed in detail in section \ref{diffractionsec}.
Therefore, we switch to
the  Green function formalism which
recovers (\ref{3.1}) in the limit of high $Q^2$.

Gluons are treated as massless and transverse. For the
factor $P$ defined in
(\ref{1.1}) one can write,
\beq\label{gluonl}
\Bigl\la P^{G}\Bigr\ra = 
\frac{N^G}{D^G},
\label{3.2}
\eeq
where 
\beqn\nonumber
N^{G}&=& m_N\,x_{Bj}\,
\int d^2r_{1G}\,d^2r_{1q\bar q}\,d^2r_{2G}\,
d^2r_{2q\bar q}\,d\alpha_q
\,d\ln(\alpha_G)\,
\widetilde \Psi^{\dagger}_{{q\bar q}G}
\left(\vec r_{2G},\vec r_{2q\bar q},
\alpha_q,\alpha_G\right)\\ & \times &
\left(\int\limits_{z_1}^\infty dz_2\, G_{{q\bar q}G}
\left(\vec r_{2G},\vec r_{2q\bar q},z_2;
\vec r_{1G},\vec r_{1q\bar q},z_1\right)
\right)\,
\widetilde \Psi_{{q\bar q}G}\left(\vec r_{1G},\vec r_{1q\bar q},
\alpha_q,\alpha_G\right)
\label{3.3}
\eeqn
\beqn\nonumber
D^{G}&=&\int d^2r_{1G}\,d^2r_{1q\bar q}\,d^2r_{2G}\,
d^2r_{2q\bar q}\,d\alpha_q
\,d\ln(\alpha_G)\,
\widetilde \Psi^{\dagger}_{{q\bar q}G}
\left(\vec r_{2G},\vec r_{2q\bar q},
\alpha_q,\alpha_G\right)\\ & \times &
\delta^{\left(2\right)}\left(\vec r_{2G}-\vec r_{1G}\right)\,
\delta^{\left(2\right)}
\left(\vec r_{2q\bar q}-\vec r_{1q\bar q}\right)\,
\widetilde \Psi_{{q\bar q}G}\left(\vec r_{1G},\vec r_{1q\bar q},
\alpha_q,\alpha_G\right)
\label{3.4}
\eeqn
Here we have introduced the Jacobi variables, 
$\vec r_{q\bar q}=\!\vec R_{\bar q}-\vec R_{q}$ and
$\vec r_{G}=\!\vec R_G-(\alpha_{\bar q}\vec R_{\bar q}+\alpha_q\vec R_{q})
/(\alpha_{\bar q}+\alpha_q)$. $\vec R_{G,q,\bar q}$ are the position vectors of
the gluon, the quark and the antiquark in the transverse plane and
$\alpha_{G,q,\bar q}$ are the longitudinal momentum fractions.

The Green function describing the propagation
of the ${q\bar q}G$ system satisfies the two dimensional Schr\"odinger 
equation \cite{KST99},
\beqn
&&
\left[\frac{\partial}{\partial z_2}
-\frac{Q^2}{2\nu}
+\frac{\alpha_q+\alpha_{\bar q}}
{2\nu\alpha_q\alpha_{\bar q}}
\Delta_\perp\left(r_{q\bar q}\right)
+\frac{\Delta_\perp\left(r_{2G}\right)}
{2\nu\alpha_G\left(1-\alpha_G\right)}
-V\left(\vec r_{2G},
\vec r_{2q\bar q},\alpha_q,\alpha_G,z_2\right)
\right]
\nonumber\\ \nonumber&\times&
G_{q\bar qG}
\left(\vec r_{2G},\vec r_{2q\bar q},z_2;
\vec r_{1G},\vec r_{1q\bar q},z_1\right)
\,=\,
\delta\left(z_2-z_1\right)\,
\delta^{\left(2\right)}\left(\vec r_{2G}-\vec r_{1G}\right)\,
\delta^{\left(2\right)}
\left(\vec r_{2q\bar q}-\vec r_{1q\bar q}\right). 
\label{3.5}\\
\eeqn
This equation accounts for the relative transverse motion of the three particles
and for the interaction between them.

In order to calculate the coherence
length relevant to shadowing, we employ the amplitude for the diffractive
dissociation $\gamma^*\to {q\bar q}G$, which is 
the  ${q\bar q}G$ wavefunction 
weighted by the cross section \cite{KST99}, in analogy to (\ref{diffamp})
\beqn\nonumber\lefteqn{
\widetilde \Psi_{{q\bar q}G}
\left(\vec r_{G},\vec r_{q\bar q},
\alpha_q,\alpha_G\right) =
\Psi^{T,L}_{{q\bar q}}\left(\vec r_{q\bar q},\alpha_q\right)
\left[
\Psi_{qG}\left(\frac{\alpha_G}{\alpha_q},\vec r_{G}
+\frac{\alpha_{\bar{q}}}{\alpha_q+\alpha_{\bar{q}}}\,
\vec r_{q\bar q}\right)\right.}
&&\\
\nonumber & - & \left.
\Psi_{\bar qG}\left(\frac{\alpha_G}{\alpha_{\bar{q}}},\vec r_{G}
-\frac{\alpha_{q}}{\alpha_q+\alpha_{\bar{q}}}\,
\vec r_{q\bar q}\right)
\right]\ 
\frac{9}{8}\ 
\left[\sigma_{q\bar q}\left(s,\vec r_{G}
+\frac{\alpha_{\bar{q}}}{\alpha_q+\alpha_{\bar{q}}}\,
\vec r_{q\bar q}\right)
\right.\\
&&\left.
+\ \sigma_{q\bar q}\left(s,\vec r_{G}
-\frac{\alpha_{q}}{\alpha_q+\alpha_{\bar{q}}}\,
\vec r_{q\bar q}\right)
-\sigma_{q\bar q}\left(s,r_{q\bar q}\right)
\right].
\label{3.6}
\eeqn

As different from the case of the $|{q\bar q}\ra$ Fock state,
where perturbative QCD can be safely applied at high $Q^2$,
the nonperturbative effects remain important for the
$|q\bar qG\ra$ component even for highly virtual photons.
Since calculations for the full three body problem are obviously quite involved,
one has to introduce appropriate simplifications.
At high $Q^2$  the size of a ${q\bar q}$ pair 
from a longitudinal photon
is always of order $\sim 1/Q$.
However, the mean quark-gluon separation at $\alpha_G \ll 1$ 
depends on the strength of gluon interaction which
is characterized in this limit by the
parameter $b_0\approx 0.65$ GeV \cite{KST99}.
For $Q^2\gg b_0^2$ the ${q\bar q}$ is small,
$r^2_{{q\bar q}} \ll r_G^2$, and one can treat the ${q\bar q}G$ system
as a color octet-octet dipole, cf.\ section \ref{diffractionsec},
\beq\label{app1}
G_{q\bar qG}\left(\vec r_{2G},
\vec r_{2q\bar q},z_2;\vec r_{1G},\vec r_{1q\bar q},z_1
\right)\ \to 
G_{q\bar q}\left(\vec r_{2q\bar q},z_2;
\vec r_{1q\bar q},z_1\right)\,
G_{GG}\left(\vec r_{2G},z_2;\vec r_{1G},z_1\right)\ .
\label{3.7}
\eeq
Such a Green function $G_{GG}$ satisfies the much simpler 
equation \cite{KST99},
\beqn\label{x}\nonumber{
\left[\frac{\partial}{\partial z_2}-\frac{Q^2}{2\nu}
+\frac{\Delta_\perp\left(r_{2G}\right)}
{2\nu\alpha_G\left(1-\alpha_G\right)}
-\frac{b_0^4\,r_{2G}^2}
{2\nu\alpha_G\left(1-\alpha_G\right)}
\right]
G_{GG}\left(\vec r_{2G},z_2;\vec r_{1G},z_1\right)}\\
\qquad=\delta\left(z_2-z_1\right)
\delta^{\left(2\right)}\left(\vec r_{2G}-\vec r_{1G}\right).
\label{3.8}
\eeqn
Correspondingly, the modified ${q\bar q}G$ wave function
simplifies too,
\beq\label{app2}
\widetilde \Psi_{{q\bar q}G}
\left(\vec r_{G},\vec r_{q\bar q},
\alpha_q,\alpha_G\right) \Rightarrow
-\,\Psi^L_{{q\bar q}}\left(\vec r_{q\bar q},
\alpha_q\right)\ \vec r_{q\bar q}\cdot\vec\nabla\,
\Psi_{qG}\left(\vec r_{G}\right)\ 
\sigma_{GG}\left(s,r_{G}\right)\ ,
\label{3.9}
\eeq
where the nonperturbative quark-gluon wave function 
has the form \cite{KST99},
\beq
\Psi_{qG}\left(\vec r_{G}\right)=
\lim_{\alpha_G\to 0}\Psi_{qG}\left(\alpha_G,r_{G}\right)=
\frac{2}{\pi}\,
\sqrt\frac{\alpha_s}{3}\ 
\frac{\vec e\cdot\vec r_{G}}{r^2_{G}}\,
\exp\left(-\frac{b_0^2}{2}\,r^2_{G}\right)\ ,
\label{3.10}
\eeq
and the color-octet dipole cross section reads,
\beq
\sigma_{GG}\left(s,r_{G}\right)
=\frac{9}{4}\,
\sigma_{q\bar q}\left(s,r_{G}\right),
\label{3.11}
\eeq
as it was already discussed in section \ref{diffractionsec}.

With the approximations given above, the factor $\la P^G\ra$
in (\ref{1.1}) for the gluon coherence length is calculated
in Appendix ~A with the result
\beqn
\left<P^G\right>&=&\frac{2}{3\ln(\alpha_G^{max}/\alpha_G^{min})}\ 
\int\limits_{\delta^{min}}^{\delta^{max}}\frac{d\delta}{\delta}
\left[\frac{5}{8\left(1+\delta\right)}
+\frac{7}{8\left(1+3\delta\right)}\right.\\
&-&\left.\frac{\delta}{\delta^2-1}\left(
\psi\left(2\right)-\psi\left(\frac{3}{2}+\frac{1}{2\delta}\right)\right)
\right],
\label{3.12}
\eeqn
where
\beq
\psi\left(x\right)=
\frac{d\ln\Gamma\left(x\right)}{dx}\quad,\quad
\hspace{2cm}
\delta=\frac{2\,b_0^2}{Q^2\,\alpha_G}\ .
\label{3.14}
\eeq

Both the numerator and denominator in (\ref{3.12})
diverge logarithmically for $\alpha_G^{min}\to 0$,
as it is characteristic for radiation of vector bosons. 
To find an appropriate lower cut off, 
note that the mass of the $q\bar qG$ system is
approximately given by
\beq
M^2_{q\bar qG}\approx \frac{2b_0^2}{\alpha_G}+Q^2,
\eeq
where (\ref{3.1}) was used
with $\la p_T^2\ra \approx b_0^2$.
Demanding $M^2_{q\bar qG}<0.2s$  leads to
$\alpha_G^{min}=2b_0^2/(0.2s-Q^2)$. 
Furthermore
we work in the approximation of $\alpha_G\ll 1$ and also 
have to choose an upper cut off.
We use 
\beq\label{limitst}
\frac{2b_0^2}{0.2s-Q^2}\le\alpha_G\le\frac{2b_0^2}{Q^2},
\eeq
which means that we take only masses 
$2Q^2\leq M^2_{q\bar qG}\leq 0.2\,s$ into account.
The two limits become equal  at
$x_{Bj}\approx 0.1$.

Our results for $\la P^G\ra = \la l_c^G\ra/l_c^{max}$ 
are depicted in fig.\ \ref{l-Q2}. 
The approximations made above break down at low $Q^2$. Calculations are
therefore done only at high virtualities $Q^2\ge 1$ GeV$^2$.
The coherence length for the $q\bar qG$ state is much shorter than 
both $l_c^T$ and $l_c^L$ for $|{q\bar q}\ra$ fluctuations. This
conclusion corresponds to delayed onset of gluon shadowing
shifted to smaller $x_{Bj}$ predicted in \cite{KST99}.



\def\Hs{\hat{s}}
\def\Ht{\hat{t}}
\def\Hu{\hat{u}}
\def\bb{{\bf b}}
\def\bs{{\bf s}}
\def\eps{\varepsilon}
\def\sNN{\sigma^{in}_{N\!N}}
\def\pp#1{p^\Vert_{#1}}
\def\pP#1{p^+_{#1}}
\def\pT#1{{\bf p}_{\bot{#1}}}
\def\abs#1{\left\vert{#1}\right\vert}
\def\dsdMdt#1{\frac{d^2\sigma{#1}}{dM^2 d\Ht}}
\def\dsdMdy#1{\frac{d^2\sigma{#1}}{dM dy}}
\def\xGmax{{x_{G\,max}}}
\def\pGmax{{p_{G\,max}}}
\def\qqG{q\!\rightarrow\!qG}
\def\qGllX{qG\!\rightarrow\!l^+\!l^-\!X}
\def\gtsim{\lower-0.45ex\hbox{$>$}\kern-0.77em\lower0.55ex\hbox{$\sim$}}

\subsection{The bremsstrahlungs contribution to 
low mass dileptons}\label{dilepton}

An advantage of the light-cone approach to 
the DY process \cite{boris,kst,KST99}, which is explained
in section \ref{dipoledy},
is that the dipole cross
section contains the effects of higher order corrections as well as
nonperturbative effects. Therefore, the light-cone approach can be
applied in the low mass region, $M\le 1$ GeV, where pQCD is questionable.
Although the term DY is normally only used for high mass dileptons, we will also
refer to low mass dileptons, which are produced via the bremsstrahlungs
mechanism, as DY dilepton since they are produced by the same mechanism.

Low mass dileptons have been measured recently in proton nucleus ($pA$)
\cite{pBepAu} and in nucleus-nucleus ($AB$) \cite{CERES,HELIOS}
collisions at the CERN SpS collider. At a laboratory energy of $158~A$ GeV
approximately three times more dileptons were detected, than one expects from
the calculation of hadron decays into $e^+e^-$ in the final state. Furthermore,
the observed shape of the spectrum is different from the calculated. 
Most theories locate the origin
for the observed dileptons in the hot and dense phase of hadrons
\cite{Brown}-\cite{Peters}, which is rather
late in the time evolution of a heavy ion
collision. The dilepton enhancement is a highly interesting effect, because it
might point at a restauration of chiral symmetry.

In this section, we estimate the contribution of the DY process to the dilepton
cocktail, using the light-cone approach. These dileptons arise from the 
very early 
stages of the heavy ion reaction in which partons are the relevant
degrees of freedom.
We calculate (i) the direct production $N\!N %
\rightarrow l^+l^-X$ in the light cone approach and in a parton 
model and (ii), following an idea of J.\ H\"ufner \cite{drei},
the lepton production via a gluonic Compton process
$G\!N \rightarrow l^+l^-X$ from prompt gluons. These prompt gluons are radiated
in a nucleon-nucleon collision and may lead to an enhancement of the dilepton 
yield via the gluonic Compton process, see fig.\ \ref{FigDiags}b, where another
nucleon scatters off the previously released gluon and radiates a virtual
photon.
Prompt gluons have
been recently identified as as an important source for charmonium
suppression in heavy ion collisions \cite{Charm} and therefore the question
arises whether they might also be responsible for a part of the dilepton
enhancement. However, in \cite{drei} this mechanism could be
excluded.



\begin{figure}[t]
\centerline{\scalebox{0.8}{\includegraphics{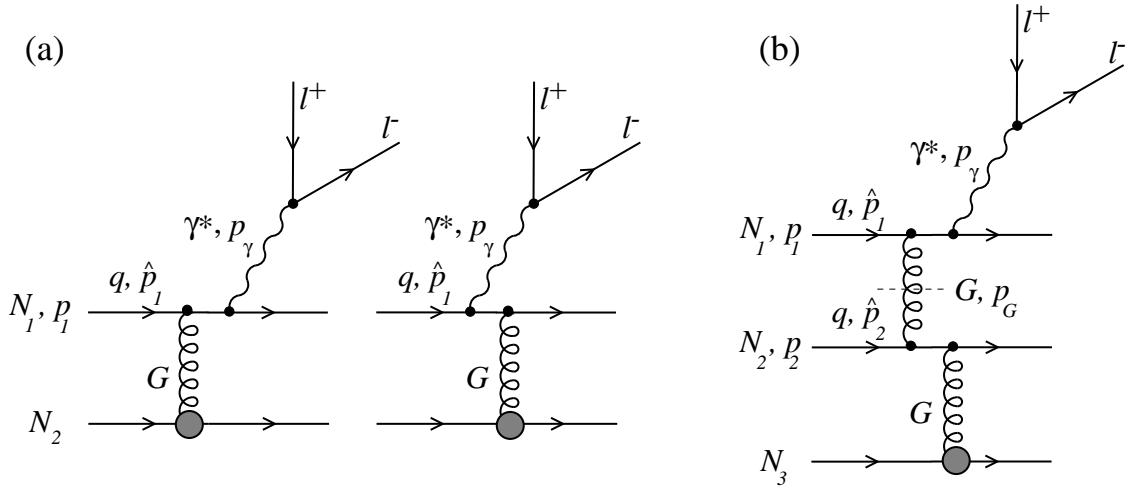}}}
\caption{
  \label{FigDiags}
  Diagrams for the two mechanisms for $l^+l^-$ production considered
  in this paper. (a) Dilepton production in $N\!N$ collisions which
  can be viewed as bremsstrahlung in $qq$ scattering (light cone
  approach) or as gluonic Compton scattering ($Gq \rightarrow %
  \gamma^*q$) from the gluon cloud of the second nucleon (parton
  approach). (b) Dileptons from prompt gluons $Gq \rightarrow %
  \gamma^*q$, where the on-mass-shell gluon G (crossed by a dashed
  line) is produced in another $N\!N$ collision.}
\end{figure}

First we discuss the direct production of lepton pairs within the light cone
approach, fig.\ \ref{FigDiags}a.
We employed the older parametrization (\ref{sasha}) from \cite{KST99}
in our calculations, which works very well at the low $M^2$ values we are
interested in. The improvement (\ref{improved}) modifies the DCS only small
$\rho$, corresponding to large virtuality, and will not alter the result.
The partonic cross section (\ref{dytotal}) has to be embedded into
the hadronic process. The cross section of direct ($D$) dilepton production
in nucleon-nucleon ($N\!N$)
takes the form
\BE
  \label{ds_D}
  \frac{d^2\sigma^{D}_{N\!N}}{dM dy} = 
    \frac{\alpha_{em}}{3\pi M} \sum_f e_q^2 \int_{x_1}^{1}d\alpha 
    \frac{x_1}{\alpha^2} \left[
        q_f       \left(\frac{x_1}\alpha\right)
      + q_{\bar f}\left(\frac{x_1}\alpha\right)
   \right]
   \frac{d\sigma(qN\rightarrow\gamma^* qN)}{d(\ln\alpha)}
\, +\, \Bigl\{y\Rightarrow - y\Bigr\},
\EE
where $y$ is the rapidity of the lepton pair in the c.m. frame and
$x_1=(\sqrt{x_F^2+4M^2/s}+x_F)/2$. The first term corresponds to 
radiation from the projectile quarks
and the second term to radiation from the target quarks (see fig.~%
\ref{FigDiags}a), which corresponds to replacement $y\to -y$ in the
first term. 
Since the dominant contribution to the cross section comes from large
values of $x_1/\alpha$, we neglect antiquarks in the projectile,
$q_{\bar f}\left(x_1/\alpha\right)\equiv 0$.
We parameterize the valence quark distribution in the 
form
\BA
  q_u(x) &=& \frac{C_u}{\sqrt x} (1-x)^3, \\
  q_d(x) &=& \frac{C_d}{\sqrt x} (1-x)^4,
\EA
where $C_{u,d}$ are defined by the normalization to the number of 
valence quarks $N_{u,d}$ in a nucleon
\BE
  \int_0^1\!\!dx \,q_{u,d}(x) = N_{u,d}.
\EE
In the case of the proton-nucleus ($pA$) and nucleus-nucleus ($AB$) 
collisions the cross section of the direct production is given
by the integration of the nuclear densities $\rho_{A,B}$ over
the impact parameter
\BE
  \label{AB_D}
  \dsdMdy{_{A\!B}^{D}} = 
    \int\!\! d^2b \int\!\! d^2s
    \int_{-\infty}^\infty\!\! dz_A\, \rho_A(\vec s        ,z_A)
    \int_{-\infty}^\infty\!\! dz_B\, \rho_B(\vec b\!-\!\vec s,z_B)
    \, \dsdMdy{_{N\!N}^D}
    = AB \dsdMdy{_{N\!N}^D}.
\EE
In this expression we neglect small shadowing effects. To obtain
the rate of the dilepton events per interaction one has to divide
expression \Ref{AB_D} by the total cross section, which can be 
calculated using the Glauber expressions
\BA
  \sigma_{pA} &=&
    \int d^2b\left[1-\exp\left(-\sNN T_A(\vec b)\right)\right], \\
  \sigma_{AB} &=&
    \int d^2b\left[1-\exp\left(-\sNN \int d^2s
    \,T_A(\vec s)\,T_B(\vec b-\vec s)\right)\right],
\EA
where $\sNN \approx 30~{\rm mb}$ denotes the inelastic $N\!N$ cross
section and we used the standard Woods-Saxon parameterization \cite{Jager}
for the nuclear density $\rho_{A,B}$ in 
\BE
  T_{A,B}(\vec b) = \int^{\infty}_{-\infty} \!\!dz\,\rho_{A,B}(\vec b,z).
\EE

There are still two sources of uncertainty:
\begin{itemize}
\item The dipole cross section $\sigma_{q\bar q}(s,\rho)$ cannot be
      calculated perturbatively at large transverse separations,
      which are important in the small mass region, however. 
      We employ the phenomenological expression (\ref{sasha}).
\item The light-cone wavefunction  $\Psi_{\gamma^* q}(\alpha,\rho)$
      depends on the mass $m_f$ of the quark. To illustrate its influence
      we calculate the cross sections for the
      two extreme cases of the constituent quark mass $m_f = 150$~MeV
      and $m_f = 300$~MeV.
\end{itemize}

Our results for proton (p-Be, p-Au) and heavy ion (Pb-Au) scattering
are shown in figs.~\ref{FigBeAu},\ref{FigPbAu} under the label ``light
cone''. The border lines of the shadowed areas correspond to different choices
of the constituent mass $m_f$ ($150$~MeV and $300$~MeV
for the upper and lower lines, respectively). We  believe, that also
a different dipole cross section cannot produce values, which are
significantly out of the shaded area, because $\sigma_{q\bar q}$
is well constrained to describe the structure function $F_2$ and
hadronic cross sections.

\begin{figure}[t]
\centerline{
  \scalebox{0.4}{\includegraphics{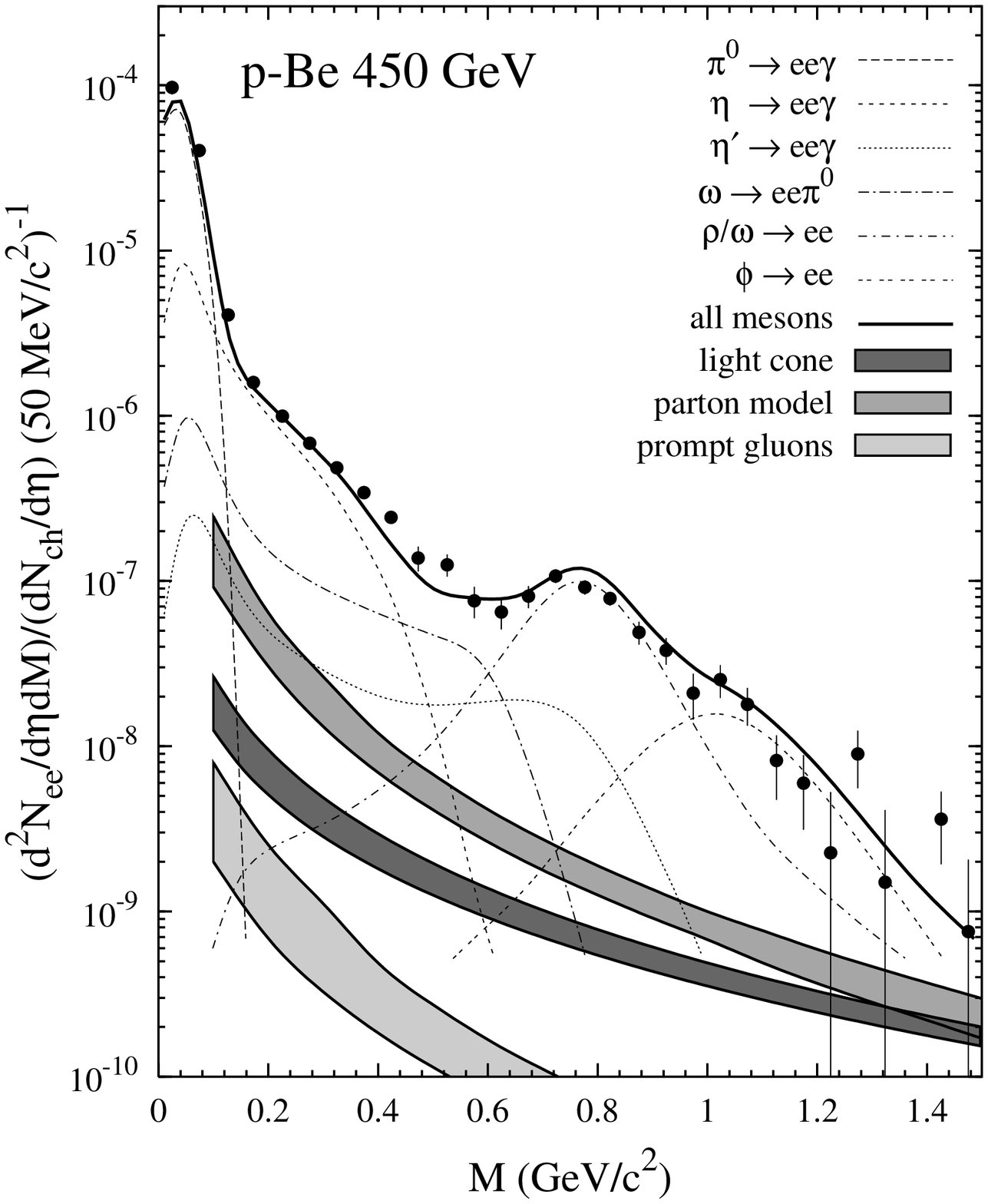}}
  \scalebox{0.4}{\includegraphics{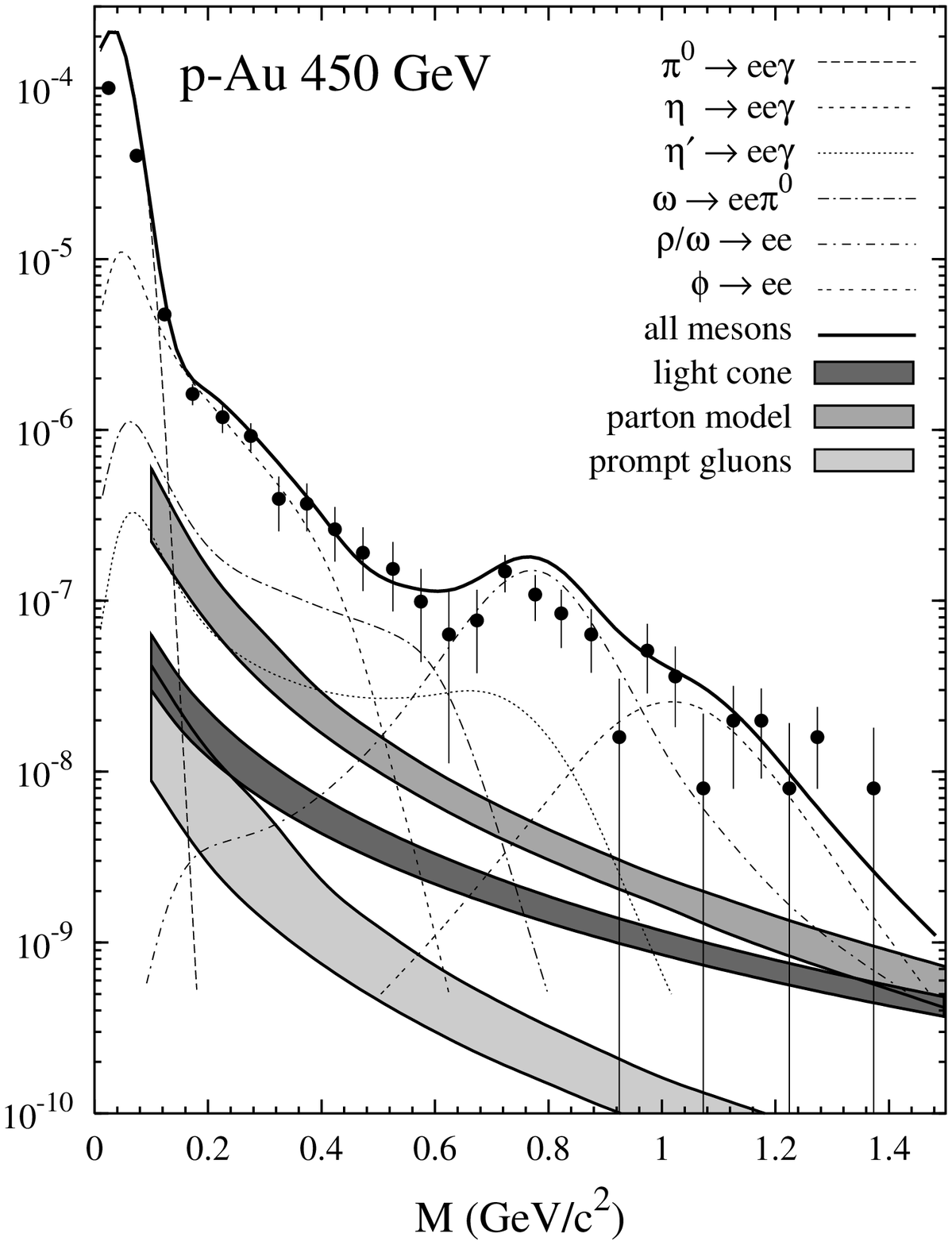}}
}
\caption{
  \label{FigBeAu}
  Data \cite{pBepAu} on dilepton events in p-Be and p-Au scattering
  together with predictions from different mechanisms. The various
  curves correspond to meson decay channels. The shaded areas represent
  our results for direct production (in the light cone and the parton
  model approaches) and via prompt gluons:
  The upper limits correspond to the quark mass $m_f=150$~MeV and the
  lower ones to $m_f=300$~MeV. The curves for the parton model and for prompt
  gluons were calculated by Yu.\ Ivanov.
}
\end{figure}

\begin{figure}[t]
  \scalebox{0.5}{\includegraphics{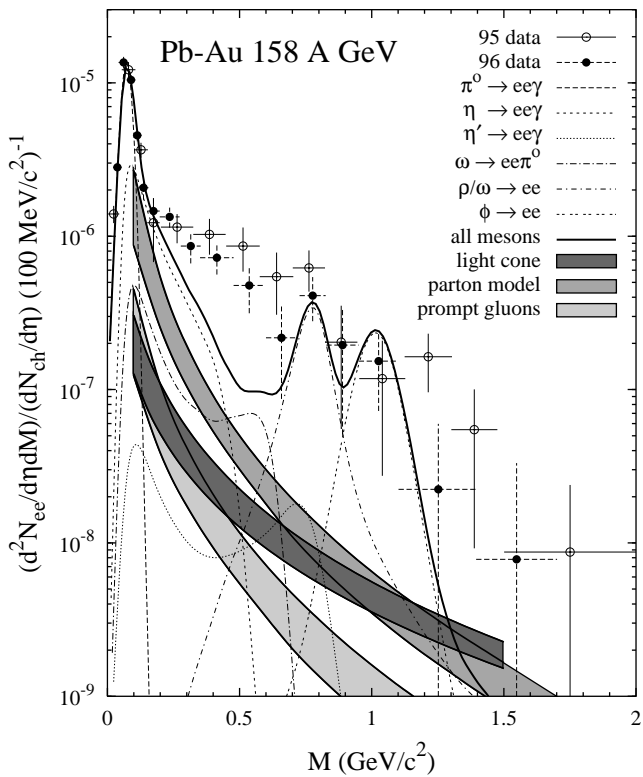}}\hfill
  \raise17mm\hbox{\parbox[b]{2.4in}{
    \caption{
      \label{FigPbAu}
      Data \cite{CERES,HELIOS} on dilepton events in Pb-Au scattering.
      The different curves correspond to meson decay channels. The 
      shaded areas are our results for direct production in the light
      cone approach and via prompt gluons:
      the upper limits correspond to the quark mass $m_f=150$~MeV and
      the lower ones to $m_f=300$~MeV. Both processes contribute
      separately to the dilepton production and therefore their
      contributions have to be added.
      The curves for the parton model and for prompt
      gluons were calculated by Yu.\ Ivanov.
    }
  }
}
\end{figure}

We compare these results with predictions of a parton model,
which treats direct radiation of the lepton pair
as Compton scattering of a target gluon on a beam quark (or {\em
vice versa}) in pQCD.  
This formulation is in principle identical to the light-cone approach, but all
nonperturbative and higher order effects which are parametrized in the DCS are
not taken into account. The amplitude for the subprocess $\qGllX$ is well known
and be found e.g.\ in \cite{ApQCD}. 
The main uncertainties in this case originate from:
\begin{itemize}
\item Possible failure of pQCD calculations for the soft
      Compton amplitude.
\item Poor knowledge of the gluon distribution function,
      especially at a soft scale.
\end{itemize}

We use the following gluon distribution function,
\BE
  \label{fG}
  f_G(x) = \frac{C_G}{x^{1.1}} (1-x)^5,
\EE
where $C_G$ is defined from the condition that gluons carry
half of the nucleon momentum
\BE
  \int_0^1\!\!dx \,x \,f_G(x) = \frac12.
\EE

The results for proton (p-Be, p-Au) and heavy ion (Pb-Au)
scattering are shown in figs.~\ref{FigBeAu}, \ref{FigPbAu}.
At low $M$ the results of pQCD (labeled as ``parton model'')
are higher than those from the light cone calculations but 
start to agree for $M \gtsim 1$~GeV. The reason for this
is that the saturation of the dipole cross section becomes 
important at small $M$, and the cross section decays weaker
than $1/M^3$. Although the parton model results are supposed
to coincide with the predictions of the light-cone approach
different approximations are used and the amount of disagreement
between the results may serve as a measure of theoretical 
uncertainty. Nevertheless, we believe that light-cone prediction
is more reliable.

The data in figs.~\ref{FigBeAu} and \ref{FigPbAu} 
have been measured in a pseudorapidity interval of
$2.1<\eta<2.65$, while 
our calculations have been performed for rapidity $y=2.4$.
Note that the data are also subject to $p_T$ cuts, which exclude very small
transverse momenta, $p_T<50$ MeV. 
These cuts are not included in our calculation. 
In the region of  interest ($M = 0.2$--$0.8$~GeV) however
both approaches underestimate the experimental data by at least a
factor of 10.
We therefore believe, 
that also a more careful calculation would not change our results.


The pQCD calculation of the Compton amplitude is similar to
that for an interaction with on mass shell gluons.
The second process, labeled ``prompt gluons'', considers gluons
which are radiated in an elementary $N\!N$ collision and which 
convert into a virtual photon via gluonic Compton scattering on
another nucleon. To estimate their contribution to the dilepton 
pair production we start with the elementary Compton subprocess
$G\,q\rightarrow\gamma^*\,q$ (see the upper part of Fig.~\ref{FigDiags}b
where the on-mass-shell gluon is marked by a dashed line). 
To calculate the prompt gluon spectrum we use the perturbative
evaluation for the cross section of gluon radiation \cite{kst}
\BE
  \label{dsdadk}
  \frac{d^2\sigma(\qqG)}{d\alpha dk^2} = 
    \frac{3\alpha_s C}{\pi}
    \frac{2 m_f^2 \alpha^4 k^2
         + \left[1+(1-\alpha)^2\right]
           (k^4+\alpha^4 m_f^4)}{\left(k^2+\alpha^2 m_f^2\right)^4}
    \left[\alpha+\frac94\frac{1-\alpha}{\alpha}\right],
\EE
where $k^2$ is the transverse momentum of the gluon, 
$C$ is the factor of the dipole
approximation for the cross section of a $q\hat q$ pair with
a nucleon (in this energy range $C \approx 3$ \cite{kz}) and
$\alpha$ is the fraction of the quark light cone momentum
carried by the gluon, as usual.

Finally, we combine the  cross section
for the elementary process with the distribution function $q_f(x_1)$
for the quark and the distribution $d n_G / d x_G$ of the gluon to the
cross section for the prompt (``P'') process on the nucleon
\BE
  \label{ds_My}
  \dsdMdy{^P} = \dsdMdy{^P_+} + \dsdMdy{^P_-} ,
\EE
with
\BA
  \dsdMdy{^P_+} &=& \sum_f
     \int_0^1      \!\!dx_1 q_f(x_1) 
     \int_0^\xGmax \!\!dx_G \,\frac{d n_G}{d x_G}\,
     \dsdMdy{(\qGllX)} , \\
  \dsdMdy{^P_-} &=& \dsdMdy{^P_+}\left(y \Rightarrow -y \right) .
\EA
Here, $dn_G/dx_G$ is the gluon distribution over the longitudinal momentum
fraction, the on mass shell gluon carries away from its parent quark, fig.\
\ref{FigDiags}.
The calculations for the parton model
and the prompt gluons were done by Yu.\ Ivanov and we refer to \cite{drei} for
an expression for $dn_G/dx_G$ and for details of the calculation.

For nucleus-nucleus collisions the geometric factor has to take
into account that points of gluon creation and interaction are
different
\BA
  \label{AB_P}
  \dsdMdy{_{A\!B}^P} &=&
    \int\!\! d^2b \int\!\! d^2s
    \int_{-\infty}^\infty\!\! dz_A\, \rho_A(\vec s        ,z_A)
    \int_{-\infty}^\infty\!\! dz_B\, \rho_B(\vec b\!-\!\vec s,z_B) \\
    & \times & \sNN \,
      \left[ \int_{-\infty}^{z_A} dz \rho_A(     \vec s,z) \dsdMdy{^P_-}
           + \int_{z_B}^\infty dz \rho_B(\vec b\!-\!\vec s,z) \dsdMdy{^P_+}
      \right]. \nonumber
\EA
Again, expression \Ref{AB_P} has to be divided by the total inelastic
cross section to compare with experiment. 


The results of our calculations are summarized in Figs.~\ref{FigBeAu}
and \ref{FigPbAu}: the mechanisms considered contribute to the 
observed spectrum less than 10\% and therefore are unimportant on
the present level of discussion. On this level also the theoretical
uncertainties, for instance in the choice of the value for constituent
quark mass or in the evaluation via two mechanisms (light cone {\em vs.}
parton model) are not yet of importance.

\subsection{DY shadowing in proton-nucleus collisions}\label{dyshadow}

In section \ref{dipoledy} we introduced the light-cone approach to the DY
process at low $x_2$, in which DY dilepton production 
is viewed in the rest frame of the target and appears as bremsstrahlung
from an incident quark. In the case of a nuclear target, this quark undergoes
multiple rescatterings and in the simplest case radiates 
electromagnetic bremsstrahlung every
time it is hit. As it is well known, this naive expectation is wrong at high
energies. Landau and Pomeranchuk \cite{Landau,Landau2}
were the first who noticed that the
cross section for bremsstrahlung of a high energetic charge in an amorphous
medium is suppressed compared to the Bethe-Heitler cross section. 
Somewhat later, a quantum mechanical treatment of this effect, 
today known as the
Landau-Pomeranchuk-Migdal (LPM) effect, was given by Migdal \cite{Migdal}. 
The LPM effect was measured recently at SLAC \cite{a1,a2,a3}.

At low $x_2$, multiple rescattering of the quark inside the nucleus leads to an
LPM suppression of bremsstrahlung from that quark. This suppression is the
nuclear shadowing effect for the DY process, which was measured for the first
time by the E772 collaboration \cite{E772}.

What is the physical mechanism behind this suppression? The answer is that the
quark needs a rather long time to recreate its electromagnetic field.
This coherence time is related to the longitudinal momentum transfer by the
uncertainty relation
\beq
l_c=\frac{1}{q_L}
=\frac{\alpha(1-\alpha)E_q}{(1-\alpha)M^2+\alpha^2m_f^2+p_T^2},
\eeq
where $E_q$ is the energy of the projectile quark. If the quark is hit again
before it has recreated its field, it cannot radiate another photon. Note
however that the Fourier modes of the field with high transverse momentum $p_T$
are
recreated earlier as the one with low $p_T$. A large transverse momentum
corresponds to a small size of the $\gamma^*q$-fluctuation. The small size
fluctuation are therefore less shadowed. The low $p_T$ components of the field
correspond to large $\gamma^*q$ fluctuations which are strongly shadowed.

Since the photon carries the momentum fraction $x_1$ away from the proton, the
coherence length for DY can be written as
\beq
l_c=\frac{1}{m_Nx_2}
\frac{(1-\alpha)M^2}{(1-\alpha)M^2+\alpha^2m_f^2+p_T^2}.
\eeq
The role of $x_{Bj}$ in DIS is played by $x_2$ in DY. One can define a mean
coherence length for DY in analogy to the one for DIS, see section
\ref{meancoh}. We will however not pursue the concept of the 
mean coherence length
for DY in this work. This has to be postponed to future studies.
 Like in low $x_{Bj}$ DIS, the coherence length for DY
can become substantially longer than a nuclear radius.  

In the case of an infinitely long coherence length $l_c\to\infty$, the 
whole target acts like a single scattering center and the
partonic
DY cross section can again be written, cf.\ (\ref{dylctotal}), 
in factorized light-cone form  
\beq\label{dyeikonal}
\frac{d\sigma^{qA}_{{T,L}}}{d\ln\alpha}
=\int d^2\rho\, |\Psi^{T,L}_{\gamma^* q}(\alpha,\rho)|^2
    \Sigma_{q\bar q}(s,\alpha\rho),
\eeq
where
\beq\label{Sigma}
\Sigma_{q\bar q}(s,\alpha\rho)=2\int d^2b
\left(1-\exp\left(-\frac{\sigma_{q\bar q}(s,\alpha\rho)}{2}T(b)\right)\right)
\eeq
is the eikonalized dipole cross section. One immediately sees that
(\ref{dyeikonal}) is the analog of 
the eikonal formula for DIS (\ref{eikonalappr}). 
The eikonal term obviously leads to a reduction of the total cross section.
It contains the LPM suppression.
As already pointed out
in section \ref{shadowdiffr} for the case of shadowing in DIS, also
(\ref{dyeikonal}) is different from the usual Glauber formula, where 
$\sigma_{q\bar q}(s,\alpha\rho)$ is averaged in the exponent.

In the case of a finite coherence length, which is relevant for all present and
most of the future experiments, the DY cross section on the
partonic level can be
calculated with the same Green function technique as the DIS cross section 
\cite{kst}
\beqn\nonumber
\frac{d\sigma^{qA}_{{T,L}}}{d\ln\alpha} &=& 
A\,\frac{d\sigma^{qp}_{{T,L}}}{d\ln\alpha}
\,-\, \frac{1}{2} {\rm Re}\int d^2b
\int\limits_{-\infty}^{\infty} dz_1 \int\limits_{z_1}^{\infty} dz_2
\int d^2\rho_1\int d^2\rho_2\,
\\ & \times &\nonumber
\Bigl[\Psi^{T,L}_{\gamma^*q}\left(\alpha,
\rho_2\right)\Bigr]^*\,
\rho_A\left(b,z_2\right)\sigma_{q\bar{q}}\left(s,\alpha\rho_2\right)
G\left(\vec \rho_2,z_2\,|\,\vec \rho_1,z_1\right)\\
&\times&
\rho_A\left(b,z_1\right)\sigma_{q\bar{q}}\left(s,\alpha\rho_1\right)
\Psi^{T,L}_{{\gamma^*q}}\left(\alpha,\rho_1\right), 
\label{dyshadowing}
\eeqn
cf.\ section
\ref{lightcone}. The Green function $G$ fulfills a two dimensional
Schr\"odinger equation, similar to (\ref{schroedinger}), 
\beqn\nonumber
\left[{\rm i}\frac{\partial}{\partial z_2}
+\frac{\Delta_\perp\left(\rho_2\right)-\eta^2}
{2E_q\alpha\left(1-\alpha\right)}
+\frac{{\rm i}}{2}\rho_A\left(b,z_2\right)\,
\sigma_{q\bar{q}}\left(s,\alpha\rho_2\right)
\right]
G\left(\vec \rho_2,z_2\,|\,\vec \rho_1,z_1\right)\\
=
{\rm i}\delta\left(z_2-z_1\right)
\delta^{\left(2\right)}\left(\vec \rho_2-\vec \rho_1\right).
\label{sgl}
\eeqn
The two dimensional Laplacian in (\ref{sgl}) acts on the transverse
coordinate and takes into account that the quark is deflected 
due to the radiation process by a small, but
non-zero angle.
Note, that for convenience 
we included the phase factor 
$\exp(-{\rm i}q_L^{min}(z_2-z_1))$, 
describing the longitudinal motion, into the Green
function $G$, 
($q_L^{min}=\eta^2/2E_q\alpha\left(1-\alpha\right)$). This phase
factor is not contained in $W$ and appears explicitly in (\ref{propagation}).
For a calculation that can be compared with experimental data, one has to embed
the partonic cross section into the hadronic environment. 
For a proton as projectile,
this is done in the
same way as in (\ref{dylctotalhadr}),
\beq\label{dylctotalA}\nonumber
\frac{d\sigma}{dM^2dx_F}=\frac{\alpha_{em}}{6\pi M^2}
\frac{1}{x_1+x_2}\int_{x_1}^1\frac{d\alpha}{\alpha}
F_2^p\left(\frac{x_1}{\alpha}\right)
\frac{d\sigma(qp\to q\gamma^*p)}{d\ln\alpha}.
\eeq

Given a parametrization of the proton structure function $F_2^p$, 
like the one from \cite{Mils}, one is in
the position to calculate shadowing for the DY process. A comparison to
experimental data will be presented in section \ref{comparison}. 
First we investigate the behavior of shadowing and find out, how much it will
influence dilepton production at RHIC.

An analytical expression for the Green function $G$ can be obtained if one
employs the approximation $\sigma_{q\bar q}(a,\alpha\rho)=C(s)\alpha^2\rho^2$
for the dipole cross section and a uniform nuclear density, 
cf.\ section \ref{lightcone}. 
Such an approximation works remarkably, especially for heavy
nuclei. Nevertheless, it can improved if one makes use 
of the fact that the asymptotic expression (\ref{dyeikonal}) 
is easily calculated with
the realistic parametrization of the
dipole cross section (\ref{improved})
and realistic nuclear density.
One needs to use the full Green function only in the transition
region from no-shadowing to a fully developed shadowing given by
(\ref{dyeikonal}). Therefore,  the value of the uniform nuclear density is fixed
by demanding that the asymptotic shadowing ratio is the same for the realistic
Woods-Saxon form of the nuclear density \cite{Jager} and for the effective
uniform density $\rho_0$.
We have checked that the found value of $\rho_0$ is practically
independent of the value of the cross section in the interval $1-50$ mb.
The factor $C(s)$ is fixed in the same way, but now 
separately for transverse and longitudinal photons and for each
value of $\alpha$. All further analytical calculations are similar to the one
performed for shadowing in DIS, see appendix \ref{appshadow}.

\begin{figure}[t]
  \centerline{\scalebox{0.8}{\includegraphics{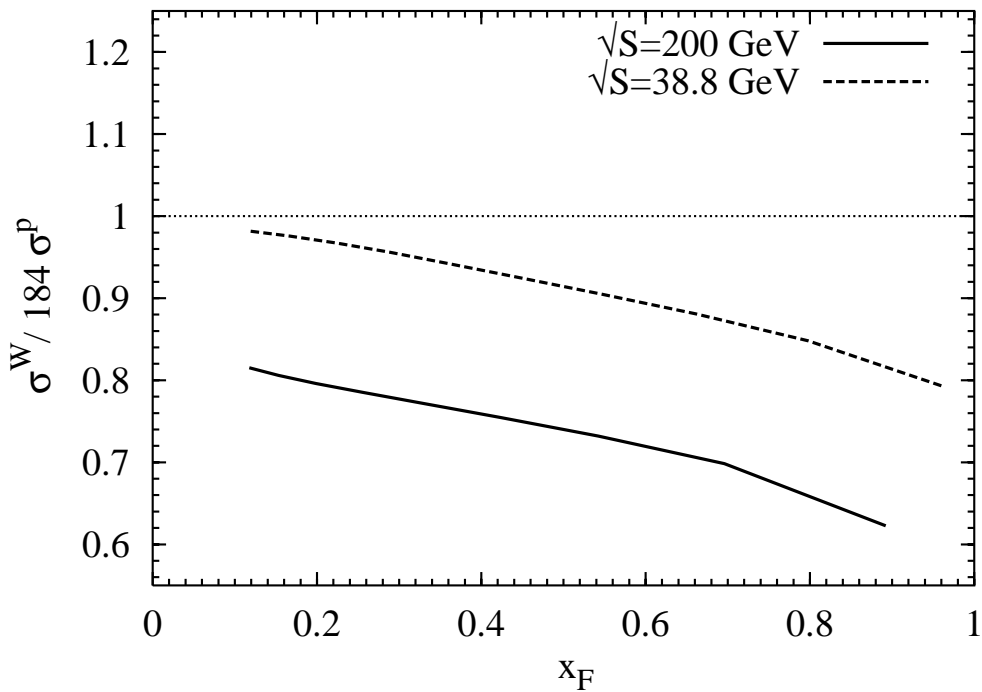}}}
    \caption{
      \label{dyxf}
      The dependence of shadowing in proton-nucleus collisions
      for the DY process on $x_F$ for FNAL (E772)
      ($\sqrt{S}=38.8$ GeV) and RHIC ($\sqrt{S}=200$ GeV) energies.
      The calculations are for tungsten. 
      Both curves are calculated for
      $M=5$ GeV and show the same points as in fig.\ \ref{dyx2}.
      At RHIC, the whole $x_F$-range is shadowed.
    }  
\end{figure}

\begin{figure}[t]
  \centerline{\scalebox{0.8}{\includegraphics{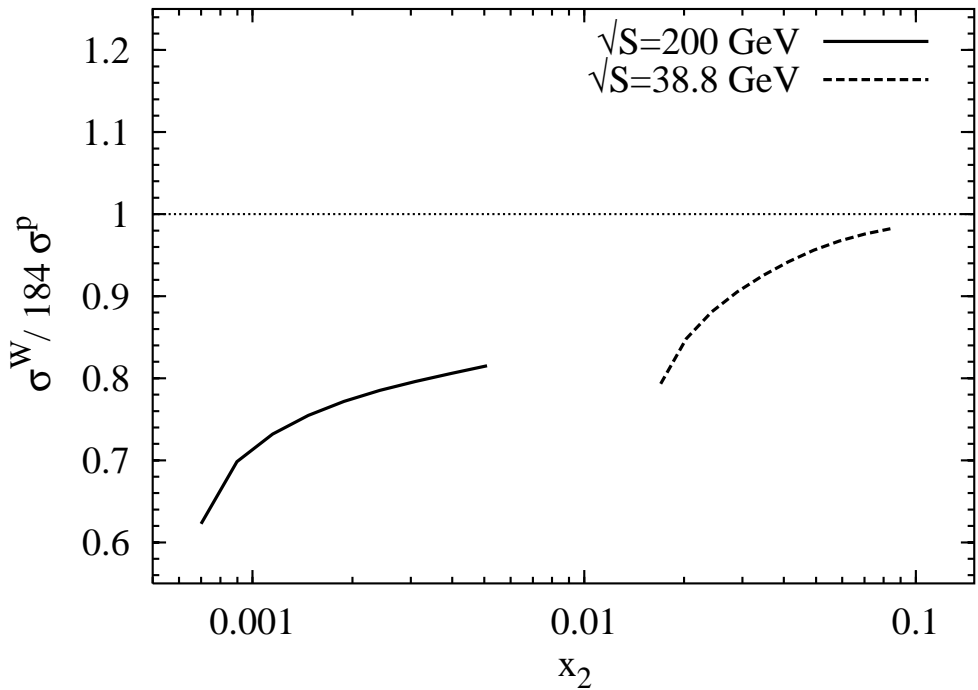}}}
    \caption{
      \label{dyx2}
      The dependence of shadowing 
      in proton-nucleus collisions
      for the DY process on $x_2$. The curves show 
      the same points as fig.\ \ref{dyxf}. 
      The calculations are for tungsten. 
      Much lower values of $x_2$ are reached at RHIC than at FNAL.
    }  
\end{figure}

Our predictions for RHIC are depicted in figs.\ \ref{dyx2} and \ref{dyxf} in
comparison with a curve calculated at the highest energy that can be reached
today at FNAL. ISR at CERN
can actually reach higher energies, $\sqrt{S}=62$ GeV, but
did never include nuclei nor performed measurements at low $x_2$.
The curves in the two figures show the same points, but in one figure,
\ref{dyx2}, displayed vs.\ $x_2$ and in the other figure, \ref{dyxf}, vs.\
Feynman $x_F$. Shadowing in DY is more complicated than in DIS, because there a
two structure function, of the projectile and of the target, involved in DY.
Calculations are performed only for values of $x_F>0.1$, because at smaller
values quarks from the target will also radiate in the same direction, which is
not taken into account in (\ref{dylctotalA}). 

One immediately recognizes that RHIC can reach much lower values
of $x_2$ than FNAL. Even more striking, the whole $x_F$ range at RHIC shows
strong shadowing. This illustrates the close connection between heavy ion
collisions at high energies and low $x$ physics. The  physics 
at RHIC is
governed by  coherence effects. 
The situation is different at SpS, where the energy is
not high enough to allow for strong coherence effects. 
For a reliable interpretation of the data from RHIC and LHC it is essential to
have good theoretical understanding of shadowing effects, because DY dilepton
serve as a reference for the study of heavy quarkonium suppression, esp.\
$J/\psi$,  which is  believed to be a signal for quark gluon plasma
formation \cite{ms}.

It is also interesting to investigate the influence of the nuclear medium on the
transverse momentum distribution of the DY pairs. The formula for the
differential cross section was derived in \cite{kst}. It reads,
\beqn\label{dynucleartrans}\nonumber
\frac{d^3\sigma^{qA}}{d(\ln\alpha)d^2p_T}&=&
\frac{\alpha_{em}}{(2\pi)^4 4E_q^2(1-\alpha)^2}
2{\rm Re}\int_{-\infty}^{\infty}dz_1\int_{z_1}^{\infty}dz_2
\int d^2b\,d^2k_T\,d^2\rho_1\,d^2\rho_2\\
\nonumber&\times&\exp\left[{\rm i}\alpha\vec p_2\cdot\vec\rho_2
-{\rm i}\alpha\vec p_1\cdot\vec\rho_1
-{\rm i}\int\limits_{z_2}^\infty dzV_{opt}(b,\rho_2,z)
-{\rm i}\int\limits_{-\infty}^{z_1}dzV_{opt}(b,\rho_1,z)\right]\\
&\times&
\left[\widehat{\cal O}_{\gamma^*q}(\vec\rho_2)\right]^*
\widehat{\cal O}_{\gamma^*q}(\vec\rho_1)
G\left(\vec \rho_2,z_2\,|\,\vec \rho_1,z_1\right),
\eeqn
where $k_T$ is the transverse momentum of the quark in the final state. It is
convenient to introduce the transverse momenta relative to the direction of the
virtual photon. The transverse momentum of the incoming quark 
relative to this direction is then given by
\beq
\vec p_1=\frac{\vec p_T}{\alpha}
\eeq
and the transverse momentum of the final quark by
\beq
\vec p_2=\vec k_T-\frac{1-\alpha}{\alpha}\vec p_T.
\eeq
The interaction with the nuclear medium is contained in the Green function $G$,
which fulfills the two dimensional Schr\"odinger equation (\ref{sgl}) and in the
optical potential
\beq\label{vopt}
V_{opt}(b,\rho,z)=-\frac{{\rm i}}{2}\rho_A(b,z)\sigma_{q\bar q}(s,\rho).
\eeq
This potential describes the shadowing corrections and is also present in
(\ref{sgl}). The operators $\widehat{\cal O}$, acting on 
the variable $\vec\rho_1$ or $\vec\rho_2$ in
the Green
function $G$ in (\ref{dynucleartrans}), are given by (\ref{transdyop}) and
(\ref{longdyop}). They are the same operators which appear in the definition of
the LC wavefunctions for the transition $q\to\gamma^*q$, cf.\ section
\ref{dipoledy}.
It has been shown in \cite{kst}, that in the limit of infinitely high energy of
the incoming quark, $E_q\to\infty$, the differential cross section 
(\ref{dynucleartrans}) takes an eikonal form, similar to (\ref{dylcdiff}),
\beqn\nonumber\label{dylcpadiff}
\frac{d\sigma(qp\to q\gamma^*p)}{d\ln\alpha d^2p_T}
&=&\frac{1}{(2\pi)^2}
\int d^2\rho_1d^2\rho_2\, \exp[{\rm i}\vec p_T\cdot(\vec\rho_1-\vec\rho_2)]
\Psi^*_{\gamma^* q}(\alpha,\vec\rho_1)\Psi_{\gamma^* q}(\alpha,\vec\rho_2)\\
&\times&
\frac{1}{2}
\left\{\Sigma_{q\bar q}(\alpha\rho_1)
+\Sigma_{q\bar q}(\alpha\rho_2)
-\Sigma_{q\bar q}(\alpha(\vec\rho_1-\vec\rho_2))\right\},
\eeqn
where $\Sigma_{q\bar q}$ is given by (\ref{Sigma}). 

\begin{figure}[t]
  \centerline{\scalebox{0.8}{\includegraphics{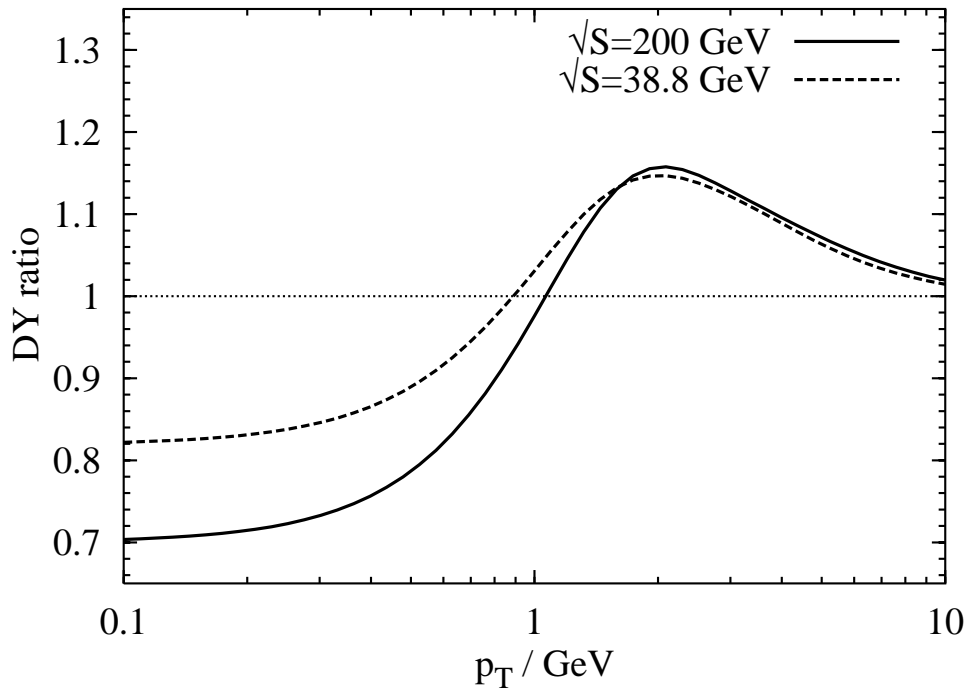}}}
    \caption{
      \label{pt}
      The transverse momentum dependence of the LPM effect for different
      energies. The curves are for tungsten and $M=5$ GeV,
      $x_F=0.7$. an approximate dipole cross section $\sigma_{q\bar
      q}(s,\rho)=C(s)\rho^2$ is used.
    }  
\end{figure}

The transverse momentum dependence of the LPM effect, calculated from
(\ref{dynucleartrans}), is depicted in fig.\ \ref{pt}. 
The calculations are for tungsten. The mass of the lepton pair is $M=5$ GeV and
Feynman $x_F=0.7$. Since (\ref{dynucleartrans}) is very complicated, the
following simplifying approximations are employed. 
First, a uniform nuclear density of
$\rho_A=0.16$ fm$^{-3}$ is used, 
which is well justified for a heavy nucleus like
tungsten, and second, the approximation 
$\sigma_{q\bar q}(s,\rho)=C(s)\rho^2$ is applied. 
The latter approximation works remarkably well, especially for heavy nuclei
\cite{kp}. 
With these approximations, some of the 12 integrations in 
(\ref{dynucleartrans}) can be performed analytically. The result of the
analytical calculations is given in appendix \ref{appdypa}.
One further integration has to be performed, in order to account for the
hadronic environment, cf.\ (\ref{dylcdiffhadr}).
To make the calculations even more realistic, an energy dependence of
the factor $C(s)$ is introduced. At FNAL energies, a value of 
$C(\sqrt{s}=38.8~{\rm GeV})=3$ is appropriate, 
while at RHIC energies one rather has $C(\sqrt{s}=200~{\rm GeV})=6$. 

Note that pairs
with low transverse momenta are suppressed at small
$p_T\le 1$ GeV, but enhanced at $p_T\ge 1$ GeV. 
This enhancement in the intermediate $p_T$ region is already
known for a long time as Cronin-effect
\cite{Cronin}. It is usually assumed that the Cronin-effect is caused by
multiple scatterings and the light-cone approach reproduces this phenomenon. 
At very large $p_T$, all
nuclear effects vanish. Furthermore, the shadowing at small $p_T$ increases with
energy, since the dipole cros section becomes larger. 

Although (\ref{dynucleartrans}) looks rather complicated, it contains
intuitively understandable physics. 
The nucleus acts as a color filter \cite{bbgg}. Small size dipoles, 
having large
$p_T$, can pass the nucleus nearly undisturbed
while dileptons 
with small transverse momentum are
shadowed, because they correspond to large arguments $\rho$ of the dipole cross
section.
Note that (\ref{dynucleartrans}) contains also the phenomenon of nuclear
broadening. Due to multiple rescatterings, the projectile quark
performs a
random walk in impact parameter space, while propagating through the nucleus.
Therefore, not the whole suppression at small $p_T$ is due to nuclear shadowing.
Part of the dileptons reappear at larger transverse momentum, which leads to the
enhancement in the intermediate $p_T$ region. Indeed, it was found in
\cite{dhk}, that in the approximation $\sigma_{q\bar q}(s,\rho)=C\rho^2$, the
transverse momentum broadening is given by
\beq
\delta\la p_T\ra^{pA}=\la p_T\ra^{pA}-\la p_T\ra^{pp}=C\la T(b)\ra,
\eeq
where $\la T(b)\ra=\int d^2b T^2(b)/A$, is the mean nuclear thickness.

First numerical calculation with 
the eikonal approximation (\ref{dylcpadiff}) are performed in
\cite{kst}, but only for quark-nucleus scattering. Our results are consistent
with the one from \cite{kst}.
Note that the transverse
momentum dependence of the LPM effect 
in QED is also investigated in \cite{sonst3}.
The case of gluon radiation off a quark is
studied in \cite{Urs,Miklos}.

Recently, also the approach by Luo, Qiu and Sterman
\cite{LQS1,LQS2} has received much attention, see e.\ g.\
\cite{Guo1}-\cite{Rainer2}. 
This approach extends the QCD factorization theorem \cite{factorization}
to higher twist operators. The double scattering contribution to
the DY process is described in terms of a matrix element, which contains all the
nonperturbative physics, and a hard part that can be calculated in perturbation
theory. However, it was already mentioned in section \ref{dysec} that the
calculation of the transverse momentum distribution is very complicated in pQCD,
since it requires the resummation of large logarithms $\ln(p_T^2/M^2)$ at
$p_T^2\ll M^2$. It is known, how to perform
this resummation for the leading twist
\cite{resumal,resumcol,Pirner}, but it still has to be investigated, how these
techniques can be generalized to higher twists.

\subsection{Comparison with data}\label{comparison}

The mean coherence length
introduced in section \ref{meancoh} alone is not sufficient
to predict any nuclear effect in the structure function $F_2$. It is merely
intended to be a tool for qualitative considerations. The Green function
technique
from section \ref{lightcone} does not need the mean coherence length as input.
Instead, all coherence effects are automatically taken into account by the Green
function (\ref{schroedinger}). Together with a realistic phenomenological 
dipole cross section  
we are now able to perform parameter free 
calculations for nuclear shadowing in the region
which is most difficult for theory: The transition region between no
shadowing at $x_{Bj}\sim 0.1$
and saturated (for the $|{q\bar q}\ra$ component) shadowing
at very small $x_{Bj}$.

In the rest frame of the nucleus shadowing in the total
virtual photoabsorption cross section $\sigma^{\gamma^*A}_{tot}$
(or in the structure function $F_2^A$) can be decomposed over 
different Fock components of the photon,
\beq
\sigma_{tot}^{\gamma^*A} = A\,\sigma_{tot}^{\gamma^*p}\,
-\, \Delta\sigma_{tot}({q\bar q})\,
-\, \Delta\sigma_{tot}({q\bar q}G)\, 
-\, \Delta\sigma_{tot}({q\bar q}2G)\, -\,...
\label{4.1}
\eeq
It was found in section \ref{meancoh} that the coherence length for the $|q\bar
qG\ra$ fluctuation is rather short at $x_{Bj}\sim 0.01$, 
compared to the mean internucleon spacing in a nucleus,
because of it's large mass, see fig.\
\ref{l-Q2}. 
Although the calculations were done only at large $Q^2$ and some data are at
quite small $Q^2$, we neglect the 
$|q\bar qG\ra$ component of the virtual photon for the present 
comparison with data. 
It is one of the most important tasks for the future, to include also the gluon,
since it is important for shadowing of the nuclear gluon distribution. The
treatment of the propagation of the 
three body system $q\bar qG$ through the nucleus is however
numerically quite involved.

Holding only the first two terms on the r.h.s.\ of (\ref{4.1}), the total
cross section for transverse and longitudinal photons on a nucleus can be
represented as
\beqn\nonumber
\sigma^{\gamma^*A}_{{T,L}} &=& 
A\,\sigma^{\gamma^*p}_{{T,L}}
\,-\, \frac{1}{2} {\rm Re}\int d^2b
\int\limits_0^1 d\alpha
\int\limits_{-\infty}^{\infty} dz_1 \int\limits_{z_1}^{\infty} dz_2
\int d^2\rho_1\int d^2\rho_2\,
\\ & \times &\nonumber
\Bigl[\Psi^{T,L}_{{q\bar q}}\left(\varepsilon,
\lambda,\rho_2\right)\Bigr]^*\,
\rho_A\left(b,z_2\right)\sigma_{q\bar{q}}\left(s,\rho_2\right)
G\left(\vec \rho_2,z_2\,|\,\vec \rho_1,z_1\right)\\
&\times&
\rho_A\left(b,z_1\right)\sigma_{q\bar{q}}\left(s,\rho_1\right)
\Psi^{T,L}_{{q\bar q}}\left(\varepsilon,\lambda,\rho_1\right), 
\label{4.1a}
\eeqn
where $\rho_A(b,z)$ is the nuclear density dependent on impact parameter $b$
and longitudinal coordinate $z$.
In (\ref{4.1a}) we have generalized our result from section \ref{lightcone} to
include also the nonperturbative interaction proposed in \cite{KST99}, 
see also section \ref{dipoledis}.
The nonperturbative wave functions for the ${q\bar q}$ component
of the photon are defined in 
(\ref{psitnpt})-(\ref{psilnpt}). The Green function
$G\left(\vec \rho_2,z_2\,|\,\vec \rho_1,z_1\right)$ describes propagation
of a nonperturbatively  interacting ${q\bar q}$ pair in an
absorptive medium. It fulfills the two dimensional Schr\"odinger-equation,
\beqn\nonumber
\left[{\rm i}\frac{\partial}{\partial z_2}
+\frac{\Delta_\perp\left(\rho_2\right)-\varepsilon^2}
{2\nu\alpha\left(1-\alpha\right)}
+\frac{{\rm i}}{2}\rho_A\left(b,z_2\right)\,
\sigma_{q\bar{q}}\left(s,\rho_2\right)
-\frac{a^4\left(\alpha\right)\rho_2^2}
{2\nu\alpha\left(1-\alpha\right)}
\right]
G\left(\vec \rho_2,z_2\,|\,\vec \rho_1,z_1\right)\\
=
{\rm i}\delta\left(z_2-z_1\right)
\delta^{\left(2\right)}\left(\vec \rho_2-\vec \rho_1\right)\ ,
\label{4.2}
\eeqn
where $a(\alpha)$ and $\lambda$ are given by (\ref{2.1a}) and 
(\ref{2.10}) respectively. The third term on the l.h.s.\ of
(\ref{4.2}) describes the absorption of a ${q\bar q}$ pair with cross section 
$\sigma_{{q\bar q}}\left(s,r\right)$
in the medium of  density 
$\rho_A\left(b,z\right)$. Note that the Green function $G$ 
includes now also the motion in longitudinal direction, 
in contrast to the Green
function
$W$ in (\ref{schroedinger}). The longitudinal motion is contained in the term 
$\eps^2/2\nu\alpha\left(1-\alpha\right)$ in (\ref{4.2}).

An explicit analytical expression for $W$ can be found 
only for a
dipole cross section  $\sigma_{{q\bar q}}(s,\rho)= C(s)\rho^2$
and a constant nuclear density $\rho_A(b,z)=\rho_0$.
Such an approximation has a reasonable accuracy, especially for heavy
nuclei. Nevertheless, it can be even more precise if one makes use 
of the fact that the asymptotic expression (\ref{eikonalappr}) 
is easily calculated with
the realistic parametrization of the 
dipole cross section (\ref{improved}) and a realistic
nuclear density.
One needs the full Green function only in the transition
region from no-shadowing to a fully developed shadowing given by
(\ref{eikonalappr}). Therefore, we fix the value of the uniform nuclear density
by demanding that the asymptotic shadowing ratio is the same for the realistic
Woods-Saxon form of the nuclear density \cite{Jager} and for the effective
uniform density $\rho_0$.
We checked that the found value of $\rho_0$ is practically
independent of the value of the cross section in the interval $1-50$ mb.
The factor $C$ is fixed in the same way, but now 
separately for transverse and longitudinal photons and for each
value of $\alpha$. More details can be found in appendix \ref{appshadow}.

We perform calculations with the parameter $v$ in (\ref{2.1a})
fixed at $v=0.5$,
because the parameters in the dipole cross section 
are fixed for this value of $v$, see section
\ref{DCS}. The dependence on $v$ is very weak.
All calculations are also performed with perturbative wavefunctions 
(\ref{psit}, \ref{psil}) and the
quark masses given on page \pageref{quarkmasses}.
The results are shown in figs.\ \ref{nmcd}-\ref{e665}.
Curves obtained with nonperturbative wavefunctions are drawn with solid lines
and the results from perturbative wavefunctions are depicted by dashes lines.
The calculations are  for the same
kinematics as the datapoints. We neglect shadowing in the
deuterium structure function $F_2^D$, which is of order $\le5$\% \cite{nmcdeut}.
The antishadowing effect which occurs at $x_{Bj}\sim 0.1$ is neglected as well. 
Shadowing is calculated
only for nuclei with $A\ge 12$, since the basic formula
is derived in the approximation $A\gg 1$.
Although the calculations
are parameter free, agreement is reasonably good. 

Fig.\ \ref{nmcd} shows a comparison of the calculations with experimental data
 for the structure function ratios of carbon and
calcium over deuterium from NMC \cite{nmc951,nmc952}.
Unlike other NMC data, these data reach down to quite
small $x_{Bj}$, where
the error bars for carbon are quite large. At the last
point in the carbon plot, the solid curve turns unexpectedly upwards.
No such behavior is visible in the perturbative calculation.
Note, that the data are at different values of $Q^2$. The leftmost
point has $Q^2=0.035$ GeV$^2$ and the rightmost point $Q^2=9.6$ GeV$^2$.
Because of the very small $Q^2$ of the point at the lowest value of $x_{Bj}$, we
attribute the decrease of shadowing to the model of the nonperturbative
interaction. 
The calcium data and part of the carbon data shown in
this figure are the
result of a reevaluation of the data from \cite{nmcold,nmcold2}.
Among other reasons, the reevaluation was necessary, because the radiative
corrections require knowledge of $F_2^D$ over a large range of $x_{Bj}$ and
$Q^2$. When new information about $F_2^D$ became available from SLAC, the
radiative corrections were performed again \cite{nmc951}.
This lead to an increase of the structure function ratio 
by at most 2.5\% for Ca/D at
the lowest value of $x_{Bj}$ \cite{nmc951}.

In fig.\ \ref{nmcc}, we compare our calculation to NMC data from
\cite{NMCfit,nmc951} for different nuclei, ranging from aluminium to lead
measured relative to carbon. 
Shadowing is always slightly larger in the
nonperturbative calculation. This is in accordance with the observation in
section \ref{DCS} that the 
choice of the quark mass squeezes the wavepacket
somewhat more than the nonperturbative interaction. Since shadowing is 
determined
by diffraction, which is an almost entirely soft process, one is 
more sensitive
to the large $\rho$ behavior of the dipole cross section and 
to details of the
nonperturbative interaction. The perturbative calculation seems to reproduce the
data better than the nonperturbative. Note however that taking into account the
antishadowing effect would shift the curves upwards. Whether this will happen 
at all values of $x_{Bj}$ or only around $x_{Bj}\sim 0.1$ is model dependent.
We prefer not to introduce new parameters.
Note that the data in fig.\ \ref{nmcc}  correspond to different muon beam
energies, $90$ GeV for the calcium data in the upper row and $200$ GeV for all
other data. 
The structure function ratio is however independent of the beam
energy. Indeed one sees from the equations in section \ref{f2sec} that
\beq
\frac{F_2^A}{F_2^D}=\frac{\sigma_T^{\gamma^*A}+\sigma_L^{\gamma^*A}}
{\sigma_T^{\gamma^*D}+\sigma_L^{\gamma^*D}},
\eeq
while for the measured lepton-target cross section one has  
\beq\label{relation}
\frac{\sigma^{lA}}{\sigma^{lD}}=
\frac{F_2^A}{F_2^D}\frac{(1+\epsilon R_A)(1+R_D)}{(1+R_A)(1+\epsilon R_D)},
\eeq
where
\beq\label{polarization}
\epsilon=\frac{4(1-y)-\frac{4m_N^2x_{Bj}^2}{Q^2}}
{4(1-y)+2y^2+\frac{4m_N^2x_{Bj}^2}{Q^2}}
\eeq
is the photon polarization parameter and
\beq\label{R}
R_A=\frac{\left(1+\frac{4m_N^2x_{Bj}^2}{Q^2}\right)\sigma_L^{\gamma^*A}
+\frac{4m_N^2x_{Bj}^2}{Q^2}\sigma_T^{\gamma^*A}}{\sigma_T^{\gamma^*A}}
\approx\frac{\sigma_L^{\gamma^*A}}
{\sigma_T^{\gamma^*A}}.
\eeq
The lepton nucleon cross section ratio is only equal to the structure function
ratio, if there are no nuclear effects in $R$. This is assumed in all NMC data
and supported by experimental observations. In the kinematical region relevant
for NMC, $R$ data show practically no $A$ dependence \cite{Muecklich}. 

The best data available today for nuclear shadowing in DIS
are for the structure function  
ratio Sn/C from NMC \cite{nmc}. These are shown
in figs.\ \ref{sncx} and \ref{sncq2}. Although our calculation describes the
existing data quite well, the dip in the $Q^2$ dependence of the
solid curve in fig.\ \ref{sncq2} looks rather suspicious. This dip is
practically invisible if one employs the older parametrization (\ref{sasha})
of the dipole cross section \cite{latest}. 
It is therefore caused by the suppression factor in
the improved parametrization (\ref{improved}). Since the dip is not present in
the perturbative calculation, it is also related to the nonperturbative
interaction. We do not believe that this indicates a physical effect. One rather
has to improve the dipole cross section and the 
nonperturbative interaction, in order to get a
better description of the $Q^2$ dependence of shadowing. We also hope to
overcome the numerical 
prescription we use for the dipole 
cross section in the future and to perform the
calculations with the exact parametrization.

We also compare our parameter free calculations to data from E665
\cite{e66592,e66595}, where much lower values of $x_{Bj}$ are covered, fig.\
\ref{e665}.  The decrease of shadowing in the solid curve at very
low $x_{Bj}$ and $Q^2$, which was already visible in fig.\ \ref{nmcd}, is also
seen in the Xe/D plot. The agreement with data is less good for the E665 data
than for the NMC data. Indeed, the two datasets contradict each other
\cite{NMCfit,Muecklich} and no calculation can reproduce both. 
The disagreement between NMC and E665 vanishes however, in the
$F_2$ ratio relative to carbon. Note also that the $q\bar qG$ component
will  become important at very low $x_{Bj}$ and calculations, taking only
the $q\bar q$ state into account, will be unreliable at $x_{Bj}\ll 0.01$.

Besides describing all data reasonably well, the success of  
the Green function technique is to
reproduce the correct shape of the curves vs.\ $x_{Bj}$. Without
treating the coherence length properly, 
shadowing would set in quite abruptly at
$x_{Bj}\sim 0.1$ and quickly saturate as one goes to lower $x_{Bj}$, cf.\ fig.\
\ref{bild}.

Nuclear shadowing is very challenging from the experimental point of view,
because of large radiative corrections. These corrections  occur, e.g.\
when
the incident lepton, see fig.\ \ref{dis1}, radiates bremsstrahlung and does not
produce a deep inelastic event \cite{Levy}.
In the older publications \cite{nmcold,nmcold2}, the 
NM collaboration calculated the
radiative corrections employing the computer program FERRAD, which relies on the
theoretical analysis by Mo and Tsai \cite{Mo}.
For the reevaluation of the shadowing data \cite{nmc951} three different codes
were tested, FERRAD, an improved version of FERRAD and
TERAD. The latter relies on the work of Akhundov, Bardin, and Shumeiko
\cite{B1,B2,B3} and was used for the published NMC result. 
A more detailed discussion on radiative corrections for NMC and
HERMES can be found in \cite{Jeroen}. 
We point out that the whole
shadowing effect in the data shown in
figs.\ \ref{nmcd}-\ref{sncq2} is calculated. Without
radiative corrections, the cross section ratio would be around
unity \cite{Jeroen,Bprivate}. However, the correctness of the calculation was
checked by comparing to the "hadron tagged" data, making sure that really a deep
inelastic event was measured.

A different way to identify the deep inelastic events was chosen by the E665
collaboration \cite{e66592,e66595}, 
where cuts in the electromagnetic calorimeter were applied. 
In \cite{e66592} the xenon 
data obtained in this way were compared to an evaluation
with hadron tagging, see fig.\ \ref{e665}. The two methods give consistent
results.
In \cite{e66595}, the E665 collaboration also applied the FERRAD code for 
a comparison with NMC.
The results are depicted by open circles in fig.\ \ref{e665}. The code wass
assumed to be reliable only in the region $y<0.7$, see (\ref{ypsilon}), and 
$x_{Bj}>0.002$ \cite{e66595}. Therefore it was not applied to all data
points, but only to those, where it was supposed to work. One recognizes a
systematic discrepancy between the two evaluation methods. The radiative
corrected cross section ratios are lower by $\sim$5\% than the points from the
calorimeter analysis. We argue however 
from the date of the publication
that the improved input for FERRAD used
in \cite{nmc951} was not taken into account by E665 in \cite{e66595}. It is of
course impossible for the author to conclude which experiment is correct.

\begin{figure}[t]
  \centerline{\scalebox{1.0}{\includegraphics{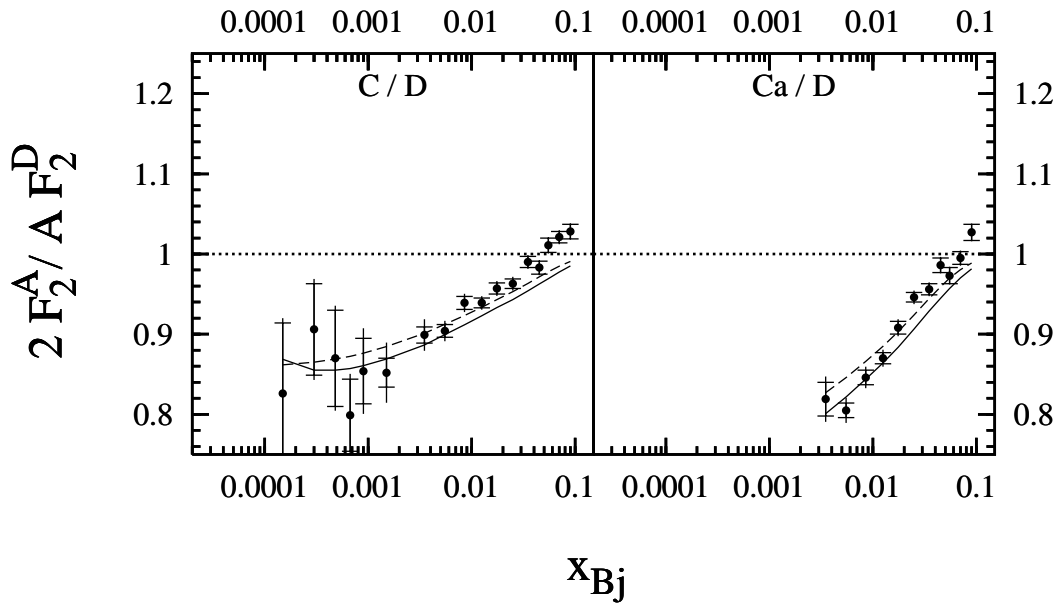}}}
    \caption{
      \label{nmcd}
      Comparison between calculations for shadowing in DIS and experimental
      data from NMC \cite{nmc951,nmc952}
      for the structure functions of
      carbon and calcium relative to deuterium.  
      The $x_{Bj}$-dependence of shadowing is shown.
      The leftmost
      point of the carbon data is measured at 
      $Q^2=0.035$ GeV$^2$ and the rightmost point at $Q^2=9.6$ GeV$^2$.
      The $Q^2$ values for calcium are
      $0.6~{\rm GeV}^2\le
        Q^2\le 9.7~{\rm GeV}^2$.
      The inner error bars show the statistical error and the
      outer error bars statistical and systematic error added in quadrature. 
      The full curves are calculated with the Green function method 
      (\ref{4.1a})
      including the nonperturbative interaction. 
      The dashed curves are calculated with the perturbative wavefunctions
      (\ref{psit},\ref{psil}).
    }  
\end{figure}

\begin{figure}[t]
  \centerline{\scalebox{1.0}{\includegraphics{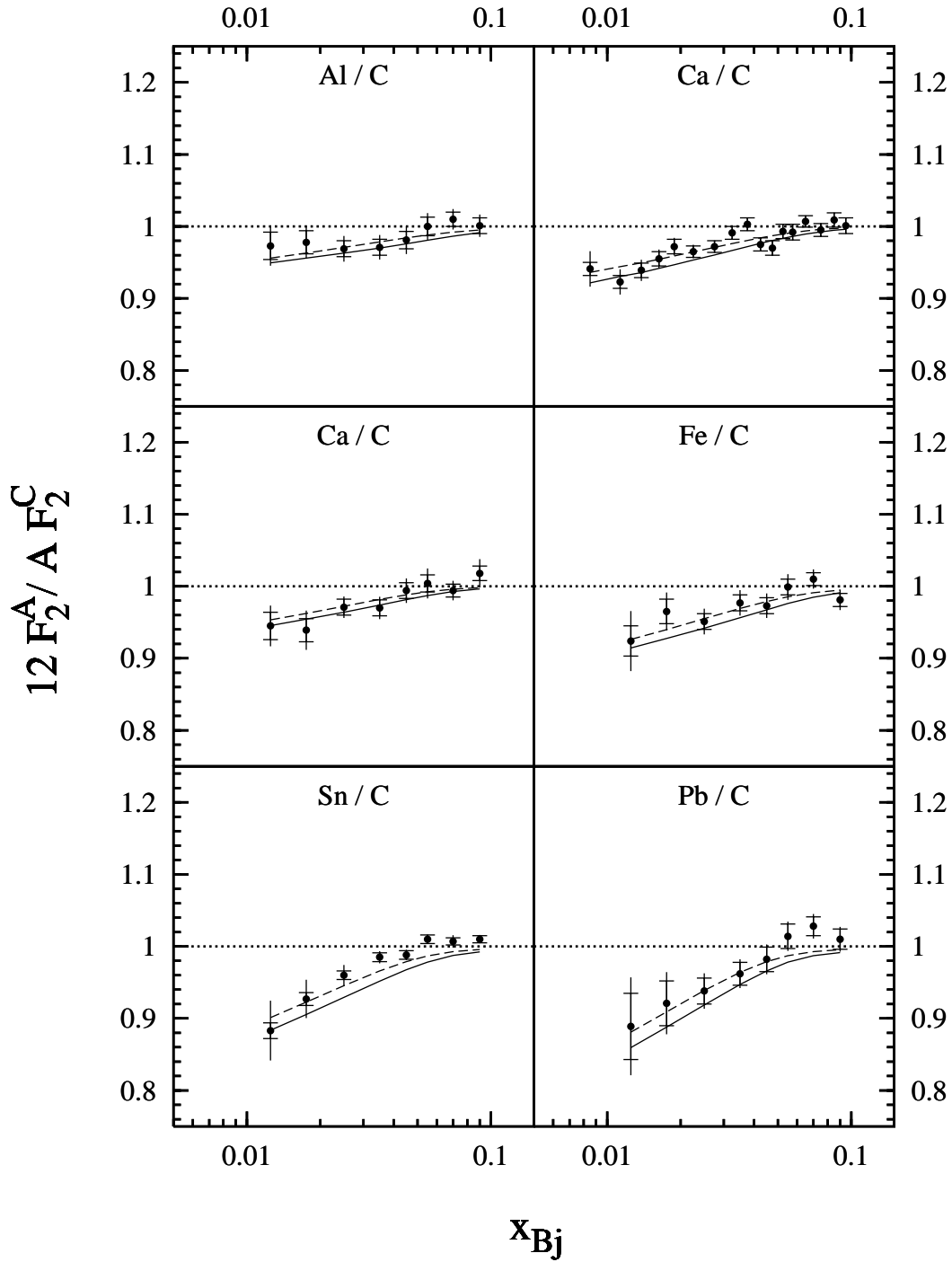}}}
    \caption{
      \label{nmcc}
      Comparison between calculations for shadowing in DIS and experimental
      data from NMC \cite{NMCfit,nmc951}
      for the structure functions of
      different nuclei relative to carbon.  
      The $x_{Bj}$-dependence of shadowing is shown.
      The $Q^2$ range covered by the data 
        is approximately $3~{\rm GeV}^2\le
        Q^2\le 17~{\rm GeV}^2$ from the lowest to the highest $x_{Bj}$ bin.
      The different curves and error bars 
      have the same meaning as in fig.~\ref{nmcd}.
    }  
\end{figure}

\begin{figure}[t]
  \centerline{\scalebox{0.9}{\includegraphics{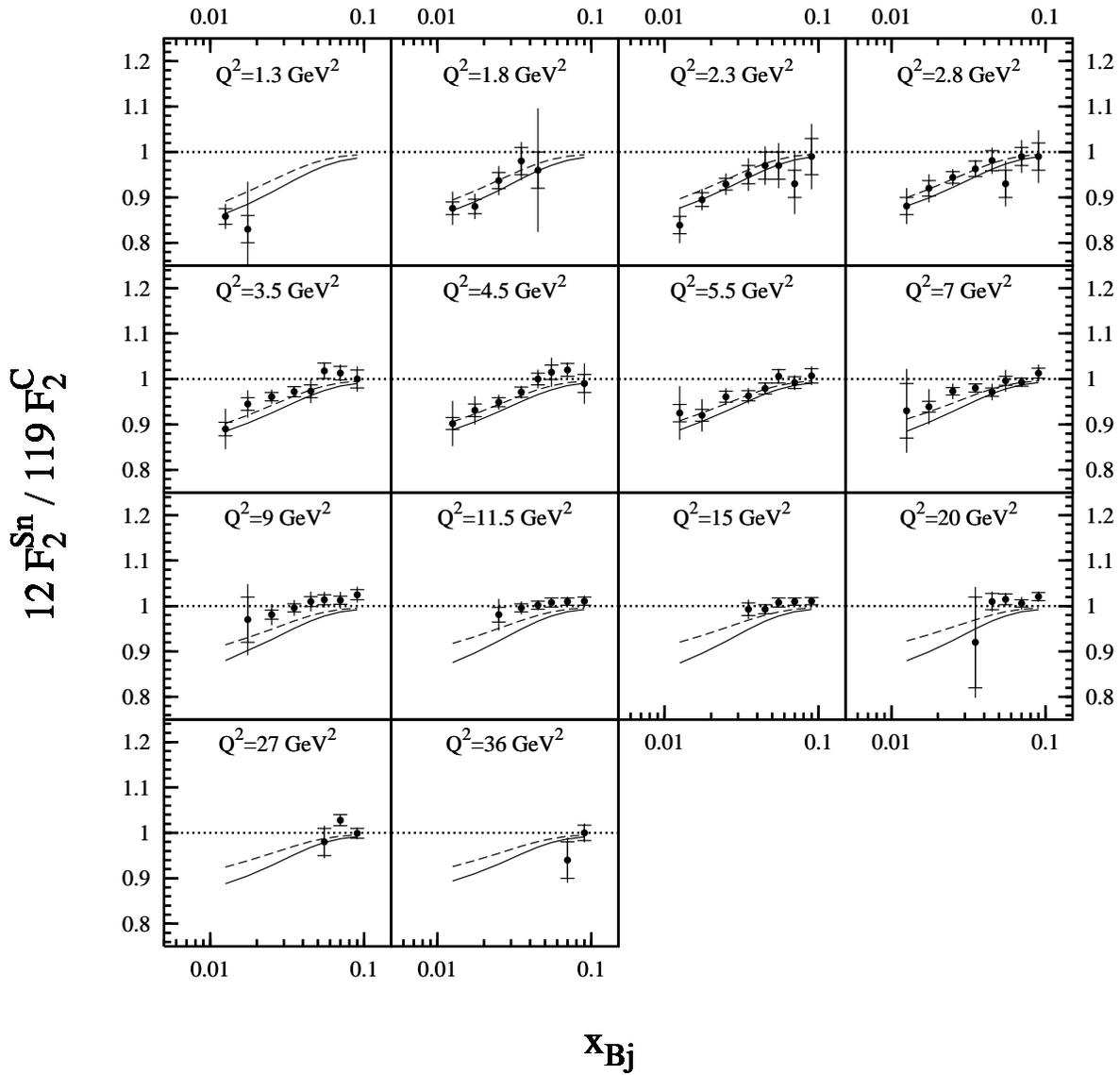}}}
    \caption{
      \label{sncx}
      The $x_{Bj}$ dependence of nuclear shadowing in DIS
      for the structure function ratio of tin relative to carbon.
      The data are from NMC \cite{nmc}.  
      The different curves and error bars 
      have the same meaning as in fig.~\ref{nmcd}.
    }  
\end{figure}

\begin{figure}[t]
  \centerline{\scalebox{0.8}{\includegraphics{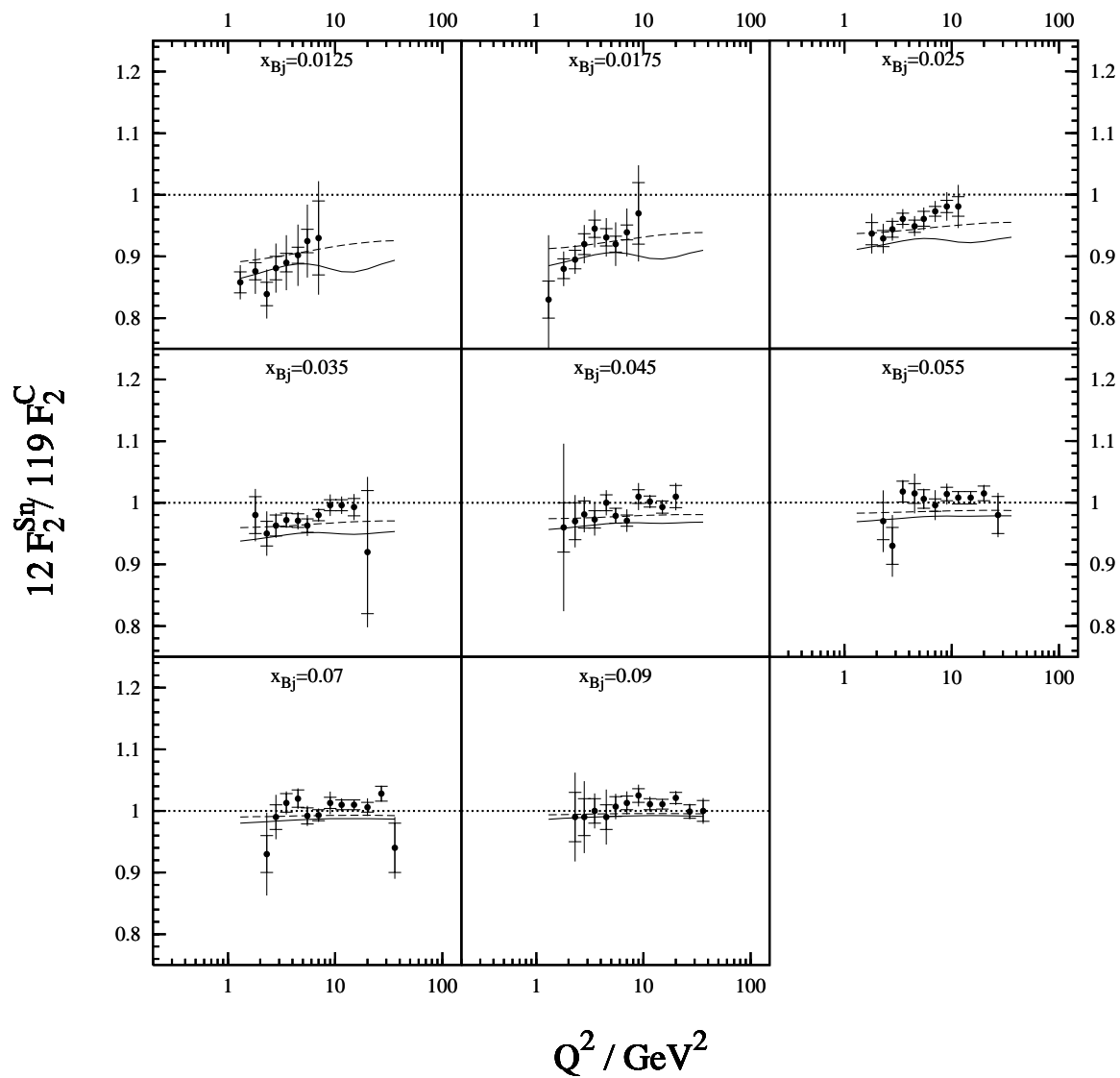}}}
    \caption{
      \label{sncq2}
      The $Q^2$ dependence of nuclear shadowing in DIS
      for the structure function ratio of tin relative to carbon.
      The data points \cite{nmc}
      are the same as shown in fig.~\ref{sncx}.  
      The different curves and error bars 
      have the same meaning as in fig.~\ref{nmcd}.
    }  
\end{figure}

\begin{figure}[t]
  \centerline{\scalebox{1.0}{\includegraphics{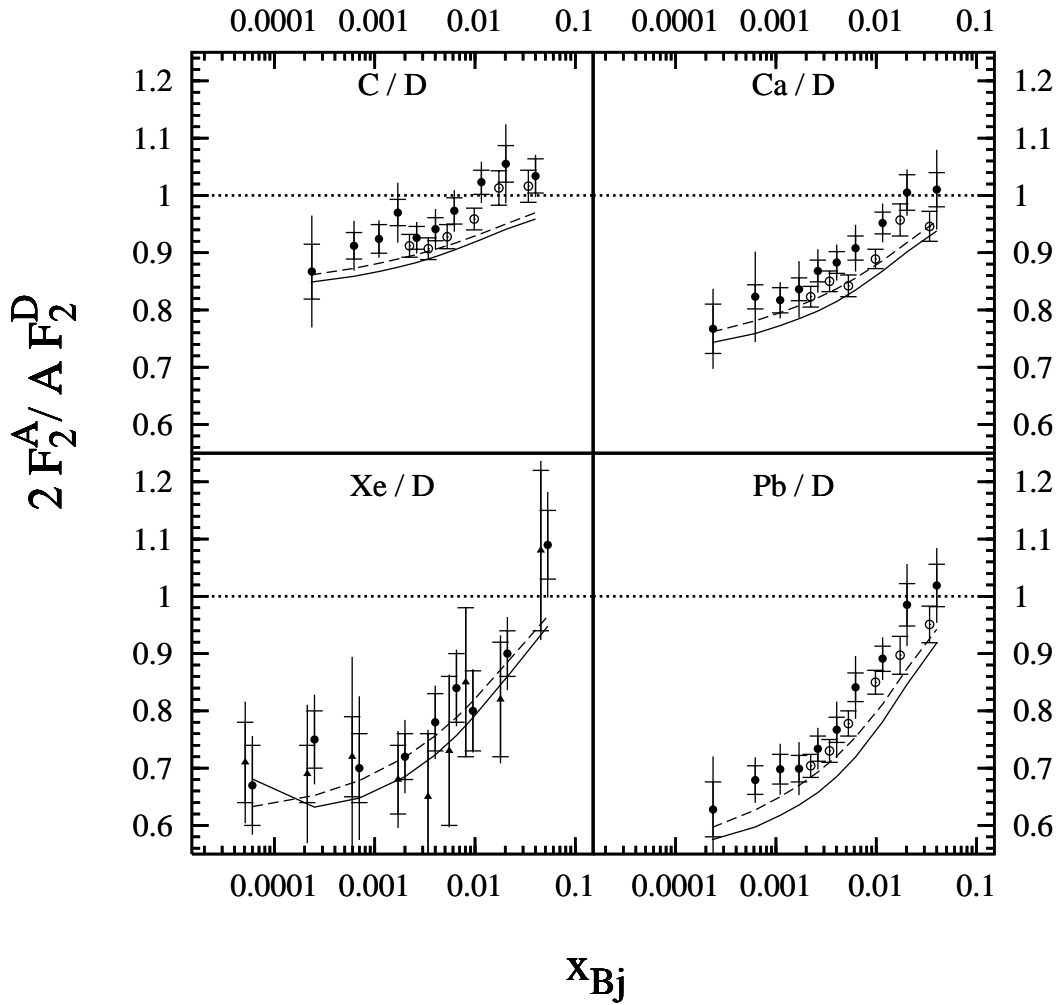}}}
    \caption{
      \label{e665}
      The $x_{Bj}$ dependence of nuclear shadowing in DIS for various nuclei. 
      The data are from E665 \cite{e66592,e66595}.
      Full circles show data taken with electromagnetic calorimeter cuts, while
      the open circles result from the FERRAD radiative 
      correction
      code. Hadronic cuts were applied for the triangles in the lower left 
      plot.
      For better visibility,
      triangles and open circles are displaced to slightly lower 
      $x_{Bj}$. 
      The $Q^2$ range of the data is $0.15~{\rm
        GeV}^2\le Q^2\le~22.5~{\rm GeV}^2$, except for the xenon data,  
        $0.03~{\rm GeV}^2\le Q^2\le 17.9~{\rm GeV}^2$, 
        from the lowest to the
        highest value of $x_{Bj}$.
        The meaning of the different curves and error bars 
      is explained below fig.~\ref{nmcd}.
    }  
\end{figure}

\clearpage

\begin{figure}[ht]
\centerline{
  \scalebox{0.45}{\includegraphics{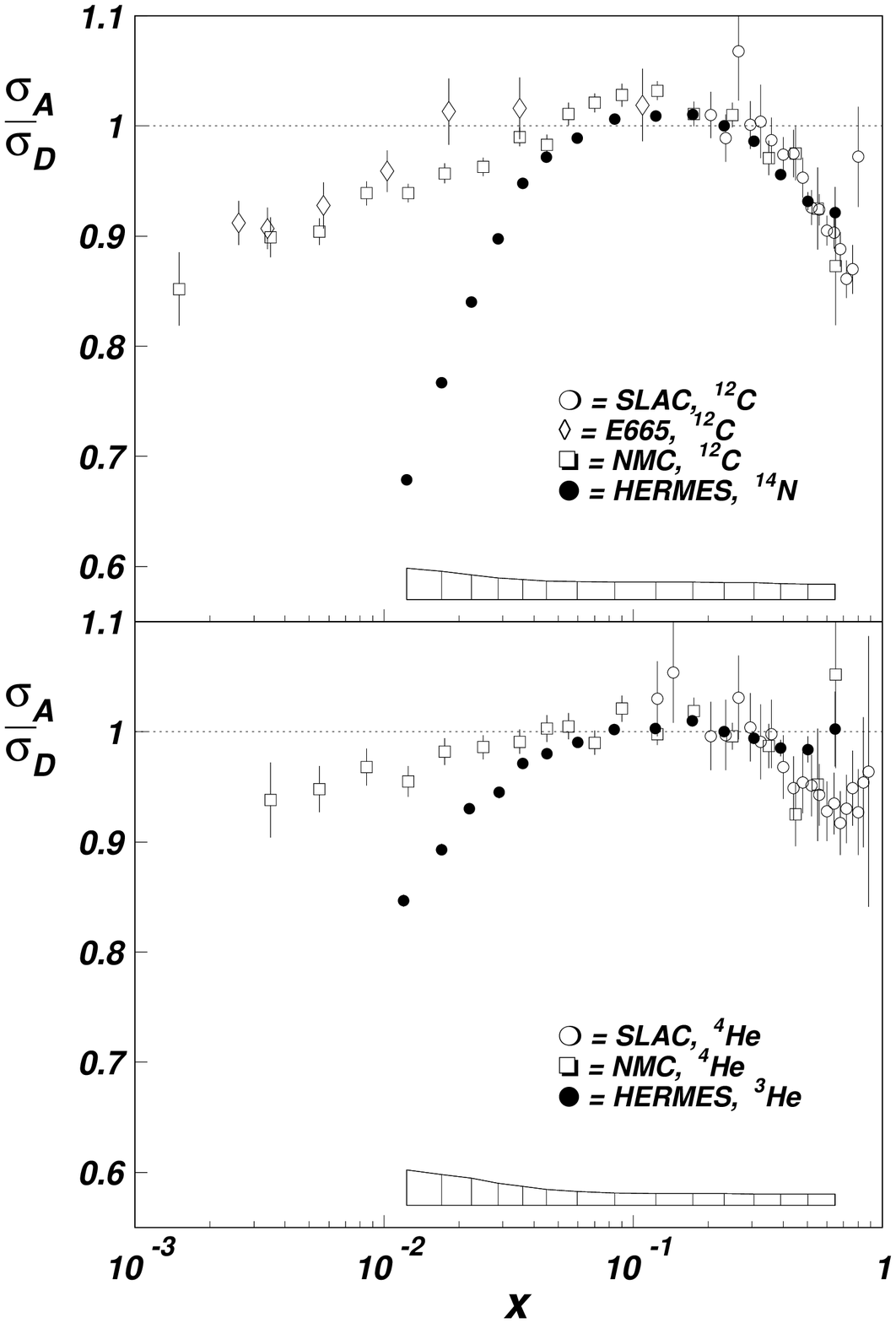}}\hfill
  \scalebox{0.45}{\includegraphics{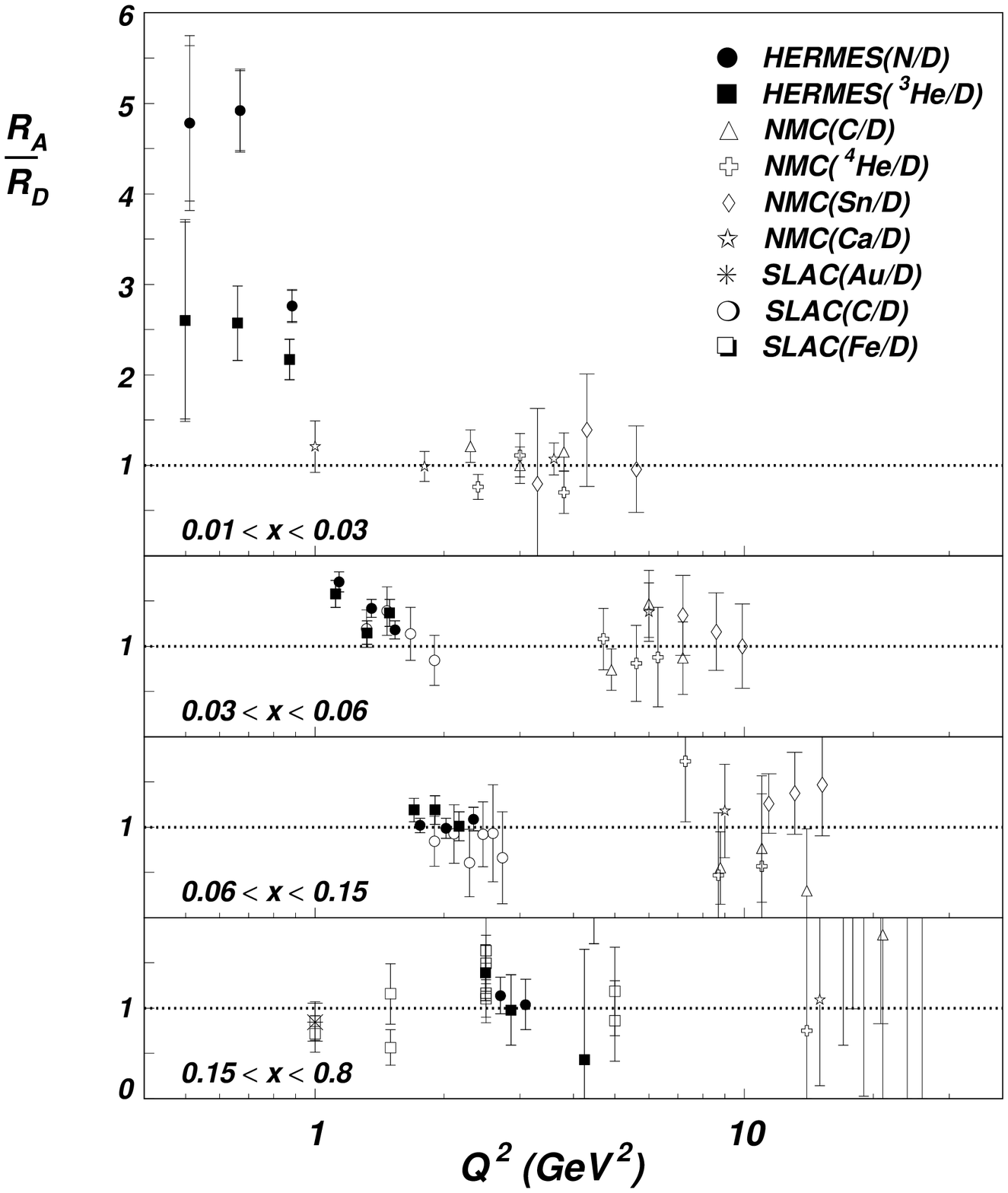}}
}
\caption{
  \label{hermesplots}
  The left figure shows the
  leptonic 
  cross section ratio measured by HERMES for helium and nitrogen. The data are
  not at variance with other experiments, 
  since the points correspond to different values of $Q^2$.
  The large enhancement in the $R$-ratio is shown on the right, 
  $R_A=\sigma_L^{\gamma^*A}/\sigma_T^{\gamma^*A}$
  The figures are from
  \cite{hermes}.
}
\end{figure}

Much attention has been drawn to
a possible $A$ dependence of $R$ and the importance of radiative corrections 
by the data released recently by the HERMES collaboration \cite{hermes}. HERMES
measured nuclear shadowing in DIS at the same values of $x_{Bj}$ as NMC, but at
smaller virtuality, $0.5~{\rm GeV}^2\le Q^2\le 2.018~{\rm GeV}^2$
and larger $y<0.85$ (\ref{ypsilon}). The 
$27.5$ GeV positron beam at HERA was used. 
The results shown in fig.\ \ref{hermesplots},
expose several striking features. The cross section ratio for nitrogen to
deuterium is suppressed much stronger than it would have been expected
extrapolating
previous NMC measurements. 
This ratio is related to the $F_2$ ratio by (\ref{relation}). The
two quantities are equal, only when $\epsilon=1$ or when there are no nuclear
effects in $R$ (\ref{R}). Since NMC could not find any nuclear effects in $R$,
all NMC data were evaluated assuming $R_A=R_D$.  
Note that at HERMES $\epsilon\sim 0.4$ at low $x_{Bj}$. The discrepancy in the
left part of fig.\ \ref{hermesplots}, can be
interpreted in terms of different shadowing for transverse and
longitudinal photons. It was concluded in \cite{hermes} 
that while $\sigma_L$ is
enhanced,  $\sigma_T$ is suppressed on nitrogen by at least
factor of two compared to a deuterium target. Then one obtains a structure
function ratio of the same order of magnitude as observed before, fig.\ 
\ref{hermesf2}.

\begin{figure}[t]
  \centerline{\scalebox{0.8}{\includegraphics{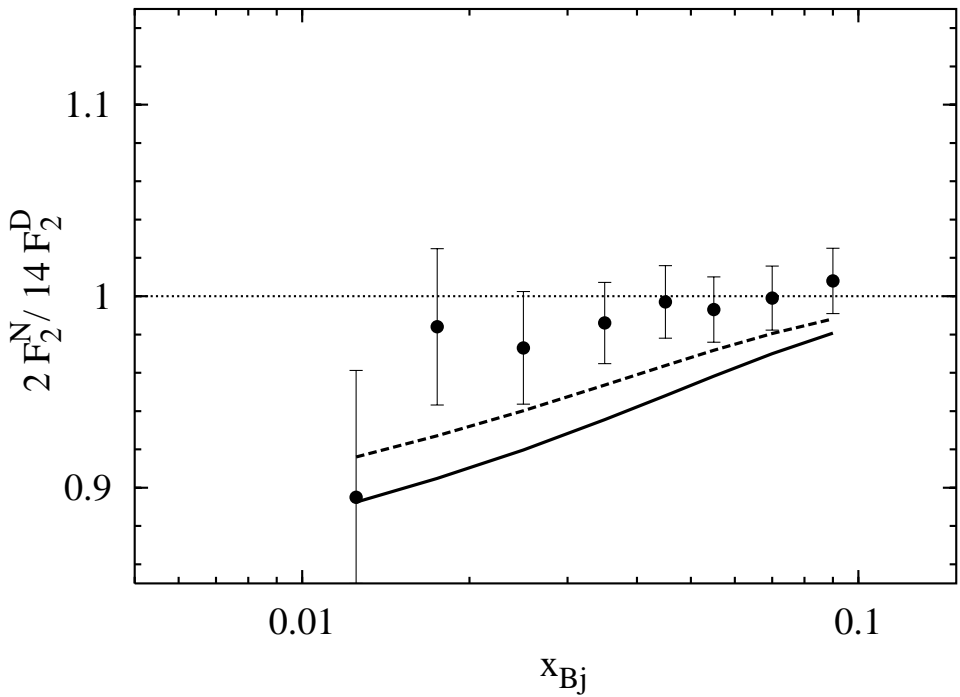}}}
    \caption{
      \label{hermesf2}
      The $F_2$ ratio for nitrogen to deuterium measured by HERMES \cite{hermes}
      compared calculations with the Green function method 
      (\ref{4.1a}).
      The full curve is calculated with nonperturbative
      and the dashed curves with perturbative wavefunctions. The $Q^2$ range of
      the data is $0.5~{\rm GeV}^2\le Q^2\le 2.018~{\rm GeV}^2$.
    }  
\end{figure}

Radiative corrections are huge at HERMES, because the positron mass is much
smaller than the muon mass and because radiative corrections become very large
at
$y\to 1$ \cite{Levy}. Furthermore, $F_2$ and $R$ are needed as input to
calculate the radiative corrections and have to be known over a wide range of
$x_{Bj}$ and $Q^2$. Therefore an iterative procedure was applied by HERMES
\cite{Jeroen,hermes}, using
existing data as input to calculate the radiatively corrected cross sections
in a first approximation. The new structure function and 
the new $R$ obtained in
this way were again used as input to the computer program. This was 
repeated until
convergence is obtained. 
To make the result reliable, three different codes were
compared, FERRAD \cite{Mo}, TERAD \cite{B1,B2,B3} and POLRAD \cite{Pol}. In
spite of the statement in \cite{e66595} 
that FERRAD is not applicable for $y>0.7$,
all codes gave the same result within 2\%. 
More details can be found in \cite{Jeroen}.

The HERMES experiment drew attention to the fact that only very few data are 
available in this kinematical region. 
Moreover, no reliable
theoretical calculations are done yet.
Approaches based on QCD evolution equations need parton distributions at a semi
hard scale as input and their applicability is doubtful at the low $Q^2$ values
of HERMES.  
More promising is the 
intuitive approach treating nuclear 
effects in the spirit of vector dominance model (VDM) \cite{bauer} 
as shadowing for the total cross section of hadronic fluctuations 
of the virtual photon (see e.\ g.\  \cite{fs}).
However, the VDM is sensible only
at small $Q^2\to 0$ not so easily generalized to longitudinal photons. 

Eventually we are able to provide theoretical predictions from the light-cone
approach, since nonperturbative effects are assumed to be parametrized in the
dipole cross section and we also included 
the nonperturbative interaction introduced in
\cite{KST99} between the $q$ and the
$\bar q$ for a more realistic description of the soft physics.
The results of our calculations for the nitrogen $F_2$-ratio 
is shown in fig.\ \ref{hermesf2} in comparison with HERMES data. 
We expect somewhat larger shadowing for $F_2$ than seen by HERMES.

\begin{figure}[t]
  \centerline{\scalebox{0.8}{\includegraphics{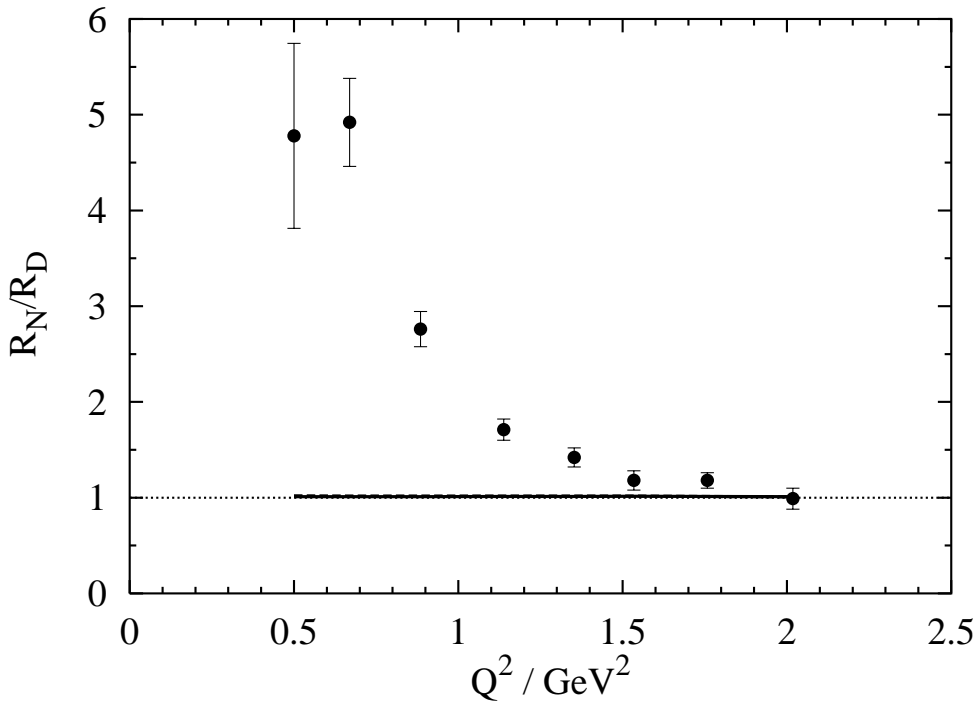}}}
    \caption{
      \label{rhermes}
      The result of calculations in the light-cone approach 
      compared to the HERMES measurement
      \cite{hermes} of
      $R_N/R_D$. The data points are for the same kinematics as those shown in
      fig.\ \ref{hermesf2}. 
      On this scale, the lines corresponding to perturbative (dashed)
      and nonperturbative (solid) wavefunctions lie on top of each other and
      predict an $R$-ratio of practically unity. The large enhancement for
      $\sigma_L^{\gamma^*N}$ measured by HERMES is clearly not reproduced in the
      light-cone approach.
    }  
\end{figure}

The most striking feature of the HERMES data for shadowing is
a dramatically rising $R=\sigma_L/\sigma_T$ ratio on nitrogen compared to
proton target at $Q^2<1~{\rm GeV}^2$ \cite{hermes}. Our predictions 
for $R_N/R_p$ are plotted in fig.\ \ref{rhermes} versus $Q^2$ at
the same values of $x_{Bj}$ as the experimental points. Like in all other
calculations, we neglect nuclear effects in deuterium.
Apparently, we do not expect any remarkable effect.
There is a priori no reason to think that the 
light-cone approach leads to equal
shadowing for longitudinal and transverse photons. It is calculated
 in section
\ref{meancoh} that the coherence length for longitudinal photons is
significantly longer than for transverse. Apparently this is compensated by the
smaller size of $q\bar q$-fluctuations from longitudinal photons, cf.\ section
\ref{dipoledis}. One has however to bear in mind, that we can calculate
shadowing for $\sigma_L$ only at small $Q^2$. In the approximation, when only
the $q\bar q$ pair is taken into account, shadowing for longitudinal photons
vanishes $\propto 1/Q^2$ at large $Q^2$. Taking also the $q\bar qG$ component
into account, shadowing for $\sigma_L$ will persist at
high $Q^2$, cf.\ section
\ref{diffractionsec}.
We also calculated $R$ for a proton target and compared to NMC data \cite{nmc1}
in the low
$x_{Bj}$ region, fig.\ \ref{sloverst}. Good agreement with the data is achieved
over the whole $Q^2$ range, without introducing any new parameters. 
One might argue that the
beginning of an increase of $R$ is visible from the low $Q^2$ points in fig.\
\ref{sloverst}. There 
are however not enough data to allow for definite conclusions. 
Note that nuclear effects for $R$ were estimated 
previously in a different approach \cite{barone,barone2}.

\begin{figure}[t]
  \centerline{\scalebox{0.8}{\includegraphics{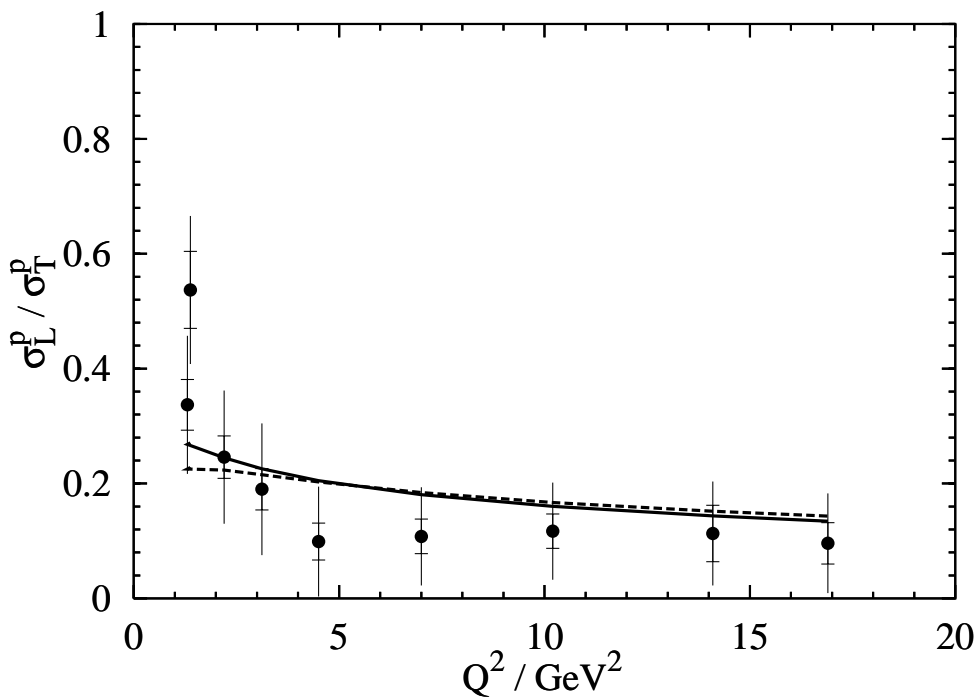}}}
    \caption{
      \label{sloverst} The ratio of
      longitudinal to transverse cross section, 
      $R=\sigma_L^{\gamma^*p}/\sigma_T^{\gamma^*p}$, for a pro\-ton 
      measured by NMC
      \cite{nmc1}. The inner error bars represent the statistical error and the
      outer error bars statistical and systematic error added in quadrature. 
      The full curve is calculated using nonperturbative
      and the dashed curves using perturbative wavefunctions. 
      From the left to the right, 
      the data are for
      the $x_{Bj}$ values $0.0080,~0.0045,~0.0125,~
      0.0175,~0.025,~0.035,~0.05,~0.07$, and $0.09$.
    }  
\end{figure}

We suppose that our results 
provide a reliable base line for nuclear effects
in this region. 
The dramatic effects revealed by the HERMES 
experiment probably cannot be explained without involving  new
nonstandard dynamics, see e.g. \cite{mbk}.
While a nuclear enhancement in the longitudinal cross section might be
qualitatively explained by higher twist contributions, the enormous
suppression for $\sigma_T$ which is implied by the $\sigma_L$ enhancement and
fig.\ \ref{hermesplots} still remains. Indeed, for the two leftmost points in
fig.\ \ref{rhermes}, one finds \cite{Jeroen} 
$\sigma_T^{\gamma^*N}/\sigma_T^{\gamma^*p}\sim 0.45$. For nitrogen, this implies
that the virtual photon splits up into something that interacts with an
effective cross section of $\sim 100$ mb, compared to the usual $\sim 25$ mb for
real photons and for $Q^2>1$ GeV$^2$.

Experimental data for shadowing in DY
are available from the collaborations E772 
\cite{E772} and
E866 \cite{E866}. We
compare only to the E772 data, although even smaller values
of $x_2$ could be reached by the E866/NuSea collaboration. The latter measured
however relative to a beryllium target. The large $A$ approximation and the 
uniform nuclear density we have to apply is not justified for such a light
nucleus. These simplifications are necessary, because of merely technical
difficulties and are not demanded by the nature of the light-cone approach,
section \ref{dyshadow}. The result of the
calculation, shown in fig.\ \ref{e772}, agrees well with the data. 

Practically all DY data are taken at masses
$M$ higher than the
$J/\psi$ mass, because of the large underground of dileptons
from other sources at small $M^2$. It is therefore impossible 
to reach values of $x_2$ as low as the
values of $x_{Bj}$ in the E665 data, fig.\ \ref{e665}, at present. 
This situation will however change in the nearest future, when RHIC starts
operating. With PHENIX, values of $x_2\sim 0.001$ and $M\sim 2$ GeV can be
reached \cite{Gprivate}.
 
\begin{figure}[t]
  \centerline{\scalebox{0.8}{\includegraphics{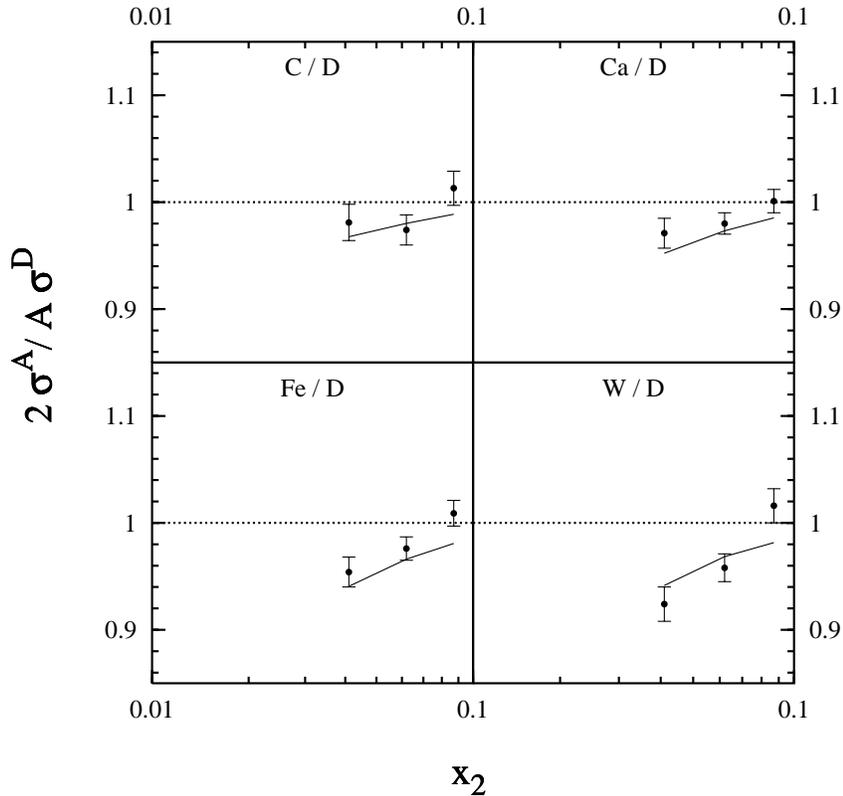}}}
    \caption{
      \label{e772}
      Comparison between calculations in the light-cone approach and
      E772 data \cite{E772} for shadowing in DY. Only perturbative
      wavefunctions are applied, since there is no real dipole in DY. 
      The calculations are performed at the mean values of the lepton pair mass.
      From the
      left to the right, these values are $5$ GeV$^2$,
      $5.7$ GeV$^2$, and $6.5$ GeV$^2$. 
    }  
\end{figure}


\section{Summary and outlook}

In this work a phenomenological approach to nuclear shadowing in DIS 
and DY in the target rest frame is developed. 
As it is well known, the partonic interpretation of high energy scattering
processes depends on the reference frame. In a frame, where the target is fast
moving, DIS looks like the scattering of virtual photons off sea quarks in the
target (at low $x_{Bj}$). 
In contrast to this, in the rest frame of the target the virtual photon
splits up into a $q\bar q$-pair which then scatters off the target gluon field,
see fig.\ \ref{lc1}.
Note that in the infinite momentum frame the $q\bar q$-pair belongs to the
target and is generated from gluon splitting, while in the target rest frame
the incoming virtual photon is decomposed into a superposition of Fock-states.
The $q\bar q$-pair has to be regarded as the lowest Fock-state of the photon. 

The DY process may be seen in a way similar to DIS: In 
the target rest frame DY dilepton production looks like 
brems\-strahl\-ung
\cite{boris}, 
see fig.\ \ref{bremsdy}. A quark from the projectile scatters off
 the target gluon field and
emits a  photon of mass $M$, which decays into a dilepton pair.
Like in DIS, the incoming quark is decomposed into a superposition of
Fock-states, but now the lowest nontrivial
state is the $q\gamma^*$-fluctuation.
The single quark and the $q\gamma^*$-state scatter with different amplitudes off
the target. This disturbes the coherence between the Fock-states and the
$\gamma^*$ is freed.

It is a special virtue of the target rest frame point of view that coherence
effects, which are so important at high energies, can be understood in a very
easy and intuitive way.

It is convenient to work in a mixed representation, where the two transverse
dimensions are treated in coordinate space while the longitudinal direction is
described in momentum representation. The advantage of this representation is
that Fock-states with fixed transverse separations between the particles are
eigenstates of the interaction 
at infinitely high energy
\cite{zkl,Borisold,Miettinen}. The total cross
section for DIS and DY can then be written in factorized light-cone form, namely
as product of a LC wavefunction and the dipole cross section, integrated over
the two transverse directions and the longitudinal momentum. 
The differential DY cross section can also be expressed in terms of 
the dipole cross section
\cite{kst}. The LC
wavefunctions describe the probabilities for the transitions
 $\gamma^*\to q\bar q$
(DIS) and $q\to\gamma^*q$ (DY), respectively. In the case of DIS, a
nonperturbative interaction \cite{KST99} between the particles in the pair is
introduced to model confinement.

Working in a representation, in which the scattering matrix is diagonal, 
makes it
particularly simple to resum multiple scattering terms.
In both cases, DIS and DY, shadowing is caused by multiple
scattering within the coherence length. The coherence length is the lifetime of
the fluctuation of the projectile and can substantially exceed the nuclear
radius at high energies. 
In DIS, large size dipoles are absorbed at the surface of the nucleus, casting a
shadow on the inner nucleons. For the DY process, shadowing is due to the LPM
suppression of bremsstrahlung, as explained in the introduction.
The conditions which have to be fulfilled in order to
observe shadowing are 
\begin{itemize}
\item{The fluctuation of the projectile has to interact with a cross
section that is large enough to make the mean free path of the fluctuation
shorter than the nuclear radius.  
Only for these fluctuations, multiple scattering is possible.}
\item{The coherence length has to be longer than the mean internucleon spacing.
Otherwise, the projectile would scatter incoherently off the different nucleons
and no shadowing is observed.}
\end{itemize}
Since shadowing is dominated by large size fluctuations, it is an essentially
nonperturbative phenomenon, even at high virtualities.
Shadowing does not vanish at high virtuality, because the large cross section of
the aligned jet configurations compensates for the small probability to create
a very asymmetric pair. Note that there are no large $q\bar q$-pairs from
longitudinal photons. Nevertheless, shadowing for longitudinal photons does not
vanish with $Q^2$ either, when higher
Fock-component of the photon, containing gluons, are
taken into account. Since longitudinal photons measure the gluon density in the
target, one expects that gluon shadowing is a leading twist effect.

The most important nonperturbative input to all equations in this work is the
dipole cross section which is essentially unknown and has to be parametrized.
The parametrization presented in this work is an improvement of the one given in
\cite{KST99}. The dipole cross section is known to rise quadratically with the
transverse separation $\rho$ at small $\rho$. It is expected to become constant
as function of $\rho$ at large separations. The energy dependence 
of the dipole cross section correlates with $\rho$. At small separations, the
dipole cross section grows with a hard pomeron intercept while at large
separations it grows much more slowly with a soft pomeron intercept.
The dipole cross section is written as function of the {\em c.m.} energy squared
$s$ of the colliding objects, rather than as function of $x_{Bj}$, in this work.
It can therefore also be applied to calculate total hadronic cross sections.

The propagation of the $q\bar q$-pair through the nucleus is described in the
light-cone approach by a Green function which fulfills a two dimensional
Schr\"odinger equation. The kinetic part of the Hamiltonian accounts for the
transverse motion of the particles in the pair which is important at finite
energies. The dipole cross section multiplied with the nuclear density enters
as imaginary potential in the Schr\"odinger equation. The optical potential
accounts for the shadowing corrections. 
The Green function contains all multiple scattering terms and treats the nuclear
form factor correctly.
Furthermore, the nonperturbative interaction from \cite{KST99} is also
implemented into the formalism. It leads to a real part of the potential that
describes the confining interaction between quark and antiquark.
The total nuclear 
cross section is represented
as the difference of the unshadowed cross section and a suppression term that
represents the shadowing correction. This interference part contains also the
longitudinal momentum transfer. Several known approximations can be obtained as
limiting cases. For infinitely high energy, an eikonal
formula similar to the Glauber formula \cite{Glauber}
is recovered. For light nuclei, where only the double scattering term is
important, one reproduces the formula of Karmanov and Kondratyuk \cite{kk}.
Shadowing vanishes as $x_{Bj}\to 0.1$, because the longitudinal momentum
transfer leads to an oscillating phase factor which makes the shadowing term
vanish. The same formalism can also be applied to shadowing in DY. 
The two dimensional Laplacian in the Schr\"odinger equation describes the
deflection of the quark after radiation of the photon. This term becomes
negligible at infinitely high energies and one obtains a similar eikonal
expression as in DIS.
The Green function method is however less intuitive in this case since there is
no physical dipole in the DY process. For the same reason, the nonperturbative
interaction is not applied to DY.

The main achievements of this work can be summarized as follows:
\begin{itemize}
\item{The Glauber-Gribov theory of multiple scattering is improved to include
the correct nuclear form factor in all higher order scattering terms.
Calculations for DIS and DY
are in good agreement with data, without introducing additional
free parameters.}
\item{The mean coherence length of a Fock state is defined. This quantity is
intended to be a tool for qualitative considerations. We estimate that shadowing
for the $q\bar qG$ Fock state of the photon is negligible at $x_{Bj}\ge 0.01$.}
\item{For proton-nucleus collisions at 
RHIC energies, considerable shadowing of DY dileptons
is predicted for the whole $x_F$-range. 
The transverse momentum dependence of the LPM effect is studied,
too. While shadowing is predicted for 
dileptons with low transverse momenta, an enhancement at larger transverse
momentum is expected.}
\item{The contribution of low mass dileptons, which are produced via the
brems\-strahlungs-mechanism, to the dilepton cocktail is estimated to be of
order 10\% only.}   
\item{The transverse momentum distribution of DY pairs in proton-proton
collisions is calculated. The result is finite at zero transverse momentum due
to the saturation of the dipole cross section. Note that first order pQCD leads
to a divergent result.}
\item{A new parametrization of the dipole cross section is presented.  It is
written as a function of the {\em c.m.} energy squared $s$ and the dipole
size $\rho$.  The parametrization describes well all HERA data up to
$Q^2\le 35 $ GeV$^2$, as well as DY data at large $x_F$ and hadronic cross
sections. No $K$ factor is introduced to describe the DY data.}
\end{itemize}
In spite of the successes of the light-cone approach, there are also some open
issues which have to be considered in the future:
\begin{itemize}
\item{The transverse momentum spectrum of DY pairs in proton-proton collisions,
calculated in the light-cone approach, looks different from what one would
expect from data at higher $x_2$. It will be interesting to compare to 
data at low $x_2$,
once they are available.} 
\item{The strong enhancement of the longitudinal cross section
$\sigma_L^{\gamma^*A}$, recently observed by HERMES
in DIS off nitrogen, is not reproduced in  
in the light-cone approach.}
\item{There are still unsolved technical problems, how to do calculations with
 the
Green function method, without the approximation of a uniform nuclear density
and a quadratically rising dipole cross section.}
\end{itemize}
This work is far from being complete. There are many interesting applications
and challenges which have to be investigated in the future. The two most natural
continuations of this work are,
\begin{itemize}
\item{to develop a framework that consistently takes into account higher Fock
states of the projectile, containing gluons. 
This is especially important in view of nuclear gluon shadowing.
First calculations are already done
in \cite{KST99}.}
\item{to investigate the suppression of gluon radiation in nucleus-nucleus
collisions and it's influence on the physics at RHIC and LHC.}
\item{to extend the Green function method to vector-meson 
production and to study the
influence of the nuclear medium on the total cross section and on the transverse
momentum distribution.}
\end{itemize}
  
\appendix
\section{Appendix}

\subsection{Calculations for DY cross sections in pp scattering}\label{appdypp}

In section \ref{dipoledy}, 
the differential DY cross section is expressed as a four-fold
Fourier integral (\ref{dylcdiff})
\beqn\nonumber
\frac{d\sigma(qp\to q\gamma^*p)}{d\ln\alpha d^2p_T}
&=&\frac{1}{(2\pi)^2}
\int d^2\rho_1d^2\rho_2\, \exp[{\rm i}\vec p_T\cdot(\vec\rho_1-\vec\rho_2)]
\Psi^*_{\gamma^* q}(\alpha,\vec\rho_1)\Psi_{\gamma^* q}(\alpha,\vec\rho_2)\\
&\times&
\frac{1}{2}
\left\{\sigma_{q\bar q}(\alpha\rho_1)
+\sigma_{q\bar q}(\alpha\rho_2)
-\sigma_{q\bar q}(\alpha(\vec\rho_1-\vec\rho_2))\right\},
\eeqn
where
\beqn\nonumber
\Psi^{*T}_{\gamma^* q}(\alpha,\vec\rho_1)\Psi^T_{\gamma^* q}(\alpha,\vec\rho_2)
&=& \frac{\alpha_{em}}{\pi^2}\Bigg\{
     m_f^2 \alpha^4 {\rm K}_0\left(\eta\rho_1\right)
     {\rm K}_0\left(\eta\rho_2\right)\\
   &+& \left[1+\left(1-\alpha\right)^2\right]\eta^2
   \frac{\vec\rho_1\cdot\vec\rho_2}{\rho_1\rho_2}
     {\rm K}_1\left(\eta\rho_1\right)
     {\rm K}_1\left(\eta\rho_2\right)\Bigg\},\\
 \Psi^{*L}_{\gamma^* q}(\alpha,\vec\rho_1)\Psi^L_{\gamma^* q}(\alpha,\vec\rho_2)
&=& \frac{2\alpha_{em}}{\pi^2}M^2 \left(1-\alpha\right)^2
  {\rm K}_0\left(\eta\rho_1\right)
     {\rm K}_0\left(\eta\rho_2\right). 
\eeqn
The Fourier-integral is inconvenient for numerical calculations, but one can
perform three of the integrations analytically for arbitrary 
$\sigma_{q\bar q}(\alpha\rho)$.

Consider the K$_0$-part first. With help of the relation
\beq\label{k0relation}
{\rm K}_0(\eta\rho)=\frac{1}{2\pi}\int d^2l\frac{{\rm e}^{{\rm i}\vec
l\cdot\vec\rho}}{l^2+\eta^2},
\eeq
one finds
\beqn\nonumber\label{k0part}
\frac{d\sigma(qp\to q\gamma^*p)}{d\ln\alpha d^2p_T}\Bigg|_{{\rm K}_0{\rm -part}}
&=&\frac{\alpha_{em}}{\pi^2}
\left[m_f^2 \alpha^4+2M^2\left(1-\alpha\right)^2\right]
\frac{1}{(2\pi)^2}\int d^2\rho_1d^2\rho_2
\frac{d^2l_1}{2\pi}\frac{d^2l_2}{2\pi}\\
\nonumber&\times&
\frac{{\rm e}^{{\rm i}\vec p_T\cdot(\vec\rho_1-\rho_2)}\,
{\rm e}^{-{\rm i}\vec l_1\cdot\vec\rho_1}\,
{\rm e}^{{\rm i}\vec l_2\cdot\vec\rho_2}}
{\left(l_1^2+\eta^2\right)\,\left(l_2^2+\eta^2\right)}\\
&\times&
\frac{1}{2}
\left\{\sigma_{q\bar q}(\alpha\rho_1)
+\sigma_{q\bar q}(\alpha\rho_2)
-\sigma_{q\bar q}(\alpha(\vec\rho_1-\vec\rho_2))\right\}.
\eeqn
Note that the term in the curly brackets consists of three contributions, which
depend either only on $\rho_1$ or on $\rho_2$ or on the difference 
$\vec\rho_1-\vec\rho_2$. Thus, the integral (\ref{k0part}) can be split into
three terms. In the integral which arises from the 
$\sigma_{q\bar q}(\alpha\rho_1)$ part, the $\rho_2$-integration is trivially
performed and leads to a two dimensional delta-function 
$\delta^{(2)}(\vec p_T-\vec l_2)$. This makes it possible two perform also the
$l_2$ integration. The integration over $l_1$ gives just the MacDonald function
K$_0$ (\ref{k0relation}).
Thus one is left with a two fold integration over $\rho_1$. Provided the DCS
depends only on the modulus of $\rho$, one can use the relation
\beq
{\rm J_0}=\frac{1}{2\pi}\int d\phi\,{\rm e}^{{\rm i}\vec
l\cdot\vec\rho}
\eeq
to perform one more integration. Here, J$_0$ is a Bessel function of first
kind. The contribution arising from $\sigma_{q\bar q}(\alpha\rho_2)$
is calculated in exactly the same way. For the 
$\sigma_{q\bar q}(\alpha(\vec\rho_1-\vec\rho_2))$-part one has to introduce the
auxiliary variable $\vec d=\vec\rho_1-\vec\rho_2$, before the procedure
described above can be applied.

The K$_1$ part is calculated in a similar way. Note that
\beq\label{k1relation}
{\rm K}_1(\eta\rho)=-\frac{1}{\eta}\frac{d}{d\rho}{\rm K}_0(\eta\rho).
\eeq
The K$_1$-part reads
\beqn\nonumber\label{k1part}
\frac{d\sigma(qp\to q\gamma^*p)}{d\ln\alpha d^2p_T}\Bigg|_{{\rm K}_1{\rm -part}}
&=&\frac{\alpha_{em}}{\pi^2}
\left[1+\left(1-\alpha\right)^2\right]
\frac{1}{(2\pi)^2}\int d^2\rho_1d^2\rho_2
\frac{d^2l_1}{2\pi}\frac{d^2l_2}{2\pi}\\
\nonumber&\times&
\frac{{\rm e}^{{\rm i}\vec p_T\cdot(\vec\rho_1-\rho_2)}\,
{\rm e}^{-{\rm i}\vec l_1\cdot\vec\rho_1}\,
{\rm e}^{{\rm i}\vec l_2\cdot\vec\rho_2}}
{\left(l_1^2+\eta^2\right)\,\left(l_2^2+\eta^2\right)}
\vec l_1\cdot\vec l_2\\
&\times&
\frac{1}{2}
\left\{\sigma_{q\bar q}(\alpha\rho_1)
+\sigma_{q\bar q}(\alpha\rho_2)
-\sigma_{q\bar q}(\alpha(\vec\rho_1-\vec\rho_2))\right\}.
\eeqn
Like the K$_0$-part, the complete integral (\ref{k1part}) is split into three
pieces, corresponding to the three terms in the curly brackets. Again, one
integration over $\rho$ is immediately performed, leading to a
$\delta$-functions, which allows to do one integration over $l$. With the
second $l$-integration, one recovers the MacDonald function K$_1$ via
(\ref{k1relation}). For the K$_1$-part, one also needs the relation
\beq
{\rm J}_1(z)=-\frac{d}{dz}{\rm J}_0(z).
\eeq
Although the calculation is slightly more cumbersome for the
$\sigma_{q\bar q}(\alpha(\vec\rho_1-\vec\rho_2))$-part, which leads to an
additional K$_0$-term in the final result, all calculations are easily
performed.

Finally, one finds
\beqn\nonumber
\frac{d\sigma(qp\to q\gamma^*p)}{d\ln\alpha d^2p_T}
&=&\frac{\alpha_{em}}{\pi^2}\left\{
\left[m_f^2 \alpha^4+2M^2\left(1-\alpha\right)^2\right]
\left[\frac{1}{p_T^2+\eta^2}{\cal I}_1-\frac{1}{4\eta}{\cal I}_2\right]
\right.\\
&+&\left.\left[1+\left(1-\alpha\right)^2\right]
\left[\frac{2\eta p_T}{p_T^2+\eta^2}{\cal I}_3-{\cal I}_1+\frac{\eta}{2}{\cal
I}_2\right]\right\},
\eeqn
with
\beqn
{\cal I}_1&=&\int_0^\infty dr r {\rm J}_0(p_Tr){\rm K}_0(\eta r)
\sigma_{q\bar q}(\alpha r)\\
{\cal I}_2&=&\int_0^\infty dr r^2 {\rm J}_0(p_Tr){\rm K}_1(\eta r)
\sigma_{q\bar q}(\alpha r)\\
{\cal I}_3&=&\int_0^\infty dr r {\rm J}_1(p_Tr){\rm K}_1(\eta r)
\sigma_{q\bar q}(\alpha r).
\eeqn
The remaining integrals are evaluated numerically with the Numerical Recipes
\cite{numrec} routines.

\subsection{Calculations for the mean coherence length}\label{appmeancoh}

In this appendix, some details of the calculation of 
the mean coherence length for the $q\bar qG$-component, section \ref{meancoh}, 
are presented.
We consider the case $Q^2\gg b_0^2$. With the approximations (\ref{app1})
to (\ref{app2}) we obtain for the denominator of (\ref{gluonl})
\beq
D^G=\int d^2r_{G}d^2r_{q\bar q}\int_0^1 d\alpha_q 
\int\limits_{\alpha_G^{min}}^{\alpha_G^{max}}\frac{d\alpha_G}{\alpha_G}
\left|\Psi^L\left(r_{q\bar q}\alpha_{q\bar q}\right)\right|^2
\left[\sigma_{q\bar q}\left(s,r_G\right)\right]^2
\left[\vec r_{q\bar q}\cdot\vec\nabla\Psi_{qG}\left(r_{G}\right)\right]^2.
\eeq
This integral diverges logarithmically for $\alpha_G^{min}\to 0$. 
To find an appropriate lower cut off, 
note that the mass of the $q\bar qG$ system is
approximately given by
\beq
M^2_{q\bar qG}\approx \frac{2b_0^2}{\alpha_G}+Q^2.
\eeq
We demand that $M^2_{q\bar qG}<0.2s$ which leads to
$\alpha_G^{min}=2b_0^2/(0.2s-Q^2)$. 
Furthermore
we work in the approximation of $\alpha_G\ll 1$ and we also 
have to choose an upper cut off.
We use 
\beq\label{limits}
\frac{2b_0^2}{0.2s-Q^2}\le\alpha_G\le\frac{2b_0^2}{Q^2}
\eeq
The two limits become equal  at
$x_{Bj}\approx 0.1$.
In our further calculation of $D^G$ we do the replacement 
$r_{q\bar qi}r_{q\bar qj}\to r_{q\bar q}^2\delta_{ij}$ and perform the
derivative. This yields
\beqn\nonumber
D^G&=&\left(\frac{2}{\pi}\sqrt{\frac{\alpha_s}{3}}\right)^2
\frac{6\alpha_{em}}{\left(2\pi\right)^2}4Q^2\pi
\int d^2r_{G}dr_{q\bar q}r^3_{q\bar q}\int_0^1 d\alpha_q 
\int\limits_{\alpha_G^{min}}^{\alpha_G^{max}}\frac{d\alpha_G}{\alpha_G}
{\rm K}_0^2\left(\varepsilon r_{q\bar q}\right)\alpha_q^2\left(1-\alpha_q\right)^2
\\
&\times&\left[\sigma_{q\bar q}\left(s,r_G\right)\right]^2
e^{-b_0^2r^2_{G}}\left(\frac{2}{r^4_{G}}+
\frac{2b_0^2}{r^2_{G}}+b_0^4\right).
\eeqn
For the integration over $r_{q\bar q}$ we use the integral representation
\beq
{\rm K}_0(x)=\frac{1}{2}\int_0^\infty\frac{dt}{t}\exp\left(-t-\frac{x^2}{4t}\right)
\eeq
of the modified Bessel function and obtain
\beq\label{Integral}
\int_0^\infty dr_{q\bar q}r^3_{q\bar q}
{\rm K}_0^2\left(\varepsilon r_{q\bar q}\right)
=\frac{2}{3\varepsilon^4}.
\eeq
Thus we have for the denominator
\beq
D^G=\frac{32\alpha_{em}\alpha_s}{3\pi^2Q^2}
\ln\frac{\alpha_G^{max}}{\alpha_G^{min}}
\int_0^\infty dr_Gr_G
\left[\sigma_{q\bar q}\left(s,r_G\right)\right]^2
e^{-b_0^2r^2_{G}}\left(\frac{2}{r^4_{G}}+
\frac{2b_0^2}{r^2_{G}}+b_0^4\right).
\eeq
Now we restrict ourselves to the a dipole cross section of the form
\beq
\sigma_{q\bar q}\left(s,r_G\right)=C(s)r_G^2
\eeq
and perform the last integration with the result
\beq
D^G=\frac{32\alpha_{em}\alpha_sC^2(s)}{\pi^2Q^2b_0^2}
\ln\frac{\alpha_G^{max}}{\alpha_G^{min}}.
\eeq
Note that the factor $C(s)$ will drop out, when one takes the ratio 
$<P^G>=N^G/D^G$.

Next we  calculate the numerator
\beqn\nonumber
N^G&=&m_Nx_{Bj}\left(\frac{2}{\pi}\sqrt{\frac{\alpha_s}{3}}\right)^2
\frac{6\alpha_{em}}{\left(2\pi\right)^2}4Q^2
\int d^2r_{1G}d^2r_{2G}d^2r_{q\bar q}\int_0^1d\alpha_q 
\int\limits_{\alpha_G^{min}}^{\alpha_G^{max}}\frac{d\alpha_G}{\alpha_G}
\alpha_q^2\left(1-\alpha_q\right)^2\\
\nonumber&\times&{\rm K}_0^2\left(\varepsilon r_{q\bar q}\right)
\sigma_{q\bar q}\left(s,r_{1G}\right)\sigma_{q\bar q}\left(s,r_{2G}\right)
\left[\vec r_{q\bar q}\cdot\vec\nabla_{r_{1G}}
\frac{\vec e\cdot\vec r_{1G}}{r^2_{1G}}e^{-\frac{b_0^2r^2_{1G}}{2}}\right]\\
&\times&
\left[\vec r_{q\bar q}\cdot\vec\nabla_{r_{2G}}
\frac{\vec e\cdot\vec r_{2G}}{r^2_{2G}}e^{-\frac{b_0^2r^2_{2G}}{2}}\right]
\int_0^\infty d\Delta z G_{GG}\left(\vec r_{2G},\vec r_{1G},\Delta z\right)
\eeqn
where
\beq
G_{GG}\!\left(\vec r_{2G},\vec r_{1G},\Delta z\right)
=\frac{b_0^2e^{-\frac{Q^2\Delta z}{2\nu}}}
{2\pi\sinh\!\left(\omega\Delta z\right)}
\exp\!\left\{-\frac{b_0^2}{2}\left[\left(r^2_{1G}+r^2_{2G}\right)
{\rm cth}\!\left(\omega\Delta z\right)
-\frac{2\vec r_{1G}\cdot\vec r_{2G}}{\sinh\!\left(\omega\Delta z\right)}
\right]\right\}
\eeq
is the solution of (\ref{x}) with $\omega=b_0^2/(\nu\alpha_G)$.

Again we can do the replacement
$r_{q\bar qi}r_{q\bar qj}\to r_{q\bar q}^2\delta_{ij}$, perform the
derivatives, sum over gluon polarizations and use (\ref{Integral}) to obtain
\beqn\nonumber
N^G&=&m_Nx_{Bj}\frac{8\alpha_{em}\alpha_s}{3\pi^4Q^2}
\int\!\! d^2r_{1G}d^2r_{2G}d\Delta z
\int\limits_{\alpha_G^{min}}^{\alpha_G^{max}}\!\!\frac{d\alpha_G}{\alpha_G}
\sigma_{q\bar q}\left(s,r_{1G}\right)\sigma_{q\bar q}\left(s,r_{2G}\right)
\frac{b_0^2e^{-\frac{Q^2\Delta z}{2\nu}}}
{\sinh\left(\omega\Delta z\right)}
\\&\times&\nonumber
\left[
4\frac{\left(\vec r_{1G}\cdot \vec r_{2G}\right)^2}{r^4_{1G}r^4_{2G}}
+b_0^4\frac{\left(\vec r_{1G}\cdot \vec r_{2G}\right)^2}{r^2_{1G}r^2_{2G}}
+2b_0^2\frac{\left(\vec r_{1G}\cdot \vec r_{2G}\right)^2}{r^2_{1G}r^4_{2G}}
+2b_0^2\frac{\left(\vec r_{1G}\cdot \vec r_{2G}\right)^2}{r^4_{1G}r^2_{2G}}
\right.\\&&\left.
-\frac{b_0^2}{r^2_{1G}}-\frac{b_0^2}{r^2_{1G}}-\frac{1}{r^2_{1G}r^2_{2G}}
\right]
\exp\left(-\beta\left(r^2_{1G}+r^2_{2G}\right)
+2\gamma\vec r_{1G}\cdot \vec r_{2G}\right)
\eeqn
where
\beq
\beta=\frac{b_0^2}{2}\left(1+\coth\left(\omega\Delta z\right)\right),
\eeq
\beq
\gamma=\frac{b_0^2}
{2\sinh\left(\omega\Delta z\right)}.
\eeq
With a cross section like $\sigma_{q\bar q}\left(s,r\right)=C(s)r^2$ 
the integrations over $r_{G}$ are easily
performed with the result
\beqn\nonumber
N^G&=&m_Nx_{Bj}\frac{8\alpha_{em}\alpha_sC^2(s)}{3\pi^2Q^2b_0^2}
\int d\Delta z
\int\limits_{\alpha_G^{min}}^{\alpha_G^{max}}\frac{d\alpha_G}{\alpha_G}
\frac{e^{-\frac{Q^2\Delta z}{2\nu}}}
{\sinh\left(\omega\Delta z\right)}
\left\{\frac{10}{\left(1+\coth\left(\omega\Delta z\right)\right)^2}
\right.\\
&&\left.
+\frac{12}{\sinh^2\left(\omega\Delta z\right)
\left(1+{\rm cth}\left(\omega\Delta z\right)\right)^3}
-8\sinh^2\!\left(\omega\Delta z\right)
\ln\!\left(\frac{1+\coth\left(\omega\Delta z\right)}{2}\right)\!\!
\right\}\\
\nonumber
&=&\frac{8\alpha_{em}\alpha_sC^2(s)}{3\pi^2Q^2b_0^2}
\int_0^1 dy 
\int\limits_{\delta^{min}}^{\delta_G^{max}}\frac{d\delta}{\delta^2}
\left\{4y^{\frac{1}{\delta}-2}\left(1-y^2\right)\ln\left(1-y^2\right)\right.\\
&&\qquad\qquad\qquad\qquad\qquad\qquad\qquad\qquad\qquad\left.
+5y^{\frac{1}{\delta}}\left(1-y^2\right)
+12y^{\frac{1}{\delta}+2}
\right\}.
\eeqn
In the last step we have introduced the new variables $y=e^{-\omega\Delta z}$
and $\delta=2b_0^2/(Q^2\alpha_G)$. 
According to (\ref{limits}) the limits for $\delta$ are
\beq
1\le\delta\le 0.2\frac{s}{Q^2}-1.
\eeq
For the integral containing the logarithm it
is convenient to do one more substitution, $x=y^2$. Then one finds
\beqn
\int_0^1 dy 
y^{\frac{1}{\delta}-2}\left(1-y^2\right)\ln\left(1-y^2\right)
&=&
\frac{1}{2}\lim_{\eta\to 0}\frac{\partial}{\partial\eta}
\int_0^1 dx x^{\frac{1}{2\delta}-\frac{3}{2}}
\left(1-x\right)^{1+\eta}\\
&=&\frac{2\delta^2}{\delta^2-1}\left(
\psi\left(\frac{3}{2}+\frac{1}{2\delta}\right)
-\psi\left(2\right)\right).
\eeqn
Now only one integration is left in $N^G$
\beq
N^G=\frac{64\alpha_{em}\alpha_sC^2(s)}{3\pi^2Q^2b_0^2}\!\!
\int\limits_{\delta^{min}}^{\delta_G^{max}}\!\!\frac{d\delta}{\delta}
\left[\frac{5}{8\left(1+\delta\right)}
+\frac{7}{8\left(1+3\delta\right)}-\frac{\delta}{\delta^2-1}\left(
\psi\left(2\right)-\psi\left(\frac{3}{2}+\frac{1}{2\delta}\right)\right)
\right]
\eeq
and we end up with the result (\ref{3.12}) for the factor
$\la P^G\ra=N^G/D^G$ at $Q^2\gg b_0^2$,
\beq
\la P^G\ra=\frac{2}{3\ln\left(\frac{\alpha_G^{max}}{\alpha_G^{min}}\right)}
\int\limits_{\delta^{min}}^{\delta_G^{max}}\frac{d\delta}{\delta}
\left[\frac{5}{8\left(1+\delta\right)}
+\frac{7}{8\left(1+3\delta\right)}-\frac{\delta}{\delta^2-1}\left(
\psi\left(2\right)-\psi\left(\frac{3}{2}+\frac{1}{2\delta}\right)\right)
\right]
\eeq

\subsection{Calculations for DY cross sections in pA scattering}\label{appdypa}

The calculation of the transverse momentum distribution of DY pairs in pA
collisions  requieres rather involved analytical and numerical
calculations. In order to be able to perform some of the integrations
in (\ref{dynucleartrans})
analytically we employ the approximations of a uniform nuclear density $\rho_A$
and a dipole cross section $\sigma_{q\bar q}(\alpha\rho)=C\alpha^2\rho^2$. With
these approximations, it is possible to write down an analytical expression for
the Green function,
\beqn\nonumber
G\left(\vec \rho_2,z_2\,|\,\vec \rho_1,z_1\right)&=&
\frac{a{\rm e}^{-{\rm i}q_L\Delta z}}
{2\pi\sinh\left(\omega\Delta z\right)}\\
&\times&
\exp\left\{-\frac{a}{2}\left[\left(\rho_1^2+\rho_2^2\right)
\coth\left(\omega\Delta z\right)
-\frac{2\vec \rho_1\cdot\vec \rho_2}{\sinh\left(\omega\Delta z\right)}
\right]\right\}.
\eeqn
Here,
\beq a=(-1+{\rm i})\sqrt{\rho_A E_q\alpha^3\left(1-\alpha\right)C/2}
\eeq
and
\beq \omega=-(1+{\rm i})\sqrt{\rho_A C\alpha/(2E_q\left(1-\alpha\right))}
\eeq
is the complex oscillator frequency.

We decompose (\ref{dynucleartrans}) in six integrals which
correspond to the integration regions
\begin{enumerate}
\item{${\cal I}_1:\quad z_1\le z_2\le -L=-\sqrt{R_A-b^2},$}
\item{${\cal I}_2:\quad z_1\le -L\le z_2\le L,$}
\item{${\cal I}_3:\quad z_1\le -L$ and $L\le z_2,$}
\item{${\cal I}_4:\quad -L\le z_1\le z_2\le L,$}
\item{${\cal I}_5:\quad -L\le z_1\le L\le z_2,$}
\item{${\cal I}_6:\quad L\le z_1\le z_2.$}
\end{enumerate}
The differential cross section is then given by
\beq
\frac{d^3\sigma^{qA}}{d(\ln\alpha)d^2p_T}=\sum\limits_{j=1}^6 {\cal I}_j.
\eeq

For convenience we introduce the notation
\beq
H_1=1+\left(1-\alpha\right)^2,
\eeq
\beq
H_0=2M^2\left(1-\alpha\right)^2
\eeq
and employ the auxiliary functions
\beqn
G(z,t)&=&\frac{\eta^2}{\eta^2\cosh(\omega z)+2at\sinh(\omega z)},\\
\beta(z,t)&=&\frac{1}{2a}\frac{2at\cosh(\omega z)+\eta^2\sinh(\omega z)}
{\eta^2\cosh(\omega z)+2at\sinh(\omega z)},\\
F(L,z)&=&2C\alpha^2\rho_A(2L-z).
\eeqn
The calculations are too long, to be shown in detail. The result is
\beqn
{\cal I}_1&=&\frac{\alpha_{em}}{\pi}
\frac{\left(H_0+H_1p_T^2\right)}
{\left(\eta^2+p_T^2\right)^2}R_A^2,\\
\nonumber
{\cal I}_2 & = & -\frac{2\alpha_{em}}{\pi}2{\rm Re}
\frac{{\rm i}q_L^{min}}{\eta^2(\eta^2+p_T^2)}
\int\limits_0^{R_A}L dL
\int\limits_0^{2L} dz
\left[
H_0+
H_1p_T^2\,G(z,0)
\right]G(z,0)
\\
&\times&
\exp\left(-{\rm i}q_L^{min}z-
p_T^2\beta(z,0)
\right),\\
\nonumber
{\cal I}_3 & = & -\frac{2\alpha_{em}}{\pi}2{\rm Re}
\frac{1}{\eta^2(\eta^2+p_T^2)}
\int\limits_0^{R_A}L dL\int\limits_0^\infty dt
\left[
H_0+
H_1p_T^2\,G(2L,t)
\right]G(2L,t)
\\
&\times&
\exp\left(-{\rm i}q_L^{min}2L-t-
p_T^2\beta(2L,t)\right),\\
\nonumber
{\cal I}_4 & = & \frac{2\alpha_{em}}{\pi}2{\rm Re}\, 
\frac{{(q_L^{min})}^2}{\eta^4}
\int\limits_0^{R_A}L dL
\int\limits_0^{2L} dz
\int\limits_0^z dz^\prime
G(z^\prime,0)\\
\nonumber&\times&
\left[
H_0+
H_1G(z^\prime,0)
\left(\frac{F(L,z)}{1+F(L,z)\beta(z^\prime,0)}+\frac{p_T^2}
{\left(1+F(L,z)\beta(z^\prime,0)\right)^2}
\right)
\right]\\
&\times&\frac{
\exp\left(-{\rm i}q_L^{min}z^\prime-\frac{\beta(z^\prime,0) p_T^2}
{1+F(L,z)\beta(z^\prime,0)}\right)}
{1+F(L,z)\beta(z^\prime,0)},\\
\nonumber
{\cal I}_5 & = & -\frac{2\alpha_{em}}{\pi}2{\rm Re}\, 
\frac{{\rm i}q_L^{min}}{\eta^4}
\int\limits_0^{R_A}L dL
\int\limits_0^{2L} dz
\int\limits_0^\infty dt\,
G(z,t)\\
\nonumber&\times&
\left[
H_0+
H_1G(z,t)
\left(\frac{F(L,z)}{1+F(L,z)\beta(z,t)}+\frac{p_T^2}
{\left(1+F(L,z)\beta(z,t)\right)^2}
\right)
\right]\\
&\times&\frac{
\exp\left(-{\rm i}q_L^{min}z-t-\frac{\beta(z,t) p_T^2}
{1+F(L,z)\beta(z,t)}\right)}
{1+F(L,z)\beta(z,t)},\\
{\cal I}_6&=&\frac{2\alpha_{em}}{\pi}
\int\limits_0^{R_A}L dL\int\limits_0^\infty dt\left[
\frac{H_0}{\eta^2}t+H_1(1-t)
\right]\frac{\exp\left(-t-\frac{p_T^2}{\eta^2+F(L,0)t}t\right)}
{\eta^2+F(L,0)t}.
\eeqn

Note that in ${\cal I}_1$ and ${\cal I}_6$, the integration over the
longitudinal coordinates $z_1$ and $z_2$ does not extend over the nucleus, which
is located between $-L$ and $L$. Nevertheless, ${\cal I}_1$ and ${\cal I}_6$ are
different from zero. One can however check, that the cross section vanishes in
the limit 
$\rho_A \rightarrow 0$. 
In particular one finds
${\cal I}_1+{\cal I}_3+{\cal I}_6=-{\cal I}_2={\cal I}_4=-{\cal I}_5$.

\subsection{Calculations for shadowing in DIS}\label{appshadow}

In the approximation where only the $q\bar q$-Fock-component of the virtual
photon is taken into account, it was found that nuclear shadowing in DIS is
given by the formula
\beqn\label{eq}
\nonumber
\sigma_{tot}^{\gamma^*A} & = & A\sigma^{\gamma^*p}-\frac{1}{2}{\rm Re}
\sum_{f=1}^{N_f}\int d^2b\inf dz_1
\int_{z_1}^\infty dz_2 \int_0^1d\alpha \int d^2\rho_1\int d^2\rho_2\\
&\times&
\nonumber 
\Psi_{T,L}^*\left(\varepsilon,\lambda,\rho_2\right)
\rho_A\left(b,z_2\right)\sigma_{q\bar{q}}\left(s,\rho_2\right)\\
&\times&
G\left(\vec \rho_2,z_2\,|\,\vec \rho_1,z_1\right)
\rho_A\left(b,z_1\right)\sigma_{q\bar{q}}\left(s,\rho_1\right)
\Psi_{T,L}\left(\varepsilon,\lambda,\rho_1\right),
\eeqn
where summation over the polarisations ($T,L$)
 of the virtual photon is understood.
The Greenfunction $G$ describes the propagation of the $q\bar q$-pair 
through the nucleus.
With the nonperturbative interaction which was introduced to model confinement
effects, $G$ fullfills the equation
\beqn\nonumber\label{scheq}
\left[i\frac{\partial}{\partial z_2}
+\frac{\Delta_\perp\left(\rho_2\right)-\varepsilon^2}
{2\nu\alpha\left(1-\alpha\right)}
+\frac{i}{2}\rho_A\left(b,z_2\right)\,
\sigma_{q\bar{q}}\left(s,\rho_2\right)
-\frac{a^4\left(\alpha\right)\rho_2^2}
{2\nu\alpha\left(1-\alpha\right)}
\right]
G\left(\vec \rho_2,z_2\,|\,\vec \rho_1,z_1\right)\\
=
i\delta\left(z_2-z_1\right)
\delta^{\left(2\right)}\left(\vec \rho_2-\vec \rho_1\right),
\eeqn
where 
\beq
\varepsilon^2=\alpha\left(1-\alpha\right)Q^2+m_f^2
\eeq
is the so-called extension parameter 
which depends on the flavor $f$ via the quark mass.
The strength of the potential between
the particles is characterized by the parameter
\beq
a^2\left(\alpha\right)=v^{1.15}(112MeV)^2
+\left(1-v\right)^{1.15}(165MeV)^2\alpha\left(1-\alpha\right).
\eeq

Since (\ref{eq}) is too complicated for a direct numerical evaluation, we
introduce the following simplifying approximations. First, we chose a dipole
cross-section of the functional shape
\beq\label{a1}
\sigma_{q\bar{q}}\left(s,\rho\right)=C(s)\rho^2,
\eeq
where the coefficient $C(s)$ does not depend on $\rho$. Second, we work with a
uniform nuclear density
\beq\label{a2}
\rho_A\left(b,z\right)=\rho_{A0}\Theta\left(R_A^2-b^2-z^2\right).
\eeq
These two approximations allow us to solve (\ref{scheq}) analytically, because 
(\ref{scheq}) reduces to the Schr\"odinger-equation of a two dimensional
harmonic oscillator, where $z$ plays the role of time. The solution is known as
\beqn\nonumber
G\left(\vec \rho_2,z_2\,|\,\vec \rho_1,z_1\right)&=&
\frac{B\left(\alpha\right){\rm e}^{-{\rm i}q_L\Delta z}}
{2\pi\sinh\left(\omega\Delta z\right)}\\
&\times&
\exp\left\{-\frac{B\left(\alpha\right)}{2}\left[\left(\rho_1^2+\rho_2^2\right)
\coth\left(\omega\Delta z\right)
-\frac{2\vec \rho_1\cdot\vec \rho_2}{\sinh\left(\omega\Delta z\right)}
\right]\right\}.
\eeqn
Note however that the frequency of the oscillator has a complex value,
\beq
\omega={\rm i}\frac{B\left(\alpha\right)}{M}.
\eeq
Furthermore we introduced the notation
\beq
\Delta z= z_2-z_1
\eeq
and
\beq
B^2\left(\alpha\right)
=-{\rm i}\rho_AMC(s)+a^4\left(\alpha\right).
\eeq
The quantity
\beq
M=\nu\alpha\left(1-\alpha\right)
\eeq
plays the role of the reduced mass of the pair and the minimal
longitudinal momentum
transfer when the photon splits is given by 
\beq
q_L=\frac{\varepsilon^2}{2M}.
\eeq

With these simplifications (\ref{eq}) reads
\beqn\nonumber\label{new}
\sigma^{\gamma^*A} & = & A\sigma^{\gamma^*p}-\frac{1}{2}\,{\rm Re}
\sum_{f=1}^{N_f}\int d^2b\int_{-l/2}^{l/2} dz_1
\int_0^{l/2-z_1} d(\Delta z) \int_0^1d\alpha \int d^2\rho_1\int d^2\rho_2\\
\nonumber&\times&
{\rm e}^{-{\rm i}q_L\left(z_2-z_1\right)}\rho_{A0}^2
C^2(s)\rho_1^2\rho_2^2\\
&\times&
\left[\Phi_{0}\left(\varepsilon,\lambda,\rho_1\right)
\Phi_{0}\left(\varepsilon,\lambda,\rho_2\right)+
\vec\Phi_{1}\left(\varepsilon,\lambda,\rho_1\right)\cdot
\vec\Phi_{1}\left(\varepsilon,\lambda,\rho_2\right)\right]\\
\nonumber&\times&
\frac{B}{2\pi\sinh\left(\omega\Delta z\right)}
\exp\left\{-\frac{B}{2}\left[\left(\rho_1^2+\rho_2^2\right)
\coth\left(\omega\Delta z\right)
-\frac{2\vec \rho_1\cdot\vec \rho_2}{\sinh\left(\omega\Delta z\right)}
\right]\right\},
\eeqn
with $l=2\sqrt{R_A^2-b^2}$.
For simplicity we do write out the dependence of $B$ on $\alpha$ explicitly.
From the physical point of view it is reasonable to distinguish between
shadowing for transverse and longitudinal photons, as indicated by the
subscripts $T$ and $L$ in  (\ref{eq}). For calculational purposes it is however
more convenient 
to use the functions
\beqn\lefteqn{
\Phi_{0}\left(\varepsilon,\lambda,\rho_1\right)
\Phi_{0}\left(\varepsilon,\lambda,\rho_2\right)
=\frac{6\alpha_{em}Z_f^2}{\left(4\pi\right)^2}
\left(m_f^2+4Q^2\alpha^2\left(1-\alpha\right)^2\right)}\\
\nonumber&\times&\int\limits_0^\infty dudt\frac{\lambda^2}
{\sinh\left(\lambda t\right)\sinh\left(\lambda u\right)}
\exp\left[-\frac{\lambda\varepsilon^2\rho_1^2}{4}\coth\left(\lambda t\right)-t
-\frac{\lambda\varepsilon^2\rho_2^2}{4}\coth\left(\lambda u\right)-u\right],\\
\lefteqn{
\vec\Phi_{1}\left(\varepsilon,\lambda,\rho_1\right)\cdot
\vec\Phi_{1}\left(\varepsilon,\lambda,\rho_2\right)
=\frac{6\alpha_{em}Z_f^2}{\left(2\pi\right)^2}
\left(1-2\alpha\left(1-\alpha\right)\right)
\frac{\vec\rho_1\cdot\vec\rho_2}{\rho_1^2\rho_2^2}}\\
\nonumber&\times&\int\limits_0^\infty dudt
\exp\left[-\frac{\lambda\varepsilon^2\rho_1^2}{4}\coth\left(\lambda t\right)-t
-\frac{\lambda\varepsilon^2\rho_2^2}{4}\coth\left(\lambda u\right)-u\right]
\eeqn
instead. The strength of the nonperturbative interaction is now characterized by
the dimensionless parameter
\beq
\lambda=\frac{2a^2\left(\alpha\right)}{\varepsilon^2}.
\eeq
Note that in the limit $\lambda\to 0$ one obtains the expressions for the
perturbative wave-functions. 
We will not write out the $\alpha$ dependence of $\lambda$ explicitly in the
following.
In particular, $\Phi_0$ goes to the $K_0$-part of
the light-cone wave-functions and $\Phi_1$ to the $K_1$-part.

The total cross section $\gamma^*A$ can now be decomposed into three terms. 
\beq
\sigma^{\gamma^*A}=A\sigma^{\gamma^*p}-\sigma^{\gamma^*A}_0- 
\sigma^{\gamma^*A}_1. 
\eeq
The shadowing correction is given by the last two terms, where the indices $0$
and $1$ refer to the $\Phi_0$ and $\Phi_1$-part of the wavefunctions,
respectivly.

First we consider the $\Phi_0$-part. Since the integrand in (\ref{new}) depends
on $z_1$ only via the upper limit of the integration over $z_2$, we are able to
integrate two times by part, with the result 
\beqn\nonumber\label{zwei}
\sigma^{\gamma^*A}_0 & = &\frac{1}{2}\,{\rm Re}
\sum_{f=1}^{N_f}\int_0^{D} dl\int_0^1d\alpha \int d^2\rho_1\int d^2\rho_2
\int_0^\infty dudt\\
\nonumber&\times&
\frac{\pi}{12}\left(l^3-3D^2l+2D^3\right){\rm e}^{-{\rm i}q_Ll}
C^2(s)\rho_{A0}^2
\frac{6\alpha_{em}Z_f^2}{\left(4\pi\right)^2}
\left(m_f^2+4Q^2\alpha^2\left(1-\alpha\right)^2\right)\\
&\times&
\frac{\lambda^2{\rm e}^{-u-t}}
{\sinh\left(\lambda t\right)\sinh\left(\lambda u\right)}
\frac{B}{2\pi\sinh\left(\omega l\right)}
\rho_1^2\rho_2^2\exp\left(-\gamma\rho_1^2-\beta\rho_2^2+
2\delta\vec \rho_1\cdot\vec \rho_2\right).
\eeqn
The polynomial in $l$ results from making the boundary terms due to the
integration by parts vanish. We also introduced the nuclear diameter
$D=2R_A$
and the notation
\beqn\label{abkz}
\gamma&=&\frac{B\coth\left(\omega l\right)}{2}
+\frac{\lambda\varepsilon^2\coth\left(\lambda t\right)}{4},\\
\beta&=&\frac{B\coth\left(\omega l\right)}{2}
+\frac{\lambda\varepsilon^2\coth\left(\lambda u\right)}{4},\\
\delta&=&\frac{B}{2\sinh\left(\omega l\right)}.
\eeqn
The exponential in (\ref{zwei}) is obtained when one inserts the expressions for
the Green function and the nonperturbative wave-functions into  (\ref{new}).
Now one notes that the integrations over $\rho_1$ and $\rho_2$ are simply
Gaussian and can immediately be performed,
\beqn\nonumber\label{Final}
\sigma^{\gamma^*A}_0 & = &\frac{\alpha_{em}}{128}C^2(s)\rho_{A0}^2\,{\rm Re}
\sum_{f=1}^{N_f}Z_f^2\int_0^{D} dl\int_0^1d\alpha 
\int_0^\infty dudt\\
\nonumber&\times&
{\rm e}^{-{\rm i}q_Ll}\left(m_f^2+4Q^2\alpha^2\left(1-\alpha\right)^2\right)
\frac{\lambda^2{\rm e}^{-u-t}}
{\sinh\left(\lambda t\right)\sinh\left(\lambda u\right)}\\
&\times&
\frac{B}{\sinh\left(\omega l\right)}
\left[\frac{1}{\left(\gamma\beta-\delta^2\right)^2}
+\frac{2\delta^2}{\left(\gamma\beta-\delta^2\right)^3}\right].
\eeqn
Thus we have reduced the number of integrations to four. Substituting
(\ref{abkz}) in (\ref{Final}) one obtains the somewhat lengthy expression
\beqn\label{final2}
\sigma^{\gamma^*A}_0 & = & 32\alpha_{em}C^2(s)\rho_{A0}^2R_A^4
{\rm Re}
\sum_{f=1}^{N_f}Z_f^2\int_0^{1} dL\int_0^1d\alpha 
\int_0^\infty dudt\\
\nonumber&\times&
{\rm e}^{-{\rm i}q_LDL}
\left(m_f^2+4Q^2\alpha^2\left(1-\alpha\right)^2\right)
\left(L^3-3L^2+2\right)
{\rm e}^{-u-t}
\lambda^2B\\
\nonumber&\times&
\left\{
\frac{\sinh\left(\omega DL\right)
\sinh\left(\lambda t\right)\sinh\left(\lambda u\right)}
{\begin{array}{c@{\quad}c}
\left[\sinh\left(\omega DL\right)
\left(4B^2\sinh\left(\lambda t\right)\sinh\left(\lambda u\right)
+\lambda^2\varepsilon^4\cosh\left(\lambda t\right)\cosh\left(\lambda u\right)
\right)\right.\hspace{1cm}\\
\hspace{5cm}\left.+2B\lambda\eps^2\cosh\left(\omega DL\right)
\sinh\left(\lambda(u+t)\right)\right]^2&\\
\end{array}}
\right.\\
\nonumber&+&
\left.
\frac{8B^2\sinh^2\left(\lambda t\right)\sinh^2\left(\lambda u\right)}
{\begin{array}{c@{\quad}c}
\left[\sinh\left(\omega DL\right)
\left(4B^2\sinh\left(\lambda t\right)\sinh\left(\lambda u\right)
+\lambda^2\varepsilon^4\cosh\left(\lambda t\right)\cosh\left(\lambda u\right)
\right)\right.\hspace{1cm}\\
\hspace{5cm}\left.+2B\lambda\eps^2\cosh\left(\omega DL\right)
\sinh\left(\lambda(u+t)\right)\right]^3&\\
\end{array}}
\right\}.
\eeqn

The $\Phi_1$-part can be treated in the same way. Once again we integrate two
times by part and obtain
\beqn\nonumber
\sigma^{\gamma^*A}_1 & = &\frac{1}{2}\,{\rm Re}
\sum_{f=1}^{N_f}\int_0^{D} dl\int_0^1d\alpha \int d^2\rho_1\int d^2\rho_2
\int_0^\infty dudt\\
\nonumber&\times&
\frac{\pi}{12}\left(l^3-3D^2l+2D^3\right){\rm e}^{-{\rm i}q_Ll}
C^2(s)\rho_{A0}^2
\frac{6\alpha_{em}Z_f^2}{\left(2\pi\right)^2}
\left(1-2\alpha\left(1-\alpha\right)\right)\\
&\times&
{\rm e}^{-u-t}
\frac{B}{2\pi\sinh\left(\omega l\right)}
\vec\rho_1\cdot\vec\rho_2\exp\left(-\gamma\rho_1^2-\beta\rho_2^2+
2\delta\vec \rho_1\cdot\vec \rho_2\right)
\eeqn
with the same notations as in (\ref{abkz}). The Gaussian integral over
$\rho_1$ and $\rho_2$ is performed with the result 
\beqn\nonumber
\sigma^{\gamma^*A}_1 & = &\frac{\alpha_{em}C^2(s)\rho_{A0}^2}{32}\,{\rm Re}
\sum_{f=1}^{N_f}Z_f^2\int_0^{D} dl\int_0^1d\alpha 
\int_0^\infty dudt\\
\nonumber&\times&
\left(l^3-3D^2l+2D^3\right){\rm e}^{-{\rm i}q_Ll}
\left(1-2\alpha\left(1-\alpha\right)\right)\\
&\times&
{\rm e}^{-u-t}
\frac{B}{\sinh\left(\omega l\right)}
\frac{\delta}{\left(\gamma\beta-\delta^2\right)^2},
\eeqn
and after substitution of the expressions in (\ref{abkz}) one finally obtains
\beqn\label{fnpt1}
\sigma^{\gamma^*A}_1 & = & 64\alpha_{em}C^2(s)\rho_{A0}^2R_A^4
{\rm Re}
\sum_{f=1}^{N_f}Z_f^2\int_0^{1} dL\int_0^1d\alpha 
\int_0^\infty dudt\\
\nonumber&\times&
{\rm e}^{-{\rm i}q_LDL}
\left(1-2\alpha\left(1-\alpha\right)\right)
\left(L^3-3L^2+2\right)
{\rm e}^{-u-t}
B^2\\
\nonumber&\times&
\frac{
\sinh^2\left(\lambda t\right)\sinh^2\left(\lambda u\right)}
{\begin{array}{c@{\quad}c}
\left[\sinh\left(\omega DL\right)
\left(4B^2\sinh\left(\lambda t\right)\sinh\left(\lambda u\right)
+\lambda^2\varepsilon^4\cosh\left(\lambda t\right)\cosh\left(\lambda u\right)
\right)\right.\hspace{1cm}\\
\hspace{5cm}\left.+2B\lambda\eps^2\cosh\left(\omega DL\right)
\sinh\left(\lambda(u+t)\right)\right]^2&\\
\end{array}}.
\eeqn

Taking the limit $\lambda\to 0$, one obtains the corresponding expressions for
the perturbative wave-functions,
\beqn\label{fpt0}
\sigma^{\gamma^*A}_0 & = & 32\alpha_{em}C^2(s)\rho_{A0}^2R_A^4
{\rm Re}
\sum_{f=1}^{N_f}Z_f^2\int_0^{1} dL\int_0^1d\alpha 
\int_0^\infty dudt\\
\nonumber&\times&
{\rm e}^{-{\rm i}q_LDL}
\left(m_f^2+4Q^2\alpha^2\left(1-\alpha\right)^2\right)
\left(L^3-3L^2+2\right)
{\rm e}^{-u-t}
\lambda^2B\\
\nonumber&\times&
\left\{
\frac{\sinh\left(\omega DL\right)
ut}
{
\left[\sinh\left(\omega DL\right)
\left(4B^2ut
+\varepsilon^4
\right)+2B\eps^2\cosh\left(\omega DL\right)
(u+t)\right]^2
}
\right.\\
\nonumber&+&
\left.
\frac{8B^2u^2t^2}
{
\left[\sinh\left(\omega DL\right)
\left(4B^2ut
+\varepsilon^4
\right)+2B\eps^2\cosh\left(\omega DL\right)
(u+t)\right]^3
}
\right\},
\eeqn

\beqn\label{fpt1}
\sigma^{\gamma^*A}_1 & = & 64\alpha_{em}C^2(s)\rho_{A0}^2R_A^4
{\rm Re}
\sum_{f=1}^{N_f}Z_f^2\int_0^{1} dL\int_0^1d\alpha 
\int_0^\infty dudt\\
\nonumber&\times&
{\rm e}^{-{\rm i}q_LDL}
\left(1-2\alpha\left(1-\alpha\right)\right)
\left(L^3-3L^2+2\right)
{\rm e}^{-u-t}
B^2\\
\nonumber&\times&
\frac{
u^2t^2}
{
\left[\sinh\left(\omega DL\right)
\left(4B^2ut
+\varepsilon^4
\right)+2B\eps^2\cosh\left(\omega DL\right)
(u+t)\right]^2
}.
\eeqn

To make our simplifications (\ref{a1}) and  (\ref{a2}) more realistic, 
we determine $C(s)$ by demanding,
\beqn\nonumber
\lefteqn{\frac{2\int d^2b d^2r
\left|\Psi_{T,L}\left(\varepsilon r\right)\right|^2
\left[1-\exp\left(-\frac{C_{T,L}r^2}{2}
T\left(b\right)\right)\right]}
{\int d^2r
\left|\Psi_{T,L}\left(\varepsilon r\right)\right|^2
C_{T,L}r^2
}}\\
& = & \frac{2\int d^2b d^2r
\left|\Psi_{T,L}\left(\varepsilon r\right)\right|^2
\left[1-\exp\left(-\frac{\sigma_{q\bar q}\left(s,r\right)}{2}
T\left(b\right)\right)\right]}
{\int d^2r
\left|\Psi_{T,L}\left(\varepsilon r\right)\right|^2
\sigma_{q\bar q}\left(s,r\right)},
\eeqn
where
\beq
T\left(b\right)=\int\limits_{-\infty}^{\infty} dz\, \rho_A\left(b,z\right)
\eeq
is the nuclear thickness. This means that $C(s)$ is chosen such that in the
limit of infinite coherence length, the same value of the cross section
as obtained with the
realistic parametrization (\ref{sasha}) is reproduced. This procedure is
performed separately for transverse and longitudinal photons and for each value
of $\alpha$.
We use realistic parametrizations for $\rho_A(b,z)$ with parameter values from 
\cite{Jager}.

The value for the mean nuclear density $\rho_{A0}$ is determined in a similar
way from
\beq
 \int d^2b\left[1-\exp\left(-\frac{\sigma_0}{2}
2\sqrt{R_A^2-b^2}\rho_{A0}\right)\right]
 =  \int d^2b
\left[1-\exp\left(-\frac{\sigma_0}{2}
T\left(b\right)\right)\right]
\eeq
We tried values between $1$ mb and $50$ mb for $\sigma_0$ and found the result
for $\rho_{A0}$ practically independent of this value.



\pagestyle{empty}\clearpage
\vspace*{1cm}\clearpage
\centerline{\Large\bf Acknowledgements}
\vspace{2cm}

First of all, I would like to 
express my gratitude to my adviser Boris Kopeliovich. During all the
time, in which this work was performed, he provided me with constant
encouragement and stimulating discussion. He always found the time to explain
the physics of processes which seemed hopelessly complicated. 
It stands to reason that his guidance was essential for a successful
accomplishment of this research work.

Specially, I would like to thank J\"org H\"ufner and Sasha Tarasov for sharing
their views and illuminating discussions,
which were indispensable for making
this work a success. In particular without Sasha's 
help, many of the equations in
this thesis could not have been turned into numbers.

I am grateful to Andreas Sch\"afer for his interest in my work and his
willingness to referee this thesis and to take part in the defense. 

I am indepted to
 B.\ Povh and H.\ Weidenm\"uller for the hospitality at the MPI f\"ur
Kernphysik in Heidelberg. They gave me the possibility to stay in close contact
to my advisers. Furthermore, without the computers at the MPI, many calculations
in this thesis could not have been done. I thank Hans-Christian Pauli for being
one of my supervisors.

I also would like to thank all members of the theoretical nuclear and hadronic
physics group for the pleasant atmosphere in the institute
and for all kinds of conversation, not only about physics. I am especially
grateful to Dagmar Isert and Kai Schwenzer for reading the manuscript of this
thesis and making suggestions for improvements. I also would like to thank Yuri
Ivanov for his help with several undocumented features of the software.

This work was supported by the Gesellschaft f\"ur Schwerionenforschung GSI,
Darmstadt, grant HD H\"UF T and by the Graduiertenkolleg {\em Physikalische
Systeme mit vielen Freiheitsgraden}.

\end{document}